%% file: main.tex
\RequirePackage{snapshot}
\documentclass[a4paper,11pt]{article}
\usepackage{jheppub}
\pdfoutput=1 
\usepackage{amsmath,amssymb}
\usepackage{mathtools}
\usepackage{amsthm}
\usepackage{graphicx}
\usepackage{caption}
\usepackage{subcaption}
\usepackage[export]{adjustbox}
\usepackage[utf8]{inputenc}
\usepackage{slashed}
\usepackage{bbold}
\usepackage{tikz}
\usetikzlibrary{arrows,shapes}
\usepackage[most]{tcolorbox}
\usepackage{pdfpages}
\usepackage[mathscr]{euscript}
\usepackage{hyperref}
\usepackage{float}
\usepackage{amsfonts, bm}
\usepackage[printonlyused,nohyperlinks]{acronym}
\usepackage{mathrsfs}
\usepackage{afterpage}
\usepackage{stackengine}

\graphicspath{{figs/}}

\newcommand{\eqs}[2][0.3]{\includegraphics[width=#1\linewidth, valign=c]{#2}}

\newcommand{\edit}[1]{{#1}}

\newcommand{\Tr}[1]{\text{Tr}{\,#1}}

\newcommand{\K}[1]{k_{\perp #1}}

\newcommand{\sq}[2]{\s{Q}_{#1,#2}}
\newcommand{\q}[2]{{Q_{#1,#2}}}

\newcommand{\m}{M}
\newcommand{\prop}[1]{(#1^2-\m^2)}
\newcommand{\e}{\epsilon}

\newcommand{\tq}[1]{t^{#1}}
\newcommand{\s}[1]{\slashed{#1}}
\newcommand{\gs}{g_{s}}
\newcommand{\g}{e}
\newcommand{\as}{\alpha_{s}}

\newcommand\numberthis{\addtocounter{equation}{1}\tag{\theequation}}

\title{Local form factor subtraction for three-loop QCD corrections to electroweak production in quark-antiquark annihilation}

\author[]{{\small Rayan Haindl}}

\affiliation[]{
Institut f{\"u}r Theoretische Teilchenphysik, Karlsruhe Institute of Technology (KIT), Wolfgang-Gaede Stra{\ss}e 1, 76131 Karlsruhe, Germany
}
\affiliation[b]{}

\emailAdd{rayan.haindl@kit.edu}

\abstract{
We extend a local subtraction framework to three-loop QCD corrections for the production of multiple electroweak bosons in quark-antiquark annihilation. 
We derive two-loop Ward identities that ensure the factorisation of most collinear singularities from the hard-scattering process in the sum over integrands. 
Infrared and ultraviolet singularities are removed point-by-point in loop momentum space using a minimal set of counterterms, which can be integrated analytically in terms of known master integrals.
Additional counterterms eliminate non-factorising terms arising from loop momentum shifts and one-loop corrections to the gluon three-point function.
We identify previously unknown non-factorising loop polarisation effects in the single-collinear regions, which pose challenges for local integrability and require further investigation.
The techniques presented here are a first crucial step in formulating a systematic approach for constructing finite integrands for general electroweak amplitudes at three-loop order.

}

\begin{document}

\maketitle 
\flushbottom

\acrodef{sm}[SM]{Standard Model}
\acrodef{bsm}[BSM]{Beyond Standard Model}
\acrodef{bphz}[BPHZ]{Bogoliubov–Parasiuk–Hepp–Zimmermann}
\acrodef{mssm}[MSSM]{Minimal Supersymmetric Standard Model}
\acrodef{np}[NP]{New Physics}
\acrodef{qft}[QFT]{Quantum Field Theory}
\acrodef{qcd}[QCD]{Quantum Chromodynamics}
\acrodef{pqcd}[pQCD]{perturbative Quantum Chromodynamics}
\acrodef{qed}[QED]{Quantum Electrodynamics}
\acrodef{ew}[EW]{electroweak}
\acrodef{lhc}[LHC]{Large Hadron Collider}
\acrodef{hl}[HL]{High-Luminosity}
\acrodef{cms}[CMS]{Compact Muon Solenoid}
\acrodef{atlas}[ATLAS]{A Toroidal LHC Apparatus}
\acrodef{heft}[HEFT]{Higgs Effective Field Theory}
\acrodef{vbf}[VBF]{vector boson fusion}
\acrodef{ggf}[ggF]{gluon-gluon fusion}
\acrodef{dis}[DIS]{deep-inelastic scattering}
\acrodef{dy}[DY]{Drell-Yan}
\acrodef{ccdy}[CCDY]{charged-current Drell-Yan}
\acrodef{ncdy}[NCDY]{neutral-current Drell-Yan}
\acrodef{lo}[LO]{leading order}
\acrodef{nlo}[NLO]{next-to-leading order}
\acrodef{nnlo}[NNLO]{next-to-next-to-leading order}
\acrodef{n3lo}[N$^3$LO]{next-to-next-to-next-to-leading order}
\acrodef{n3ll}[N$^3$LL]{next-to-next-to-next-to-leading logarithmic}
\acrodef{fcnc}[FCNCs]{flavour-changing neutral currents}
\acrodef{p2b}[P2B]{Projection to Born}
\acrodef{rs}[RS]{regularisation scheme}
\acrodef{cdr}[CDR]{Conventional Dimensional Regularisation}
\acrodef{kln}[KLN]{Kinoshita-Lee-Nauenberg}
\acrodef{uv}[UV]{ultraviolet}
\acrodef{ir}[IR]{infrared}
\acrodef{mhv}[MHV]{Maximally Helicity Violating}
\acrodef{ps}[PS]{pinch surface}
\acrodef{1pi}[1PI]{one-particle irreducible}
\acrodef{2pi}[2PI]{two-particle irreducible}
\acrodef{1pr}[1PR]{one-particle reducible}
\acrodef{rge}[RGE]{renormalisation group equation}
\acrodef{mi}[MI]{master integral}
\acrodef{mis}[MIs]{master integrals}
\acrodef{ibp}[IBP]{integration by parts}
\acrodef{li}[LI]{Lorentz invariance}
\acrodef{ms}[MS]{minimal subtraction}
\acrodef{msb}[$\overline{\textrm{MS}}$]{modified minimal subtraction}
\acrodef{pdfs}[PDFs]{parton distribution functions}
\acrodef{ffs}[FFS]{form-factor subtraction}
\acrodef{ltd}[LTD]{Loop-Tree Duality}

\section{Introduction}

The calculation of higher-order corrections to scattering amplitudes in perturbative \ac{qcd} is paramount for reaching the target precision of current and future collider experiments. Given that the dominant experimental uncertainties, currently at order of a few percent at ATLAS~\cite{ATLAS:2019pzw,ATLAS:2017bje} and CMS~\cite{CMS:2021xjt,CMS:2016lmd}, are expected to decrease further with the start of the High Luminosity phase of the LHC (HL-LHC), a concerted effort towards percent-level phenomenology is key to both verifying the structure of the \acl{sm} and unlocking possible \acl{np} effects. \edit{Better understanding of both the Higgs potential and Yukawa sector are crucial in this endeavour.}

There has been steady progress in improving both analytic and numerical control of loop amplitude calculations. Powerful tools have been developed that exploit the algebraic properties of iterated integrals~\cite{Remiddi:1999ew,Vollinga:2004sn,Goncharov:2010jf,Duhr:2011zq,Duhr:2012fh,Duhr:2019tlz,Henn:2013pwa,Broedel:2017kkb,Ablinger:2018sat,Broedel:2021zij,Abreu:2022mfk,Duhr:2014woa,Gehrmann:2024tds} including algorithmic methods related to the elliptic sector~\cite{Duhr:2019rrs,Walden:2020odh,Duhr:2024uid,Frellesvig:2024rea,DHoker:2025szl,Marzucca:2025eak}, while steady progress has been made in understanding Feynman graphs related to curves of higher genus~\cite{Huang:2013kh,Georgoudis:2015hca,Doran:2023yzu,Marzucca:2023gto}. There has been continued improvement in reducing Feynman diagrams to a finite set of \ac{mis} through optimised \ac{ibp} reduction techniques~\cite{Anastasiou:2004vj,Smirnov:2008iw,Smirnov:2013dia,Smirnov:2014hma,Smirnov:2019qkx,Lee:2012cn,Lee:2013mka,Studerus:2009ye,vonManteuffel:2012np,Maierhofer:2017gsa,Maierhofer:2018gpa,Badger:2024sqv}, in Feynman parameter space using projective geometry~\cite{Artico:2023jrc},
finite field reconstruction techniques~\cite{Peraro:2016wsq,Peraro:2019svx,Klappert:2019emp,Klappert:2020aqs,Klappert:2020nbg}, (numerical) unitarity-based methods~\cite{Anastasiou:2006jv,Britto:2004nc,Bern:2004ky,Bern:1995db,Bern:1994cg,Kosower:2011ty,Ita:2015tya,Larsen:2015ped,Abreu:2023bdp}, 
algebraic geometry techniques~\cite{Mastrolia:2016dhn,Bohm:2018bdy,Bendle:2019csk}, or methods based on properties due to intersection theory~\cite{Caron-Huot:2012awx,Frellesvig:2019uqt,Mastrolia:2018uzb, Frellesvig:2019kgj, Mizera:2019ose, Frellesvig:2020qot}. 
There is a sophisticated toolkit for computing master integrals using differential equations~\cite{Kotikov:1990kg,Gehrmann:1999as,Caffo:1998du,Czakon:2013goa,Abreu:2020jxa,Casconi:2019xof,Frellesvig:2019byn,Hidding:2020ytt,Moriello:2019yhu,Liu:2022chg,Hidding:2022ycg,Czakon:2020vql,Lee:2023dtc}, solved e.g. in terms of a numerically efficient basis of pentagon functions~\cite{Gehrmann:2018yef,Abreu:2023rco}, and we note recently developed analytic techniques for the parametric integration of massive two-loop four-point Feynman integrals in the high-energy region~\cite{Zhang:2024fcu}. Many numerical methods
have also emerged in the past decades, one of the standard approaches being sector decomposition~\cite{Hepp:1966eg,Roth:1996pd,Binoth:2000ps,Anastasiou:2005pn,Anastasiou:2008rm,Lazopoulos:2007ix,Smirnov:2008py,Borowka:2017idc,Carter:2010hi,Borowka:2016ehy,Borowka:2016ypz}, which provide an algorithmic and automated way of dealing with dimensionally regulated singularities,
Mellin-Barnes integration~\cite{Smirnov:1999gc,Tausk:1999vh,Anastasiou:2005cb,Czakon:2005rk,Smirnov:2009up,Gluza:2016fwh,Dubovyk:2018rlg,Dubovyk:2019krd}, and direct parametric integration of Feynman integrals in the Schwinger parametrisation~\cite{Panzer:2014caa} using quasi-finite bases~\cite{vonManteuffel:2014qoa}.

The last two decades have seen incredible progress in the automation of \ac{nlo} computations for all the important hard-scattering processes relevant for the LHC~\cite{Ossola:2006us,Ellis:2007br,Giele:2008ve,Berger:2008sj,Ellis:2011cr,Bredenstein:2010rs,Hirschi:2011pa,Cascioli:2011va}. 
Replicating this success at higher orders remains a formidable challenge, given that both the analytic complexity and computational demands grow rapidly with the number of loops, particle masses and kinematic invariants. 
Extending the current frontier of $2\to 3$ \ac{nnlo} and $2\to 2$ \ac{n3lo} processes~\cite{Huss:2022ful} to include additional loops and external legs, particularly for processes involving multiple mass scales, is a crucial step towards advancing the precision physics program.

In recent years, significant progress has been made in this direction, including calculations of two-loop \ac{qcd} amplitudes with four or more scales~\cite{Agarwal:2020dye,Bronnum-Hansen:2021olh,Jones:2018hbb,Chen:2020gae,Borowka:2016ehy,Becchetti:2023wev,Coro:2023mwl}, as well as mixed strong-electroweak corrections~\cite{Bonetti:2020hqh,Heller:2020owb,Bonciani:2021zzf,Becchetti:2021axs,Bonetti:2022cnr}, two-loop five-point amplitudes with up to one massive particle~\cite{Chawdhry:2020for,Agarwal:2021grm,Badger:2021nhg,Abreu:2021oya,Chawdhry:2021mkw,Agarwal:2021vdh,Badger:2021ega,Abreu:2021asb,Badger:2022ncb,Agarwal:2023suw,Badger:2024sqv,Badger:2024gjs,Badger:2024awe}, and the first calculation of the two-loop finite remainders for $gg \to t\bar{t}\,+\text{ jet}$ at leading colour~\cite{Badger:2024gjs}. Additionally, important milestones have been reached in the calculation of three-loop massless $2\to 2$ amplitudes~\cite{Caola:2020dfu,Caola:2021rqz,Bargiela:2021wuy}, three-loop form factors with massive particles~\cite{Fael:2022rgm,Schonwald:2024jtw,Datta:2024cen,Fael:2024vko}, four-loop massless form factors~\cite{Henn:2016men,Lee:2022nhh,Chakraborty:2022yan}, as well as recent calculations of three-loop four-point Feynman integrals with masses~\cite{Henn:2020lye,Canko:2021xmn,Long:2024bmi}, including the first calculation of three-loop \ac{mis} for the production of two off-shell vector bosons with different masses~\cite{Canko:2024ara}.

Despite these advances, many existing techniques face a rapid increase in complexity with the number of loops and kinematic scales, necessitating the development of alternative methods to mitigate computational bottlenecks.
Direct numerical integration in momentum space remains a tantalising solution, and has received renewed interest in recent years. The aim is to circumvent traditional obstacles to higher-order calculations, including the generation of large systems of \ac{ibp} identities and the challenges involved in evaluating multi-scale master integrals, particularly in understanding the space of special functions of multi-loop Feynman integrals.

At the turn of the century, Soper developed numerical techniques for the integration of \ac{nlo} corrections to infrared-safe observables~\cite{Soper:1998ye}, and subsequently implemented for three-jet quantities in $e^+e^-$ annihilation at \ac{nnlo}~\cite{Soper:1999xk}. A few years later, Nagy and Soper~\cite{Nagy:2003qn} developed a local momentum-space subtraction scheme for generic one-loop \ac{qcd} amplitudes, with analytically calculable and process-independent counterterms regulating both infrared and ultraviolet divergences on a graph-by-graph basis. This was eventually combined with Monte Carlo integration of the finite remainder on a suitably deformed contour directly in loop-momentum space~\cite{Gong:2008ww} and using Feynman parameters~\cite{Binoth:2005ff,Nagy:2006xy}. Subsequent years has seen fundamental progress in combining local subtraction methods with numerical integration techniques, formulated at the level of the amplitude~\cite{Becker:2010ng,Assadsolimani:2009cz,Becker:2012aqa}, applied in the leading colour approximation for electron-positron annihilation up to seven jets at NLO~\cite{Becker:2011vg}. A contour deformation applicable to multi-loop integrals has been proposed in ref.~\cite{Becker:2012bi}. 
Recent advancements in the ``local unitarity" method~\cite{Capatti:2020xjc,Capatti:2021bsm,Capatti:2022tit,AH:2023kor} and groundbreaking work on a new methodological approach dubbed \ac{ltd} ``causal unitarity"~\cite{Ramirez-Uribe:2024rjg,LTD:2024yrb} have led to the development of locally finite representations of differential cross sections, facilitating the cancellation of final-state infrared singularities between real and virtual contributions.
Methods to organise infrared singularities in parametric space~\cite{vonManteuffel:2014qoa,vonManteuffel:2015gxa} and using Landau equations~\cite{Gambuti:2023eqh} to determine finite bases of Feynman integrals have also been explored.

Building on this foundational work, we developed a local subtraction method for general electroweak amplitudes in $e^+e^-$-annihilation at two-loop order~\cite{Anastasiou:2020sdt}. The approach is based on factorisation theorems for wide-angle or large-momentum transfer scattering processes~\cite{Sen:1982bt,Catani:1998bh,Collins:2011zzd,Feige:2014wja,Erdogan:2014gha,Sterman:2002qn,Ma:2019hjq,Dixon:2008gr}, in which infrared singularities are factorised from the underlying hard-scattering process into universal, i.e. process independent jet and soft functions. This allows us to use the simplest $2 \to 1$ process to construct a minimal number of universal amplitude-level form factor subtraction terms, without referring to individual Feynman graphs, that are independent of the number of final-state particles and the external mass scales. The subtraction defines a locally finite representation of the loop amplitude, in which both infrared and ultraviolet singularities are removed point-by-point in loop momentum space. A key advantage of this approach is that the complexity of the problem scales primarily with the number of loops rather than the number of external particles, and is independent of the external scales. The counterterms can be evaluated analytically in $D=4-2\e$ dimensions using known \ac{ibp} identities and master integrals.
This is followed by numerical integration in $D=4$ dimensions of the finite remainder, facilitated by recent developments in \ac{ltd}~\cite{Capatti:2019ypt,Capatti:2019edf,Capatti:2020ytd,Kromin:2022txz,Runkel:2019yrs,JesusAguilera-Verdugo:2020fsn,Aguilera-Verdugo:2020set} based on foundational works of refs.~\cite{Catani:2008xa,Bierenbaum:2010cy,Buchta:2015wna}, and combined with a way to treat threshold singularities, for example through contour deformation~\cite{Becker:2012bi}, or threshold subtraction~\cite{Kilian:2009wy,Kermanschah:2021wbk,Kermanschah:2024jbp}. The method was extended in subsequent work to $q\bar{q}$-initial states in \acs{qcd}~\cite{Anastasiou:2022eym} and Higgs production through gluon-fusion
at \ac{nnlo}~\cite{Anastasiou:2024xvk,Karlen:2024zgo}.
In a recent breakthrough, the $n_f$-contribution to triboson production at two loops was computed for up to three different external masses for the first time, using threshold subtraction~\cite{Kermanschah:2024utt}. This makes a strong case for developing the \acl{ffs} method further and improve its versatility by including higher loop orders as well as colourful final states in the future.

Since so few three-loop amplitudes are known, especially for processes with external masses, the foremost aim of this paper is to determine how far the \ac{ffs} method developed originally for two-loop electroweak amplitudes can be pushed to higher loop orders, and identify possible challenges that hinder local integrability. While phenomenological applications at this loop order are still limited, our work in ref.~\cite{Anastasiou:2020sdt} was motivated in large part by a desire to calculate yet unknown two-loop corrections to the process $q\bar{q} \to W^{\pm}W^{\mp}Z$ and eventually compute the complete \ac{nnlo} \ac{qcd} corrections to triboson production at the LHC. Already the \ac{nlo} \ac{qcd} correction to $WWZ$ production at the LHC is found to be about $100\%$~\cite{Nhung:2013jta}, and we anticipate that \ac{nnlo} corrections, as well as \ac{n3lo} corrections to diboson and triboson production processes to be significant~\cite{Bozzi:2012mh,Azzi:2019yne,Kallweit:2020gcp,Campbell:2022uzw,Denner:2024ufg}.
Indeed, multiboson production is of intense phenomenological interest since it is an important background in \acl{np} searches through anomalous gauge couplings~\cite{Green:2016trm}.

The challenge is to construct a locally finite representation of the loop integrand amenable to numerical integration.
Ward identities are the basic mechanism by which local factorisation is achieved, in which virtual collinear gluons acquire a longitudinal, or scalar polarisation. Section~\ref{sec:ward} is dedicated to reviewing the tree-level identities. For a triple-gluon vertex contracted with a longitudinal polarisation of one of its external gluons the result can be written in terms of ``scalar" contributions, where one of the hard propagators has been cancelled, and terms related to ghost-gluon vertices~\cite{Anastasiou:2022eym}. In section~\ref{sec:ward_1L} we review their application to one-loop corrections to the quark propagator and quark-antiquark-gluon vertex. Then, in section~\ref{sec:ward_2L} we derive the Ward identities relevant for two-loop subgraphs in quark-antiquark annihilation at three loops, which constitutes one of the main results of this paper. In this context, we identify non-factorising \textit{shift-integrable} contributions in two different loop momentum variables which integrate to zero but hinder integrability.

Importantly, factorisation theorems \textit{a priori} do not guarantee \textit{local} factorisation mainly because certain symmetries, e.g. gauge symmetry, are obscured at the level of the Feynman integrand. This requires the modification of the traditional representation of scattering amplitudes, obtained directly from Feynman rules, through the addition of local infrared counterterms to ensure integrability. In ref.~\cite{Anastasiou:2020sdt} we identified ``loop polarisation" contributions of the one-loop jet function to the collinear regions whereby virtual collinear gluons acquire arbitrary polarisation that spoil local factorisation. This was remedied by exploiting the symmetries of the problematic integrands under loop momentum shifts, which was later extended to QCD corrections of the one-loop quark function in ref.~\cite{Anastasiou:2022eym}. At the three-loop order similar non-factorising loop polarisation terms appear in one-loop corrections to the triple-gluon vertex, and in sections~\ref{sec:gluon_self_energy} and~\ref{sec:gluon_triangle} we derive locally modified versions of the gluon self-energy and triangle subgraphs to eliminate these contributions. 

The rest of this paper is organised as follows. In section~\ref{sec:setup} we review the construction of the locally finite remainder in the \ac{ffs} method, and establish the notation. 
In section~\ref{sec:UV} we discuss regularisation in the ultraviolet regions, and in appendix~\ref{app:uv} provide the expressions for Ward identity preserving ultraviolet counterterms that enable collinear factorisation. In section~\ref{sec:shift} we develop local infrared counterterms that eliminate shift terms due to scalar contributions to the Ward identities for general electroweak integrand. The counterterms are written in terms of standard three-loop Feynman graphs multiplied by non-standard colour factors.
Finally, in section~\ref{sec:ghosts} we show that local factorisation is achieved only up to non-cancelling loop polarisation terms. Additionally, we identify shift-integrable terms due to ghosts contributions to the Ward identities, which appear for the first time the three-loop order, and remove them using local counterterms. We include several appendices with technical material and definitions omitted throughout the main text.

\input{setup}

\input{ward}

\input{ward_1L}

\input{ward_2L}

\input{UV}

\input{shift}

\input{gluon_self_energy}

\input{gluon_triangle}

\input{ghost}

\section{Conclusions}

In this paper, we have made crucial steps in applying the local factorisation framework developed in refs.~\cite{Anastasiou:2020sdt,Anastasiou:2022eym} for the first time to the production of multiple off-shell electroweak bosons in quark-antiquark scattering at the three-loop order. A key achievement has been the derivation of local Ward identities for both the two-loop corrections to the quark-antiquark-gluon vertex and the one-loop gluon triangle subgraph, which constitute some of the main results of this paper. 

Ward identities are the key mechanism by which local factorisation is achieved in the \acl{ffs} method. Through the systematic use of Ward identities on the amplitude integrand, infrared singularities are shown to factorise from a hard-scattering function containing the dependence on the mass-scales and momenta of the final-state electroweak bosons. This enables the use of the simplest process, the $2\to 1$ form factor, to remove initial-state infrared singularities directly from the loop integrand. All counterterms contain quadratic denominators, so that the resulting integrals can be deformed in the complex plane to avoid threshold singularities. Moreover, we have established how the gluon three-point subgraph connects with other two-loop corrections to quark-antiquark-gluon vertex through Ward identities. In addition, we have derived Ward identity-preserving ultraviolet counterterms for two-loop Green’s functions, which are a crucial element in maintaining integrability in mixed UV-collinear regions, a necessary step for managing the interplay of singularities at three loops and beyond. 

As has been shown in ref.~\cite{Anastasiou:2022eym}, the triple-gluon vertex contracted with a scalar polarisation of an external gluon can be written as a sum of two scalar terms (for which a hard propagator is cancelled) and two ghost-gluon vertices multiplied by a momentum vector each. We have generalised this to the one-loop gluon three-point function, which is equivalent to a local implementation of the well known Ward-Slavnov-Taylor identities. Using local infrared counterterms that correspond to symmetric integration of the gluon triangle in the single-collinear regions, we were able to 1) combine bubble-type integrands with one-loop corrections to the external legs, and 2) remove locally non-factorising contributions from the gluon triangle subgraph.

At three loops, the amplitude exhibits locally non-factorising contributions of two kinds: those that cancel after individual loop momentum shifts and those involving virtual gluon momenta that do not satisfy local Ward identities. To address the former, we have developed local infrared counterterms for both the amplitude and the form factor subtraction terms, in a non-trivial extension of the approach introduced ref.~\cite{Anastasiou:2022eym}. The challenge was to avoid spurious contributions of counterterms in different collinear regions, made possible by the introduction of a fictitious fermion propagator with a judicious choice of mass regulator.  The shift counterterms are a critical component of the \acl{ffs} approach and underscore its versatility at higher loops. In particular, we have derived new two-loop shift-integrable counterterms that cancel shift mismatches in two loop momentum variables. The counterterms for scalar contributions to shifted integrands are represented in terms of standard Feynman diagrams multiplied by non-standard colour factors. We have also identified familiar one-loop factorisable and shift-integrable terms in diagrams containing a regularised one-loop quark jet function. Additionally, we have encountered for the first time shift integrands due to ghosts contributions to the Ward identities propagating in the loop. We note that the corresponding shift counterterms, which eliminate these terms directly from the three-loop integrand, do not allow a representation in terms of regular Feynman diagrams. 

While the introduction of zero integrals via integrand symmetrisation and shift terms --- required to enable local Ward identity cancellations --- 
does not pose a conceptual issue, it is expected to impact numerical implementations. In particular, these additional terms may increase the evaluation time of integrands and potentially affect numerical stability due to large local cancellations.
We aim to address these challenges in future work. 

Another significant challenge we encountered is the appearance of loop polarisation terms in ghost contributions to single-collinear regions. 
At two loops ghost contributions factorise separately and are cancelled by the corresponding form factor counterterms containing a triple-gluon vertex. However, at three loops the gluon triangle subgraph, combined with subgraphs containing two triple-gluon vertices, introduces non-factorised ghost contributions that we could not eliminate using symmetric integration. 

Loop polarisations also occur when a virtual collinear gluon connects to a jet subgraph. While counterterms have been developed for two-loop QED corrections in electron-positron annihilation~\cite{Anastasiou:2020sdt} and later for QCD corrections in quark-antiquark scattering~\cite{Anastasiou:2022eym}, extending this procedure to the two-loop quark jet function at three loops remains essential. Moreover, similar generalisations will be critical for the gluon jet function, which plays a key role in other phenomenologically significant processes, such as electroweak production through gluon fusion near the quark pair production threshold and Higgs production at two loops in \acl{heft}. In addition, the expectation is that the methods developed in this paper - including the two-loop Ward identities and modifications to the gluon triangle subgraph - can, in future work, be applied to processes with colourful final states.

Loop polarisations present a significant challenge to the local subtraction program, and whether the standard symmetrisation approach is valid for more complex processes remains an open question. Since these issues originate from the momentum-space integral representation of the scattering amplitude, exploring alternative formulations, such as Schwinger, Feynman-parameter, or Mellin-Barnes representations, may provide new insights, albeit at the cost of additional integrations over auxiliary variables and potentially obscuring gauge symmetries as well as complicating numerical integration. However, we anticipate that further refinements in our approach will help address these challenges in future work.

In summary, the techniques introduced in this work, including three-loop shift counterterms, local modifications to the gluon triangle subgraph, and the derivation of two-loop Ward identities, lay the groundwork for extending the approach beyond two loop electroweak amplitudes, while resolving the remaining challenges involving loop polarisation terms will be crucial for completing the local subtraction framework at three loops.

\section*{Acknowledgements}
We thank J.  Karlen,  M.  Vicini,  D.  Kermanschah and C.  Anastasiou for insightful discussions.

\appendix
\input{app_framework}
\input{app_jet}
\input{app_greens}
\input{app_shift}

\input{app_ghost}
\input{app_UV}

\newpage
\bibliographystyle{JHEP}
\bibliography{main}

\end{document}

%% file: setup.tex
\section{Framework}
\label{sec:setup}
We consider processes with multiple colourless final states $X\in \{\gamma^*,H,W,Z\}$ being produced at wide angles in quark-antiquark scattering, 
\begin{equation}
q(p_1) + \bar{q}(p_2)  \to X(Q)\,,
\label{eq:basic-processes}
\end{equation}
where $p_1^2=p_2^2=0$ and momentum conservation implies $p_1+p_2 = Q$, with ${Q=\sum_{i=1}^n q_i}$. 
Our starting point is the $2\to n$ (off-shell) scattering amplitude $\mathcal{M}$, generated directly from the \acs{qcd} Lagrangian in Feynman gauge. In this text, we adopt the Feynman rule conventions of ref.~\cite{sterman_1993}. 

The electroweak amplitude $\mathcal{M}$ admits a perturbative expansion in the bare \acs{qcd} coupling $\as=\gs^2/4\pi$ as follows,
\begin{align}
\begin{split}
\mathcal{M}
&=
\mathcal{M}^{(0)} + \frac{\as}{2\pi}\,\mathcal{M}^{(1)}
+ \left(\frac{\as}{2\pi}\right)^2 \mathcal{M}^{(2)} 
+ \left(\frac{\as}{2\pi}\right)^3 \mathcal{M}^{(3)} +\mathcal O(\as^4)
\,, 
\end{split}
\label{eq:M_series}
\end{align}
where $\mathcal{M}^{(0)}$ is the tree-level amplitude while $\mathcal{M}^{(1)}$, $\mathcal{M}^{(2)}$ and $\mathcal{M}^{(3)}$ denote the one-, two- and three-loop corrections, respectively. For clarity, we have suppressed the dependence of the amplitude components on their (loop) momenta. 
It will sometimes be useful to consider a form of the amplitude with external spinors removed,
\begin{align}
    \mathcal{M}^{(L)}
    (p_1,p_2,\ell_1,\ldots, \ell_L;\{q_1,\ldots,q_n\})
    = \bar{v}(p_2)\,\widetilde{\mathcal{M}}^{(L)}
    (p_1,p_2,\ell_1,\ldots, \ell_L;\{q_1,\ldots,q_n\})
    \,u(p_1)\,, 
\end{align}
The truncated amplitude $\widetilde{\mathcal{M}}^{(L)}$ of loop order $L$ is a matrix in spinor space, though spinor indices are suppressed for simplicity. We will also use this notation to denote a truncated off-shell amplitude with external fermion propagators removed. 

We shall denote by $M^{(L)}$ the integrated $L$-loop component in $D=4-2\epsilon$ dimensions over the set of loop momenta 
$\ell_1,\ldots,\ell_L$, calculated in the physical region where ${s\equiv 2p_1\cdot p_2 >0}$,
\begin{align}
M^{(L)}(p_1,p_2;
\{q_1,\ldots,q_n\})
= 
\int_{\ell_1,\ldots,\ell_L}\,
\mathcal M^{(L)}
(p_1,p_2,\ell_1,\ldots,\ell_L;\{q_1,\ldots,q_n\})
\,,
\label{eq:Mk_int}
\end{align}
with normalisation
\begin{align}
\int_{\ell_1,\ldots,\ell_L} \equiv \prod_{i=1}^L\, \mu_0^{2\epsilon} \int \frac{\mathrm{d}^D \ell_i}{(2\pi)^D}\,,
\label{eq:integration_measure}
\end{align}
Infrared and ultraviolet singularities will appear as poles in the dimensional regularisation parameter $\epsilon$ after integration. We note that analytic continuation is performed by restoring the causal %
prescription to the denominators,
\begin{align}
    P_i \to P_i + 
    i0^+\,,
    \quad s \to s+ 
    i0^+\,,
\end{align}
where the $P_i$ represent linear combinations of the loop momenta 
$\ell_1,\ldots,\ell_L$
and the external momenta $p_1,p_2,\{q_1,\ldots,q_n\}$.

The aim is to develop local counterterms for three-loop corrections to the electroweak amplitude that eliminate both infrared and ultraviolet divergences directly at the integrand-level. To this end, we re-write the loop-momentum space representation of the integrand in the form,
\begin{align}
\begin{split}
&M^{(L),R} = \int_{\ell_1,\ldots, \ell_L}\,
\mathcal H^{(L),R}
(\ell_1,\ldots,\ell_L)
+ \int_{\ell_1,\ldots,\ell_L}\,
\mathcal M^{(L),R}_\text{singular}
(\ell_1,\ldots,\ell_L)
\,,
\end{split}
\label{eq:Mfin_Msing}
\end{align}
where we have suppressed the dependence on the external momenta. Here, the function $\mathcal H^{(L),R}$ denotes the hard finite remainder, which is amenable to numerical integration in $D=4$ dimensions (at lowest order, $\mathcal H^{(0),R} \equiv \mathcal{M}^{(0)}$).
The singular function $\mathcal M^{(L),R}_\text{singular}$ is constructed out of a set of counterterms that are local in loop-momentum and coordinate space, and match the amplitude of loop order $L$ in all \acs{ir}- and \acs{uv}-singular regions, up to finite terms. By construction, the set of counterterms is easily integrable in {${D=4-2\epsilon}$} dimensions using standard integration by parts and reduction to master integrals techniques. The superscript $``R"$ in the quantities in eq.~\eqref{eq:Mfin_Msing} denotes regularisation in the \acl{uv} regions, through local UV counterterms. The method of their construction is described in section~\ref{sec:UV_1L}. 

Following the notation of refs.~\cite{Anastasiou:2020sdt,Anastasiou:2022eym,Anastasiou:2024xvk} the infrared regions can be regulated order-by-order in the strong coupling by iterative subtraction of a set of local form-factor counterterms $\mathcal{F}$,
\begin{align}
    \mathcal{H}^{(L),R} &=
    \Delta\mathcal{M}^{(L),R} -\sum_{i=0}^{L-1}\mathcal{F}^{(L-i),R}
    (\ell_1,\ldots,\ell_{L-i})
    \left[ 
    \mathbf P_1\,
    \widetilde{\mathcal{H}}^{(i),R}
    \,\mathbf P_1 
    \right]\,, \quad L>0\,.
\label{eq:Mfin-iterative}
\end{align}
Here, the hard scales, of the order of the
external invariants, are separated from the region of the loop momentum space in which
the amplitude becomes divergent in the soft and collinear limits. The local vertex $\mathbf{P}_1 \, \widetilde{\mathcal{H}}^{(i),R}\, \mathbf{P}_1$ denotes a truncated, renormalised hard-scattering function $\widetilde{\mathcal{H}}^{(i),R}$ enclosed by a pair of Dirac projectors, defined as,
\begin{align}
  \mathbf P_1 &\equiv \frac{\slashed p_1 \slashed p_2} {2p_1 \cdot p_2}\,,  
\quad 
\mathbf P_1^2 = \mathbf P_1\,,
\label{eq:P1projector}
\end{align}
which satisfy $\mathbf P_1\, u(p_1) = u(p_1)$ and $\bar{v}(p_2) \mathbf P_1 = \bar{v}(p_2)$.
Importantly, the projectors act as the identity on soft and collinear lines. Their role is to restore gauge invariance to the lower-order amplitude embedded within the local vertex, and prevent spurious singularities at two-loop order and beyond (c.f. the discussion in appendix~I.1 of ref.~\cite{Haindl:2022hcb}).

The basic construction of eq.~\eqref{eq:Mfin-iterative} relies on the factorisation of infrared poles into a product of process-independent soft and a jet functions, and a perturbative short-distance function in which all lines are off-shell. This type of factorisation is characteristic of wide-angle or large-momentum transfer scattering processes~\cite{Sterman:2002qn,Ma:2019hjq,Dixon:2008gr}, and the universality of the soft and jet functions is key in using the simplest process, the form factor, to regulate infrared divergences of the full amplitude.
For the processes under study the infrared divergences occur either for virtual lines that become collinear to the incoming (anti)quark line, or have vanishing momenta (ie. become soft) and attach to either the incoming quark-antiquark pair or to collinear lines connected to them. 

Starting at two-loops, the integrand is defined up to additive infrared counterterms $\delta^{(L)}$ through,
\begin{align}
\begin{split}
    \Delta \mathcal{M}^{(1),R}(l) &\equiv
    \mathcal{M}^{(1),R}(l)\,,\\
    \Delta\mathcal{M}^{(L),R}
    (\ell_1,\ldots,\ell_L)
    &= \mathcal{M}^{(L),R}
    (\ell_1,\ldots,\ell_L)
    + \delta^{(L)}
    (\ell_1,\ldots,\ell_L)
    \,, \quad
    L > 1\,,
\end{split}
\label{eq:M_mod}
\end{align}
which do not affect the result after integration, given
\begin{align}
    \int_{\ell_1,\ldots,\ell_L}\,\delta^{(L)}
    (\ell_1,\ldots,\ell_L)
    =0\,.
\end{align}
The purpose of $\delta^{(L)}$ is twofold: a) to remove locally non-factorisable contributions to the collinear regions that cancel by a loop-momentum shift and b) to remove non-longitudinal, leading power polarisations of a virtual gluon connecting lower-order jet subgraphs to the hard sub-amplitude. Treatment of the former is extended to three-loop order in section~\ref{sec:shift}. At present, the latter is only known to two-loop order.

The validity of eq.~\eqref{eq:Mfin-iterative} has so far been demonstrated for electroweak production in $q\bar{q}$-scattering up to two loops~\cite{Anastasiou:2020sdt,Anastasiou:2022eym} and gluon fusion mediated by a heavy quark loop below the production threshold of a heavy quark pair~\cite{Anastasiou:2024xvk}.

%% file: ward.tex
\section{Leading regions and factorisation}
\label{sec:ward}

The \acl{ffs} method utilises power counting arguments to identify the leading regions in loop momentum space in fixed-angle scattering that produce infrared singularities. Such regions are characterised by the location of pinches, in which poles due to vanishing propagators coalesce from opposite sides of the integration contour. Such pinch surfaces can occur for configurations where two internal lines become parallel to a lightlike direction of an external particle (collinear pinches) or when all four components of the loop momentum vanish (soft or infrared pinches).

In covariant gauges the reduced diagram corresponding to the leading regions has longitudinally polarised massless vector particles that flow out of the hard subdiagram and connect to jet lines. Gauge invariance ensures that these unphysical polarisations decouple in the \emph{sum} of diagrams. In ref.~\cite{Anastasiou:2020sdt}  we have shown that in the collinear regions $k \,||\, p_1$ and $k \,||\, p_2$ we can approximate the virtual jet-line by making the replacements,
\begin{align}
    \eqs[0.24]{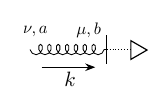} \, :\,\frac{-i \delta^{ab}}{k^2+i\epsilon} g^{\mu\nu}\,\to\,  
    \frac{-i\delta^{ab}}{k^2+i\epsilon}
    \frac{2\eta_1^\nu \, k^\mu} {d_1}
    \,, \quad k \,||\, p_1\,,
\label{eq:g_k_p1}
\end{align}
and
\begin{align}
    \eqs[0.24]{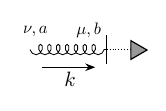}\, :\,\frac{-i\delta^{ab}}{k^2+i\epsilon} g^{\mu\nu} \, \to \,
    \frac{-i\delta^{ab}}{k^2+i\epsilon}
    \frac{2\eta_2^\nu \, k^\mu} {d_2}\,, \quad k \,||\, p_2\,,
\label{eq:g_k_p2}
\end{align}
with quadratic denominators
\begin{align}
    d_1(k,\eta_1) = -(k-\eta_1)^2 +\eta_1^2\,, \quad
    d_2(k,\eta_2) = (k+\eta_2)^2 -\eta_2^2\,.
\end{align}
Here, $\eta_i$ is an auxiliary vector chosen to have a large rapidity separation from $p_1$ ($p_2$) in the collinear limit $k\,||\,p_1$ ($k\,||\,p_2$), with $g_\perp^{\mu\nu}k_\nu =g_\perp^{\mu\nu}\eta_{i\,\nu}=0$, to avoid producing additional pinches.
Above, we have used a empty (shaded) triangle to denote the approximation in the collinear region $k \,||\, p_1$ ($k \,||\, p_2$). The case with a collinear photon is the same, up to the colour matrix $\delta^{ab}$. The approximations in eqs.~\eqref{eq:g_k_p1} and~\eqref{eq:g_k_p2} are key to showing local factorisation of the divergent collinear dynamics from the hard region (independent of the number of electroweak final states) of the loop-momentum space at the amplitude level, and allow us to subtract the divergent part using a small set of form factor counterterms. For completeness, we repeat the derivations of eqs.~\eqref{eq:g_k_p1} and~\eqref{eq:g_k_p2} in appendix~\ref{app:framework}.

As in refs.~\cite{Anastasiou:2020sdt,Anastasiou:2022eym} we find it useful to characterise the leading singular regions at the $L$-loop order by an ordered tuple $(A_{\ell_1},A_{\ell_2}\ldots)$ with $A_{\ell_i}\in\{1_{\ell_i},2_{\ell_i},H_{\ell_i},H_{\ell_i\to\infty}\}$ for loop momenta $\ell_i$ with $i=1,\ldots,L$. The implied ordering determines the sequence in which limits are taken. For instance, the ``mixed"-collinear region\footnote{Throughout this text we assign the loop momentum labels $q$, $k$ and $l$ up three-loop order.} $(1_l,2_q,H_k)$ implies $l$ first becomes lightlike to the incoming quark momentum $p_1$, after which we apply the $q\,||\, p_2$ limit while the loop momentum $k$ remains hard (i.e. of of the order of the typical hard momentum transfer of the process). The subscript $\ell_i\to\infty$ is used to identify ultraviolet regions of the loop momentum space.

\subsection{Tree-level Ward identities}

\begin{figure}[!ht]
    \centering
\begin{align*}
&\hspace{0.4mm}
\eqs[0.23]{g_collinear_k_p2} =\frac{-i\delta^{ab}}{k^2 + i\epsilon}k_\mu
\qquad
\eqs[0.23]{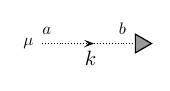} =\frac{i\delta^{ab}}{k^2 + i\epsilon}k_\mu
\\&\hspace{0.4mm}
\eqs[0.26]{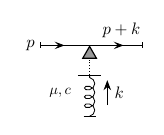}
=-i \gs\tq{c} \s{k}
\quad
\eqs[0.26]{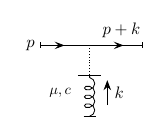} 
= \gs\tq{c}
\\&\hspace{0.4mm}
\eqs[0.26]{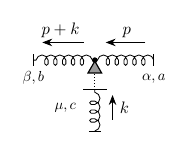}
= k_\mu\,C^{\alpha\mu\beta}_{acb}(-p,-k,p+k)
\quad
\eqs[0.26]{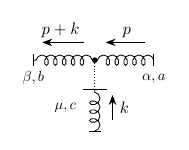} 
= -i\gs\,f^{acb} g^{\alpha\beta}
\end{align*}
    \caption{Modified Feynman rules used to show local collinear factorisation in the region where the external gluon with virtual momentum $k$ becomes collinear to an external antiquark with momentum $p_2$, using the diagrammatic notation of ref.~\cite{Anastasiou:2022eym}. The multiplicative term $2\eta_2^\nu/d_2$ associated to the collinear approximation (c.f. eq.~\eqref{eq:g_k_p2}) is not shown, and will be implicit throughout this paper.}
    \label{fig:Feynman_rules_Ward}
\end{figure}

\begin{figure}[!t]
\centering
\begin{align*}
\eqs[0.22]{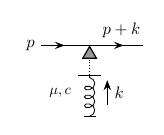} &=  
\eqs[0.22]{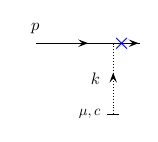}  
-\eqs[0.22]{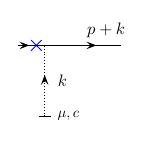}
\end{align*}
\caption{A graphical representation of the abelian Ward identity for quark lines, eq.~\eqref{eq:qq_Ward}. A ghost ending at a quark line indicates an insertion of the collinear momentum $k$ at the vertex, where the adjacent quark propagator is cancelled, denoted by a cross. 
}
\label{fig:qqg_Ward}
\end{figure}

Ward identities form a crucial element in this methodology.
The modified Feynman rules that we will use in the diagrammatic representation of the tree- and loop-level Ward identities are summarised in figure~\ref{fig:Feynman_rules_Ward} for the $k \,||\, p_1$ limit. 
Consider then the quark-antiquark-gluon vertex in which the external gluon with momentum $k$ acquires a longitudinal polarisation, according to the collinear approximation of eq.~\eqref{eq:g_k_p1}. The abelian-type tree-level Ward identity for fermion lines, follows from a simple partial fractioning relation,
\begin{align}
    i S_0(p')(-i\gs\, \slashed{k})iS_0(p) = e_0\left[iS_0(p)-iS_0(p')\right]\,, \quad p' = p+k\,,
    \label{eq:qq_Ward}
\end{align}
where $S_0$ is the the free fermion propagator,
\begin{align}
    \eqs[0.26]{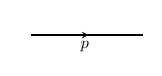}
    \equiv iS_0(p)= \frac{i}{\slashed{p}+i\epsilon}\,.
    \label{eq:electron_prop}
\end{align}
This is shown diagrammatically in figure~\ref{fig:qqg_Ward} using the modified Feynman rules of figure~\ref{fig:Feynman_rules_Ward}. If instead the vertex ends on an external quark line, we have
\begin{align}
    i S_0(p')(-i\gs\, \slashed{k})u(p) = e_0\left[1-S_0(p')\s{p}\right]u(p)\,, \quad p' = p+k\,,
    \label{eq:qq_Ward_spinor}
\end{align}
where the second term in square brackets vanishes due to the massless Dirac equation.

\begin{figure}[!t]
\centering
\begin{align*}
\eqs[0.26]{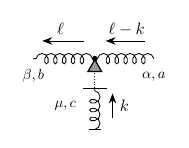} &=  
\eqs[0.26]{ggg_ward_RHS_1}  
+\eqs[0.26]{ggg_ward_RHS_2}
\\&+\eqs[0.26]{ggg_ward_RHS_4}  
-\eqs[0.26]{ggg_ward_RHS_3}
\end{align*}
\caption{A pictorial representation of the QCD Ward identity for the three-gluon vertex, eq.~\eqref{eq:gg_Ward}. Ghost lines ending at a gluon line indicate an insertion of the momentum $k$ at the vertex, where the adjacent gluon line is cancelled. 
}
\label{fig:3g_Ward}
\end{figure}

For the QCD Ward identity we consider the contraction of the triple-gluon vertex with a longitudinally polarised gluon\footnote{We mention that ref.~\cite{Anastasiou:2024xvk} uses an alternative approach in which eq.~\eqref{eq:ggg} is decomposed into three pairs of terms, each of which can be interpreted as an scalar-scalar-gluon interaction vertex.
},
\begin{align}
    \frac{-i}{l^2}\frac{-i}{(l-k)^2} k_\mu C^{\mu\beta\alpha}_{cba}(-k,l,k-l) = 
    f^{cba}\left[Q_0^{\alpha\beta}(k,l)+O_0^{\alpha\beta}(k,l)\right]\,,
    \label{eq:gg_Ward}
\end{align}
where we have defined,
\begin{align}
\begin{split}
    &\gs\,C^{\mu\beta\alpha}_{cba}(-k,l,k-l) \equiv
    \eqs[0.22]{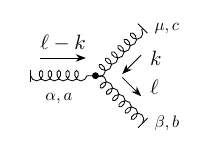}
    \\&\qquad
    =
    \gs\,f^{cba}\left[-g^{\mu\beta}(k+l)^\alpha +g^{\beta\alpha}(2l-k)^\mu +
    g^{\alpha\mu}(2k-l)^\beta\right]\,,
\end{split}
    \label{eq:ggg}
\end{align}
and~\cite{Anastasiou:2022eym},
\begin{align}
    Q_0^{\alpha\beta}(k,l) &= 
    \frac{g^{\alpha\beta}k\cdot(k-2l)}{l^2(l-k)^2} = g^{\alpha\beta}\left[\frac{1}{l^2}-\frac{1}{(l-k)^2}\right]\,,
    \label{eq:gg_Ward_Q0}
    \\
    O_0^{\alpha\beta}(k,l) &= %
    \frac{l^\alpha l^{\beta} - (l-k)^\alpha(l-k)^\beta}{l^2(l-k)^2}
    \label{eq:gg_Ward_O0}
    \,.
\end{align}
The term $Q_0$ is what was called the scalar part in ref.~\cite{Anastasiou:2022eym}, as it can be re-written using a partial fractioning identity in analogy to the quark-gluon vertex. The two terms in $O_0$ can be interpreted as ghost-gluon vertices multiplied by the momentum of the outgoing ghost. A diagrammatic representation of the Ward identity for the triple-gluon vertex is shown in figure~\ref{fig:3g_Ward}, with ghost and scalar terms forming the first and second lines on the right-hand side of the relation, respectively.

In our notation, a cross on a fermion or gluon line represents a cancelled propagator as in the terms on the right-hand side of eqs.~\eqref{eq:qq_Ward} and~\eqref{eq:gg_Ward_Q0},
\begin{align}
    \eqs[0.25]{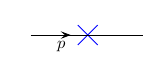} = 1\,, \qquad
    \eqs[0.25]{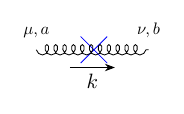} = \delta^{ab}g^{\mu\nu}\,,
    \label{eq:cancel_prop}
\end{align}
Thus, a cancelled fermion or gluon propagator is proportional to the unit matrix in the spin and colour space (the colour matrix depends on whether the cancelled particle transforms in the fundamental or adjoint representation of SU$(N_c)$).

\begin{figure}[!ht]
\centering
\begin{align*}
&\eqs[0.3]{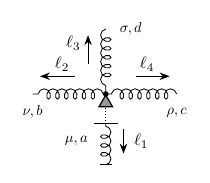} =  
\eqs[0.39]{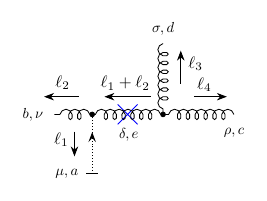}  
\\&\hspace{1cm}+\eqs[0.34]{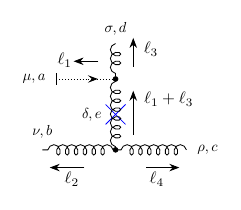}
+\eqs[0.39]{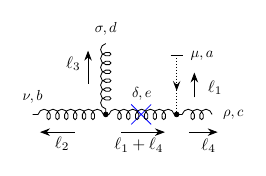}  
\end{align*}
\caption{A diagrammatic version of the QCD Ward identity for the four-gluon vertex, eq.~\eqref{eq:4g_ward}. For consistency with the modified ghost-gluon-gluon vertex we have introduced a dummy Lorentz index $\delta$.
}
\label{fig:4g_ward}
\end{figure}

At three-loop order the four-gluon vertex becomes relevant also. In the case where one of the gluons becomes collinear to an external (anti-)quark, its propagator can be replaced by one of the approximations in eqs.~\eqref{eq:g_k_p1} and~\eqref{eq:g_k_p2}, and the four-gluon vertex is contracted with a longitudinal polarisation. This can be related to a sum of triple-gluon vertices as follows~\cite{Papavassiliou:1992ia},
\begin{align}
\begin{split}
    &(-\ell_1^\mu )D_{\mu\nu\rho\sigma}^{abcd}(\ell_1,\ell_2,\ell_3,\ell_4) = -i
    \left[
    f^{abe}C^{edc}_{\nu\sigma\rho}(\ell_1+\ell_2,\ell_3,\ell_4)
    \right. \\&\qquad \left.
    +f^{ade}C^{ecb}_{\nu\sigma\rho}(\ell_2,\ell_1+\ell_3,\ell_4)
    +f^{ace}C^{ebd}_{\nu\sigma\rho}(\ell_2,\ell_3,\ell_1+\ell_4)
    \right]\,,
\end{split}
\label{eq:4g_ward}
\end{align}
and similarly for contractions with $\ell_i$, $i\in\{2,3,4\}$ (note that we can use momentum conservation $\sum_i^4 \ell_i=0$ to eliminate one of the $\ell_i$). Here we have defined,
\begin{align*}
    &\gs^2\,D^{\mu\nu\rho\sigma}_{abcd}(p_1,p_2,p_3,p_4)\equiv
    \eqs[0.22]{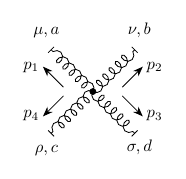}
    =-i
    \gs^2
    \,\big[f^{abe}f^{cde}(g^{\mu\rho}g^{\nu\sigma}-g^{\mu\sigma}g^{\nu\rho})
    \\&\qquad
    +\,f^{ace}f^{bde}(g^{\mu\nu}g^{\rho\sigma}-g^{\mu\sigma}g^{\nu\rho})
    +\,f^{ade}f^{bce}(g^{\mu\nu}g^{\rho\sigma}-g^{\mu\rho}g^{\nu\sigma})
    \big]\,.
    \numberthis
    \label{eq:4g}
\end{align*}
The four-gluon identity is shown in figure~\ref{fig:4g_ward}. Equation~\eqref{eq:4g_ward} can easily be derived by using the familiar Jacobi identity,
\begin{align}
    f^{abe}f^{cde}-f^{ace}f^{bde}+f^{ade}f^{bce}=0\,.
\end{align}

The three-loop electroweak amplitude also contains subgraphs which are one-loop corrections to the three-point function with external gluons. These integrands are logarithmically divergent whenever one of these gluons becomes collinear to an incoming quark line. Using the approximations of eqs.~\eqref{eq:g_k_p1} and~\eqref{eq:g_k_p2}, this leads to the insertion of a longitudinally polarised gluon at the one-loop subgraph, which is either a fermion, gluon or ghost loop. This is discussed in section~\ref{sec:gluon_triangle}.

As we will see in secs.~\ref{sec:Greens_1L} and~\ref{sec:ward_2L}, where we discuss the Ward identities valid for the QCD Green's functions at one- and two-loop order, it will also be useful to use the notation,
\begin{align}
    \eqs[0.25]{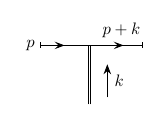} = 1 \qquad
    \eqs[0.25]{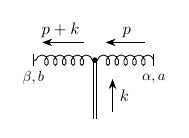} = g^{\alpha\beta}\,.
    \label{eq:insertion}
\end{align}
which represents an insertion of the momentum $k$ of the longitudinal gluon, neglecting the colour matrix of the original quark-gluon or triple-gluon vertex.

\subsection{Loop momentum flow}
\label{sec:routing}

The choice of loop-momentum flow is not unique, and one is free to utilize the shift-invariance of loop integrals to pick a different routing per diagram, as is often convenient in analytic calculations. However, a judicious choice of loop momenta assignments will ensure that cancellations of collinear singularities in the leading regions are manifest in the sum of diagrams. This will lead to a factorised integrand up to \textit{shift-integrable} contributions, first identified in ref.~\cite{Anastasiou:2020sdt} in the context of local factorisation. Such shift terms integrate to zero, i.e. cancel at the level of the integrand only by performing suitable loop-momentum shifts, 
but lead to non-factorisable contributions at the amplitude level. Therefore, shift terms have to be subtracted in the \acs{ffs} scheme to preserve manifest locality of the procedure. We will discuss the factorisation of the scalar contributions to the general three-loop electroweak annihilation amplitude (except gluon self-energy and triangle contributions, which are treated separately) up to shift-integrable terms in section~\ref{sec:shift}.

\begin{figure}
     \centering
     \begin{subfigure}[b]{0.4\textwidth}
         \centering
        \begin{align*}
        \eqs[0.7]{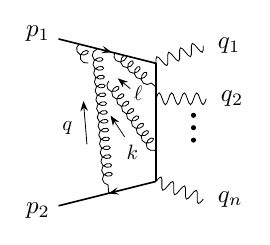}
        \end{align*}
         \caption{}
         \label{fig:flow_ab}
     \end{subfigure}
     \hfill
     \begin{subfigure}[b]{0.4\textwidth}
         \centering
        \begin{align*}
        \eqs[0.7]{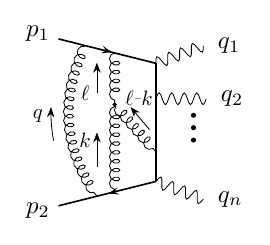}
        \eqs[0.7]{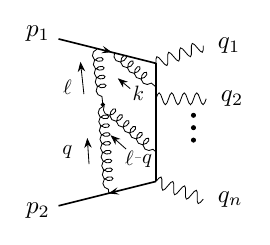}
        \end{align*}
         \caption{}
         \label{fig:flow_nab1}
     \end{subfigure}
     \hfill
     \begin{subfigure}[b]{0.4\textwidth}
         \centering
        \begin{align*}
        \eqs[0.7]{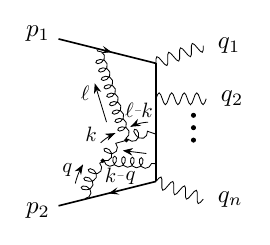}
        \eqs[0.7]{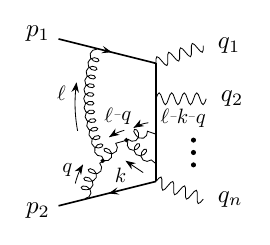}
        \end{align*}
         \caption{}
         \label{fig:flow_nab2}
     \end{subfigure}
     \begin{subfigure}[b]{0.4\textwidth}
         \centering
        \begin{align*}
        \eqs[0.7]{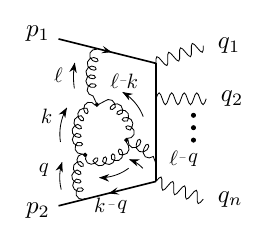}
        \end{align*}
         \caption{}
         \label{fig:flow_nab3}
     \end{subfigure}
     \begin{subfigure}[b]{0.4\textwidth}
         \centering
        \begin{align*}
        \eqs[0.7]{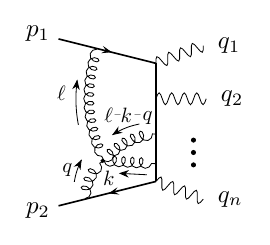}
        \end{align*}
         \caption{}
         \label{fig:flow_nab4}
     \end{subfigure}
        \caption{Loop momentum assignment for the three-loop amplitude. 
        }
        \label{fig:flow}
\end{figure}

To be specific, let $(q,k,l)$ be the tuple of loop momenta at three-loop order, assigned to the virtual gluon lines according to their proximity to the incoming antiquark $\bar{q}(p_2)$, i.e. as we move in the opposite direction to the fermion charge flow. In this way, the virtual gluon connected to the $q\bar{q}g$-vertex closest to $\bar{q}(p_2)$ is chosen to have loop momentum $q$, flowing in the direction of the incoming quark. For ``abelian" diagrams, i.e. diagrams without three-gluon vertices, the virtual gluon connected to the $q\bar{q}g$-vertex second-closest to the incoming antiquark is labelled $k$, the third-closest is assigned the variable $l$. An example is shown in figure~\ref{fig:flow_ab}

For diagrams with a single-three-gluon vertex we follow the conventions of ref.~\cite{Anastasiou:2020sdt}. The assignment is as follows: We start at the $q\bar{q}g$-vertex closest to the incoming antiquark, and choose the loop momentum $\ell_1\in(q,k,l)$ for the outgoing virtual gluon (which may or may not be part of a three-gluon vertex) according to the rule above. 
We then move opposite the fermion flow to the next $q\bar{q}g$-vertex and assign the loop momentum $\ell_2\in(q,k,l)\setminus\{\ell_1\}$ to the virtual gluon. Finally, we move in a clockwise orientation around the triple-gluon vertex, starting from the gluon connected to the $q\bar{q}g$ closest to the antiquark, and assign the remaining loop momentum $\ell_3\in(q,k,l)\setminus\{\ell_1,\ell_2\}$ to the second gluon. Two three-loop examples with a single triple-gluon vertex are shown in figure~\ref{fig:flow_nab1}. Finally, for diagrams with two or more triple-gluon vertices as well as diagrams with a quartic gluon vertex the assignment is fixed according to figs.~\ref{fig:flow_nab2}-~\ref{fig:flow_nab4}. 

Since specific graphs may contribute to multiple collinear regions, we perform a three-fold symmetrisation of the amplitude in the loop momentum variables $q$, $k$ and $l$, so the exact ordering becomes unimportant. We replace the three-loop integrand with an equivalent version,
\begin{align}
    \mathcal{M}^{(3)}(q,k,l) \to \mathcal{M}_{sym}^{(3)}(q,k,l) \equiv\frac{1}{6}\left(\mathcal{M}^{(3)}(q,k,l) + 5 \text{ permutations of }(q,k,l)\right)\,.
    \label{eq:sym}
\end{align}
Both $\mathcal{M}^{(3)}$ and $\mathcal{M}^{(3)}_{sym}$ integrate to the same value, of course. Averaging over the momenta of the virtual gluon lines, $q$, $k$ and $l$, ensures that the Ward identities
can be applied consistently to the sum of diagrams in all \acs{ir}-divergent limits. In particular, it provides a systematic way of combining integrands in all \acs{ir}- and \acs{uv}-singular regions.

%% file: ward_1L.tex
\section{One-loop Ward identities}
\label{sec:ward_1L}

In this part we will introduce the relevant one-loop two- and three-point Green's functions, and discuss their UV regularisation. The results of this section have been derived in refs.~\cite{Anastasiou:2020sdt,Anastasiou:2022eym}. Ward identities will play a key role in ensuring the factorisation of collinear singularities from the hard (UV) regions of loop-momentum space remains local. These functions can be obtained directly from the Feynman rules, unless they are part of an external jet and require additional modifications in the collinear regions. The relevant one-loop jet counterterms are summarised in 
appendix~\ref{app:jet}.
Our understanding of the Ward identities at one-loop will form the basis of our treatment of the three-loop $q\bar{q}$-annihilation amplitude in later sections, so we find it useful to review the one-loop case in some detail. 

\subsection{One-loop two- and three-point Green's functions}
\label{sec:Greens_1L}
The one-loop \acs{qcd} correction to the electroweak vertex has the following momentum-space representation,
\begin{align}
    \begin{split}
        &\g\,\Gamma_{qq\gamma}^{(1)\,\mu}(p,k,l) = \eqs[0.25]{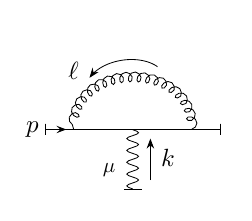} = -\g\,\gs^2 C_F V^{(1)\,\mu}(p,k,l) 
    \end{split}
\label{eq:EW_vertex_1L}
\end{align}
where $e$ is the bare electric charge and
\begin{align}
    V^{(1)\,\mu}(p,k,l) = \frac{\gamma^\nu(\s{p} + \s{l}+\s{k})\gamma^\mu(\s{p}+\s{l})\gamma_\nu}{l^2(p+l)^2(p+k+l)^2}\,.
    \label{eq:qqg_1L_V}
\end{align}
Here, the momentum $p$ is in general off-shell and $k$ is the momentum of the external photon\footnote{If we regard $\Gamma_{qq\gamma}^{(1)\,\mu}$ as a subgraph in mixed QCD-EW corrections to the electroweak amplitude, then $k$ denotes the loop momentum of a virtual photon and the amplitude diverges if the photon becomes collinear to an external quark or antiquark.} with Lorentz index $\mu$.

Similarly, for the \acs{qcd} vertex we have,
\begin{align}
\begin{split}
    &\gs\,\Gamma_{qqg}^{(1)\,\mu,c}(p,k,l) = \eqs[0.25]{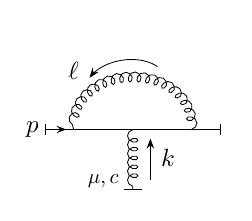} + \eqs[0.25]{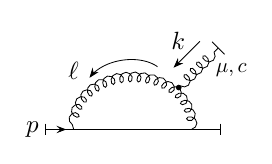} \\
    &\quad = \gs^3 \tq{c}\left[
    -T_F(2C_F-C_A)
    V^{(1)\,\mu}(p,k,l) + 
    2T_F^2C_A
    W^{(1)\,\mu}(p,k,l) \right]\,,
\end{split}
\label{eq:QCD_vertex_1L}
\end{align}
where $T_F=\frac{1}{2}$ is defined by $\Tr{(\tq{a}\tq{b})}=T_F\delta^{ab}$, and
$C_F$ and $C_A$ denote the quadratic Casimirs of the fundamental and adjoint representations of SU$(N_c)$,
\begin{align}
    C_F = \frac{N_c^2-1}{2N_c}\,,\quad C_A = N_c\,.
    \label{eq:casimirs}
\end{align}

The \acs{qcd} vertex correction has a term with crossed ladder structure, denoted by $\Gamma_{qqg}^{(1,XL)}$, and a term containing a three-gluon vertex, denoted by $\Gamma_{qqg}^{(1,3V)}$, corresponding to the first and second diagrams on the right hand side of eq.~\eqref{eq:QCD_vertex_1L}, respectively. It is useful to decompose the one-loop three-gluon vertex function $W^{(1)\,\mu}$ in eq.~\eqref{eq:QCD_vertex_1L} into a scalar contribution $Q^{(1)\,\mu}$ and a contribution associated to ghosts $O^{(1)\,\mu}$,
\begin{align}
    W^{(1)\,\mu}(p,k,l) = Q^{(1)\,\mu}(p,k,l) + O^{(1)\,\mu}(p,k,l)\,,
    \label{eq:qqg_1L_W}
\end{align}
with
\begin{align}
    Q^{(1)\,\mu}(p,k,l) &= 2(1-\epsilon)\frac{(k-2l)^\mu(\s{p}+\s{l})}{l^2(p+l)^2(l-k)^2}\,, 
    \label{eq:Q1}
\end{align}
and
\begin{align}
    O^{(1)\,\mu}(p,k,l) &= -\frac{\gamma^\mu(\s{p}+\s{l})(\s{l}-2\s{k})+(\s{l}+\s{k})(\s{p}+\s{l})\gamma^\mu}{l^2(p+l)^2(l-k)^2}\,.
    \label{eq:O1}
\end{align}

The one-loop self-energy correction is given by,
\begin{align}
    \Pi_{qq}^{(1)}(p,l) = \eqs[0.25]{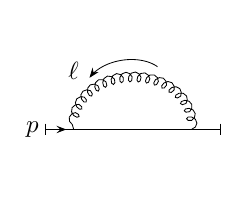} = \gs^2 C_F S^{(1)}(p,l)\,,
    \label{eq:QCD_selfenergy_1L}
\end{align}
with
\begin{align}
    S^{(1)}(p,l)=2(1-\epsilon)\frac{\s{p}+\s{l}}{l^2(p+l)^2}\,.
    \label{eq:qq_1L_S}
\end{align}

The QED Ward identity at one-loop follows directly from the discussion of section~\ref{sec:ward}. It is the relation,
\begin{align}
    k_\mu \Gamma^{(1)\,\mu}_{qq\gamma}(p,k,l)=
    \Pi^{(1)}_{qq}(p,l)-\Pi^{(1)}_{qq}(p+k,l)\,,
    \label{eq:Ward_QED_1L}
\end{align}
For the \acs{qcd} three-point function we have instead~\cite{Anastasiou:2022eym},
\begin{align}
\begin{split}
    k_\mu\,  \Gamma_{qqg}^{(1)\,\mu}(p,k,l) &= 
    \Pi^{(1)}_{qq}(p,l)-\Pi^{(1)}_{qq}(p+k,l)
    - \Pi^{(1)\,\text{shift}}_{qq}(p,k,l) 
    \\&+ \gs^2\frac{C_A}{2}k_\mu O^\mu(p,k,l)\,,
\end{split}
\label{eq:Ward_QCD_1L}
\end{align}
with \textit{shift} propagator
\begin{align}
\begin{split}
    \Pi^{(1)\,\text{shift}}_{qq}(p,k,l) &= 
    -\gs^2\frac{C_A}{2} 
    k_\mu [V^{(1)\,\mu}(p,k,l) + Q^{(1)\,\mu}(p,k,l)]
    \\ &=
    \frac{C_A}{2C_F}\left[\Pi^{(1)}_{qq}(p+k,l-k)-\Pi^{(1)}_{qq}(p+k,l)\right]\,.
\end{split}   
\label{eq:shift_1L}
\end{align}
To obtain the second line, we have used the relations,
\begin{align}
    k_\mu V^{(1)\,\mu}(p,k,l) &=
    S^{(1)}(p+k,l)-S^{(1)}(p,l)\,,
    \label{eq:V1_ward}
    \\
    k_\mu Q^{(1)\,\mu}(p,k,l) &=
    S^{(1)}(p,l)-S^{(1)}(p+k,l-k)
    \,,
    \label{eq:Q1_ward}
\end{align}
which can easily be verified using a simple partial fraction identity.
The shift term integrates to zero but is important in cancelling locally non-factorising contributions in the amplitude~\cite{Anastasiou:2022eym}.

Graphically, we can represent the one-loop QCD Ward identity, eq.~\eqref{eq:Ward_QCD_1L}, as
\begin{align}
\begin{split}
    &k_\mu\,  \Gamma_{qqg}^{(1)\,\mu,c}(p,k,l)=
    \eqs[0.22]{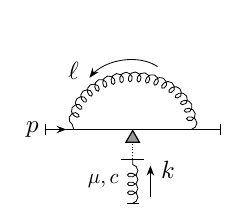}
    +\eqs[0.22]{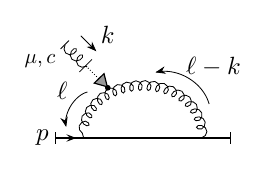}
    \\&\qquad= 
    \eqs[0.24]{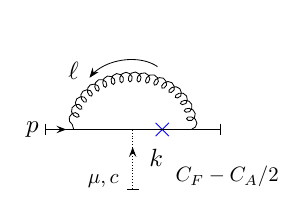}
    -\eqs[0.24]{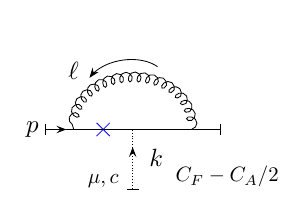}
    +\eqs[0.24]{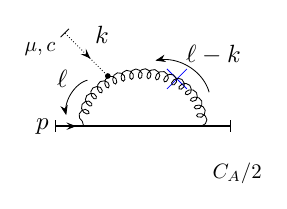}
    \\&\qquad-\eqs[0.24]{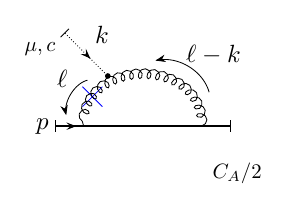}
    + \text{ ghosts}\,,
\end{split}
\label{eq:Ward_QCD_1L_graph}
\end{align}
where on the right-hand side we have explicitly written the colour factors below each graph.
As we have explained in section~\ref{sec:ward} the shaded triangle denotes a longitudinal gluon polarisation, i.e. contraction at the vertex by the momentum of the external collinear gluon. Cancelled propagators, as on the right-hand side of eqs.~\eqref{eq:V1_ward} and~\eqref{eq:Q1_ward}, are represented by blue crosses (c.f. eq.~\eqref{eq:cancel_prop}). Ghost (dotted) lines connected to a quark or gluon propagator represent the insertion of the collinear momentum $k$ at the vertex and carry the Lorentz index $\mu$ and colour index $c$ of the longitudinal gluon, according to the modified Feynman rules introduced in figure~\ref{fig:Feynman_rules_Ward}. 

The shift propagator arises from the parts of the second and fourth graphs on the right-hand side of eq.~\eqref{eq:Ward_QCD_1L_graph} proportional to the non-abelian colour factor $C_A/2$.
\begin{align}
\begin{split}%
    &\Pi^{(1)\,\text{shift}}_{qq}(p,k,l) =
    \frac{C_A}{2C_F}\Biggl[\eqs[0.26]{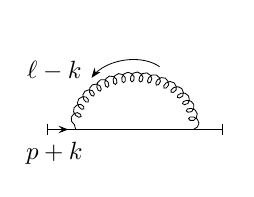}
    -\eqs[0.26]{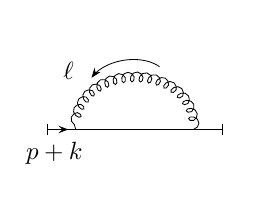}\Biggr]\,,
\end{split}
\label{eq:shift_1L_graph}
\end{align}
where
\begin{align}
    \int_\ell\, \Pi^{(1)\,\text{shift}}_{qq}(p,k,l) =0\,.
\end{align}
For the sake of clarity, we remark on a subtlety in the pictorial notation above. Since the virtual gluon line with loop momentum $\ell$ is cancelled there is an insertion of a scalar polarised gluon (represented by a ghost line in eq.~\eqref{eq:Ward_QCD_1L_graph})
directly at the quark-gluon vertex. This implies an incoming quark momentum of $p+k$. 

The ghost term in eq.~\eqref{eq:Ward_QCD_1L}, $k_\mu O^\mu$, has been shown to factorise independently~\cite{Anastasiou:2022eym} from the scalar contributions in the single-collinear regions. %
\edit{In section~\ref{sec:ghosts} we will investigate to what extent this statement applies to the two-loop QCD vertex.}
Pictorially, we can represent the ghost contribution to the QCD Ward identity for the one-loop quark propagator as follows,
\begin{align}
\begin{split}
    &\gs^3\,\frac{C_A}{2}\tq{c} k_\mu O^\mu(p,k,l) =
    \eqs[0.22]{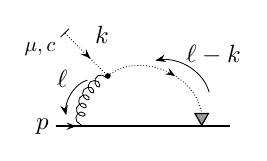}
    +\eqs[0.22]{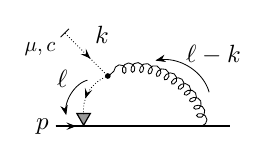} \\
    &\qquad = \eqs[0.22]{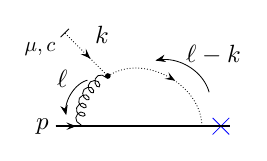}
    -\eqs[0.22]{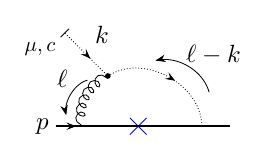}
    +\eqs[0.22]{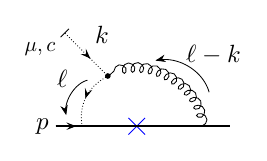} \\
    &\qquad -\eqs[0.22]{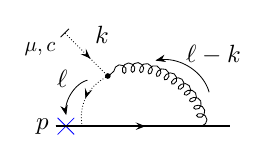}
\end{split}
\label{eq:ghosts_ward_1L}
\end{align}
Summing the second and third graphs on the second line of eq.~\eqref{eq:ghosts_ward_1L} we obtain an integrand proportional to the loop momentum vector of the scalar polarised gluon. This can be cast in terms of the standard abelian-type Ward identity, multiplied by an additional bubble integrand. Indeed, in ref.~\cite{Anastasiou:2020sdt} it was shown that this combination, a difference of two self-energy subgraphs, is equivalent to the following pictorial rule,
\begin{align}
\begin{split}
    &\eqs[0.22]{qqg_QCD_1L_3V_ward_ghost_RHS_diag4} 
    -\eqs[0.22]{qqg_QCD_1L_3V_ward_ghost_RHS_diag2} =
    -\gs^3 \tq{c} \frac{C_A}{2} \frac{i}{\s{p}}\,\frac{\s{k}}{l^2(l-k)^2}\,\frac{i}{\s{p}+\s{k}} \\
    &\qquad\qquad =
    \frac{1}{2}\Biggl[\eqs[0.19]{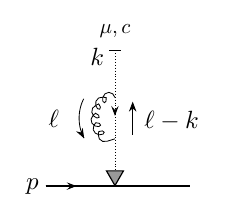} 
    +\eqs[0.21]{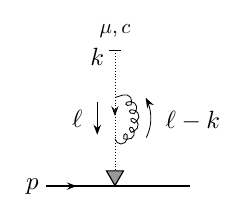}
    \Biggr]
    \equiv 
    \eqs[0.19]{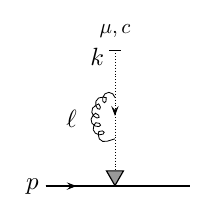}
    \,,
\end{split}
\label{eq:qqg_1L_ward_ghost_id}
\end{align}
which simplifies the bookkeeping of ghost contributions to the amplitude in the collinear regions. We will encounter similar identities at the three-loop level. For simplicity, we will use the shorthand shown on the right-hand side of the second line, where the symmetrisation of the bubble integrand under the exchange $\ell \to k-\ell$ is implied.
Clearly, the second line in eq.~\eqref{eq:qqg_1L_ward_ghost_id} satisfies the standard fermion-line identity, c.f. figure~\ref{fig:qqg_Ward},
\begin{align}
\begin{split}
    &\eqs[0.19]{qqg_1L_ghost_identity}=
\eqs[0.2]{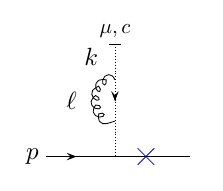}
-\eqs[0.2]{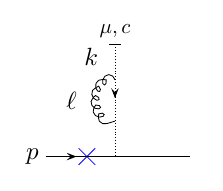}
\\&\qquad\equiv i \gs\,\tq{c}\,
\widetilde{\Pi}(k,l)\left(\frac{1}{\s{p}}-\frac{1}{\s{p}+\s{k}}\right)
\,,
\end{split}
\label{eq:qqg_1L_ward_ghost_id_2}
\end{align}
with symmetrised ghost self-energy,
\begin{align}
    \widetilde{\Pi}(k,l) = \gs^2\,\frac{C_A}{2}\,\frac{1}{l^2(l-k)^2}
    \label{eq:ghost_self_energy}
\end{align}

\subsection{One-loop jet subgraph and loop polarisations}
\label{sec:LP_1L}

At two loops the Feynman rules for two- and three-point Green's functions have to be modified by terms that integrate to zero, both to ensure that the Ward identity is preserved locally, and to remove non-factorisable contributions due to unphysical polarisations of the virtual gauge boson momenta. %
Jet subgraphs $\mathcal{J}$ contain terms that spoil local factorisation in the limit where the external (virtual) gluon, which connects to the hard subdiagram $\mathcal{H}$ from which the final state electroweak bosons are emitted, becomes collinear to the jet. Such contributions, which we called \textit{loop polarisations} in ref.~\cite{Anastasiou:2020sdt}, are proportional to the loop momentum vector of the jet function and do not satisfy standard Ward identities. 

In refs.~\cite{Anastasiou:2020sdt, Anastasiou:2022eym} for the one-loop quark jet function it was shown that these problematic regions can be removed entirely by an appropriate symmetrisation of the integrand, including in light-cone components transverse to the collinear direction. This is equivalent to adding corresponding counterterms that do not affect the result after integration but ensures local factorisation. 

We briefly review the regularisation of the one-loop jet function in appendix~\ref{app:jet} since at \acs{n3lo} loop polarisations occur when $(a)$ both $\mathcal{J}$ and the corresponding hard-scattering subdiagram are at one-loop order and (b) when $\mathcal{H}$ is the Born-level subdiagram and the jet subgraph is at two-loop order. At three-loop order we have two-loop jet subgraphs that exhibit loop polarisation terms in both single and double-collinear regions of the loop momentum space. It remains unclear how the procedure outlined in refs.~\cite{Anastasiou:2020sdt,Anastasiou:2022eym} for the one-loop quark jet function can be generalised to higher loop orders, and we defer the treatment of the two-loop quark jet function to future work.

\subsection{One-loop ultraviolet counterterms}
\label{sec:UV_1L}

The \acs{uv} counterterms can be obtained by performing a Taylor expansion around the large loop momentum and truncating the series at the order corresponding to a logarithmic divergence after integration \cite{Anastasiou:2020sdt}. In this way, the power counting is respected at the integrand level, leading to a locally finite amplitude in all \acs{uv} regions. At the same time, the poles in the dimensional regularisation parameter $\epsilon$ coincide with those of any renormalisation scheme, such as $\overline{\text{MS}}$. This renormalisation procedure is originally based on the \acs{bphz} formula~\cite{Bogoliubov:1957gp,Hepp:1966eg,Zimmermann:1969jj}, adjusted to satisfy local Ward identities. The finite terms are chosen carefully to ensure the local factorisation of the amplitude in the mixed collinear and ultraviolet regions $(1_k,H_{l\to\infty})$ and $(2_k,H_{l\to\infty})$, as has been discussed in detail in refs.~\cite{Anastasiou:2020sdt,Anastasiou:2022eym}. We note that since our convention is to \textit{subtract} \acs{uv} counterterms, they have the opposite sign to ref.~\cite{Anastasiou:2022eym}.

The one-loop QCD vertex counterterm has the contributions,
\begin{align}
    \begin{split}
        &\gs\,\Gamma_{qqg,\text{ UV}}^{(1,XL)\,\mu,c}(p,k,l) 
        = -\gs^3\,\tq{c}\,\left(C_F-\frac{C_A}{2}\right) V_\text{UV}^{(1)\,\mu}(l)\,,
    \end{split}
\label{eq:QCD_vertex_1L_XL_UV}
\end{align}
and
\begin{align}
\begin{split}
    &\gs\,\Gamma_{qqg,\text{ UV}}^{(1,3V)\,\mu,c}(p,k,l) 
    = \gs^3\, \tq{c}\, \frac{C_A}{2}\, W_\text{UV}^{(1)\,\mu}(l) \,.
\end{split}
\label{eq:QCD_vertex_1L_3V_UV}
\end{align}
The kinematic parts in eqs.~\eqref{eq:QCD_vertex_1L_XL_UV} and~\eqref{eq:QCD_vertex_1L_3V_UV} are given by
\begin{align}
    V_\text{UV}^{(1)\,\mu}(l) &= 2(1-\e)\left(\frac{\gamma^\mu}{(l^2-\m^2)^2}-\frac{2l^\mu\s{l}}{(l^2-\m^2)^3}\right)\,,
    \label{eq:V_1L_UV}\\
    W_\text{UV}^{(1)\,\mu}(l) &= 
    Q_\text{UV}^{(1)\,\mu}(l)+ O_\text{UV}^{(1)\,\mu}(l)\,,
    \label{eq:W_1L_UV}
\end{align}
with
\begin{align}
    Q_\text{UV}^{(1)\,\mu}(l) &= -2(1-\e)\frac{2l^\mu \s{l}}{(l^2-\m^2)^3}\,, 
    \label{eq:Q_1L_UV}\\
    O_\text{UV}^{(1)\,\mu}(l) &=- \frac{2\gamma^\mu}{(l^2-\m^2)^2}\,.
    \label{eq:O_1L_UV}
\end{align}
The term proportional to $\gamma^\mu$ in the definition of $W^\mu_\text{UV}$ is associated to ghosts. The label ``$(1)$" in the diagrammatic representation of the UV counterterms refers to the loop order, while the labels ``$(XL)$" and ``$(3V)$" denote the crossed ladder and three-gluon vertex terms, respectively. The mass regulator $\m$ is added to the denominators of eqs.~\eqref{eq:V_1L_UV} and~\eqref{eq:W_1L_UV} to ensure finiteness in the infrared regions (i.e. it plays the role of a renormalisation scale). We will use this construction also for the \acs{uv} counterterms at two-loop order. 

The sum of eqs.~\eqref{eq:QCD_vertex_1L_XL_UV} and \eqref{eq:QCD_vertex_1L_3V_UV} yields,
\begin{align}
    \begin{split}
        &\gs\,\Gamma_{qqg,\text{ UV}}^{(1)\,\mu,c}(l) = \eqs[0.23]{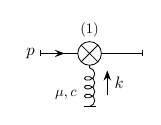}  \\
    &\qquad = -\gs^3\, \tq{c} \left[C_A\left((1-\e) \frac{4l^\mu \s{l}}{(l^2-\m^2)^3}+\e \frac{\gamma^\mu}{(l^2-\m^2)^2}\right)\right. \\
    &\qquad \qquad\left. +C_F\,2(1-\e)\left(\frac{\gamma^\mu}{(l^2-\m^2)^2}-\frac{2l^\mu\s{l}}{(l^2-\m^2)^3}\right)\right]
    \end{split}
\label{eq:QCD_vertex_1L_UV}
\end{align}
Note that the corresponding UV counterterm for the electroweak vertex of eq.~\eqref{eq:EW_vertex_1L} can be obtained from eq.~\eqref{eq:QCD_vertex_1L_UV} by setting $C_A \to 0$,
\begin{align}
    \begin{split}
        &e_{0}\Gamma_{qq\gamma,\text{ UV}}^{(1)\,\mu}(l) = \eqs[0.23]{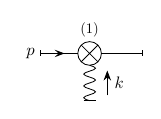} = 
        -e\,\gs^2\, C_F\, V_\text{UV}^{(1)\,\mu}(l)\,,
    \end{split}
\label{eq:EW_vertex_1L_UV}
\end{align}

The \acs{uv} counterterm for the fermion self-energy is given by
\begin{align}
\begin{split}
    &\Pi_{qq,\text{ UV}}^{(1)}(p,l) = \eqs[0.23]{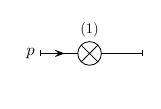} =
    \gs^2\, C_F\, S_\text{UV}^{(1)}(p,l)\,,
\end{split}
\label{eq:QCD_selfenergy_1L_UV}
\end{align}
with
\begin{align}
    S_\text{UV}^{(1)}(p,l)= 2(1-\e)\left(\frac{\s{l}+\s{p}}{(l^2-\m^2)^2}-\frac{2(l\cdot p) \s{l}}{(l^2-\m^2)^3}\right)\,.
    \label{eq:S_1L_UV}
\end{align}
We note that it can be written in terms of a contraction of the \acs{uv} vertex function,
\begin{align}
\begin{split}
    S_\text{UV}^{(1)}(p,l) =
    2(1-\e)\frac{\s{l}}{(l^2-\m^2)^2}
    + p_\mu V^{(1)\,\mu}_{\text{UV}}(l)
    \,,
\end{split}
\label{eq:EW_selfenergy_1L_UV}
\end{align}
where the first term on the right-hand side diverges linearly in the UV region but vanishes after integration over the large loop momentum $l$. Locally, it ensures that the subtracted amplitude converges in the UV region, while it plays an important role in cancelling shift mismatches in contributions where the tree-level jet function connects to the one-loop hard function \cite{Anastasiou:2022eym}.

It is easy to see that the cancellation of the integrated vertex and self-energy contributions is exact in the electroweak case while in \acs{qcd} we receive an additional term from ghosts, proportional to $C_A$. To be specific, we have
\begin{align}
    \int_l \, \Gamma_{qqg,\text{ UV}}^{(1)\,\mu}(l) &= \frac{\as}{2\pi}\left(C_F \widetilde{Z}^{(1)}_{1,F} + C_A \widetilde{Z}^{(1)}_{1,A}\right) (-i \gs \gamma^\mu)\,, \\
    \int_l \, \Pi_{qq,\text{ UV}}^{(1)}(p,l) &= -\frac{\as}{2\pi}C_F \widetilde{Z}^{(1)}_{1,F} (-i \s{p})\,,
\end{align}
where the wavefunction renormalisation constants $\widetilde{Z}^{(1)}_{1,F}$ and $\widetilde{Z}^{(1)}_{1,A}$ are given by
\begin{align}
    \widetilde{Z}^{(1)}_{1,F} &= \left(\frac{4\pi \mu^2}{\m^2}\right)^\e\frac{1}{2}(1-\epsilon)\Gamma(\e)\,, \\
    \widetilde{Z}^{(1)}_{1,A} &= \left(\frac{4\pi \mu^2}{\m^2}\right)^\e \frac{3-\e}{2}\Gamma(\e)\,.
\end{align}

It is important to note that this prescription for the \acs{uv} counterterms is not unique, but has been chosen to respect the \acs{qcd} Ward identity \emph{locally}, up to terms that vanish after integration\footnote{Alternative versions that match the \ac{uv}-divergent behavior of the amplitude exactly but differ by finite terms after integration are of course possible (c.f. prescription by Nagy and Soper~\cite{Nagy:2003qn}).},
\begin{align}
\begin{split}
    k_\mu\,  \Gamma_{qqg,\text{ UV}}^{(1)\,\mu}(l) &= 
    \Pi^{(1)}_{qq,\text{ UV}}(p,l)-\Pi^{(1)}_{qq,\text{ UV}}(p+k,l)
    - \Pi^{(1)\,\text{shift}}_{\text{ UV}}(k,l) \\&-\gs^2 C_A \frac{\s{k}}{(l^2-\m^2)^2}\,,
\end{split}
\label{eq:Ward_QCD_1L_UV}
\end{align}
where $\Pi^{(1)\,\text{shift}}_{\,\text{UV}}$ compensates for the ultraviolet behaviour of the shift propagator defined in eq.~\eqref{eq:shift_1L},
\begin{align}
\begin{split}
    \Pi^{(1)\,\text{shift}}_{\,\text{UV}}(k,l) %
    &=\gs^2C_A (1-\epsilon)\left(\frac{4(l\cdot k)\s{l}}{(l^2-\m^2)^3}-\frac{\s{k}}{(l^2-\m^2)^2}\right)\,.
\end{split}   
\label{eq:shift_1L_UV}
\end{align}
The first two terms on the right-hand side of eq.~\eqref{eq:Ward_QCD_1L_UV} define the \textit{abelian} part of the Ward identity 
\cite{Anastasiou:2020sdt},
\begin{align}
\begin{split}
    k_\mu\,  \Gamma_{qq\gamma,\text{ UV}}^{(1)\,\mu}(l) &= 
    \Pi^{(1)}_{qq,\text{ UV}}(p,l)-\Pi^{(1)}_{qq,\text{ UV}}(p+k,l)\,,
\end{split}
\label{eq:Ward_QED_1L_UV}
\end{align}
which is a local version of the relation $Z_1 = Z_2$ between wavefunction renormalisation constants in QED.

The UV counterterms for the modified quark jet function, provided in eq.~\eqref{eq:jet_mod_p1}, have been calculated in ref.~\cite{Anastasiou:2022eym} and we will not reproduce them here, as they are not of interest for the analysis that follows.

%% file: ward_2L.tex
\section{Two-loop Ward identities}
\label{sec:ward_2L}

Now that we have understood the one-loop Ward identities, we are ready to discuss the two-loop \acs{qcd} corrections to the two and three-point function, both for an external off-shell electroweak boson and gluon. The latter will be necessary ingredients for extensions of the \acs{ffs} to the two-loop \acs{qcd} amplitude with gluons in the final state as well as  three-loop \acs{qcd} or mixed \acs{qcd}-\acs{ew} corrections to the electroweak annihilation amplitude. We ignore gluon self-energy and triangle contributions here, which are discussed separately in section~\ref{sec:gluon_self_energy}.

Following the strategy of the previous section, the approximated integrands in the single-\acs{uv} region can be written in terms of the one-loop counterterms of eqs.~\eqref{eq:QCD_vertex_1L_UV}, \eqref{eq:EW_vertex_1L_UV} and \eqref{eq:QCD_selfenergy_1L_UV}. In the double-\acs{uv} region we will be careful in constructing explicit integrand representations that respect local factorisation of collinear singularities, using appropriate Ward identities. The relevant counterterms are collected in appendix~\ref{app:uv}.

It will be useful to decompose two-loop \acs{qcd} corrections to the quark self-energy as follows,
\begin{align}
    &\Pi_{qq}^{(2)}(p,l,k) = 
    \sum_X\,\Pi_{qq}^{(2,X)}(p,l,k)\,, \quad X\in\{UL,XL,3V\}
    \,,
    \label{eq:qq_QCD_2L}
\end{align}
where $p$ denotes the incoming fermion momentum while $l$ and $k$ are the loop momenta assigned to gluon lines. 
The two-point functions with superscripts $``(UL)"$, $``(XL)"$ and $``(3V)"$ represent the \ac{1pi} uncrossed ladder (or planar), crossed ladder and three-gluon vertex contributions, respectively.

Likewise, for convenience, we decompose the two-loop \acs{qcd} corrections to the quark-antiquark-gluon vertex into several contributions,
\begin{align}
\begin{split}
    &
    \Gamma_{qqg}^{(2)\,\mu,c}(p,q,k,l) 
    \\&\qquad= 
    \sum_X\,\Gamma_{qqg}^{(2,X)\,\mu,c}(p,q,k,l)\,, \quad X\in\{UL,XL,3V,4V,UL-3V,XL-3V,d3V\}\,.
\end{split}
\label{eq:qqg_QCD_2L}
\end{align}
These include the uncrossed, crossed and three-gluon vertex topologies already mentioned, but with an external off-shell gluon connecting to the quark line.
In addition, we have the four-gluon vertex contribution, labelled by the superscript $``(4V)"$, and the usual uncrossed ladder, crossed ladder and three-gluon vertex diagrams where an additional off-shell gluon attaches to an internal gluon line, denoted by the superscripts $``(UL-3V)"$, $``(XL-3V)"$ and $``(d3V)"$, respectively. The terms on the right-hand side of eqs.~\eqref{eq:qq_QCD_2L} and~\eqref{eq:qqg_QCD_2L} are defined in appendix~\ref{sec:Greens_2L}.

Analogously to the one-loop case, the three- and two-point Green's functions for the two-loop QED vertex and propagator subgraphs satisfy the following Ward identity,
\begin{align}
q_\mu \Gamma^{(2)\,\mu}_{qq\gamma} (p,q,k,l) = \Pi^{(2)}_{qq}(p,k,l) - \Pi^{(2)}_{qq}(p+q,k,l)\,,
\label{eq:Ward_QED_2L}
\end{align}
where $q_\mu$ is the momentum of a scalar polarised external photon, $p$ is the momentum of the incoming quark and $k,\,l$ denote virtual gluon momenta. The QED vertex function $\Gamma^{(2)\,\mu}_{qq\gamma}$ is defined below eq.~\eqref{eq:QED_vertex_2L}. We remark that eq.~\eqref{eq:Ward_QED_2L} applies topology-wise as
\begin{align}
    q_\mu V^{(2,X)\,\mu}(p,q,k,l)=
    S^{(2,X)}(p,k,l) - S^{(2,X)}(p+q,k,l)\,,
    \label{eq:Ward_V_2L}
\end{align}
where $V^{(2,X)\,\mu}$ and $S^{(2,X)}$ are defined in appendix~\ref{sec:Greens_2L}, with $X\in \{UL,XL,3V\}$. The relation~\eqref{eq:Ward_V_2L} follows almost directly from the tree-level Ward identity, eq.~\eqref{eq:qq_Ward}, by applying a simple partial fraction identity on each graph, leading to two terms each with a differently cancelled propagator. Then, we are left with the difference of two quark self-energy corrections where the incoming quark momentum in one graph is shifted by the momentum $q$ of the collinear gluon.

The Ward identity for the QCD vertex has a more complicated structure since it involves contributions to the quark self-energy with a shift mismatch, with non-trivial cancellations between different colour coefficients. In analogy to the one-loop Ward QCD Ward identity, at two loops we have,
\begin{align}
\begin{split}
    q_\mu\,  \Gamma_{qqg}^{(2)\,\mu}(p,q,k,l) &= 
    \Pi^{(2)}_{qq}(p,k,l)-\Pi^{(2)}_{qq}(p+q,k,l)
    - \Pi^{(2)\,\text{shift}}_{qq}(p,q,k,l) 
    \\&+ \text{ ghosts}\,,
\end{split}
\label{eq:Ward_QCD_2L}
\end{align}
with $\Gamma_{qqg}^{(2)\,\mu,c}\equiv \tq{c}\,\Gamma_{qqg}^{(2)\,\mu}$.
We decompose the two-loop shift propagator $\Pi^{(2)\,\text{shift}}_{qq}$ as follows,
\begin{align}
    &\Pi^{(2)\,\text{shift}}_{qq}(p,q,k,l)  = 
    \Pi_{qq}^{(2,UL)\,\text{shift}}+ \Pi_{qq}^{(2,XL)\,\text{shift}} + \Pi_{qq}^{(2,3V)\,\text{shift}}
    \,,
    \label{eq:qq_QCD_2L_shift}
\end{align}
where we have suppressed the arguments of the functions on the right-hand side for readability. We remark that the Ward identity for two-loop one-particle reducible diagrams follows directly from the discussion in section~\ref{sec:Greens_1L} for one-loop subgraphs. Below, we will derive each of the terms on the right-hand side of eq.~\eqref{eq:qq_QCD_2L_shift} in turn. 

We reiterate that shift terms integrate to zero but lead to non-factorisable contributions at the amplitude level, and have to be subtracted in the \acs{ffs} scheme. We illustrate the two-loop QCD Ward identity using the planar topologies, which in QCD include graphs where the collinear gluon with scalar polarisation attaches to a virtual gluon line. The result is,
\begin{align*}
    &q_\mu \left(\Gamma^{(2,UL)\,\mu,c}_{qqg}(p,q,k,l)+\Gamma^{(2,UL-3V)\,\mu,c}_{qqg}(p,q,k,l)\right)
    =\eqs[0.2]{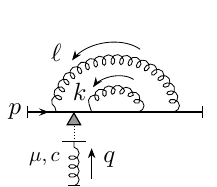}
    +\eqs[0.2]{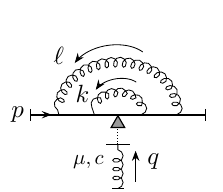}
    \\&\qquad+\eqs[0.2]{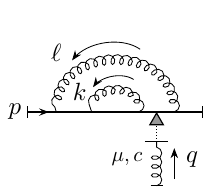}
    +\eqs[0.24]{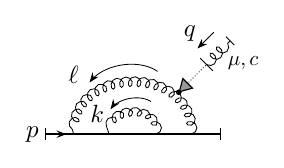}
    +\eqs[0.24]{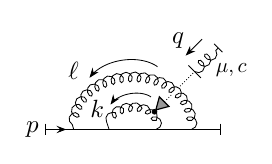}
    \\&\qquad=
    \tq{c}\bigg[\Pi_{qq}^{(2,UL)}(p,k,l)
    -\Pi_{qq}^{(2,UL)}(p+q,k,l)
    -\Pi_{qq}^{(2,UL)\,\text{shift}}(p,q,k,l)
    \\&\qquad
    +i\gs^4\,q_\mu\left(
    \frac{C_AC_F}{2}\,
    O^{(2,UL)\,\mu}(p,q,k,l)
    -\frac{C_A^2}{4}\,O^{(2,UL)\,\mu}_2(p,q,k,l)
    \right)\bigg]\,.
\numberthis
\label{eq:qqg_ward_2L_UL}
\end{align*}
The first two terms on the second line denote the usual abelian-like Ward identity (c.f. eq.~\eqref{eq:Ward_QED_2L}). Above, we have collected ghost terms according to their colour coefficients, with $O^{(2,UL)}\equiv O^{(2,UL)}_1+O^{(2,UL)}_2$, where $O^{(2,UL)}_1$ and $O^{(2,UL)}_2$ are defined below eq.~\eqref{eq:QCD_vertex_2L_UL-3V}.

The shift subtraction term  $\Pi_{qq}^{(2,UL)\,\text{shift}}$ in eq.~\eqref{eq:qqg_ward_2L_UL} is given by,
\begin{align*}
   &\Pi_{qq}^{(2,UL)\,\text{shift}}(p,q,k,l) = 
   i\gs^4\, \frac{C_AC_F}{2}\left[S^{(2,UL)}(p+q,k,l-q)-S^{(2,UL)}(p+q,k,l)\right]
   \\&\qquad-i\gs^2\left(C_F-\frac{C_A}{2}\right)
   \frac{\gamma^\alpha(\s{p}+\s{l}+\s{q})\Pi_{qq}^{(1)\,\text{shift}}(p+l,q,k)(\s{p}+\s{l})\gamma_\alpha}{l^2(p+l)^2(p+l+q)^2}\,,
   \numberthis
\label{eq:shift_2L_UL}
\end{align*}
where $\Pi_{qq}^{(1)\,\text{shift}}$ was defined in eq.~\eqref{eq:shift_1L}. The terms on the right-hand side of eq.~\eqref{eq:shift_2L_UL} correspond to quark self-energy contributions with non-standard colour factors. The term on the second line is equivalent to a shift subtraction of the ``inner" loop.
Graphically, we represent eq.~\eqref{eq:shift_2L_UL} as,
\begin{align}
    \begin{split}
        &\Pi_{qq}^{(2,UL)\,\text{shift}}(p,q,k,l) = 
        \frac{C_A}{2C_F}\Biggl[
        \eqs[0.2]{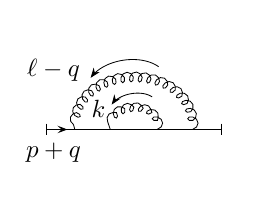}
        -\eqs[0.2]{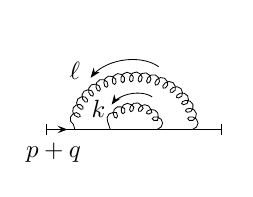}
        \Biggr]
        \\&\qquad
        +\frac{C_A}{2C_F}\left(1-\frac{C_A}{2C_F}\right)
        \Biggl[
        \eqs[0.2]{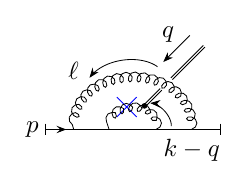}
        -\eqs[0.2]{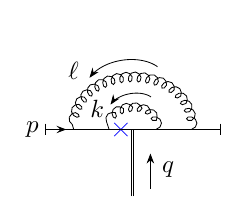}
        \Biggr]\,.
    \end{split}
\label{eq:shift_2L_UL_graph}
\end{align}
Here, a double-line ending at a gluon or quark line indicates the inflow of momentum $q$ at the vertex, according to the notation defined in eq.~\eqref{eq:insertion}.

Equation~\eqref{eq:qqg_ward_2L_UL} can be derived using similar relations to eqs.~\eqref{eq:V1_ward} and~\eqref{eq:Q1_ward} in the one-loop case. In terms of the kinematic functions introduced in the previous sections, the contraction of the planar contributions to the QCD vertex, denoted by the left-hand side of eq.~\eqref{eq:qqg_ward_2L_UL}, is given by,
\begin{align}
\begin{split}
    &q_\mu \left(\Gamma^{(2,UL)\,\mu}_{qqg}+\Gamma^{(2,UL-3V)\,\mu}_{qqg}\right) 
    \\&\qquad =
    i \gs^4\,q_\mu\left[
    C_F^2 V^{(2,UL)\,\mu} 
    + \frac{C_A^2}{4}\left(V_3^{(2,UL)\,\mu}-W^{(2,UL)\,\mu}_2\right)
    \right.\\&\qquad \quad \left.
    - \frac{C_AC_F}{2}\left(V^{(2,UL)\,\mu}+V^{(2,UL)\,\mu}_3 - W^{(2,UL)\,\mu}\right)
    \right]\,,
\end{split}
\label{eq:qqg_ward_2L_UL_decomposed}
\end{align}
where we have suppressed the kinematic dependence for the sake of readability. The components of the uncrossed ladder vertex function $V^{(2,UL)\,\mu}=\sum_{i=1}^3V^{(2,UL)\,\mu}_i$ are defined in eqs.~\eqref{eq:EW_vertex_2L_UL_V1}~-~\eqref{eq:EW_vertex_2L_UL_V3}. As usual, it proves salient to decompose the non-abelian terms $W^{(2,UL)\,\mu}=W^{(2,UL)\,\mu}_1+W^{(2,UL)\,\mu}_2$ into scalar and ghost parts, defined below eq.~\eqref{eq:QCD_vertex_2L_UL-3V}. The scalar terms will enter the definition of the shifted quark self-energy, eq.~\eqref{eq:shift_2L_UL}.

Similarly, the crossed ladder type diagrams satisfy the relation,
\begin{align}
\begin{split}
    &q_\mu \left(\Gamma^{(2,XL)\,\mu}_{qqg}(p,q,k,l)+\Gamma^{(2,XL-3V)\,\mu}_{qqg}(p,q,k,l)\right)
    \\&\qquad=
    \Pi_{qq}^{(2,XL)}(p,k,l)
    -\Pi_{qq}^{(2,XL)}(p+q,k,l)
    -\Pi_{qq}^{(2,XL)\,\text{shift}}(p,q,k,l)
    \\&\qquad
    -i\gs^4\,q_\mu
    \frac{C_A}{2}\left(C_F-\frac{C_A}{2}\right)
    O^{(2,XL)\,\mu}(p,q,k,l)\,,
\end{split}
\label{eq:qqg_ward_2L_XL}
\end{align}
where we have collected the ghost terms $O^{(2,XL)\,\mu}=O^{(2,XL)\,\mu}_1+O^{(2,XL)\,\mu}_2$, with $O^{(2,XL)\,\mu}_i$, $i\in\{1,2\}$, defined in eqs.~\eqref{eq:O1_2L_XL} and~\eqref{eq:O2_2L_XL}. Again, the first two terms on the right-hand side of eq.~\eqref{eq:qqg_ward_2L_XL} form the abelian part of the Ward identity, while the third term, which denotes the shifted quark-self energy contribution for the crossed ladder topology, reads
\begin{align*}
    &\Pi_{qq}^{(2,XL)\,\text{shift}}(p,q,k,l) =
    i\gs^4\,\frac{C_A}{2}\left(C_F-\frac{C_A}{2}\right)\Bigg[
    S^{(2,XL)}(p+q,k,l-q)-S^{(2,XL)}(p+q,k,l)
   \\&\qquad\left.+
   \frac{\gamma_\alpha(\s{p}+\s{k})V^{(1)\,\alpha}(p,k,l)}{(k-q)^2(p+k)^2}
   -\frac{\gamma_\alpha(\s{p}+\s{k}+\s{q})V^{(1)\,\alpha}(p,k+q,l)}{k^2(p+k+q)^2}
   \right]\,. \numberthis%
    \label{eq:shift_2L_XL}
\end{align*}
The one-loop vertex function $V^{(1)\,\alpha}$ appearing on the second line was defined in eq.~\eqref{eq:qqg_1L_V}. We give a diagrammatic representation of the uncrossed ladder shift term as follows,
\begin{align}   
    \begin{split}
        &\Pi_{qq}^{(2,XL)\,\text{shift}}(p,q,k,l) =
        \frac{C_A}{2C_F}\Biggl[
        \eqs[0.2]{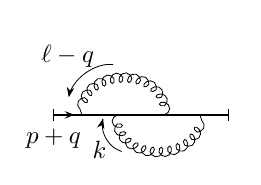}
        -\eqs[0.2]{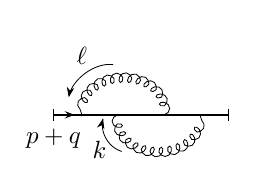}
        \Biggr.
        \\&\qquad\Biggl.
        +\eqs[0.2]{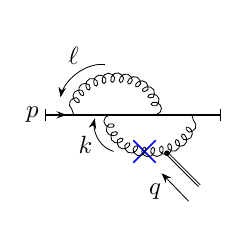}
        -\eqs[0.2]{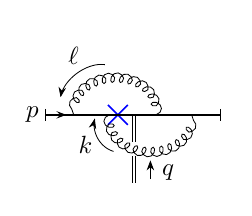}
        \Biggr]\,.
    \end{split}
\label{eq:shift_2L_XL_graph}
\end{align}
It is easy to see that the first graph on the second line is related to the second graph by a loop-momentum shift $l \to l+q$. The first two terms in square brackets has the ``standard" shift relation, while the second line is akin to the ``inner" shift of the uncrossed ladder topology, eq.~\eqref{eq:qqg_ward_2L_UL} . The derivation of eq.~\eqref{eq:qqg_ward_2L_XL} is completely analogous to that of the uncrossed ladder Ward identity, eq.~\eqref{eq:qqg_ward_2L_UL}.

Finally, we have that
\begin{align}
\begin{split}
    &q_\mu \left(\Gamma^{(2,3V)\,\mu}_{qqg}(p,q,k,l)+\Gamma^{(2,d3V)\,\mu}_{qqg}(p,q,k,l)+\Gamma^{(2,4V)\,\mu}_{qqg}(p,q,k,l)\right)
    \\&\qquad=
    \Pi_{qq}^{(2,3V)}(p,k,l)
    -\Pi_{qq}^{(2,3V)}(p+q,k,l)
    -\Pi_{qq}^{(2,3V)\,\text{shift}}(p,q,k,l)
    \\&\qquad
    +i\gs^4\,q_\mu
    \frac{C_A^2}{4}
    O^{(2,d3V)\,\mu}(p,q,k,l)\,,
\end{split}
\label{eq:qqg_ward_2L_3V}
\end{align}
with ghost contribution $O^{(2,d3V)\,\mu}$ defined in eq.~\eqref{eq:O1_2L_d3V}. Here, we have defined the shift term,
\begin{align*}
    &\Pi_{qq}^{(2,3V)\,\text{shift}}(p,q,k,l) =
    i\gs^4\,\frac{C_A^2}{4}\Bigg[
    S^{(2,3V)}(p+q,k,l)-S^{(2,3V)}(p+q,k,l-q)
   \\&\qquad\left.+
   \frac{\gamma_\alpha(\s{p}+\s{k}+\s{q})W^{(1)\,\alpha}(p,k+q,l)}{k^2(p+k+q)^2}
   -\frac{\gamma_\alpha(\s{p}+\s{k})W^{(1)\,\alpha}(p,k,l)}{(k-q)^2(p+k)^2}
   \right]\,, \numberthis
    \label{eq:shift_2L_3V}
\end{align*}
which has the same structure as the shift contribution to the crossed-ladder topology, eq.~\eqref{eq:shift_2L_XL}. Diagrammatically, we represent the cubic gluon vertex shift propagator by
\begin{align}
\begin{split}
    &\Pi_{qq}^{(2,3V)\,\text{shift}}(p,q,k,l) =
    \frac{C_A}{2C_F}\Biggl[
    \eqs[0.2]{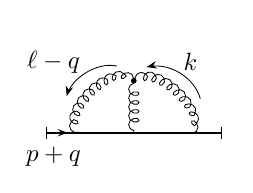}
    -\eqs[0.2]{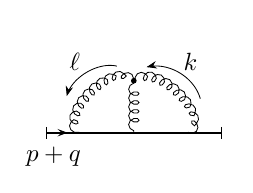}\Biggr.
    \\&\qquad
    \Biggl.
    +\eqs[0.2]{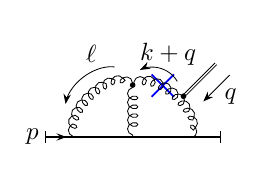}
    -\eqs[0.2]{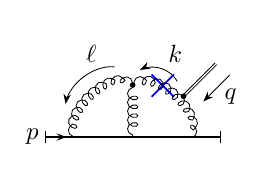}
    \Biggr]\,.
    \label{eq:shift_2L_3V_graph}
\end{split}
\end{align}

Note that, as usual, symmetrisation under the exchange of the loop momenta $k\leftrightarrow l$ in eq.~\eqref{eq:qqg_ward_2L_3V} is implied and so their assignment is unimportant. As we will show below, the derivation of \eqref{eq:shift_2L_3V} is non-trivial, due to contributions from the four-gluon vertex correction to the QCD vertex. In section~\ref{sec:ward} we have seen that at tree-level the four-gluon vertex decomposes into a sum over longitudinal gluon insertions on the triple-gluon vertex (c.f. figure~\ref{fig:4g_ward}). 

We write the left-hand side of eq.~\eqref{eq:qqg_ward_2L_3V} in terms of the kinematic functions introduced in appendix~\ref{sec:Greens_2L} and collect the contributions according to their colour coefficients,
\begin{align}
    &q_\mu \left(\Gamma^{(2,3V)\,\mu}_{qqg}+\Gamma^{(2,d3V)\,\mu}_{qqg}+\Gamma^{(2,4V)\,\mu}_{qqg}\right) 
    \label{eq:qqg_ward_2L_3V_kinem}
    \numberthis
    \\&\qquad=
    i\gs^4\,q_\mu\left[-\frac{C_AC_F}{2}
    V^{(2,3V)\,\mu}
    +\frac{C_A^2}{4}\left(
    V^{(2,3V)\,\mu}
    + Q^{(2,d3V)\,\mu}
    + W^{(2,4V)\,\mu}
    \right)
    \right] + \text{ghosts}\,,
    \nonumber
\end{align}
where $V^{(2,3V)\,\mu}$ was defined below eq.~\eqref{eq:EW_vertex_2L_3V}, $W^{(2,4V)\,\mu}$ in eq.~\eqref{eq:QCD_vertex_2L_4V_W}, and the contributions to $Q^{(2,d3V)\,\mu} = Q_1^{(2,d3V)\,\mu}+Q_2^{(2,d3V)\,\mu}$ were defined in eqs.~\eqref{eq:Q1_2L_d3V} and~\eqref{eq:Q2_2L_d3V}. Above, we have suppressed the kinematic dependence for readability. The first term in eq.~\eqref{eq:qqg_ward_2L_3V_kinem} proportional to $C_AC_F$ has the usual abelian-like structure,
\begin{align}
    q_\mu V^{(2,3V)\,\mu}(p,q,k,l) =
    S^{(2,3V)}(p,k,l)-
    S^{(2,3V)}(p+q,k,l)\,,
    \label{eq:qqg_ward_2L_V_3V}
\end{align}
with $S^{(2,3V)}$ defined in eq.~\eqref{eq:S_QCD_2L_3V}. Next, we have
\begin{align}
\begin{split}
    q_\mu Q_1^{(2,d3V)\,\mu}(p,q,k,l) &=
    \frac{\gamma_\alpha (\s{p}+\s{k})W^{(1)\,\alpha}(p,k,l)}{(k-q)^2(p+k)^2}
    -S^{(2,3V)}(p,k,l)\,, 
    \label{eq:qqg_ward_2L_Q1_d3V}
\end{split}\\
\begin{split}
    q_\mu Q_2^{(2,d3V)\,\mu}(p,q,l,k) &=
    -\frac{W^{(1)\,\alpha}(p+l,q-l,k+q-l) (\s{p}+\s{l})\gamma_\alpha}{l^2(p+l)^2} \\
    &+S^{(2,3V)}(p+q,k,l-q)\,,
    \label{eq:qqg_ward_2L_Q2_d3V}
\end{split}\\
\begin{split}
    q_\mu W^{(2,4V)\,\mu}(p,q,k,l) &=
    \frac{W^{(1)\,\alpha}(p+l,q-l,k+q-l) (\s{p}+\s{l})\gamma_\alpha}{l^2(p+l)^2}\\
    &-\frac{\gamma_\alpha (\s{p}+\s{k}+\s{q})W^{(1)\,\alpha}(p,k+q,l)}{k^2(p+k+q)^2}
    \,.
    \label{eq:qqg_ward_2L_W_4V}
\end{split}
\end{align}
Note that we have chosen to exchange $k$ and $l$ in the second equation. Equation~\eqref{eq:qqg_ward_2L_W_4V} can be derived by applying the Ward identity for quartic gluon vertices, eq.~\eqref{eq:4g_ward}. It is easy to see that combining eqs.~\eqref{eq:qqg_ward_2L_V_3V} - \eqref{eq:qqg_ward_2L_W_4V} we obtain the shift counterterm of eq.~\eqref{eq:shift_2L_3V}.

In the next section we describe the mechanism by which we construct local UV counterterms for the electroweak and QCD vertices that preserve the Ward identities presented above. This is to ensure that collinear singularities are factorised at the integrand level. Many of the explicit expressions for the UV integrands are provided in appendix~\ref{app:uv}. Then, in section~\ref{sec:shift} we derive local infrared counterterms to remove shift contributions from the general three-loop electroweak amplitude. As we have seen, the QCD vertex receives additional contributions due to ghost terms, which we will discuss in section~\ref{sec:ghosts}.

%% file: UV.tex
\section{Ultraviolet subtractions}
\label{sec:UV}

The diagrams contributing to the three-loop electroweak amplitude still require QCD and electroweak renormalisation. For $q\bar{q}$ scattering into off-shell electroweak bosons at three-loop order, \acs{uv} divergences occur in corrections to the fermion self-energy (up to three loops), the quark-antiquark-gluon vertex (up to two loops, where the external virtual gluon may additionally become collinear to an external fermion) and the electroweak vertex (up to three loops, with an external off-shell electroweak boson). 

In appendix~\ref{sec:Greens_2L} we provide the relevant two- and three-point \acs{qcd} Green's functions at two-loop order, corresponding to UV divergent subgraphs of the three-loop electroweak amplitude. In the \ac{ffs} prescription the three-loop amplitude is rendered locally finite in the ultraviolet regions by,
\begin{align}
    \mathcal{M}^{(3),R}_\text{UV-finite}(q,k,l) = \mathcal{M}^{(3)}(q,k,l) - \mathcal{M}^{(3)}_\text{UV}(q,k,l)\,,
\end{align}
with ultraviolet counterterm
\begin{align}
\begin{split}
    \mathcal{M}^{(3)}_\text{UV}(q,k,l) &= 
    \mathscr{R}_\text{single-UV}\, \mathcal{M}^{(3)}(q,k,l)
    +\mathscr{R}_\text{double-UV}\, \mathcal{M}^{(3)}(q,k,l)
    \\&+\mathscr{R}_\text{triple-UV}\, \mathcal{M}^{(3)}(q,k,l)\,.
\end{split}
\end{align}
Each of the terms on the right-hand side approximate the amplitude in the regions where one, two or three loop momenta, all assigned to gluon lines according to the conventions introduced in section~\ref{sec:ward}, become infinitely large. The first term, $\mathscr{R}_\text{single-UV}\, \mathcal{M}^{(3)}(q,k,l)$, removes ultraviolet singular contributions from one-loop subgraphs, by the method discussed in section~\ref{sec:ward_1L},
\begin{align}
    \mathscr{R}_\text{single-UV}\, \mathcal{M}^{(3)}(q,k,l) = \left(\mathscr{R}_{k\to\infty}
    +\mathscr{R}_{l\to\infty}
    +\mathscr{R}_{q\to\infty}\right)
    \mathcal{M}^{(3)}(q,k,l)\,.
\end{align}
Here, $\mathscr{R}_{\ell\to\infty}$, $\ell\in\{q,k,l\}$ can be understood as an operator which generates an infrared-finite one-loop ultraviolet counterterm for large loop momentum $\ell$ according to the mechanism discussed in section~\ref{sec:UV_1L}. 

All double-UV and triple-UV counterterms are defined as free of UV subdivergences. This is achieved through consecutive subtractions as follows,
\begin{align}
\begin{split}
&\mathscr{R}_\text{double-UV}\, \mathcal{M}^{(3)}(q,k,l) \\
&\qquad= (\mathscr{R}_{k,l\to\infty}
+\mathscr{R}_{k,q\to\infty}+\mathscr{R}_{q,l\to\infty})(1 - \mathscr{R}_\text{single-UV} )\, \mathcal{M}^{(3)}(q,k,l)\,,
\end{split}
\end{align}
and
\begin{align}
\begin{split}
&\mathscr{R}_\text{triple-UV}\, \mathcal{M}^{(3)}(q,k,l)
= \mathscr{R}_{q,k,l\to\infty}\,(1 - \mathscr{R}_\text{double-UV} )\, \mathcal{M}^{(3)}(q,k,l)\,.
\end{split}
\end{align}
In general, the operator $\mathscr{R}_{\ell_1,\ell_2,\ldots\to\infty}$ furnishes a UV counterterm for large loop momenta $\ell_1,\ell_2,\ldots$ by performing a uniform rescaling $\ell_i \to \Lambda \ell_i$ and expanding around $1/\Lambda = 0$, keeping only linearly and logarithmic divergent terms (truncating at the order that corresponds to a logarithmic divergence after loop integration). As a result, we obtain tadpole-type integrands whose mass-regulated denominators are 
$(\ell_i^2-\m^2)^\nu$, $[(\ell_i\pm \ell_j)^2-\m^2]^\nu$, $[(\ell_i\pm \ell_j\pm \ell_k,\ldots)^2-\m^2]^\nu$, with $i\neq j \neq k$ and $\nu\in \mathbb{N}^+$.

\edit{For the electroweak amplitude of loop order $L$ there is, \textit{a priori}, some ambiguity in choosing finite terms for the UV approximations of subgraphs of loop order $\leq L-1$, though the corresponding counterterms have to be consistent with local collinear factorisation.} The UV approximations of fermion self-energy and vertex subgraphs constructed in this way respect local Ward identities, eq.~\eqref{eq:Ward_QCD_2L}, for any values of the loop momenta, even away from the limit where the loop momenta are large compared to the external scales. Thus, they are suitable for constructing UV counterterms at higher orders, for which the local cancellation of collinear singularities between individual diagrams, and therefore infrared factorisation for the sum of diagrams is preserved. The explicit results in the single- and double-UV regions are provided in appendix.~\ref{app:uv}.

\edit{On the other hand, the triple-UV counterterms for three-loop subgraphs in the hard part can be obtained straightforwardly using the method described above. For the sake of conserving space, we do not provide their explicit integrands in this paper.}

%% file: shift.tex
\section{Shift-integrable collinear singularities for the three-loop amplitude}
\label{sec:shift}

\begin{figure}[h!]
        \centering
        \begin{subfigure}[a]{0.2\textwidth}
        \centering
        \begin{align*}
        \eqs[1.1]{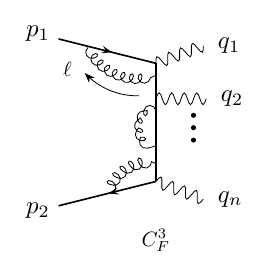}
        \end{align*}
        \caption{}
        \label{}
        \end{subfigure}
        \centering
        \begin{subfigure}[a]{0.2\textwidth}
        \centering
        \begin{align*}
        \eqs[1.1]{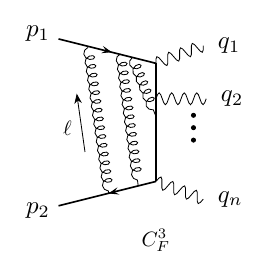}
        \end{align*}
        \caption{}
        \label{}
        \end{subfigure}
        \centering
        \begin{subfigure}[a]{0.2\textwidth}
        \centering
        \begin{align*}
        \eqs[1.1]{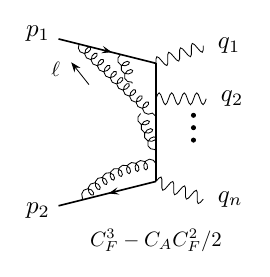}
        \eqs[1.1]{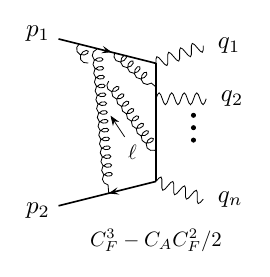}
        \end{align*}
        \caption{}
        \label{}
        \end{subfigure}
        \centering
        \begin{subfigure}[a]{0.2\textwidth}
        \centering
        \begin{align*}
        \eqs[1.1]{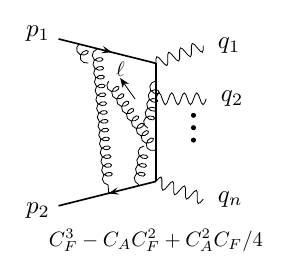}
        \eqs[1.1]{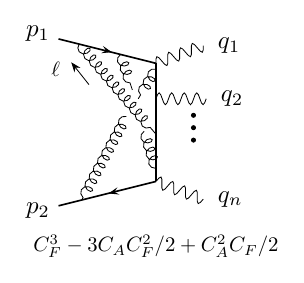}
        \eqs[1.1]{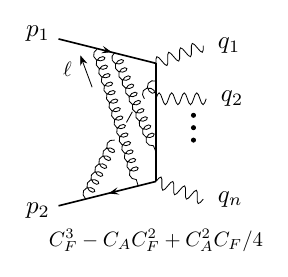}
        \eqs[1.1]{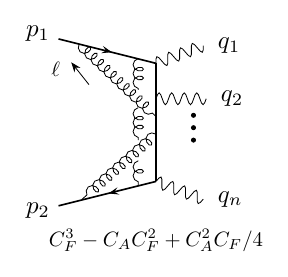}
        \end{align*}
        \caption{}
        \label{}
        \end{subfigure}
        \centering
        \begin{subfigure}[a]{0.2\textwidth}
        \centering
        \begin{align*}
        \eqs[1.1]{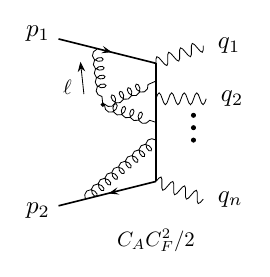}
        \eqs[1.1]{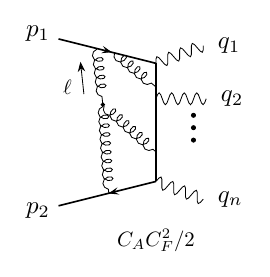}
        \eqs[1.1]{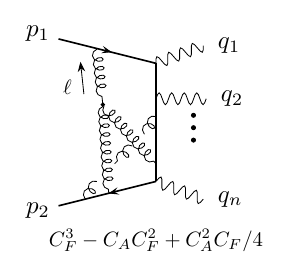}
        \eqs[1.1]{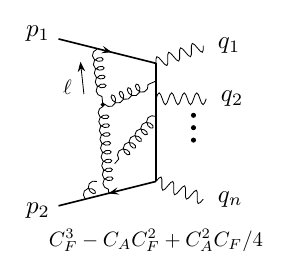}
        \end{align*}
        \caption{}
        \label{}
        \end{subfigure}
        \centering
        \begin{subfigure}[a]{0.2\textwidth}
        \centering
        \begin{align*}
        \eqs[1.1]{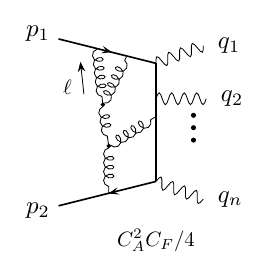}
        \eqs[1.1]{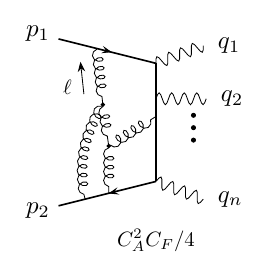}
        \end{align*}
        \caption{}
        \label{}
        \end{subfigure}
        \centering
        \begin{subfigure}[a]{0.2\textwidth}
        \centering
        \begin{align*}
        \eqs[1.1]{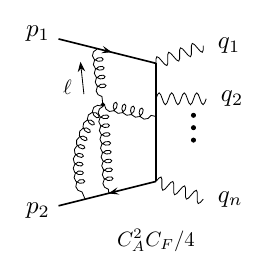}
        \end{align*}
        \caption{}
        \label{}
        \end{subfigure}
        \caption{Representative diagrams for the generic three-loop electroweak amplitude. The corresponding colour factors are displayed below each graph. The diagrams have (a) no ladder structure $(NL)$, (b) uncrossed ladder structure $(UL)$, (c) mixed crossed ladders $(XL-NL)$ or $(XL-UL)$ (from left to right), (d) three-loop crossed ladders $(XL)$, (e) mixed triple-gluon vertex corrections $(3V-NL)$, $(3V-UL)$ and $(3V-XL)$ (from left to right), (f) two triple-gluon vertices $(d3V)$ and (g) a quartic gluon-vertex $(4V)$.}
        \label{fig:3L_topo_sample}
\end{figure}

For the general electroweak amplitude, diagrams that contribute to the single-collinear regions have at least one virtual gluon attaching directly to the incoming quark or antiquark. In this section we investigate the local factorisation of the three-loop amplitude without gluon self-energy or triangle corrections in the single-collinear region denoted by $(1_l,H_k,H_q)$ (the region $(2_l,H_k,H_q)$ is analogous). 
Following the notation for the topologies introduced in section~\ref{sec:Greens_2L}, we show representative three-loop diagrams in figure~\ref{fig:3L_topo_sample}. These include ladder-like graphs which may be planar or crossed, and combinations thereof including three-gluon vertices. At three-loop order we must also include graphs with a quartic-gluon vertex. As usual, all three loop momenta flow through gluon lines, following the momentum routing convention introduced in section~\ref{sec:routing}.

We will show that through repeated applications of the QCD Ward identities in the $l\,||\, p_1$ limit, introduced in section~\ref{sec:ward}, the \textit{scalar} contributions (c.f. figure~\ref{fig:3g_Ward}) factorise locally up to shift-integrable terms, which we denote by $\mathcal{M}_\text{shift}^{(3)}$. These terms are not in local factorised form but vanish by applying appropriate loop-momentum shifts in the hard sub-graph, and therefore integrate to zero,
\begin{align}
   \int_{k,q}\, \mathcal{M}_\text{shift}^{(3)}(p_1,p_2,q,k,l;
   \{q_i\}_{i=1}^n
   ) = 0\,.
\end{align}
However, individually the terms contributing to $\mathcal{M}_\text{shift}^{(3)}$ are infrared divergent. They can be made integrable in $D=4$ dimensions through a local, additive \textit{shift counterterm} $\delta_\text{shift}^{(3)}$ as follows,
\begin{align}
    \Delta\mathcal{M}_\text{shift}^{(3)} =
    \mathcal{M}_\text{shift}^{(3)} + \delta_\text{shift}^{(3)}\,, \quad
    \int_{k,q}\, \delta_\text{shift}^{(3)}(p_1,p_2,q,k,l;
    \{q_i\}_{i=1}^n
    ) = 0\,,
    \label{eq:M3_shifted}
\end{align}
which follows the construction in eq.~\eqref{eq:M_mod}. In the single-collinear region $(1_l,H_k,H_q)$ the sum of shifted integrands is identically zero after application of the Ward identities,
\begin{align}
     \underset{l = -z_{l,1}p_1}{\lim}\,\Delta\mathcal{M}_\text{shift}^{(3)}(p_1,p_2,q,k,l;
     \{q_i\}_{i=1}^n
     ) =0\,.
\end{align}

\begin{figure}[!t]
\centering
\begin{align*}
&\eqs[0.3]{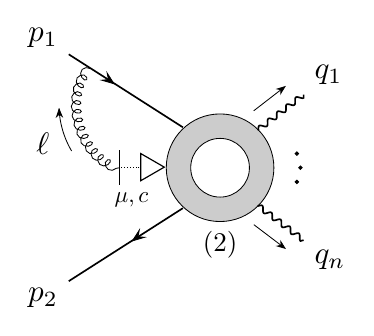} =
-\eqs[0.3]{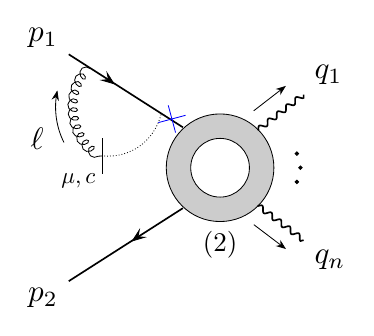}
+ \mathcal{M}_\text{shift}^{(3)}(q,k,l) +\text{ghosts}
\end{align*}
\caption{Factorisation of the scalar contributions to the three-loop integrand (without gluon self-energy or triangle corrections and $p_1$ jet subgraphs) in the single-collinear region $(1_l,H_k,H_q)$. The one- and two-loop hard sub-amplitude is represented by a grey disk and the labels ``$(1)$" and ``$(2)$", respectively, to denote the loop order. A grey blob denotes the truncated $n$-point electroweak Born amplitude. 
}
\label{fig:3L_scalar_factorisation}
\end{figure}

The behaviour outlined above is shown symbolically in figure~\ref{fig:3L_scalar_factorisation} for diagrams without one- or two-loop jet subgraphs on the incoming quark leg. In the figure, a scalar polarised gluon with loop momentum $l$, Lorentz index $\mu$ and colour index $c$ connects to the lower order two-loop hard subgraph, represented by the function $\widetilde{\mathcal{M}}^{(2)}_{\mu,c}(k,q)$ of order $\gs^5$ in the QCD coupling parameter, with an additional quark-antiquark-gluon, triple gluon or quartic gluon vertex. Ghost contributions, not explicitly shown in the figure, are discussed in section~\ref{sec:ghosts} and yield non-factorisable loop polarisation terms. 

The first term on the right-hand side of figure~\ref{fig:3L_scalar_factorisation} represents the sum of integrands where the $l\,||\,p_1$-singularity is manifestly factorised from the two-loop finite remainder $\widetilde{\mathcal{M}}^{(2)}$, due to scalar contributions to the Ward identity,
\begin{align}
    -\eqs[0.24]{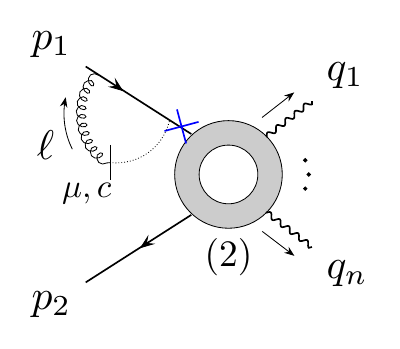} \equiv \bar{v}(p_2)
    \,\widetilde{\mathcal{M}}^{(2)}(p_1+l,p_2,q,k;\{q_1,\ldots,q_n\})\,
    \mathcal{C}(z_{1,l},p_1,l)u(p_1)\,,
    \label{eq:M3_fact}
\end{align}
with incoming quark momentum $p_1+l\simeq (1-z_{1,l})p_1$.
Here, the singularity in the limit $l\,||\,p_1$ is entirely contained in the collinear factor $\mathcal{C}(z_{1,l},p_1,l)$ defined by,
\begin{align}
    \mathcal{C}(z_{1,l},p_1,l) = -iC_F\,\frac{2(1-z_{1,l})}{z_{1,l}}\frac{1}{l^2(p_1+l)^2}\,,
    \label{eq:collinear_factor}
\end{align}
which exhibits a soft singularity for $l^\mu \to 0$ corresponding to a pole in $z_{1,l} = 0$. Both collinear and soft singularities are removed locally through a one-loop form-factor counterterm with a two-loop hard-scattering vertex $\widetilde{\mathcal{H}}^{(2),R}$, based on the construction in eq.~\eqref{eq:Mfin-iterative}. The singularities of the factorised two-loop integrand are removed according to refs.~\cite{Anastasiou:2020sdt,Anastasiou:2022eym}.

\afterpage{\clearpage}

Let us clarify the symbolic notation used in figure~\ref{fig:3L_scalar_factorisation} and later in this section. We use a grey blob to represent the truncated $n$-point (in general off-shell) electroweak Born amplitude, of zeroth order in the QCD coupling $\alpha_s$,
\begin{align}
\begin{split}
    &\widetilde{\mathcal{M}}^{(0)}(p,\bar{p},
    \{q_1,\ldots,q_n\}
    ) \equiv
    \eqs[0.21]{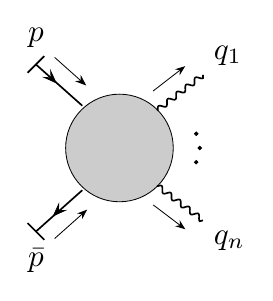}
    = \sum_{\sigma(1,\ldots,n)}
    \eqs[0.3]{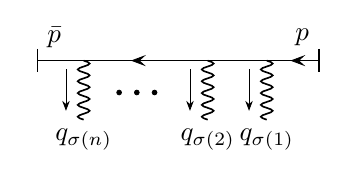}
    \\&\quad= -i e^n
    \sum_{\sigma(1,\ldots,n)}\frac{\s{\epsilon}_{\sigma(n)} (\s{p}-\sum_{j \in \sigma(1,\ldots,n)\setminus \sigma(n)}\s{q}_j)\s{\epsilon}_{\sigma(n-1)}\cdots \s{\epsilon}_{\sigma(2)} (\s{p}-\s{q}_{\sigma(1)})\s{\epsilon}_{\sigma(1)}}{(p-\sum_{j \in \sigma(1,\ldots,n)\setminus \sigma(n)}q_j)^2\cdots (p-q_{\sigma(1)})^2}\,,
    \label{eq:born}
\end{split}
\end{align}
where $\epsilon_i = \epsilon^{\mu_i}(q_i)$ is the polarisation of the off-shell photon with momentum $q_i$. The incoming fermion momenta $p$ and $\bar{p}$ may in general be off-shell, e.g. if $\widetilde{\mathcal{M}}^{(0)}$ corresponds to the hard-scattering sub-amplitude. Here, we sum over all permutations of the external photons with momenta 
$\{q_1,\ldots,q_n\}$, 
though we shall no longer write this explicitly for the rest of this paper.

Due to the factorisation behaviour shown on the right-hand side of figure~\ref{fig:3L_scalar_factorisation}, it is easiest to classify contributions to the three-loop amplitude, divergent in the region $(1_l,H_k,H_q)$, by the possible attachments of a virtual gluon with longitudinal polarisation and momentum $l^\mu \simeq -z_{1,l} p_1^\mu$, adjacent to the incoming quark with momentum $p_1$, on the truncated two-loop amplitudes as follows,
\begin{align}
    \eqs[0.27]{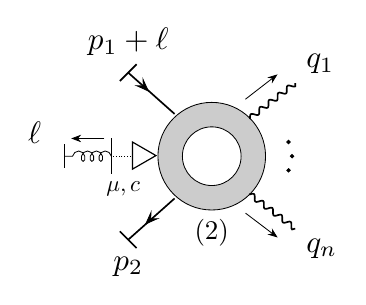} = (-l_\mu)
    \widetilde{\mathcal{M}}^{(2)\,\mu,c}(p_1+l,p_2,k,q;
    \{q_1,\ldots, q_n\}
    )
    \,,
    \label{eq:M3_l_p1}
\end{align}
where we use the $p_1$ collinear approximation, eq.~\eqref{eq:g_k_p1}. We shall refer to such attachments of a longitudinally polarised gluon, equivalent to the contraction on the right-hand side of eq.~\eqref{eq:M3_l_p1}, as \textit{collinear insertions}. In the above relation, the function ${\widetilde{\mathcal{M}}^{(2)\,\mu,c}}$ represents the sum of all two-loop diagrams, including all QCD vertices at which an external longitudinally polarised gluon with momentum $l$ attaches, but excluding the diagram where $l$ directly attaches to the incoming quark (i.e. excluding wave-function corrections to the three-loop amplitude). Diagrams where $l$ attaches to the antiquark leg are also divergent in the single-collinear region $(2_l,H_k,H_q)$. Both the external gluon and fermion with momentum {${p_1+l}$} are generically off-shell, though they approach the mass-shell in the collinear limit $l^\mu \to -z_{1,l}\,p_1^\mu$ (c.f. eq.~\eqref{eq:loop_lightcone}).

\begin{figure}[h]
        \centering
        \begin{subfigure}[a]{0.3\textwidth}
        \centering
        \begin{align*}
        \eqs[1]{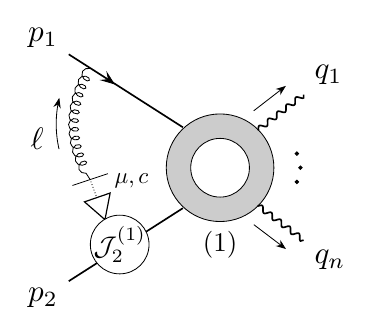} 
        \end{align*}
        \caption{}
        \label{fig:M3_l_p1_LP_1L_p2_fig1}
        \end{subfigure}
        \qquad
        \centering
        \centering
        \begin{subfigure}[a]{0.3\textwidth}
        \centering
        \begin{align*}
        \eqs[1]{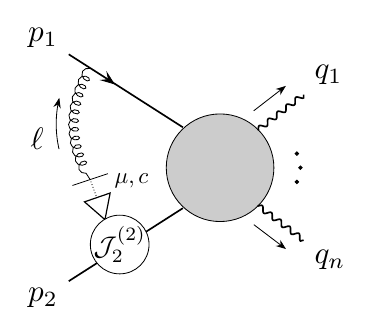} 
        \end{align*}
        \caption{}
        \label{fig:M3_l_p1_LP_1L_p2_fig2}
        \end{subfigure}
\caption{
Three-loop diagrams with a $l\,||\, p_1$ divergence containing one- and two-loop jet function subgraphs on the incoming antiquark with momentum $p_2$, denoted by $\mathcal{J}^{(1)}_2$ and $\mathcal{J}^{(2)}_2$, respectively. Counterterms for the one-loop jet function
are not shown.
}
\label{fig:M3_l_p1_jet_p2}
\end{figure}

We note that $\widetilde{\mathcal{M}}^{(2)\,\mu,c}$ in eq.~\eqref{eq:M3_l_p1} includes jet subgraph corrections to the incoming anti-quark, which require integrand modifications to treat loop polarisation terms that spoil local factorisation when $l$ becomes collinear to $p_2$. 
These diagrams are shown in figure~\ref{fig:M3_l_p1_jet_p2}. 
Local counterterms that restore the integrand-level Ward identities for diagrams with one-loop jet subgraphs have been developed in refs.~\cite{Anastasiou:2020sdt,Anastasiou:2022eym} and are summarised in appendix~\ref{app:jet}. In particular, they eliminate the shift-mismatch term generated by the one-loop jet function and the entire divergence in the region $(1_l,H_k,H_q)$ is due to integrable ghost contributions to the Ward identity, c.f. eq.~\eqref{eq:jet_mod_p1_ward}.

The two-loop $p_1$ jet function satisfies,
\begin{align}
    q_\mu\, \mathcal{J}_1^{(2)\,\mu,c}(p_1,q,k,l)u(p_1) = \tq{c}\,\left[
    \Pi^{(2)}_{qq}(p_1,k,l) - \Pi^{(2)\,\text{shift}}_{qq}(p_1,q,k,l)\right]u(p_1) + \text{ ghosts}\,,
    \label{eq:jet_p1_2L_ward}
\end{align}
with
\begin{align}
     \mathcal{J}_1^{(2)\,\mu,c}(p_1,q,k,l) = 
     \tq{c}\,\Pi^{(2)}_{qq}(p_1,k,l)\frac{\s{p}_1+\s{q}}{(p_1+q)^2}\gamma^\mu 
     + \Gamma_{qqg}^{(2)\,\mu,c}(p_1,q,k,l)\,.
\end{align}
where the two-loop quark propagator $\Pi^{(2)}_{qq}$ and $q\bar{q}g$-vertex $\Gamma_{qqg}^{(2)\,\mu,c}$ were defined in appendix~\ref{sec:Greens_2L}, while the shift subtraction term $\Pi^{(2)\,\text{shift}}_{qq}$ was defined in section~\ref{sec:ward_2L}. The $p_2$-jet function follows from complex conjugation. Here, $\Pi^{(2)}_{qq}(p_1,k,l)$ is a scaleless integral which vanishes in dimensional regularisation due to the cancellation of IR and UV poles\footnote{Throughout this text, we identify scaleless integrals that can be discarded in the \ac{ffs} scheme. As already noted in ref.~\cite{Anastasiou:2020sdt}, although this procedure mixes UV and IR poles, it does not present a problem, as the universal structure of UV and IR divergences in gauge theory amplitudes is well understood.}.
We remark that any modifications to treat loop polarisations of the two-loop quark jet function in the region 
$(1_q,H_k,H_l)$ 
will have to be designed to leave the Ward identity in region 
$(2_q,H_k,H_l)$ 
intact, though finding the appropriate counterterms is beyond the scope of this paper. Ideally, as at one-loop, the counterterms are chosen so that the first two terms on the right-hand side of eq.~\eqref{eq:jet_p1_2L_ward} vanish identically when $q$ becomes collinear to $p_2$. We note that the same potential modifications applied to the quark jet function at the amplitude level will have to be applied to the form factor counterterms as well.

We come to the main goal of this section, the treatment of shift-integrable terms due to scalar contributions to the single-collinear regions. We find it useful to decompose the truncated two-loop $n$-point amplitude into five topologies,
\begin{align}
\begin{split}
    &\eqs[0.22]{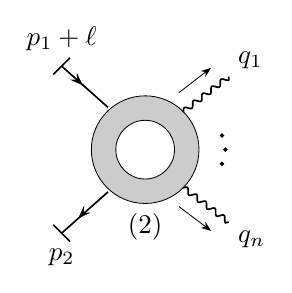} \equiv %
    \widetilde{\mathcal{M}}^{(2)}(p_1+l,p_2,k,q;
    \{q_{1},\ldots, q_{n}\}
    ) 
    \\&\qquad\qquad  = 
    \widetilde{\mathcal{M}}^{(2,NL)} +\widetilde{\mathcal{M}}^{(2,UL)}+\widetilde{\mathcal{M}}^{(2,XL)} +\widetilde{\mathcal{M}}^{(2,3V)}
    \,.
\end{split}
\end{align}
For simplicity, we have not shown the sum over permutations of the external photons and suppressed the arguments on the second line. We emphasise that $\widetilde{\mathcal{M}}^{(2)}$ contains self-energy corrections to the off-shell quark legs, though in this text we ignore two-loop jet subgraphs that lead to loop polarisation terms in the region under study. As in section~\ref{sec:Greens_2L} we use the superscripts ``$(NL)$", ``$(UL)$", ``$(XL)$" and ``$(3V)$" to denote diagrams without ladder structure, with two-loop uncrossed (planar) ladder structure, crossed (non-planar) ladder structure and triple-gluon vertices, respectively. 
It will prove equally convenient to decompose the three-loop shift counterterm by
\begin{align}
\begin{split}
    &\delta_{\text{shift}}^{(3)}(p_1,p_2,q,k,l,\{q_1,\ldots,q_n\})
    \\&\qquad=\sum_X\,\delta_{\text{shift}}^{(3,X)}(p_1,p_2,q,k,l,\{q_1,\ldots,q_n\})\,, \quad
    X\in \{\mathcal{J},NL,UL,XL,3V,O,\Delta\}
    \,.
\end{split}
    \label{eq:M3_shift_CNT}
\end{align}
The contribution $\delta_{\text{shift}}^{(3,\mathcal{J})}$ is due to three-loop diagrams containing one-loop (anti)quark jet functions. \edit{As we will see in section~\ref{sec:gluon_triangle}, the term $\delta_{\text{shift}}^{(3,\Delta)}$ removes shift mismatches originating from the gluon self-energy and triangle integrands.
The function $\delta_{\text{shift}}^{(3,O)}$, which we will define in section~\ref{sec:ghosts}, removes shifted terms due to ghost contributions in the single-collinear regions, which appear for the first time at this order.
The explicit diagrammatic representations for $\delta_{\text{shift}}^{(3,\Delta)}$ and $\delta_{\text{shift}}^{(3,O)}$ are given in eqs.~\eqref{eq:shift_DEL} and~\eqref{eq:shift_O}, respectively. We note that shift terms due to scalar contributions to the Ward identities can be cancelled by counterterms that consist of standard three-loop Feynman diagrams with a specific choice of loop momentum routing and multiplied by non-standard colour factors. As we will see in section~\ref{sec:ghosts} this is not the case for shifted ghost terms.}

The classification of the remaining terms is based on the topologies of the two-loop amplitude. For instance, $\delta_\text{shift}^{(3,UL)}$ is the local infrared counterterm added to the three-loop amplitude to cancel shift-integrable contributions derived from collinear insertions on two-loop graphs with uncrossed gluon ladder structure. 

Below, we will use the subscript $``1"$ to denote counterterms valid in the $l\,||\, p_1$ collinear limit. The complementary $l\,||\, p_2$ limit, denoted by a subscript $``2"$ follows straightforwardly from the analysis presented in this section\footnote{Throughout this paper, we will frequently switch between the single-collinear regions $l\,||\, p_1$ and $q\,||\, p_2$ in our discussions, the difference being in their approximations, eqs.~\eqref{eq:g_k_p1} and~\eqref{eq:g_k_p2}.}. Their sum covers both singular regions,
\begin{align}
    \delta_{\text{shift}}^{(3)} \equiv \delta_{\text{shift,1}}^{(3)} +\delta_{\text{shift,2}}^{(3)}\,,
    \label{eq:shift_CNT_sum}
\end{align}
where the decomposition in eq.~\eqref{eq:M3_shift_CNT} is implied for each function on the right-hand side. We will use this convention throughout the text.

Since there is no conceptual difference in the application of Ward identities for $2 \to n\geq 2$ electroweak amplitudes, we will often discuss the representative di-photon production case. 
Moreover, the calculations will be very similar, and will not derive each term on the right-hand side of eq.~\eqref{eq:M3_shift_CNT}, but sometimes simply state the result.

Finally, we note that the form factor counterterms will require the same shift modifications. Their counterterms follow straightforwardly from the $2\to 1$ amplitudes with a generic electroweak hard-scattering vertex, and are provided 
in appendix~\ref{sec:shift_FF}.

\subsection{One-loop shift-integrable integrands}
\label{sec:shift_jet}

We begin by considering the $l\,||\,p_1$ singular contributions to the three-loop amplitude component containing one-loop jet subgraph corrections to the quark leg with momentum $p_1$, where the outgoing loop momentum $l$ attaches to the one-loop hard sub-amplitude. 
In the limit where the external gluon $l$ becomes collinear to $p_1$ the jet function $\mathcal{J}_1^{(1)}$ exhibits unphysical loop polarisations that spoil standard Ward identities and require extra counterterms to restore local factorisation. 
Loop polarisation terms\footnote{From now on, the quark jet function modifications will be implied, and we will use $\mathcal{J}_i^{(1)}$ and $\Delta \mathcal{J}_i^{(1)}$ for $i=1,2$ interchangeably to refer to the regularised version.} can be removed using the prescription~\cite{Anastasiou:2020sdt,Anastasiou:2022eym} in eq.~\eqref{eq:jet_mod_p1}.

By applying the collinear approximation in eq.~\eqref{eq:g_k_p1} for the $l\,||\,p_1$ limit, we contract the integrand with a longitudinal polarisation $l_\mu$, which is represented graphically as follows,
\begin{align*}
    &\lim_{l\simeq -z_{1,l}p_1}\,
    \frac{\eta_{\mu\nu}}{l^2}\,\bar{v}(p_2)\widetilde{\mathcal{M}}^{(1)\,\mu,c}(p_1+l,p_2,q;
    \{q_1,\ldots,q_n\}
    )
    \frac{\s{p}_1+\s{l}}{(p_1+l)^2}\,\mathcal{J}_{1}^{(1)\,\nu,c}(p_1,l,k) u(p_1)
    \\&\quad=
    \frac{2\eta^\nu}{d_1(-l,\eta_1)}\,\bar{v}(p_2)(-l_\mu)\widetilde{\mathcal{M}}^{(1)\,\mu,c}(p_1+l,p_2,q;
    \{q_1,\ldots,q_n\}
    )\frac{\s{p}_1+\s{l}}{(p_1+l)^2}\,\mathcal{J}_{1}^{(1)\,\nu,c}(p_1,l,k)u(p_1)
    \\&\quad \equiv\eqs[0.25]{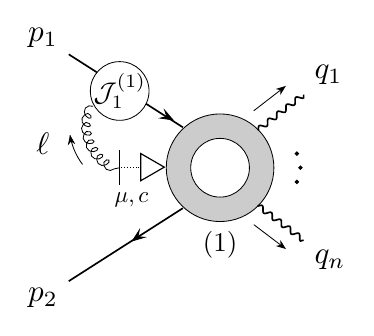}\,.
\numberthis
\label{eq:M3_jet_sum}
\end{align*}
Here, $\widetilde{\mathcal{M}}^{(1)\,\mu,c}$ denotes the truncated one-loop amplitude with an additional $q\bar{q}g$- or triple-gluon vertex. For the analysis that follows, it is convenient to express the limit as a sum of three terms through,
\begin{align}
    \widetilde{\mathcal{M}}^{(1)\,\mu,c} =\widetilde{\mathcal{M}}^{(1,A)\,\mu,c}
    +\widetilde{\mathcal{M}}^{(1,B)\,\mu,c}
    +\widetilde{\mathcal{M}}^{(1,C)\,\mu,c}\,,
    \label{eq:M3_jet_sum_decomp}
\end{align}
where we have suppressed the kinematic dependence in all functions for legibility.

\begin{figure}[!t]
        \centering
        \begin{subfigure}[a]{0.28\textwidth}
        \centering
        \begin{align*}
        \eqs[1]{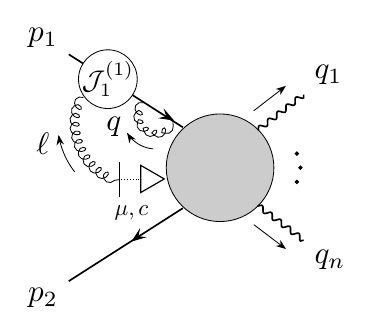} 
        \end{align*}
        \caption{}
        \label{fig:M3_jet_a}
        \end{subfigure}
        \begin{subfigure}[a]{0.28\textwidth}
        \centering
        \begin{align*}
        \eqs[1]{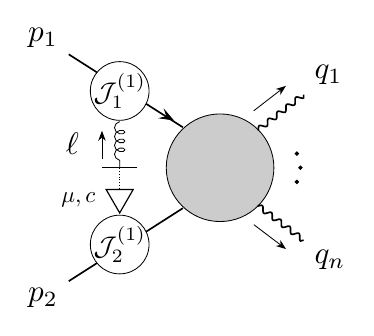}
        \end{align*}
        \caption{}
        \label{fig:M3_jet_b}
        \end{subfigure}
\caption{
One-loop jet function corrections $\mathcal{J}_1^{(1)}$ with a singularity in the $l\,||\, p_1$ limit that contain either a self-energy subgraph on the external quark leg, adjacent to $\mathcal{J}_1^{(1)}$, or an additional quark jet function $\mathcal{J}_2^{(1)}$ on the $p_2$ leg. Jet function counterterms are not shown.
}
\label{fig:M3_jet_special}
\end{figure}

This class of diagrams includes graphs with an additional self-energy subgraph on the quark leg, represented by collinear insertions on the one-loop component $\widetilde{\mathcal{M}}^{(1,A)}$, as well as graphs $\widetilde{\mathcal{M}}^{(1,B)}$ where the collinear gluon $l$ connects to a $p_2$ jet function $\mathcal{J}_2^{(1)}$. This is shown in figure~\ref{fig:M3_jet_special}.
Using both the $p_1$ and $p_2$ jet function prescriptions, eqs.~\eqref{eq:jet_mod_p1} and~\eqref{eq:jet_mod_p2}, we can cancel the $l\,||\, p_1$ singularity for both the first set of integrands, which factorises in terms of the hard Born sub-amplitude, and the second set of integrands which, up to integrable ghost contributions, vanishes in the strict collinear limit, eq.~\eqref{eq:jet_mod_p2}. We note that the sum of integrands represented by the graph in figure~\ref{fig:M3_jet_a} exhibit a linear divergence in the double-collinear region $(1_l,1_q,H_k)$ due to the repeated $p_1+l$ propagator. As at two-loop, we replace the quark self-energy, which carries loop momentum $q$ in the figure, with the symmetrised version shown in eq.~\eqref{eq:S_mod} to make this locally factorisable. 

\begin{figure}[!t]
        \centering
        \begin{align*}
        &\sum_\text{insertions}
        \eqs[0.22]{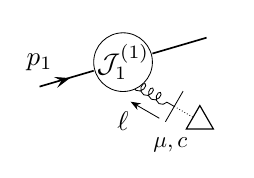} \otimes
        \eqs[0.23]{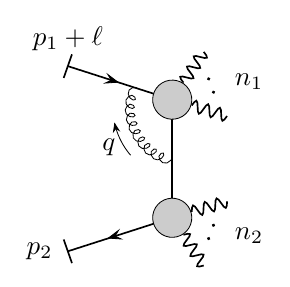}
        \end{align*}
        \caption{
        Example of a set of three-loop $2\to n$ integrands containing a (modified) one-loop quark jet function that are one-loop factorisable or shift-integrable in the limit $l\,||\,p_1$. Jet function counterterms are not shown.
        }
        \label{fig:M3_shift_jet}
\end{figure}

Next, we have collinear insertions of the $l$ loop, flowing out of the modified one-loop jet function $\Delta \mathcal{J}_1^{(1)}$, on one-loop sub-graphs $\widetilde{\mathcal{M}}^{(1,C)}$ which contain at least one $qqg$-vertex on the fermion line from which the final state photons emerge. An example for this set of integrands is shown in figure~\ref{fig:M3_shift_jet}.
In the figure, we use the symbol $``\otimes"$ to denote a matrix product in spinor space, while $n_i$, $i=1,2$ are fixed and denote the number of outgoing photons from each tree-level sub-amplitude, represented by a grey blob, with $0\leq n_i < n$ with $i=1,2$ and $n=n_1+n_2$.
In the region
$(1_l,H_k,H_q)$ each sum over insertions will factorise from a hard-scattering one-loop graph, up to shift-integrable terms that are removed by local counterterms that integrate to zero in the \acs{ffs} scheme. In this case, the shift-mismatch is in one loop-momentum variable only. We shall refer to such terms as \textit{one-loop shift-integrable}. Similarly, integrands that have collinear singularities locally factorised from the one-loop hard sub-amplitude will be called \textit{one-loop factorisable}.

The problem is equivalent to the shift mismatches encountered for the two-loop electroweak amplitude and investigated in ref.~\cite{Anastasiou:2022eym}, and the construction of the shift counterterm is straightforward,
\begin{align}
    \delta_{\text{shift},1}^{(3,\mathcal{J})}(\xi_1;p_1,p_2,q,k,l;
    \{q_1,\ldots,q_n\}
    ) = 
    \frac{C_A}{2C_F}\eqs[0.27]{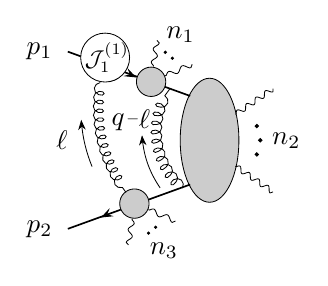} 
    - (q\to q+l)\,.
    \label{eq:shift_jet_CNT}
\end{align}
However, since it will prove indicative of more complicated integrands, we provide its derivation in appendix~\ref{app:shift_jet}. Again, $0\leq n_i \leq n$ with $i\in\{1,2,3\}$ and $n=\sum_{i=1}^3 n_i$ denote the number of outgoing photons from each Born sub-amplitude.
We restrict the values to $n_1+n_2\geq 1$ and $n_2+n_3\geq 1$ to avoid additional quark self-energy subgraphs on the $p_1$- and $p_2$-legs. We emphasise that eq.~\eqref{eq:shift_jet_CNT} is reminiscent of the shift-mismatches encountered for the two-loop electroweak amplitude, c.f. section~5 in ref.~\cite{Anastasiou:2022eym}, which are also one-loop shift-integrable. 

We remark on an important subtlety in the application of eq.~\eqref{eq:M3_shifted} in removing shift mismatches from the amplitude. Clearly, the shift counterterm $\delta_{\text{shift},1}^{(3,\mathcal{J})}$, which is designed for the $l\,||\, p_1$ limit, contains some graphs that become divergent when $l$ is collinear to the antiquark $p_2$, spoiling factorisation in that region. Inspired by a similar procedure introduced in ref.~\cite{Anastasiou:2024xvk} we can eliminate this overlap by replacing the pinched fermion propagator $1/(l-p_2)^2$ in those integrands as follows, 
\begin{align}
    iS_0(l-p_2)
    \to i\bar{S}_0(l-p_2;p_1,\xi_1)=\frac{i}{p_1\cdot p_2} \frac{(\s{l}-\s{p}_2)p_1\cdot \xi_1}{(l-\xi_1)^2-\xi_1^2}
    \equiv \eqs[0.16]{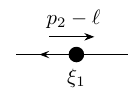}\,,
    \label{eq:shift_mod_quark}
\end{align}
where $\xi_1$ is an auxiliary vector satisfying $\xi_1\cdot p_1,\,\xi_1^2 \neq 0$. It is easy to see that the new fermion propagator reproduces the correct behaviour when $l\simeq -z_{1,l}p_1^\mu$, while guaranteeing finiteness in the $l\,||\, p_2$ limit. Thus, the shift counterterm $\delta_{\text{shift},1}^{(3,\mathcal{J})}$ defined in eq.~\eqref{eq:shift_jet_CNT} acquires an explicit $\xi_1$ dependence. \edit{Lastly, we find it convenient to introduce the graphical notation on the right-hand side, when we provide the shift counterterms for the form factors in appendix~\ref{sec:shift_FF}. In particular, it helps us distinguish it from a similar rule introduced later in this section.}

A similar replacement should be made for the counterterm $\delta_{\text{shift},2}^{(3,\mathcal{J})}$ designed to remove one-loop shift-integrable singularities in the $l\,||\,p_2$ limit,
\begin{align}
    iS_0(l+p_1) \to i\bar{S}_0(l+p_1;p_2,-\xi_2) = \frac{i}{p_1\cdot p_2}\frac{(\s{l}+\s{p}_1)p_2\cdot \xi_2}{(l+\xi_2)^2-\xi_2^2}\equiv \eqs[0.16]{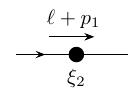}\,,
    \label{eq:shift_mod_antiquark}
\end{align}
where $\xi_2$ is an auxiliary vector chosen to have a large rapidity separation from $p_2$.

In eq.~\eqref{eq:shift_jet_CNT} and later on in the text, whenever a gluon line flows into a tree-level sub-amplitude it bisects\footnote{If the insertion is on the entire $n$-point sub-amplitude, we ignore the first term on the right-hand side of eq.~\eqref{eq:born_qqg}, since that would correspond to a wavefunction correction.} the set of outgoing photon momenta as follows,
\begin{align*}
    &\eqs[0.23]{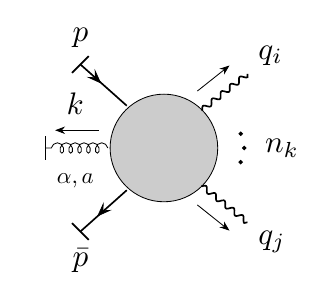}
    =
    \eqs[0.3]{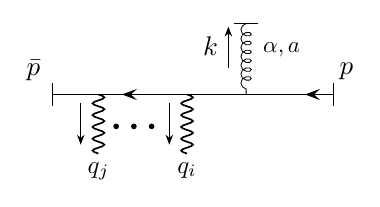}
    +\sum_{s=i}^{j}
    \eqs[0.3]{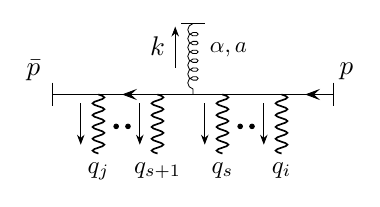}
    \\&\qquad 
    \numberthis
    \label{eq:born_qqg}
    = \tq{a}\Bigg[
    \widetilde{\mathcal{M}}^{(0)}(p,\bar{p},k;
    \{q_i,\ldots,q_j\})\frac{\s{p}-\s{k}}{(p-k)^2}\gamma^\alpha
    \\&\qquad
    +\sum_{s=i}^{j}\widetilde{\mathcal{M}}^{(0)\,\alpha}_s(p,\bar{p},k;
    \{q_i,\ldots,q_s\},\{q_{s+1},\ldots,q_j\})
    \Bigg]
    \,,
\end{align*}
with
\begin{align}   
\begin{split}
    &\widetilde{\mathcal{M}}^{(0)\,\alpha}_s(p,\bar{p},k;
    \{q_i,\ldots,q_s\},\{q_{s+1},\ldots,q_j\})
    \\&\qquad=-e^{j-i+1}\gs
    \frac{\s{\epsilon}_{j} (\s{p}-\s{k}-\sq{i}{j-1})
    \cdots
    \s{\epsilon}_{s+1}(\s{p}-\s{k}-\sq{i}{s})
    \gamma^\alpha
    (\s{p}-\sq{i}{s}) 
    \cdots
    \s{\epsilon}_{i}}{(p-k-\q{i}{j-1}))^2\cdots(p-k-\q{i}{s})^2(p-\q{i}{s})^2\cdots (p-q_{i})^2}\,,
\end{split}
\label{eq:M0_partition_def}
\end{align}
where $i<j \leq n$ and we have defined $Q_{m,n} \equiv \sum_{r=m}^n q_r$. Above, $\tq{a}\widetilde{\mathcal{M}}^{(0)\,\alpha}_s$ denotes the truncated tree-level integrand of order $\gs$ with outgoing photon momenta $\{q_i,\ldots, q_j\}$ and an additional $qqg$-vertex (with outgoing gluon momentum $k$, Lorentz index $\alpha$ and colour index $a$) between the adjacent electroweak vertices with momenta $q_s$ and $q_{s+1}$. Thus, we use an index $s$ to mark a partition of the outgoing momenta $\{q_i,\ldots q_j\}$ into two different sets, $\{q_{i},\ldots,q_{s}\}$ and $\{q_{s+1},\ldots,q_{j}\}$, where $s$ corresponds to the number of electroweak vertices between the incoming quark with momentum $p$ and the $qqg$-vertex with outgoing gluon momentum $l$. As usual, the sum over permutations of the external photon momenta is implicit. In the sum of integrands shown in eq.~\eqref{eq:born_qqg} the boundary term $n_k=0$ is equal to a single $qqg$-vertex with no photon emissions.

\subsection{Two-loop shift-integrable integrands}

\begin{figure}[!t]
        \centering
        \begin{align*}
        &\sum_\text{insertions}
        \eqs[0.14]{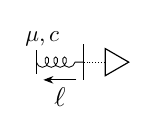}
        \otimes
        \eqs[0.2]{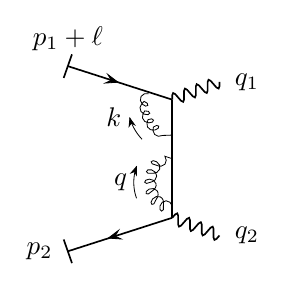}
        =
        -\eqs[0.2]{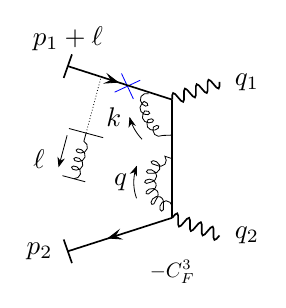}
        +\eqs[0.2]{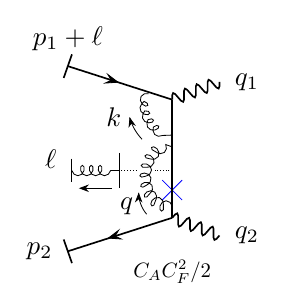}
        \\&\qquad
        -\eqs[0.2]{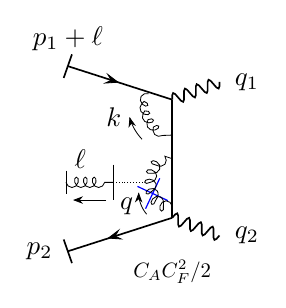}
        +\eqs[0.2]{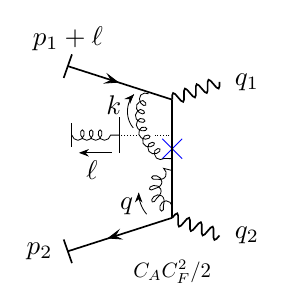}
        -\eqs[0.2]{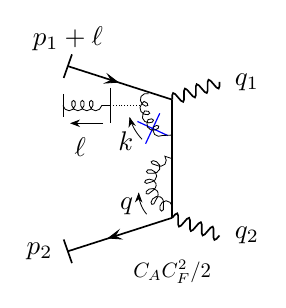}
        \end{align*}
        \caption{
        Collinear insertions on a representative set of integrands without ladder structure.
        The colour factors shown below each diagram are those of the remainder, after cancellations in the sum of integrands have been carried out. 
        This class of diagrams leads to four two-loop shift-integrable integrands.
        }
        \label{fig:M2_NL_shift_2_diag1_result}
\end{figure}

Next, we investigate collinear insertions of the longitudinal gluon $l$ on two-loop subgraphs $\widetilde{\mathcal{M}}^{(2,NL)\,\mu,c}$ without ladder structure that do not contain self-energy corrections to the external quark lines (which are part of the definition of the one- and two-loop quark jet functions with loop polarisation terms in the $l\,||\,p_1$ and $l\,||\,p_2$ limits).
The result for a representative set of integrands in $2\to 2$ electroweak production is shown in figure~\ref{fig:M2_NL_shift_2_diag1_result}. Repeated application of the QCD Ward identities yields an integrand in which the $l\,||\, p_1$ collinear divergence is factorised from a \textit{two-loop} hard subgraph, equivalent to the first diagram in figure~\ref{fig:M2_NL_shift_2_diag1_result}.
Additionally, we obtain four terms that have a shift mismatch in two loop momenta. These integrands are not factorised for fixed loop momentum $q$ and $k$ but cancel after integration, or locally after appropriate shifts in the loop momenta. We call these contributions \textit{two-loop shift integrable}.

To remove this problematic region we add a local infrared counterterm to the three-loop amplitude, designed to cancel these shift-integrable terms in the $l\,||\, p_1$ limit, of the form,
\begin{align*}
        &\delta_{\text{shift},1}^{(3,NL)}(\xi_1;p_1,p_2,q,k,l;\{q_1,\ldots,q_n\}) = 
        \frac{C_A}{2C_F}\Bigg[\eqs[0.23]{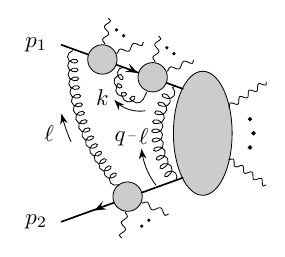} - (q\to q+l)\Bigg]
        \\&\quad
        +\Bigg[\frac{C_A}{2C_F}\Bigg(\eqs[0.23]{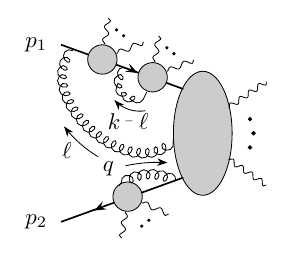}
        +\eqs[0.23]{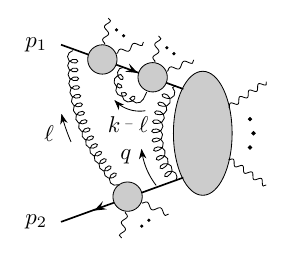}\Bigg)
        -C_A^2
        \eqs[0.23]{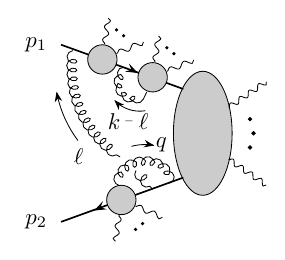} 
        \\&\quad
        - (k\to k+l)\Bigg]\,.
    \numberthis
    \label{eq:shift_NL}
\end{align*}
The first two lines are reminiscent of the shift subtraction shown earlier, where each integrand is multiplied by the non-standard colour factor $C_A/2C_F$. The last term, which is proportional to the colour factor $\sum_{a,b,c}\tq{a}\tq{c}\tq{a}\tq{b}\tq{b}\tq{c} = -C_F^2(C_A-2C_F)/2$ and is multiplied by $C_A^2$, is new. As we will see, such a term is typical of two-loop shift-integrable contributions. It cancels non-factorised remainders that occur, for example, after applying the QCD Ward identities to the first and second diagrams on the second line of eq.~\eqref{eq:shift_NL}. Here, and later in the text, we make use of the combination
\begin{align}
    \frac{1}{N_c} = C_A - 2C_F\,,
    \label{eq:subleading_colour}
\end{align}
which is easily verified by plugging in eq.~\eqref{eq:casimirs}. It signals contributions which are subleading in $N_c$. To clarify the graphical notation used in this section, we provide the explicit functional form of the first term in eq.~\eqref{eq:shift_NL} 
in appendix~\ref{app:shift}.

As a reminder, we exclude graphs in \eqref{eq:shift_NL} that contribute to the quark jet function and exhibit unphysical loop polarisations in the limit $l\,||\,p_1$, while the prescription of eq.~\eqref{eq:shift_mod_quark} is implied in all graphs containing a $1/(l-p_2)^2$ propagator.
\edit{Here, we encounter the additional problem that the function $\delta^{(3,NL)}_{\text{shift,1}}$ has a residual $q\,||\,p_2$ singularity in all graphs with a $k$-shift and where $q$ is adjacent to $p_2$. This singularity integrates to zero but is locally non-factorised, and therefore a hindrance to local factorisation. In order to overcome this issue, we replace the pinched propagator by,}
\begin{align}
    iS_0(q-p_2;l)\to i\bar{S}_0(q-p_2;l) = \frac{i(\s{q}-\s{p}_2)}{(p_2-q)^2-l^2} 
    \equiv \eqs[0.16]{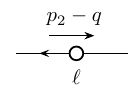}
    \,.
    \label{eq:ggg_mod_quark_def}
\end{align}
The modified fermion propagator $\bar{S}_0(q-p_2;l)$ 
matches the original in the strict $l\,||\,p_1$ collinear limit, but suppresses the counterterm in the region where $q$ becomes collinear to $p_2$ as $l^2$ acts as an infrared regulator. At the same time, and importantly, the mixed-collinear region $(1_l,2_q,H_k)$ remains intact.
From here on, both the regularisation in eq.~\eqref{eq:shift_mod_quark} and eq.~\eqref{eq:ggg_mod_quark_def} shall be implied for contributions to the shift counterterm $\delta^{(3)}_{\text{shift,1}}$.
Similarly, we use
\begin{align}
    iS_0(l+p_1;q)\to i\bar{S}_0(l+p_1;q) = \frac{i(\s{l}+\s{p}_1)}{(l+p_1)^2-q^2} 
    \equiv \eqs[0.16]{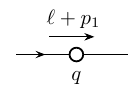}
    \,.
    \label{eq:ggg_mod_antiquark_def}
\end{align}
for the $l\,||\,p_2$ shift counterterms.

Next, we investigate shift-mismatch terms generated by $l\,||\, p_1$ collinear insertions on planar two-loop sub-graphs, $\widetilde{\mathcal{M}}^{(2,UL)}$. Again, in this text we ignore graphs that contain two-loop jet subgraphs on the antiquark leg, which require treatment of loop polarisation terms. For this class of integrands, we consider the representative two-loop diagram with two off-shell photons where the gluon lines with momentum $k$ and $q$ are adjacent to both the incoming quark and antiquark,
\begin{align*}
    &\sum_\text{insertions}
    \eqs[0.12]{l_p1}\otimes
    \eqs[0.2]{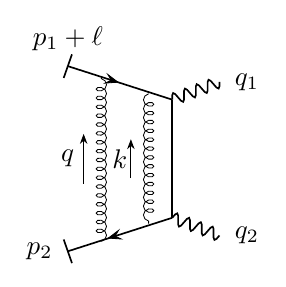}
    =
    -\eqs[0.2]{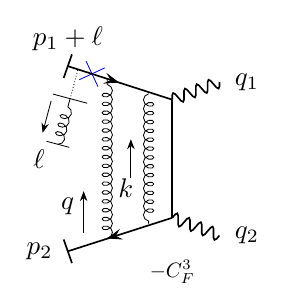}
    +\eqs[0.2]{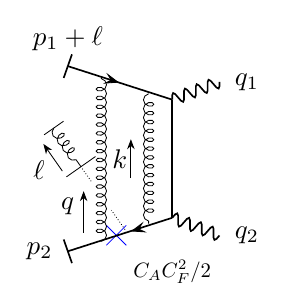}
    \\&\quad
    -\eqs[0.2]{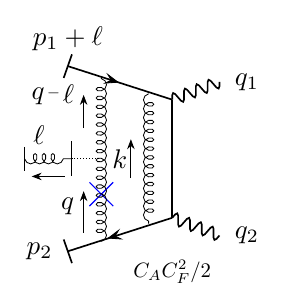}
    +\eqs[0.2]{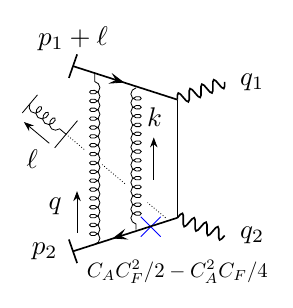}
    -\eqs[0.2]{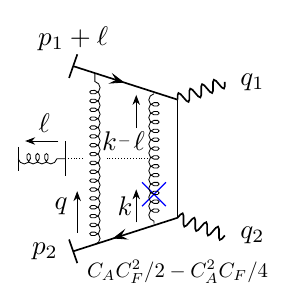}\,.
    \label{eq:M2_UL_shift_diag1}
    \numberthis
\end{align*}
The right-hand side shows the result of the sum of integrands after Ward identities have been applied to each collinear insertion and cancellations have been carried out. Below each diagram we show the colour factor that multiplies the corresponding integrand, which is not necessarily equivalent to the colour factor of the Feynman diagram itself\footnote{For instance, the second diagram on the right-hand side of eq.~\eqref{eq:M2_UL_shift_diag1} has colour factor $\sum_{a,b,c}\tq{a}\tq{b}\tq{c}\tq{c}\tq{a}\tq{b} = C_F^3-C_AC_F^2/2$, but the term proportional to $C_F^3$ is cancelled by the diagram where the gluon with longitudinal polarisation and momentum $l$ directly attaches to the external antiquark leg.}.

Again, we have applied eq.~\eqref{eq:g_k_p1} and used the Dirac equation on the quark spinor $u(p_1)$.
The first term on the right-hand side of eq.~\eqref{eq:M2_UL_shift_diag1}, proportional to the abelian colour factor $C_F^3$, contributes to the singularity, but is factorised at the integrand-level from the hard two-loop subgraph. \edit{Clearly, we can associate the colour factor $C_F^2$ with the hard-scattering process.}
In analogy to the sub-graphs without ladder structure, we obtain two sets of shift mismatch terms, one for each loop momentum $k$ and $q$ of the two-loop hard subgraph. The shift mismatch in the ``outer" loop, with loop momentum $q$ or $q-l$, is proportional to the colour factor $C_AC_F^2/2$, while that of the ``inner" loop, with loop momentum $k$ or $k-l$, is proportional to 
$C_FC_A(C_F-C_A/2)/2$.
The strategy is to first construct a shift counterterm for the set of non-factorised integrands on the right-hand side of eq.~\eqref{eq:M2_UL_shift_diag1} and then generalise the expression to the full electroweak amplitude. The result is,
\begin{align*}
        \numberthis  \label{eq:shift_UL}
        &\delta_{\text{shift},1}^{(3,UL)}(\xi_1;p_1,p_2,q,k,l;\{q_1,\ldots,q_n\}) = 
        \frac{C_A}{2C_F}\Bigg(\eqs[0.23]{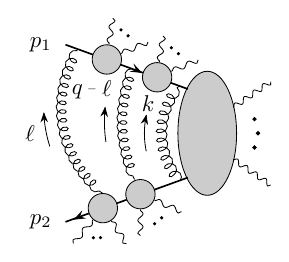} - (q\to q+l)\Bigg)
        \\&\quad
        +\Bigg[\Bigg(\frac{C_A}{2C_F}\eqs[0.23]{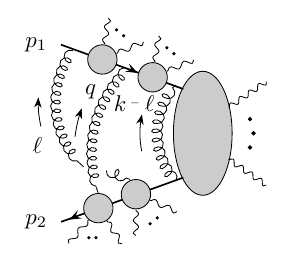}
        +\frac{C_A}{2C_F}\left(1-\frac{C_A}{2C_F}\right)
        \eqs[0.23]{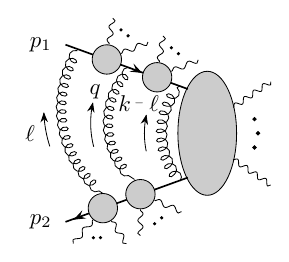} \Bigg)
        - (k\to k+l)\Bigg]\,.
\end{align*}
Here, the last diagram, multiplied by the non-standard colour factor 
$\frac{C_A}{2C_F}(1-\frac{C_A}{2C_F})$,
cancels additional non-factorised shift-integrable contributions in the $l\,||\, p_1$ limit, coming from the first diagram in square brackets. 

The shift counterterm with two-loop crossed ladder subgraphs is again two-loop shift integrable. We will not provide its derivation here, but simply state the result,
\begin{align}
    \begin{split}
        &\delta_{\text{shift},1}^{(3,XL)}(\xi_1;p_1,p_2,q,k,l;\{q_1,\ldots,q_n\}) = 
        \frac{C_A}{2C_F}\Bigg(\eqs[0.23]{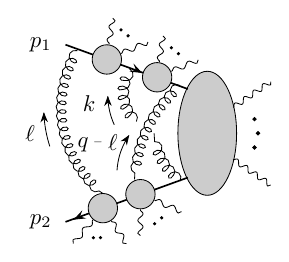} - (q\to q+l)\Bigg)
        \\&\quad
        +\Bigg[\Bigg(\frac{C_A}{2C_F}\eqs[0.23]{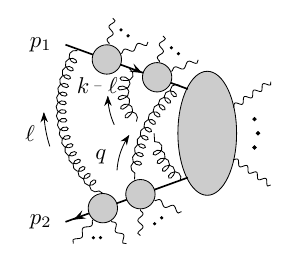}
        -C_A^2
        \eqs[0.23]{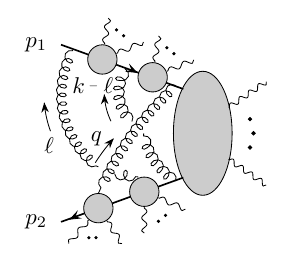} \Bigg)
        - (k\to k+l)\Bigg]\,.
    \end{split}
    \label{eq:shift_XL}
\end{align}
The sum of integrands represented by the last diagram, multiplied by $C_A^2$ in close analogy to the non-ladder type graphs in eq.~\eqref{eq:shift_NL}, is as usual required to remove non-cancelled $k$-loop shift mismatch terms.

Derivation of the infrared counterterm $\delta_{\text{shift},1}^{(3,3V)}$, which is the last remaining term in eq.~\eqref{eq:M3_shift_CNT} apart from the ghost contributions, is more intricate since it involves $l\,||\, p_1$ divergent three-loop graphs with either two triple-gluon vertices or a quartic gluon vertex. Consider the following sum of collinear insertions,
\begin{align*}
    &\sum_\text{insertions}
    \eqs[0.12]{l_p1}\otimes
    \eqs[0.2]{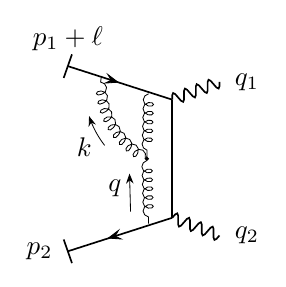}
    \to
    -\eqs[0.2]{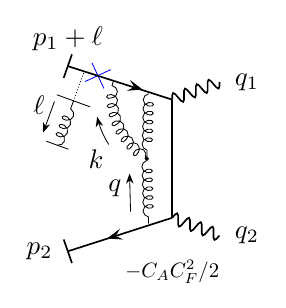}
    \\&\quad
    +\eqs[0.2]{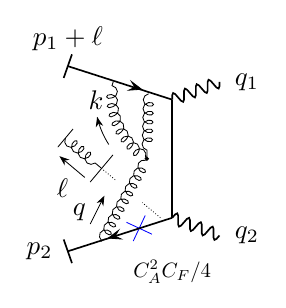}
    -\eqs[0.2]{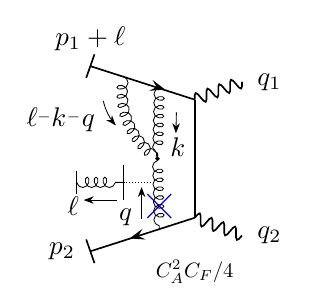}\,.
    \label{eq:M2_3V_shift_diag1}
    \numberthis
\end{align*}
The first term on the right-hand side is manifestly factorised from the two-loop hard subgraph, this time proportional to the colour factor $C_AC_F^2/2$. Clearly, the other two integrands, shown on the second line, cancel only up to a \textit{simultaneous} shift in both loop momenta, $q\to q+l$ and $k\to k-q$.

To obtain the right-hand side of eq.~\eqref{eq:M2_3V_shift_diag1}, we have used, in addition to the by now familiar Ward identities for quark and gluon lines, the quartic-gluon identity shown graphically in figure~\ref{fig:4g_ward}, which for the example above yields,
\begin{align}
\begin{split}
    &\eqs[0.2]{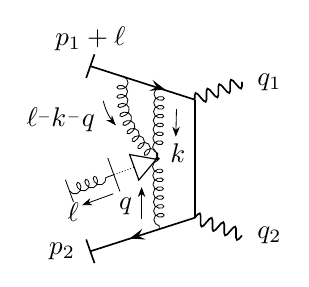}
    =
    \eqs[0.2]{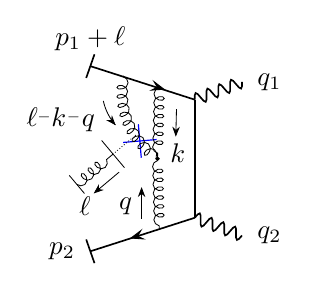}
    +\eqs[0.2]{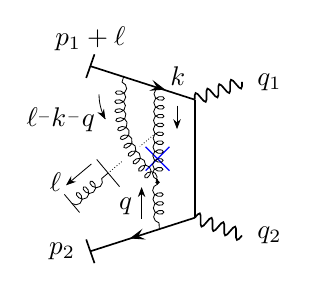}
    -\eqs[0.2]{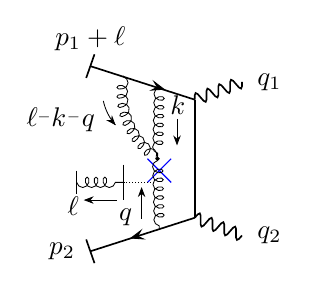}\,.
\end{split}
\end{align}
As usual, we follow the momentum flow convention of section~\ref{sec:routing}. The first and third integrands on the right-hand side are proportional to the colour factor $C_A^2C_F/4$ in the sum over attachments in eq.~\eqref{eq:M2_3V_shift_diag1}. The second diagram has vanishing colour factor, which can be proven using,
\begin{align}
    f^{abc} = -2i\,\Tr{\big[\big[\tq{a},\tq{b}\big],\tq{c}\big]}\,,
\end{align}
and repeated application of the well known Fierz identity,
\begin{align}
    \tq{a}_{ij}\tq{a}_{kl} = T_R\left(\delta_{il}\delta_{kj}-\frac{1}{N_c}\delta_{ij}\delta_{kl}\right)\,.
\end{align}

Using the information above, we can construct a shift counterterm for this class of diagrams as follows,
\begin{align}
    \begin{split}
        &\delta_{\text{shift},1}^{(3,3V)}(\xi_1;p_1,p_2,q,k,l;\{q_1,\ldots,q_n\}) = 
        \frac{C_A}{2C_F}
        \eqs[0.27]{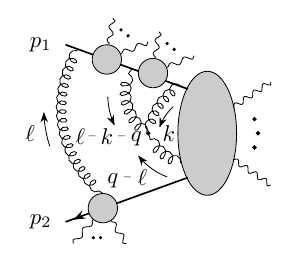}
        \\&\qquad- (q\to q+l\,,\,k \to k-q )\,.
    \end{split}
    \label{eq:shift_3V}
\end{align}
We note the unusual momentum routing of the triple-gluon vertex in the hard two-loop subgraph. Due to the ambiguous choice of loop momentum flow the shift term in the $q\,||\, p_2$ limit will look slightly different. However, this ambiguity in the double-collinear region poses no problem otherwise, since the term is cancelled exactly by the shift counterterm in the single-collinear limit already.

To summarise, we have derived local infrared counterterms that remove non-factorising shift mismatches generated by ``scalar" contributions in the single-collinear regions. Counterterms are supplemented by a shift in the fermion propagator according to the prescription in eq.~\eqref{eq:shift_mod_quark} to avoid double-counting of divergent integrands that would spoil local factorisation. In the double- and mixed-collinear regions limits are applied consecutively and the problem reduces either to the factorisable one-loop case for integrands containing one-loop jet functions, c.f. sec.~\ref{sec:shift_jet}, or the two-loop case, which yield one-loop shift-integrable contributions at worst. In the limit where three loop momenta become collinear the shift integrands factorise and pose no problem to integrability. 

For instance, in the region $(1_l,2_q,H_k)$ we apply the $q\,||\, p_2$ limit
to the two-loop $l\,||\, p_1$-finite remainder $\widetilde{\mathcal{M}}^{(2)}(p_1+l,p_2,q,k;\{q_1,\ldots,q_n\})$, c.f. eq.~\eqref{eq:M3_fact}, as well as the three-loop shift contribution $\mathcal{M}^{(3)}_\text{shift}$.
Shift mismatch terms of the $l\,||\,p_1$ limit yield non-factorisable, one-loop shift-integrable contributions in the $q\,||\,p_2$ limit, and the shift counterterm $\delta_{\text{shift},1}^{(3)}$ already guarantees integrability. We note for $q\,||\,p_2$ collinear insertions on the quark line with momentum $p_1+l$ we apply the Ward identity for on-shell quark lines, c.f. eq.~\eqref{eq:qq_Ward_spinor}, as follows,
\begin{align*}
    \eqs[0.2]{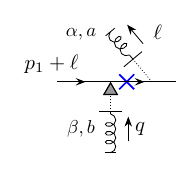} &= i\tq{a}\tq{b}\,
    S_0(p_1+q)
    \,\s{q}\,
    S_0(p_1+l)
    = i\tq{a}\tq{b}\bigg[S_0(p_1+l)-S_0(p_1+l) \frac{\s{p}_1\s{l}}{(p_1+l)^2}\bigg]
    \\&\equiv \eqs[0.2]{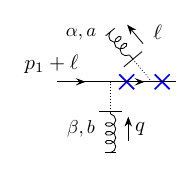}\,,
    \numberthis
\end{align*}
where the second term in square brackets vanishes in the $l\,||\,p_1$ limit as $\s{p}_1\s{l} \simeq -z_{1,l}\s{p}_1^2=0$. The expression after the second equality is obtained through a simple partial fractioning identity. Thus, the residual shift mismatch in the region $(1_l,2_q,H_k)$ can be removed locally using the standard two-loop shift counterterm ${\delta}_{\text{shift},2}$ derived in ref.~\cite{Anastasiou:2022eym}, up to a collinear factor,
\begin{align}
\begin{split}
    &\delta_{\text{shift},1,2}^{(3)}(p_1,p_2,q,k,l;\{q_1,\ldots,q_n\}) \\&\qquad = \mathcal{C}(z_{1,l},p_1,l)
    \bar{v}(p_2)\widetilde{\delta}_{\text{shift},2}^{(2)}(p_1+l,p_2,q,k;\{q_1,\ldots,q_n\})u(p_1)\,,
\end{split}
\end{align}
with $\mathcal{C}(z_{1,l},p_1,l)$ defined in eq.~\eqref{eq:collinear_factor} and
\begin{align}
\begin{split}
    &\widetilde{\delta}_{\text{shift},2}^{(2)}(p_1+l,p_2,q,k;\{q_1,\ldots,q_n\}) 
    = 
    \frac{C_A}{2C_F}\eqs[0.27]{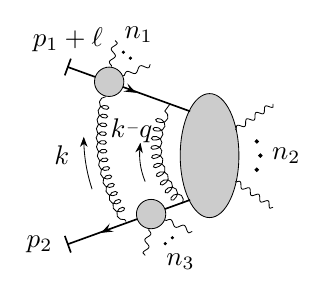} 
    - (k\to k+q)\,.
\end{split}
\end{align}
Above, the subscript $``1,2"$ denotes the order in which collinear limits are applied in the region $(1_l,2_q,H_k)$.

%% file: gluon_self_energy.tex
\section{Gluon self-energy corrections}
\label{sec:gluon_self_energy}

In this section, we discuss the treatment of subgraphs containing one and two-loop corrections to the gluon-self energy. The one-loop gluon propagator in Feynman gauge is given by,
\begin{align}
\begin{split}
    &\frac{-i}{l^2}\,
    \Pi_{gg}^{(1)\,\mu\nu,ab}(k,l)
    \equiv
    \eqs[0.26]{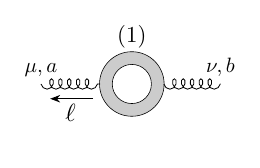} 
    \\&\qquad =
    \frac{-i}{l^2}\left[\Pi^{(1)\,\mu\nu,ab}_q + \Pi^{(1)\,\mu\nu,ab}_g +\Pi^{(1)\,\mu\nu,ab}_{gh}
    \right]
\end{split}
\label{eq:gg_1L}
\end{align}
where $\Pi^{(1)}_q$, $\Pi^{(1)}_g$ and $\Pi^{(1)}_{gh}$ denote the quark, gluon and ghost loop contributions, respectively. The gluon tadpole is a scaleless integral and therefore vanishes in dimensional regularisation, so is immediately ignored. We note that we have absorbed one of the gluon propagators $-i/l^2$ into the definition of $\Pi_{gg}$, which will be a useful convention later. The two-point function is proportional to unity in colour space,
\begin{align}
\Pi_{gg}^{(1)\,\mu\nu,ab}=\delta^{ab}\,\Pi_{gg}^{(1)\,\mu\nu}\,.
\label{eq:colour_sum}
\end{align}
The integrands are given by, 
\begin{align}
\begin{split}
    &\frac{-i}{l^2}\,
    \Pi^{(1)\,\mu\nu,ab}_q(k,l)\equiv
    \eqs[0.25]{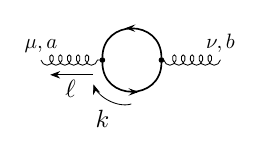} 
    \\&\qquad
    = 
    -\gs^2\, %
    n_f T_F\,
    \delta^{ab} \,
    \frac{\Tr{\left[ \slashed{k}  \, \gamma^\mu \left(\s{l} -\s{k}
    \right)\gamma^\nu  \right] }}{(l^2)^2k^2(l-k)^2}\,,
\end{split}
\label{eq:gg_floop}
\end{align}
\begin{align}
\begin{split}
    &\frac{-i}{l^2}\,
    \Pi^{(1)\,\mu\nu,ab}_g(k,l)\equiv
    \eqs[0.25]{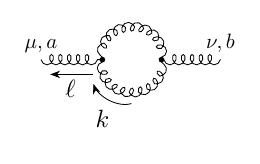} 
    \\&\qquad 
    = -\gs^2\,\frac{C_A}{2}\,\delta^{ab}\,
    \frac{C^{\mu\alpha\beta}(l,k-l,-k)C^{\nu}{}_{\beta\alpha}(-l,k,l-k)}{(l^2)^2k^2(l-k)^2}\,,
\end{split}
\label{eq:gg_gloop}
\end{align}
\begin{align}
\begin{split}
    &\frac{-i}{l^2}\,
    \Pi^{(1)\,\mu\nu,ab}_{gh}(k,l)\equiv
    \eqs[0.25]{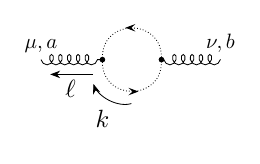} 
    = -\gs^2\,
    C_A\,
    \delta^{ab}\,
    \frac{k^\mu(l-k)^\nu 
    }{(l^2)^2k^2(l-k)^2}\,.
\end{split}
\label{eq:gg_ghloop}
\end{align}
The gluon loop contains a symmetry factor of $1/2$. Later, when we investigate one-loop gluon triangle subgraphs, it will be useful to consider an equivalent version of the ghost loop contribution symmetrised under the exchange $k\to l-k$, which we denote by $\Pi^{(1)}_{2,\,sym}$. This can be implemented through a local counterterm $\delta_{gh}$ as follows,
\begin{align}
\begin{split}
    &\frac{-i}{l^2}\,
    \Pi^{(1)\,\mu\nu,ab}_{2,\,sym}(k,l)\equiv
    \frac{-i}{l^2}\left[
    \Pi^{(1)\,\mu\nu,ab}_{2}(k,l)
    +\delta_{gh}^{\mu\nu,ab}(k,l)
    \right]
    \\&\qquad
    =\frac{1}{2}\Bigg[
    \eqs[0.25]{gg_1L_ghloop_diag1} 
    +\eqs[0.25]{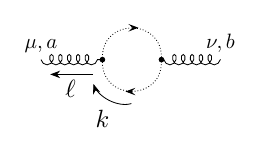}\Bigg] 
    \\&\qquad
    = -\gs^2\,
    \frac{C_A}{2}\,
    \delta^{ab}\,
    \frac{k^\mu(l-k)^\nu 
    +(l-k)^\mu k^\nu
    }{(l^2)^2k^2(l-k)^2}\,.
\end{split}
\label{eq:gg_ghloop_sym}
\end{align}
with
\begin{align}
    \delta_{gh}^{\mu\nu,ab}(k,l)=
    i\gs^2\,\frac{C_A}{2}\,\delta^{ab}\,
    \frac{k^\mu l^\nu -k^\nu l^\mu }{l^2k^2(l-k)^2}\,, \qquad \int_k\,\delta_{gh}(k,l)=0\,.
    \label{eq:gg_ghloop_sym_CNT}
\end{align}
Note that $\delta_{gh}^{\mu\nu,ab}$, unlike the symmetrised ghost loop, is not symmetric under the exchange of its Lorentz indices. Equation~\eqref{eq:gg_ghloop_sym} defines the symmetrised one-loop gluon propagator,
\begin{align}
\begin{split}
    &
    \Pi_{gg,\,sym}^{(1)\,\mu\nu,ab}=
   \Pi^{(1)\,\mu\nu,ab}_q + \Pi^{(1)\,\mu\nu,ab}_g +\Pi^{(1)\,\mu\nu,ab}_{gh,\,sym}\,.
\end{split}
\label{eq:gg_1L_sym}
\end{align}

It is easy to see that the doubled propagator in eq.~\eqref{eq:gg_1L} leads to enhanced soft and collinear singularities. Following the discussion in section~\ref{sec:ward}, in the soft limit $l^\mu\to \delta l^\mu$, $\delta \to 0$, the integrand of one-loop gluon propagator scales as
\begin{align}
    \frac{-i}{l^2}\,\Pi_{gg}^{(1)\,\mu\nu,ab}(k,l)
    \sim \delta^{-4}\,,
\end{align}
for generic non-lightlike values of the loop momentum $k$. The power singularity exhibited in the soft region would suggest we can no longer rely on the leading order approximation for the construction of the counterterm, eqs.~\eqref{eq:g_k_p1} and~\eqref{eq:g_k_p2}. Similarly, naive power counting in the collinear region $l\,||\,p_1$ leads to an apparent linear divergence.
On the collinear pinch surface the polarisation of the virtual gluon with momentum $l^\mu$ may be proportional to $k^\mu$, which is arbitrary and spoils factorisation in the collinear limit. 

In ref.~\cite{Anastasiou:2020sdt} we argued these issues can easily be resolved through a simple Passarino-Veltman tensor reduction, which removes one power of $l^2$ in the denominator of the photon propagator. Applied to the gluon self-energy, tensor reduction yields,
\begin{align}
\begin{split}
    \frac{-i}{l^2}\,
    \Pi_{gg}^{(1)\,\mu\nu,ab}(k,l)
    &= \gs^2 \delta^{ab}\frac{1}{l^2}\left(g^{\mu\nu}-\frac{l^\mu l^\nu}{l^2}\right) 
    \Gamma_{2}^\epsilon
    \int_{k}\frac{1}{k^2(l-k)^2}\,,
\end{split}
\label{eq:gg_1L_TR}
\end{align}
where we define the $\e$-dependent coefficient,
\begin{align}
\Gamma_{2}^\epsilon =
    n_f \Gamma_{n_f}^\e + C_A \Gamma_A^\e\,,
\end{align}
with
\begin{align}
    \Gamma_{n_f}^\e=\frac{2(1-\e)}{3-2\e}\,, \quad
    \Gamma_A^\e = - \frac{5-3\e}{3-2\e}\,.
\end{align}
As we have argued in ref.~\cite{Anastasiou:2020sdt}, the longitudinal components in eq.~\eqref{eq:gg_1L_TR} lead to a scaleless integral in $l$ in the sum over diagrams of the two-loop amplitude. This can be shown through repeated application of the abelian Ward identity, eq.~\eqref{eq:qq_Ward}.
Thus, the problem reduces to the one-loop subtraction, and we make the replacement,
\begin{align}
\begin{split}
    &\mathcal{M}^{(2)}_{2}(k,l)%
    \to
    \mathcal{M}^{\prime\, (2)}_{2}(k,l) =
    i \gs^2\,
    \frac{\Gamma_{2}^\epsilon}{k^2(l-k)^2}\,
    \mathcal{M}^{(1)}(l)\,.
\end{split}
\label{eq:M2_gg_TR}
\end{align}
The primed notation $ \mathcal{M}^{\prime\, (2)}_{2}$ indicates that we have performed a tensor reduction, i.e. the integrated result is the same. 
Thus, a suitably modified integrand $\mathcal{M}^{\prime\, (2)}_{2}$ can be obtained directly from the corresponding one-loop amplitude $\mathcal{M}^{(1)}(l)$ that is free of the problematic behavior in the infrared regions discussed above.
The one-loop amplitude $\mathcal{M}^{(1)}$ is rendered finite in the infrared and ultraviolet according to the prescription discussed in refs.~\cite{Anastasiou:2020sdt,Anastasiou:2022eym}.
Regularisation of $\mathcal{M}^{\prime\, (2)}_{2}$ in the \ac{uv} regions is then straightforward, leading to the following expression,
\begin{align}
    \begin{split}
    &\mathcal{M}^{\prime\, (2)}_{2}(k,l) =
    i \gs^2\,
    \Gamma_{2}^\epsilon
    \left[ 
    \left(\frac{1}{k^2(l-k)^2} -
    \frac{1}{\prop{k}^2}\right)
    +\frac{1}{\prop{k}^2}
    \right]
    \\&\qquad \times
    \left(\mathcal{H}^{(1)}(l) + \mathcal{M}^{(1)}_\text{singular}(l)\right)\,.
    \end{split}
\label{M2_gg_TR_reg}
\end{align}
Above, the term $1/\prop{k}^2$ in round brackets removes the residual UV divergence in the region $k\to \infty$. We note that while the denominator $1/(l-k)^2$ in fact suppresses any UV divergence in the limit $l \to \infty$ it is convenient to make the replacement $\mathcal{M}^{(1)} = \mathcal{H}^{(1)}(l) + \mathcal{M}^{(1)}_\text{singular}$. 

Finally, we write
\begin{align}
    \mathcal{M}^{\prime\, (2)}_{2}(k,l) =
    \mathcal{H}^{\prime\, (2)}_{2}(k,l)
    +\mathcal{M}^{\prime\, (2)}_{2\,\text{singular}}(k,l)\,,
    \label{eq:nf_reg}
\end{align}
where~\cite{Anastasiou:2020sdt},
\begin{align}
    \mathcal H^{\prime\,(2)}_{2}(k,l) = i \gs^2\,
    \Gamma_{2}^\epsilon
    \left[
    \frac{1}{k^2 (l-k)^2} 
    - \frac{1}{\prop{k}^2}\right] 
    {\mathcal H}^{(1)}(l)  
    \,,
    \label{eq:M2_ferm_finite}
\end{align}
and
\begin{align}
    \mathcal M^{\prime\,(2)}_{2\,\text{singular}}(k,l) = i \gs^2\,
    \Gamma_{2}^\epsilon
    \left[
    \frac{1}{k^2 (l-k)^2} {\overline{\mathcal M}}^{(1)}_\text{singular}(l) 
    + \frac{1}{\prop{k}^2}
    {\mathcal H}^{(1)}(l)  
    \right]\,.
    \label{eq:M2_ferm_sing}
\end{align}
The one-loop integrands ${\mathcal M}^{(1)}_\text{singular}$ and ${\mathcal H}^{(1)}$ are defined according to eqs.~\eqref{eq:Mfin_Msing} and~\eqref{eq:Mfin-iterative} (c.f. section~4.1 of ref.~\cite{Anastasiou:2020sdt}). \edit{The $n_f$-contribution ${M}^{\prime\, (2)}_{2}$, as defined by the prescription in eq.~\eqref{eq:nf_reg}, was calculated recently using numerical techniques in ref.~\cite{Kermanschah:2024utt}, for the \ac{nnlo} virtual corrections to the 
production of three equal-mass off-shell photons and the $d\bar{d}$-initiated production of a $Z$-boson and two different-mass off-shell photons.
}

The three-loop QCD amplitude will contain two types of gluon self-energy contributions, which we can construct out of the one- and two-loop amplitudes by considering all possible ``insertions" of the self-energy subgraph on the virtual gluon lines.
As for the two-loop electroweak amplitude, we need to perform a tensor reduction of each gluon self-energy subgraph to alleviate the power divergences in the soft and collinear regions. We denote by $\mathcal{M}^{(3)}_{2,a}$ contributions which are proportional to the one-loop amplitude after tensor reduction. These include consecutive insertions of two one-loop gluon polarisation tensors, or a two-loop correction on a single gluon line. For this class of diagrams we make the replacement in straightforward analogy to eq.~\eqref{eq:M2_gg_TR} and we do not show the explicit result here.

The amplitude $\mathcal{M}^{(3)}_{2,b}$ consists of all one-loop gluon self-energy insertions on the two-loop electroweak amplitude. 
For the analysis below, we consider the diagrams contributing to the $l\,||\,p_1$ collinear limit (as usual the $q\,||\,p_2$ limit follows straightforwardly). Below we assign the loop momentum $k$ to the gluon polarisation tensor. Consider the decomposition,
\begin{align}
    \lim_{l\,||\,p_1}\mathcal{M}^{(3)}_{2,b} \sim 
    \mathcal{M}^{(3,A)}_{2,b} %
    +\mathcal{M}^{(3,B)}_{2,b}\,. %
\end{align}
The first term on the right-hand side consists of all one-loop gluon self-energy corrections to the gluon line with momentum $l$ adjacent to the incoming $p_1$ quark. Replacing the gluon polarisation tensor by its tensor reduced equivalent, eq.~\eqref{eq:gg_1L_TR}, we obtain,
\begin{align}
    \mathcal{M}^{(3,A)}_{2,b} \equiv
    \eqs[0.21]{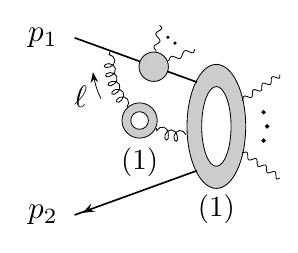} \to
    \mathcal{M}^{\prime\,(3,A)}_{2,b} =
    i \gs^2\, 
   \frac{\Gamma^\epsilon_2}{k^2(l-k)^2}
   \eqs[0.21]{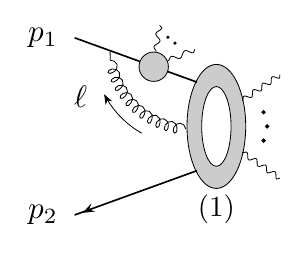}\,.
   \label{eq:M3_2_b_A}
\end{align}
The two-loop amplitude on the right-hand side is rendered locally finite in all infrared and ultraviolet regions according to refs.~\cite{Anastasiou:2020sdt,Anastasiou:2022eym}. The residual ultraviolet divergence due to the scalar bubble integrand is removed in the same way as in eq.~\eqref{M2_gg_TR_reg}.

The amplitude $\mathcal{M}^{(3,B)}_{2,b}$ contains gluon self-energy corrections to the loops $q$ and $l-q$ in diagrams with triple-gluon vertices, as well as corrections to the $q$ gluon line in uncrossed and crossed ladder type diagrams, and graphs without ladder structure,
\begin{align*}
    \mathcal{M}^{(3,B)}_{2,b} &\equiv
    \eqs[0.25]{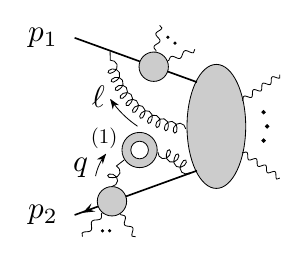}
    +\eqs[0.25]{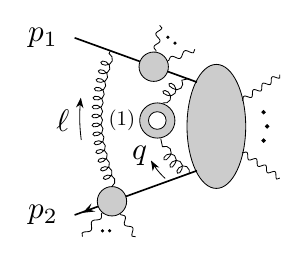}
    +\eqs[0.25]{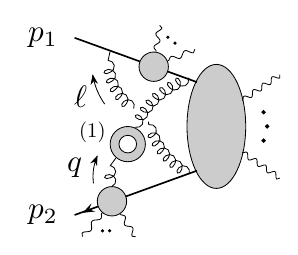}
    \\&
    +\eqs[0.25]{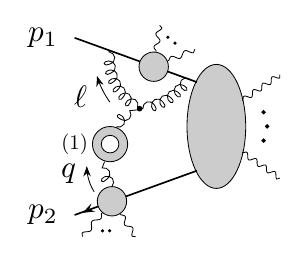}
    +\eqs[0.25]{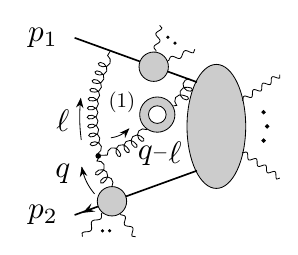}\,.
    \numberthis
    \label{eq:M3_2_b_B}
\end{align*}
Again, a tensor reduction is unavoidable due to doubled propagators which become problematic in the $q\,||\,p_2$ region or, in the case of the last set of graphs, in the mixed collinear region $(1_l,2_q,H_k)$. On the other hand, the single-collinear region $l\,||\,p_1$, while being free of power singularities, factorises only when combined with diagrams containing a gluon triangle, which we discuss in the next section. Indeed, the set of graphs with a self-energy correction on the gluon line with momentum $q-l$ do not factorise on their own. An additional subtlety involves two-loop jet subgraphs containing a gluon self-energy correction, which, after tensor reduction, require the usual one-loop modifications summarised in section~\ref{sec:LP_1L}.

%% file: gluon_triangle.tex
\section{Gluon triangle}
\label{sec:gluon_triangle}

Next, we discuss the regularisation and factorisation of one-loop corrections to the QCD three-point subgraph with gluon external states, given by,
\begin{align}
\begin{split}
    &
    \frac{-i}{l^2}\,\frac{-i}{q^2}\,
    \Gamma^{(1)\,\mu\nu\rho,abc}_{ggg}(q,k,l) 
    \equiv\eqs[0.25]{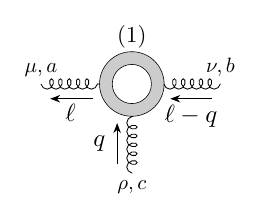}
    \\&\qquad = 
    \frac{-i}{l^2}\,\frac{-i}{q^2}\left[
    \Gamma^{(1)\,\mu\nu\rho,abc}_q
    +\Gamma^{(1)\,\mu\nu\rho,abc}_g
    +\Gamma^{(1)\,\mu\nu\rho,abc}_{gh}
    +\Gamma^{(1)\,\mu\nu\rho,abc}_{4g}
    \right]\,,
\end{split}
\label{eq:QCD_ggg}
\end{align}
where we have suppressed the momentum dependence of the functions on the second line. The grey blob represents the sum of quark, gluon and ghost loop contributions, denoted by $\Gamma^{(1)}_q$, $\Gamma^{(1)}_g$ and $\Gamma^{(1)}_{gh}$, respectively. Additionally, we have diagrams with a quartic-gluon vertex, denoted by $\Gamma^{(1)}_{4g}$, which contain a gluon self-energy subgraph. The gluon triangle is proportional to the antisymmetric structure constant $f^{abc}$,
\begin{align}
    \Gamma_{ggg}^{(1)\,\mu\nu\gamma,abc} =
    f^{abc}\,\Gamma_{ggg}^{(1)\,\mu\nu\gamma}\,.
\end{align}

The three-point gluon subgraphs lead to logarithmically divergent contributions to the three-loop amplitude (c.f. figure~\ref{fig:flow_nab3}) in the single collinear regions $(H_q,H_k,1_l)$ and $(2_q,H_k,H_l)$, when either $l$ is adjacent to the external quark line with momentum $p_1$ or $q$ is adjacent to the incoming antiquark with momentum $p_2$. For this reason we find it useful to factor out the gluon propagators $-i/l^2$ and $-i/q^2$ in eq.~\eqref{eq:QCD_ggg}.

It is well known that the three-gluon vertex satisfies a Ward-Slavnov-Taylor identity to all orders, which relates it to the ghost-gluon vertex~\cite{Slavnov:1972fg,Taylor:1971ff,Ball:1980ax}. Below, we will show how this identity is implemented locally at the one-loop order, which is the main result of this section. In the $l\,||\,p_1$ limit, where the gluon with loop momentum $l$ has longitudinal polarisation, the result is,
\begin{align}
\begin{split}
&
(-l_\mu)\,\frac{-i}{q^2} \,
\Gamma^{(1)\,\mu\nu\rho,abc}_{ggg}(q,k,l)
+\delta_{ggg,1}^{\nu\rho,abc}(q,k,l) =
\Delta\mathcal{I}^{\nu\rho,abc}_{qq,1}(q,k,l) + \Delta\mathcal{I}^{\nu\rho,abc}_{\Delta,1}(q,k,l)
\\&\qquad-i\gs\,f^{abc}\left[
\frac{-i}{(l-q)^2}\, \Pi_{gg,\,sym}^{(1)\,\nu\rho}(-k,-q) - \frac{-i}{q^2}\,\Pi_{gg,\,sym}^{(1)\,\nu\rho}(q-k,q-l)
\right]\,,
\end{split}
\label{eq:ggg_1L_ward_p1_mod}
\end{align}
with
\begin{align}
    \begin{split}
        \delta_{ggg,1}^{\nu\rho,abc} =
        \delta_{TR,1}^{\nu\rho,abc}
        +\delta_{\mathcal{U},1}^{\nu\rho,abc}
        +\delta_{\mathcal{E},1}^{\nu\rho,abc}
        +\delta_{\Delta\,gh,1}^{\nu\rho,abc}
        +\delta_{\mathcal{D},1}^{\nu\rho,abc}\,,
    \end{split}
\label{eq:delta_ggg}
\end{align}
where we have suppressed the momentum dependence. Here, we use symmetric integration, equivalent to adding a local infrared counterterm $\delta_{ggg,1}$, to remove locally non-factorisable terms from the triangle integrand in the single-collinear region $l\,||\,p_1$. The functions $\delta_{TR,1}$, $\delta_{\mathcal{U},1}$, $\delta_{\mathcal{E},1}$, $\delta_{\Delta\,gh,1}$ and $\delta_{\mathcal{D},1}$ are defined later in this section in eqs.~\eqref{eq:delta_ggg_TR}, \eqref{eq:delta_ggg_U}, \eqref{eq:delta_ggg_E}, \eqref{eq:delta_ggg_gh} and~\eqref{eq:delta_ggg_D}, respectively. Below, we will derive each of the five terms in turn. The integrands $\Delta\mathcal{I}^{\nu\rho,abc}_{qq,1}$ and $\Delta\mathcal{I}^{\nu\rho,abc}_{\Delta,1}$ are defined at the end of this section, shown graphically in eqs.~\eqref{eq:ggg_gloop_ward_I_qq} and~\eqref{eq:ggg_gloop_ward_I_Delta}, respectively.

At the amplitude level, eq.~\eqref{eq:ggg_1L_ward_p1_mod} can be implemented by replacing the gluon triangle with an equivalent version,
\begin{align}
\begin{split}
    &\frac{-i g^\alpha{}_{\mu}}{l^2}\,\frac{-i}{q^2}\,\Gamma^{(1)\,\mu\nu\rho,abc}_{ggg}(q,k,l) \to
    \frac{-i g^\alpha{}_{\mu}}{l^2}\,\frac{-i}{q^2}\,\Delta\Gamma^{(1)\,\mu\nu\rho,abc}_{ggg,1}(q,k,l) =
    \\&\qquad 
    \frac{-i  g^\alpha{}_{\mu}}{l^2}\,\frac{-i}{q^2}\,\Gamma^{(1)\,\mu\nu\rho,abc}_{ggg}(q,k,l)
    +\frac{-i}{l^2}\frac{2 \eta_{1}^\alpha}{d_1(-l,\eta_1)}
    \frac{-i}{q^2}
    \delta^{\nu\rho,abc}_{ggg,1}(q,k,l)\,.
\end{split}
\label{eq:ggg_1_CNT}
\end{align}
Both sides integrate to the same result given that,
\begin{align}
    \int_k\, \delta_{ggg,1}^{\nu\rho,abc} (q,k,l) =0\,.
\end{align}
For clarity, we have written the factor $2 \eta_{1}^\alpha/d_1(-l,\eta_1)$ from the collinear approximation for the propagator $(-i g^\alpha{}_{\mu})/l^2$, eq.~\eqref{eq:g_k_p1}, multiplying $\delta_{ggg,1}$ explicitly. \edit{The modified gluon triangle subgraph $\Delta\Gamma^{(1)\,\mu\nu\rho}_{ggg,1}$ in the $l\,||\,p_1$ region also requires ultraviolet renormalisation, and we provide the corresponding local subtraction terms throughout this section. We note that the infrared counterterm $\delta^{\nu\rho}_{ggg,1}(q,k,l)/(l^2d_1(-l,\eta_1))$ gives a finite contribution to the amplitude for large $l$ or $q$ but diverges for large values of the ``internal" loop momentum $k$. Ultraviolet counterterms are derived according to the discussion in sec.~\ref{sec:UV}, and in this case they guarantee applicability of the tree-level Ward identities in mixed collinear-UV regions.}

To avoid spurious contributions from eq.~\eqref{eq:ggg_1_CNT} to other collinear regions we introduce a massive fermion propagator to diagrams where $q$ is adjacent to the incoming antiquark $p_2$,
\begin{align}
\begin{split}
    &
    iS_0(q-p_2)
    \delta^{\nu\rho,abc}_{ggg,1}(q,k,l) \to 
    i\bar{S}_0(q-p_2;l)\,
    \delta^{\nu\rho,abc}_{ggg,1}(q,k,l)\,,
\end{split}
\label{eq:ggg_mod_quark}
\end{align}
based on the prescription used for shift counterterms, eq.~\eqref{eq:ggg_mod_quark_def}.

In eq.~\eqref{eq:ggg_1L_ward_p1_mod} and later in this section, we use a subscript $``1"$ to denote the Ward identities in the $l\,||\,p_1$ limit. For the $q\,||\,p_2$ limit, with $q^\rho \simeq z_{2,q}p_2^\rho$, we have instead,
\begin{align}
\begin{split}
&
q_\rho\,\frac{-i}{l^2} \,
\Gamma^{(1)\,\mu\nu\rho,abc}_{ggg}(q,k,l)
+\delta_{ggg,2}^{\mu\nu,abc}(q,k,l) =
\Delta\mathcal{I}^{\mu\nu,abc}_{qq,2}(q,k,l) + \Delta\mathcal{I}^{\mu\nu,abc}_{\Delta,2}(q,k,l)
\\&\qquad-i\gs\,f^{abc}\left[
\frac{-i}{l^2}\, \Pi_{gg,\,sym}^{(1)\,\mu\nu}(q-k,q-l) - \frac{-i}{(l-q)^2}\,\Pi_{gg,\,sym}^{(1)\,\mu\nu}(l-k,l)
\right]\,,
\end{split}
\label{eq:ggg_1L_ward_p2_mod}
\end{align}
where this time we label each of the functions on the second line by a subscript $``2"$. It is straightforward to see that the limits are related by a change of variables $p_1 \to -p_2$ and, for example,
\begin{align}
    \delta_{ggg,2}^{\mu\nu}(q,k,l) \equiv 
    \delta_{ggg,1}^{\mu\nu}(l-q,k-q,-q)\,,
    \label{eq:switch_limits}
\end{align}
and similarly for $\mathcal{I}_{qq,2}$ and $\mathcal{I}_{\Delta,2}$. Analogous to the prescription in eq.~\eqref{eq:ggg_mod_quark} we introduce a modified fermion propagator $1/(p_1+l)^2\to 1/[(p_1+l)^2-q^2]$ multiplying the counterterm $\delta_{ggg,2}$ for graphs where $l$ is adjacent to $p_1$. Then, we denote by $\Delta\,\Gamma_{ggg}^{(1)\,\mu\nu\rho}$ the modified integrand that takes into account modifications in both single-collinear regions as follows,
\begin{align}
    \Delta\,\Gamma_{ggg}^{(1)\,\mu\nu\rho} = 
    \Gamma_{ggg}^{(1)\,\mu\nu\rho}
    +\frac{2 \eta_{1}^\mu}{d_1(-l,\eta_1)}
    \delta^{\nu\rho,abc}_{ggg,1}
    +\frac{2 \eta_{2}^\rho}{d_2(q,\eta_2)}
    \delta^{\mu\nu,abc}_{ggg,2}\,,
    \label{eq:ggg_CNT}
\end{align}
in analogy to the shift prescription in eq.~\eqref{eq:shift_CNT_sum}.

Later on, when we investigate the local factorisation of ghost contributions, we will consider two-loop corrections to the $q\bar{q}g$-vertex containing the modified gluon triangle subgraphs $\Delta \Gamma_{ggg}^{(1)}$,
\begin{align}
\begin{split}
    &\gs\,\frac{-i}{l^2}\,\Delta \Gamma_{ggg}^{(1)\,\rho,c}(q,k,l)
    \equiv \eqs[0.25]{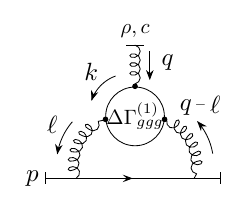}
    \\&\qquad=
    -\gs^2\, \frac{C_A}{2}\,\tq{c}
    \,\Delta\,\Gamma_{ggg}^{(1)\,\mu\nu\rho}(q,k,l)
    \frac{\gamma_\nu(\s{p}+\s{l})\gamma_\mu}{l^2(p+l)^2}
    \,.
\end{split}
\label{eq:qqg_ggg_mod}
\end{align}
Here, the limit of interest is where $q$ becomes collinear to the incoming antiquark with momentum $p_2$. We note that eq.~\eqref{eq:qqg_ggg_mod} diverges in the limit $k\to \infty$, for which we provide appropriate local ultraviolet subtraction terms in this section, but is suppressed by large values of $l$ due to the fermion propagator $1/(\s{p}+\s{l})$. The double-UV subtraction term, which renders amplitude-level contributions due to eq.~\eqref{eq:qqg_ggg_mod} finite when two loop momenta become large, is provided in eq.~\eqref{eq:qqg_ggg_mod_double-UV}.

\subsection{Fermion loop}

We begin with the quark loop contribution to the gluon three-point function, which consists of two distinguishable graphs, 
\begin{align}
\begin{split}
    &\frac{-i}{l^2}\,\frac{-i}{q^2}\,
    \Gamma^{(1)\,\mu\nu\rho,abc}_q(q,k,l) \equiv
    \eqs[0.25]{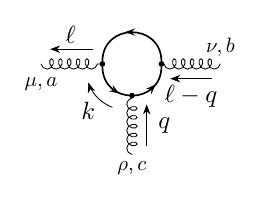}
    +\eqs[0.25]{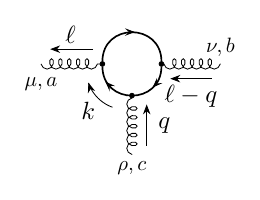}
    \\&\qquad= i\gs^3n_f\left(
    \Tr{\left[t^a t^b t^c\right]}-\Tr{\left[t^a t^c t^b\right]}
    \right) 
    \frac{-i}{l^2}\,\frac{-i}{q^2}\,
    \mathcal{K}^{\mu\nu\rho}_q(q,k,l)\,,
\end{split}
\label{eq:ggg_floop}
\end{align}
with
\begin{align}
    \mathcal{K}^{\mu\nu\rho}_q(q,k,l) = 
    -
    \frac{\Tr{\left[\s{k}\gamma^\mu (\s{l}-\s{k})\gamma^\nu(\s{q}-\s{k})\gamma^\rho\right]}}{k^2(l-k)^2(q-k)^2(l-q)^2}\,.
\end{align}
The relative sign in the kinematic part in eq.~\eqref{eq:ggg_floop} is due to the opposite fermion charge flow in the second diagram, while the trace is cyclic. The colour factor can easily be evaluated as,
\begin{align}
    \Tr{\left[t^a t^b t^c\right]}-\Tr{\left[t^a t^c t^b\right]} = i T_F f^{abc}
    \label{eq:ggg_floop_colour}
\end{align}
where 
we used the well known identity,
\begin{align}
    \Tr{\left[t^a t^b t^c\right]} =
    \frac{1}{4}d^{abc} + \frac{i}{2}T_F f^{abc}\,.
\end{align}
Here, we have introduced the symmetric structure constant $d^{abc}$,
\begin{align}
    d^{abc} = 2\,\Tr{\big[\big\{\tq{a},\tq{b}\big\}\,\tq{c}\big]}\,.
\end{align}
Clearly, the term proportional to $d^{abc}$ cancels in eq.~\eqref{eq:ggg_floop_colour}. We note that the opposite charge flow, the second diagram in eq.~\eqref{eq:ggg_floop}, is related to the crossed version of the first diagram under the simultaneous exchange $q\to k-q$ and $k\to l-k$ (or by the exchange of the two external vectors $l-q$ and $q$).

We are interested in the collinear limit $l\,||\,p_1$ (the limit $q\,||\,p_2$ is analogous) where the gluon with loop momentum $l$ is adjacent to the incoming quark and has longitudinal polarisation according to eq.~\eqref{eq:g_k_p1}, 
\begin{align}
\begin{split}
    &(-l_\mu)\,\frac{-i}{q^2} \,
    \Gamma^{(1)\,\mu\nu\rho,abc}_q(q,k,l)
    \equiv \eqs[0.25]{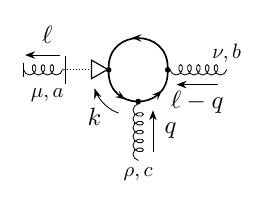}
    +\eqs[0.25]{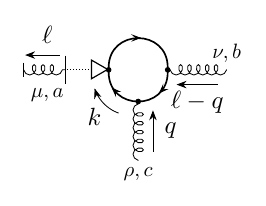}
    \\&\qquad=-i\gs f^{abc} \left[
    \frac{-i}{(l-q)^2}\Pi^{(1)\,\nu\rho}_q(-k,-q) -
    \frac{-i}{q^2}\Pi^{(1)\,\nu\rho}_q(q-k,q-l)
    \right]\,,
\end{split}
\label{eq:ggg_floop_l_p1}
\end{align}
To obtain the second line, we have applied the abelian Ward identity, shown pictorially in figure~\ref{fig:qqg_Ward}, and noticed that the result is equal to the difference of two fermion-loop bubble integrals, $\Pi^{(1)}_q$ defined in eq.~\eqref{eq:gg_floop}. 

Equation~\eqref{eq:ggg_floop_l_p1} has the equivalent, suggestive diagrammatic representation,
\begin{align*}
    \label{eq:ggg_floop_l_p1_graph}
    \numberthis
    &(-l_\mu)\,\frac{-i}{q^2} \,
    \Gamma^{(1)\,\mu\nu\rho,abc}_q(q,k,l)
    \equiv \eqs[0.25]{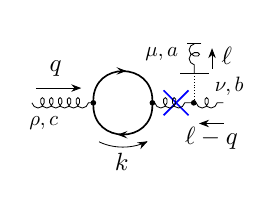}
    -\eqs[0.25]{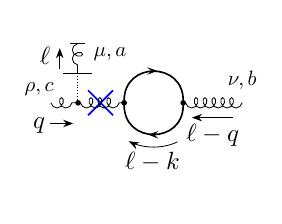}
    \\&\quad
    \to i\gs^2\, \Gamma_{n_f}^\e\Bigg[\frac{1}{k^2(q-k)^2}
    \eqs[0.2]{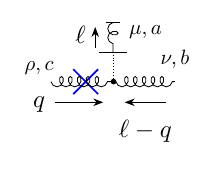}
    -
    \frac{1}{(l-k)^2(q-k)^2}
    \eqs[0.2]{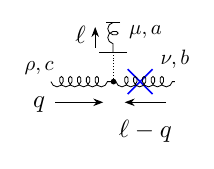}
    \Bigg]
    \\&\quad
    +\text{longitudinal terms}
    \,.
\end{align*}
On the second line, we have provided its tensor reduced equivalent.
Equation~\eqref{eq:ggg_floop_l_p1} cancels against fermion loop corrections to the external $q$ and $q-l$ loops (contributing to $\mathcal{M}^{(3,B)}_{2,b}$), which we define by,
\begin{align}
\begin{split}
    &\mathcal{G}^{(1)\,\mu\nu\rho,abc}_q(q,k,l)
    =\gs\,\frac{-i}{q^2}\frac{-i}{(l-q)^2}\Bigg[\Pi^{(1)\,\alpha\rho,cd}_{q}(-k,-q)f^{abd}C_{\alpha}{}^{\mu\nu}(-q,l,q-l)
    \\& \qquad+\Pi^{(1)\,\alpha\nu,bd}_{q}(q-k,q-l)f^{adc}C_{\alpha}{}^{\rho\mu}(q-l,-q,l)\Bigg]
    \\&\qquad 
    \equiv \eqs[0.25]{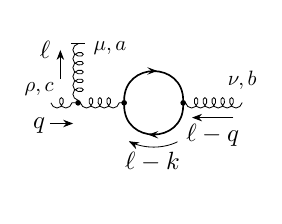}
    +\eqs[0.25]{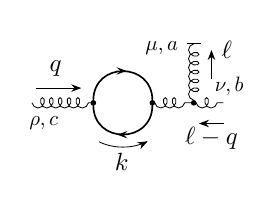}\,,
\end{split}
\end{align}
After tensor reduction, we obtain,
\begin{align*}
    \label{eq:G+T_ggg_l_p1}
    \numberthis
    &(-l_\mu)\,\left[
    \mathcal{G}^{(1)\,\mu\nu\rho,abc}_q(q,k,l)
    +\frac{-i}{q^2} \,
    \Gamma^{(1)\,\mu\nu\rho,abc}_q(q,k,l)
    \right] \\&\quad\to 
    i\gs^2\,\Gamma_{n_f}^\e\Bigg[
    \frac{1}{k^2(q-k)^2}
    \eqs[0.2]{ggg_1L_floop_ward_l_p1_RHS_fig1_TR}
    -\frac{1}{(l-k)^2(q-k)^2}
    \eqs[0.2]{ggg_1L_floop_ward_l_p1_RHS_fig2_TR}
    \\&\quad
    +\Bigg(\frac{1}{k^2(q-k)^2}+\frac{1}{(l-k)^2(q-k)^2}\Bigg)
    \Bigg(\eqs[0.2]{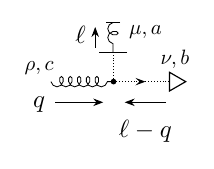}
    +\eqs[0.2]{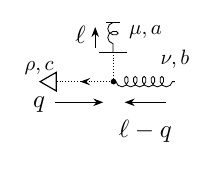}\Bigg)
    \Bigg]
    \,.
\end{align*}

The result is that fermion loop contributions to $\mathcal{M}^{(3,A)}_{2,b}$, c.f. eq.~\eqref{eq:M3_2_b_B}, when combined with the triangle graph above, are one-loop factorisable in the single collinear region $l\,||\,p_1$, up to a shift mismatch in both the scalar bubble integrand and the loop momentum $q$. 
\edit{In the sum of eq.~\eqref{eq:G+T_ggg_l_p1} the uncancelled longitudinal terms have been neglected, as they lead to scaleless integrals.}
As we will see below, both gluon and ghost loop corrections to the one-loop three-gluon vertex yield similar contributions. Factorisation of the ghost terms on the last line of eq.~\eqref{eq:G+T_ggg_l_p1} can be shown in the same way as for the two-loop amplitude. The only subtlety is that the ghost identity, eq.~\eqref{eq:qqg_1L_ward_ghost_id}, is multiplied by a scalar bubble integrand. The shift mismatch is eliminated from the amplitude by a local infrared counterterm of the form,
\begin{align*}
    &\delta_{\text{shift},1}^{(3,\Delta)}(\xi_1;p_1,p_2,q,k,l;\{q_1,\ldots,q_n\}) = 
    i\gs^2\,\frac{C_A}{2C_F}\frac{\Gamma_{2}^\epsilon}{(q-k)^2(l-k)^2}
    \eqs[0.25]{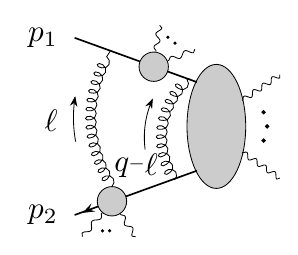} 
    \\&\qquad
    - (k\to k+l,\, q\to q+l)\,,
\numberthis
\label{eq:shift_DEL}
\end{align*}
which, up to a multiplicative one-loop factor, is the same as the two-loop shift counterterm found in ref.~\cite{Anastasiou:2022eym}. Shift subtractions add a contribution to the quark self energy given by,
\begin{align}
\begin{split}
    &\Pi^{(2)\,\text{shift}}_{\Delta}(p,q,k,l)=
    i\gs^2\,\Gamma_{2}^\epsilon\,\frac{C_A}{2C_F}
    \Bigg[
    \frac{1}{(q-k)^2(l-k)^2}
    \eqs[0.25]{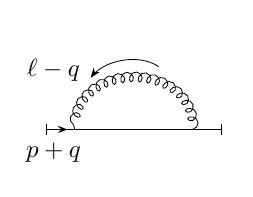}
    \\&\qquad-\frac{1}{k^2(l-k)^2}
    \eqs[0.25]{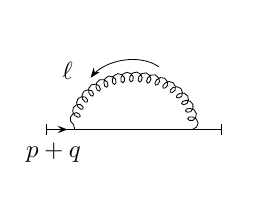}
    \Bigg]\,,
\end{split}
\label{eq:shift_2L_2}
\end{align}
which is the two-loop analogue of eq.~\eqref{eq:shift_1L_graph} with an additional one-loop scalar two-point integrand. The shift counterterm $\Pi^{(2)\,\text{shift}}_{\Delta}$ is divergent in the regions $k \to \infty$ and $k,l\to \infty$ but suppressed in the $l\to \infty$ limit.
The single-UV and double-UV singularities are removed, respectively, using the counterterms,
\begin{align}
    \Pi^{(2)\,\text{shift}}_{\Delta \text{ single-UV}}
    = i\gs^2\,\Gamma_2^\epsilon\,\frac{C_A}{2C_F}\,\frac{1}{\prop{k}^2}\,
    \Pi^{(1)\,\text{shift}}_{qq}(p,q,l)\,,
    \label{eq:shift_2L_2_singleUV}
\end{align}
and
\begin{align*}
\numberthis
\label{eq:shift_2L_2_doubleUV}
    &\Pi^{(2)\,\text{shift}}_{\Delta\,\text{double-UV}}
    = i\gs^2\,\Gamma_2^\epsilon\Bigg\{
    C_A(1-\e)\frac{1}{(l-k)^2-\m^2}
    \Bigg[
    \frac{4\,l\cdot q\, \s{l}}{\prop{l}^3\prop{k}}
    \\&\qquad
    -\frac{1}{\prop{l}^2}\Bigg(
    \frac{\s{q}}{k^2-\m^2}-\frac{2\,k\cdot q\, \s{l}}{\prop{k}^2}
    \Bigg)
    \Bigg]
    -\frac{C_A}{2C_F}\frac{1}{\prop{k}^2}
    \Pi^{(1)\,\text{shift}}_{\text{UV}}(q,l)
    \Bigg\}\,,
\end{align*}
where the one-loop shift counterterms $\Pi^{(1)\,\text{shift}}_{qq}$ and $\Pi^{(1)\,\text{shift}}_{\text{UV}}$ were defined in eqs.~\eqref{eq:shift_1L} and~\eqref{eq:shift_1L_UV}.

Above, we have demonstrated local factorisation of the (tensor reduced) amplitude $\mathcal{M}^{(3)}_{2,b}$ with gluon self-energy subgraphs when combined with the gluon triangle, which is essential for local integrability. As we will see below, we will require a local representation of the gluon loop contribution to the triangle integrand as it combines with diagrams of the $(d3V)$ topology in the single-collinear regions. We therefore avoid performing a full tensor reduction of $\Gamma^{(1)\,\mu\nu\rho}_{ggg}$ and instead add a local infrared counterterm of the form,
\begin{align*}
    &\delta_{TR,1}^{\nu\rho,abc}(q,k,l) =
    i\gs\,f^{abc}\biggl\{
    \frac{-i}{(l-q)^2}\left[\Pi^{(1)\,\nu\rho}_{gg,\,sym}(-k,-q)-
    \left(g^{\nu\rho}-\frac{q^\nu q^\rho}{q^2}\right) 
    \frac{i\,\Gamma_2^\e}{k^2(q-k)^2}\right]
    \\&-\frac{-i}{q^2}\left[
    \Pi^{(1)\,\nu\rho}_{gg,\,sym}(q-k,q-l)
    -
    \left(g^{\nu\rho}-\frac{(l-q)^\nu (l-q)^\rho}{(l-q)^2}\right) 
    \frac{i\,\Gamma_2^\e}{(l-k)^2(q-k)^2}
    \right]
    \biggr\}\,,
    \numberthis
    \label{eq:delta_ggg_TR}
\end{align*}
which defines the first term on the right-hand side of eq.~\eqref{eq:delta_ggg}.
The function $\delta_{TR,1}^{\nu\rho,abc}(q,k,l)$ replaces the bubble integrands that are a result of the $l\,||\,p_1$ Ward identity of the gluon triangle by their tensor reduced versions. Equation~\eqref{eq:delta_ggg_TR} contributes in the single-UV region as follows,
\begin{align}
\begin{split}
    &\delta_{TR,1\text{ single-UV}}^{\nu\rho,abc}(q,k,l) = i\gs\,f^{abc}\biggl\{
    \frac{1}{q^2(l-q)^2}\left(4\,n_f\,T_F\,t_1^{\nu\rho}(q,k,l) + \frac{C_A}{2}\,t_2^{\nu\rho}(q,k,l)\right)
    \\&\qquad
    - \frac{\Gamma_2^\e}{\prop{k}^2}\left[
    \frac{1}{(l-q)^2}\left(g^{\nu\rho}-\frac{q^\nu q^\rho}{q^2}\right) 
    -\frac{1}{q^2} \left(g^{\nu\rho}-\frac{(l-q)^\nu (l-q)^\rho}{(l-q)^2}\right) 
    \right]
    \biggr\}\,.
\end{split}
\label{eq:delta_ggg_TR_UV}
\end{align}
The Lorentz tensors $t_1$ and $t_2$ appearing on the first line are given by,
\begin{align}
\begin{split}
    &t_1^{\nu\rho}(q,k,l) = g^{\nu\rho} \left( \frac{l\cdot (q+k-l)}{\prop{k}^2}+ \frac{2 (k\cdot l)^2}{\prop{k}^3}\right)
    +\frac{2\, l\cdot k\left(k^\nu q^\rho +k^\rho q^\nu\right)}{\prop{k}^3}
    \\&\qquad
    + 2k^\nu k^\rho\left(\frac{l\cdot (l-2k)}{\prop{k}^3} - \frac{4\, k\cdot l\, k\cdot (l+q)}{\prop{k}^4}\right)
    -\frac{l^\nu q^\rho +l^\rho q^\nu}{\prop{k}^2}
    \\&\qquad
    + \left(l^\nu k^\rho +l^\rho k^\nu\right)\left( \frac{1}{\prop{k}^2} + \frac{2\, k\cdot (l+q)}{\prop{k}^3}\right)\,,
\end{split}
\label{eq:delta_ggg_TR_UV_t1}
\end{align}
and
\begin{align}
\begin{split}
    &t_2^{\nu\rho}(q,k,l) =
    g^{\nu\rho} \left(\frac{l\cdot(2k+3l-8q)}{\prop{k}^2}+\frac{4(k\cdot l)^2}{\prop{k}^3}\right)
    +2(3-\e)\frac{l^\nu q^\rho +l^\rho q^\nu}{\prop{k}^2}
    \\&\qquad
    -8(1-\e)k^\nu k^\rho\left(\frac{l\cdot(l-2k)}{\prop{k}^3}-\frac{4\,k\cdot l \,k\cdot (l+q)}{\prop{k}^4}\right)
    -\frac{8(1-\e)\, l\cdot k\left(k^\nu q^\rho +k^\rho q^\nu\right)}{\prop{k}^3}
    \\&\qquad
    - 4(1-\e)\left(l^\nu k^\rho +l^\rho k^\nu\right)\left(\frac{1}{\prop{k}^2}+\frac{2\, k\cdot(l+q)}{\prop{k}^3} \right)
    -\frac{2(1+\e)l^\nu l^\rho}{\prop{k}^2}\,.
\end{split}
\label{eq:delta_ggg_TR_UV_t2}
\end{align}

Similarly, the fermion-loop contribution to the triangle subgraph is made finite in the single-UV region by subtracting a local counterterm that matches the divergent behaviour. We write it as follows,
\begin{align*}
         &\Gamma^{(1)\,\mu\nu\rho,abc}_{q\text{ single-UV}}(q,k,l)=
         4\gs^3\, n_f\,T_F\,f^{abc} \frac{1}{(l-q)^2}
         \biggl[
         4\,k^\mu k^\nu k^\rho \left(\frac{2\, k\cdot (l+q)}{\prop{k}^4}+\frac{1}{\prop{k}^3}\right)
         \\&\qquad
         - 2\,\frac{k^\nu k^\rho l^\mu + k^\mu k^\rho l^\nu +k^\mu k^\rho q^\nu+k^\mu k^\nu q^\rho}{\prop{k}^3}
         -2\,\frac{k\cdot l\,g^{\nu\rho}k^\mu +k\cdot (l+q)\,g^{\mu\rho}k^\nu+k\cdot q\,g^{\mu\nu}k^\rho}{\prop{k}^3}
         \\&\qquad
         + \frac{g^{\nu\rho}(l-q-k)^\mu+g^{\mu\rho}(l+q-k)^\nu -g^{\mu\nu}(l-q+k)^\rho}{\prop{k}^2}
         \biggr]\,.
\numberthis 
\label{eq:ggg_floop_single-UV}
\end{align*}

\subsection{Ghost loop}
Next, we investigate collinear insertions on the ghost-loop contribution $\Gamma^{(1)\,\mu\nu\rho,abc}_{gh}$ to the gluon three-point function, given by
\begin{align}
\begin{split}
    &\frac{-i}{l^2}\,\frac{-i}{q^2} \,
    \Gamma^{(1)\,\mu\nu\rho,abc}_{gh}(q,k,l) \equiv
    \eqs[0.25]{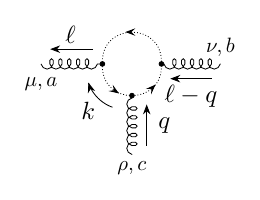}
    +\eqs[0.25]{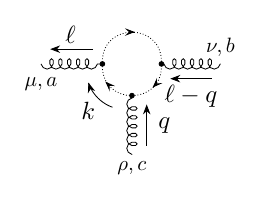}
    \\&\qquad = 
    \gs^3 \frac{C_A}{2}f^{abc}
    \frac{-i}{l^2}\,\frac{-i}{q^2} \,
    \mathcal{K}^{\mu\nu\rho}_{gh}(q,k,l)\,,
\end{split}
\label{eq:ggg_ghloop}
\end{align}
where the kinematic function $\mathcal{K}_{gh}$ reads,
\begin{align}
    \mathcal{K}_{gh}^{\mu\nu\rho}(q,k,l) = 
    \frac{k^\mu(q-k)^\rho(l-k)^\nu+(l-k)^\mu(q-k)^\nu k^\rho}{k^2(l-k)^2(q-k)^2(l-q)^2}\,.
\end{align}
In the collinear region $l\,||\,p_1$ we use the approximation of eq.~\eqref{eq:g_k_p1}. Thus, we need to consider the contraction of the gluon-ghost-ghost vertex with a longitudinally polarised gluon with loop momentum $-l^\mu \simeq z_{1,l}p_1^\mu$,
\begin{align*}
    \eqs[0.23]{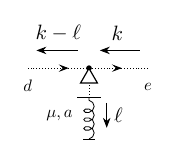} 
    &\equiv -\gs f^{ade} \frac{l\cdot k}{k^2(l-k)^2} 
    = \gs f^{ade} 
    \frac{1}{2}\left[
    \frac{1}{k^2}-\frac{1}{(l-k)^2}-\frac{l^2}{k^2(l-k)^2}
    \right]
    \\&=
    \frac{1}{2}\Bigg[
    \eqs[0.23]{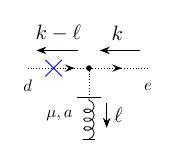}
    -\eqs[0.23]{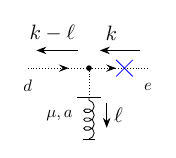}\Bigg]\,.
    \numberthis
    \label{eq:gghgh_ward}
\end{align*}
This yields an abelian-like Ward identity, shown graphically on the second line. The term proportional to $l^2 \sim \mathcal{O}(\lambda)$ vanishes in the strict collinear limit $\lambda \to 0$ and can safely be neglected in our approximation. Above, we have introduced the new Feynman rule,
\begin{align}
    \eqs[0.23]{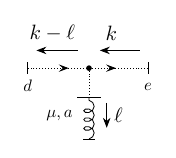} =-i\gs f^{dea}\,.
    \label{eq:ghghgh}
\end{align}
The sign is determined by moving clockwise around the three-ghost vertex starting from a ghost line whose arrow points \textit{into} the vertex. 

Applying eq.~\eqref{eq:gghgh_ward} to the contracted ghost loop, $(-l_\mu)\Gamma^{(1)\,\mu\nu\rho,abc}_{gh}$, we again write the result in terms of a difference of two self-energy integrands, 
\begin{align}
\begin{split}
    &(-l_\mu)\,\frac{-i}{q^2}\,
    \Gamma^{(1)\,\mu\nu\rho,abc}_{gh}(q,k,l)
    =-i\gs f^{abc} \frac{1}{2}\left[
    \frac{-i}{(l-q)^2}\Pi^{(1)\,\nu\rho}_{{gh},\,sym}(-k,-q) \right.
    \\&\qquad \left.
    -\frac{-i}{q^2}\Pi^{(1)\,\nu\rho}_{{gh},\,sym}(q-k,q-l)
    \right] + \mathcal{R}_1^{\nu\rho,abc}(q,k,l)
    \\&\qquad 
    \equiv 
    \frac{1}{2}\Bigg[
    \eqs[0.23]{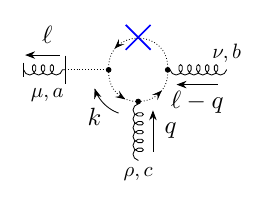}
    +\eqs[0.23]{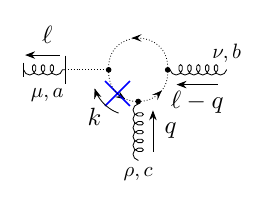}
    -\eqs[0.23]{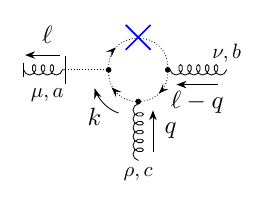}
    \\&\qquad
    -\eqs[0.23]{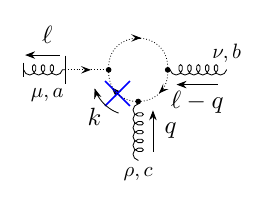}\Bigg]
    \,,
\end{split}
\label{eq:ggg_ghloop_l_p1}
\end{align}
with $\Pi^{(1)}_{{gh},\,sym}$ defined in eq.~\eqref{eq:gg_ghloop_sym}. Above, we have the remainder $\mathcal{R}_1$ defined as,
\begin{align}
    \mathcal{R}_1^{\nu\rho,abc}(q,k,l) =
    -i\gs^3f^{abc}
     \frac{C_A}{4}\frac{1}{{q^2(q-k)^2(l-q)^2}}
     \left[\frac{(q-k)^\nu l^\rho}{(l-k)^2}
     +\frac{(q-k)^\rho l^\nu}{k^2}\right]\,.
     \label{eq:ggg_ghloop_l_p1_R}
\end{align}
Below, we will show this term cancels against bubble-type contributions to the gluon loop Ward identity.

\subsection{Quartic vertex}

The three-point function $\Gamma^{(1)}_{4g}$ on the second line of eq.~\eqref{eq:QCD_ggg}, which contains a single quartic-gluon vertex, explicitly reads,
\begin{align}
\begin{split}
    &\frac{-i}{l^2}\,\frac{-i}{q^2}\,
    \Gamma^{(1)\,\mu\nu\rho,abc}_{4g}(q,k,l)
    =
    i\gs^3 
    \,\frac{-i}{l^2}\,\frac{-i}{q^2}\,\frac{1}{2}\,
    \mathcal{K}^{\mu\nu\rho,abc}_{4g}(q,k,l)
    \\&\quad
    \equiv
    \eqs[0.25]{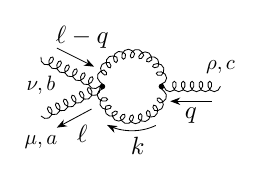}
    +\eqs[0.25]{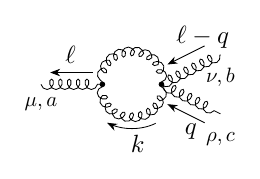}
    +\eqs[0.25]{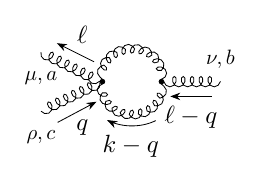}\,,
\end{split}
\label{eq:ggg_gloop_4g}
\end{align}
with
\begin{align}
\begin{split}
    &\mathcal{K}^{\mu\nu\rho,abc}_{4g}(q,k,l) = 
    f^{cde}\frac{D^{\mu\nu\beta\alpha}_{abde}(l,q-l,k-q,-k)
    C^{\rho}{}_{\beta\alpha}(-q,k,q-k)}{k^2(q-k)^2(l-q)^2}
    \\&\qquad
    +f^{aed}\frac{D^{\nu\rho\alpha\beta}_{bced}(q-l,-q,k,l-k)
    C^{\mu}{}_{\alpha\beta}(l,k-l,-k)}{k^2(l-k)^2(l-q)^2}
    \\&\qquad
    +f^{bde}\frac{D^{\rho\mu\beta\alpha}_{cade}(-q,l,k-l,q-k)
    C^{\nu}{}_{\beta\alpha}(q-l,k-q,l-k)}{(k-q)^2(l-k)^2(l-q)^2}\,.
\end{split}
\label{eq:ggg_gloop_4g_kinem}
\end{align}
where the quartic vertex $D^{\mu\nu\alpha\beta}_{abcd}$ was defined in eq.~\eqref{eq:4g}.
The factor $1/2$ multiplying $\mathcal{K}_{4g}$ in eq.~\eqref{eq:ggg_gloop_4g} accounts for the symmetry of the quartic vertex. The ``unconventional" momentum routing of the self-energy subgraph of the last diagram in eq.~\eqref{eq:ggg_gloop_4g} (neither gluon line carries momentum $k$) is useful since it allows us to combine divergent integrands in the single-collinear limits with the gluon-loop triangle graph $\Gamma_g^{(1)}$. 
The second diagram (first diagram) vanishes in the single-collinear limit $l\,||\,p_1$ ($q\,||\,p_2$),
\begin{align}
    \eqs[0.25]{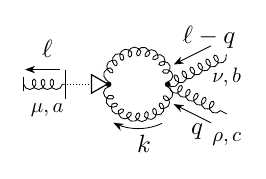} =0\,,
    \qquad
    \eqs[0.25]{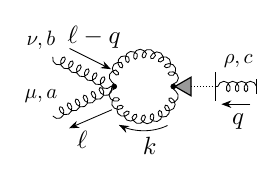} =0\,.
\end{align}
This can be checked straightforwardly by contracting the second term in eq.~\eqref{eq:ggg_gloop_4g_kinem} by the collinear momentum $l_\mu$. \edit{In fact, the scalar and ghost terms contributing to the QCD Ward identity, eq.~\eqref{eq:gg_Ward}, vanish separately}. Using the decomposition in eq.~\eqref{eq:4g_ward} for quartic-gluon vertices with a longitudinally polarised gluon, the remaining two graphs give,
\begin{align}
\begin{split}
    &\eqs[0.25]{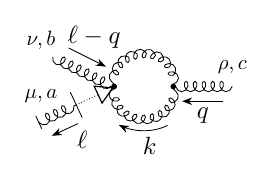}
    =-\eqs[0.25]{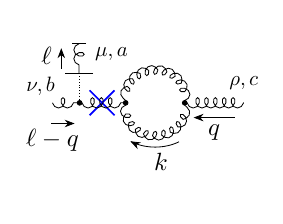}
    +\frac{1}{2}\Bigg[\eqs[0.25]{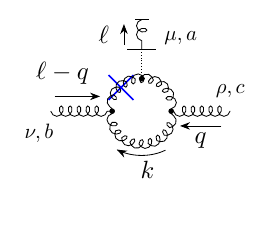}
    \\&\qquad\qquad
    -\eqs[0.25]{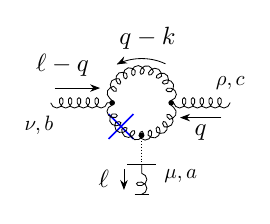}
    \Bigg]\,,
\end{split}
\label{eq:ggg_4g_ward_fig1}
\end{align}
and 
\begin{align}
\begin{split}
    &\eqs[0.25]{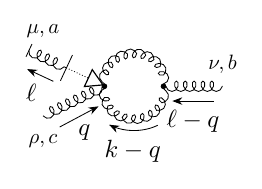}
    =\eqs[0.25]{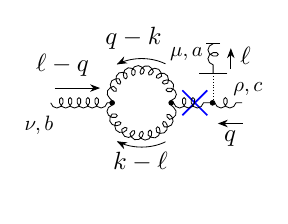}
    +\frac{1}{2}\Bigg[\eqs[0.25]{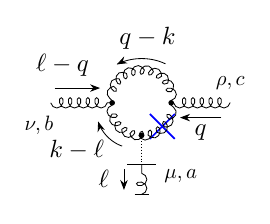}
    \\&\qquad\qquad
    -\eqs[0.25]{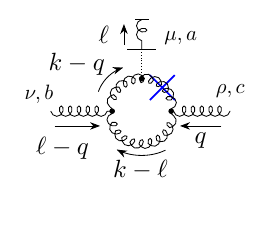}
    \Bigg]\,.
\end{split}
\label{eq:ggg_4g_ward_fig2}
\end{align}
Again, the explicit factors of $1/2$ account for the symmetry of the quartic vertex diagram. The first terms on the right-hand sides of eqs.~\eqref{eq:ggg_4g_ward_fig1} and~\eqref{eq:ggg_4g_ward_fig2} are equal to a difference of two gluon self energy graphs,
\begin{align}
\begin{split}
    &\eqs[0.25]{ggg_1L_gloop_4g_fig3_ward_l_p1_RHS_fig1}
    -\eqs[0.25]{ggg_1L_gloop_4g_fig1_ward_l_p1_RHS_fig1}
    \\&\qquad
    \equiv
    -i\gs f^{abc} \left[
    \frac{-i}{(l-q)^2}\Pi^{(1)\,\nu\rho}_g(-k,-q)
    -\frac{-i}{q^2}\Pi^{(1)\,\nu\rho}_g(q-k,q-l)
    \right] \,,
\end{split}
\label{eq:ggg_4gloop_l_p1_graph}
\end{align}
with $\Pi^{(1)}_g$ defined in eq.~\eqref{eq:gg_gloop}. In analogy to the fermion loop contribution, eq.~\eqref{eq:ggg_floop_l_p1_graph}, the integrand in eq.~\eqref{eq:ggg_4gloop_l_p1_graph} factorises up to a shift term when combined with gluon loop corrections to the $q$ and $l-q$ legs in the $l\,||\,p_1$ limit. As we will see below, the other graphs contributing to eqs.~\eqref{eq:ggg_4g_ward_fig1} and~\eqref{eq:ggg_4g_ward_fig2} can be made to vanish locally against similar self-energy contributions to the $l\,||\,p_1$ collinear limit of the gluon loop.

The quartic vertex contribution to the gluon triangle is made finite in the single-UV region $k\to \infty$ by subtracting the following counterterm,
\begin{align}
\begin{split}
        &\Gamma^{(1)\,\mu\nu\rho,abc}_{4g\text{ single-UV}}(q,k,l)
        =
         \gs^3\, \frac{9C_A}{4}\,f^{abc}\, 
         \frac{g^{\mu\nu}(2l-q)^\rho -g^{\mu\rho}(l+q)^\nu+g^{\nu\rho}(2q-l)^\mu}{(l-q)^2\prop{k}^2}\,.
\end{split}
\label{eq:ggg_4g_single-UV}
\end{align}

\subsection{Gluon loop}
Finally, we analyse the gluon-loop contributions, which are more involved. They will have to be summed with both the quartic-vertex integrand and the ghost triangle discussed previously. In addition we will need to introduce a local infrared counterterm that integrates to zero, but removes terms from the gluon triangle integrand that spoil local factorisation. Explicitly, the gluon loop contribution reads,
\begin{align}
\begin{split}
    &\frac{-i}{l^2}\,\frac{-i}{q^2}\,
    \Gamma^{(1)\,\mu\nu\rho,abc}_g(q,k,l) \equiv
    \eqs[0.25]{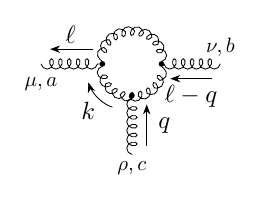}
    \\&\qquad= 
    -\gs^3 \frac{C_A}{2}f^{abc}
    \,\frac{-i}{l^2}\,\frac{-i}{q^2} \,
    \mathcal{K}^{\mu\nu\rho}_g(q,k,l)\,,
\end{split}
\label{eq:ggg_gloop}
\end{align}
with kinematic part,
\begin{align}
    \mathcal{K}_g^{\mu\nu\rho}(q,k,l) = \frac{C^{\mu\alpha\beta}(l,k-l,-k)C_{\epsilon\alpha}{}^{\nu}(k-q,l-k,q-l)C_{\beta}{}^{\epsilon\rho}(k,q-k,-q)}{k^2(l-k)^2(q-k)^2(l-q)^2}\,.
    \label{eq:ggg_gloop_kinem}
\end{align}
In the single-collinear region $l\,||\,p_1$ we contract the three-point function with {${(-l^\mu) \simeq z_{1,l}\,p_{1}^\mu}$} according to the approximation eq.~\eqref{eq:g_k_p1}, and write the result as a sum of four terms as follows,
\begin{align}
\begin{split}
    &(-l_\mu)\,\frac{-i}{q^2}\,\Gamma^{(1)\,\mu\nu\rho,abc}_g(q,k,l) \equiv
    \eqs[0.25]{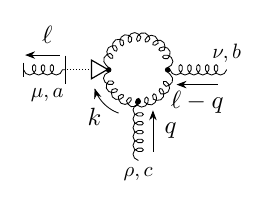}
    \\&\qquad
    =\mathcal{B}_{g,1}^{\nu\rho,abc}(q,k,l)
    +\mathcal{D}_1^{\nu\rho,abc}(q,k,l)
    +\mathcal{I}^{\nu\rho,abc}_{qq,1}(q,k,l) + \mathcal{I}^{\nu\rho,abc}_{\Delta,1}(q,k,l) \,.
\end{split}
    \label{eq:ggg_gloop_l_p1}
\end{align}
Below, we will explain each of the functions on the second line in turn. They are obtained through repeated application of the Ward identity for triple-gluon vertices, shown graphically in figure~\ref{fig:3g_Ward}. The analogous Ward identity in the $q\,||\,p_2$ limit can be obtained using eq.~\eqref{eq:switch_limits}.

Here, $\mathcal{B}_{g,1}$ represents all graphs where one of the internal gluon propagators, either $1/k^2$ or $1/(l-k)^2$, has been cancelled, and are therefore similar to two-point bubble insertions, eq.~\eqref{eq:gg_gloop}. We will use symmetric integration in the loop momentum $k$ to remove terms that vanish after integration. Graphically, the function $\mathcal{B}_{g,1}$ is equal to,
\begin{align*}
    &\mathcal{B}_{g,1}^{\nu\rho,abc}(q,k,l) =
    \eqs[0.25]{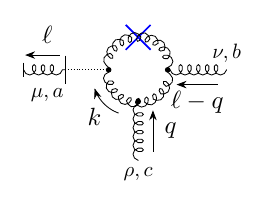}
    -\eqs[0.25]{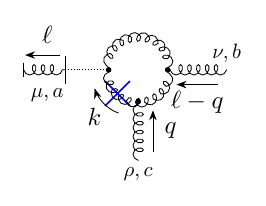}
    -\eqs[0.25]{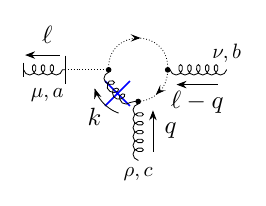}
    \\&\quad 
    +\eqs[0.25]{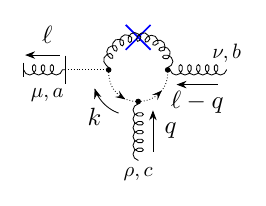}
    +\eqs[0.25]{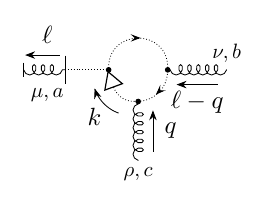}
    +\eqs[0.25]{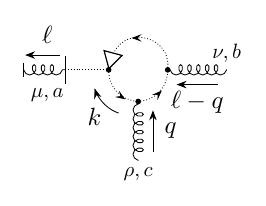}\,.
    \numberthis
    \label{eq:ggg_gloop_ward_B}
\end{align*}
The first (second) term combines with the second and third graphs on the right-hand side of eq.~\eqref{eq:ggg_4g_ward_fig1} (eq.~\eqref{eq:ggg_4g_ward_fig2}), which are contributions to the $l\,||\,p_1$ Ward identity due to graphs with a quartic gluon vertex. Their sums simplify to,
\begin{align*}
    &\eqs[0.25]{ggg_1L_gloop_B_cw_diag1}
    +\frac{1}{2}\Bigg[
    \eqs[0.25]{ggg_1L_gloop_4g_fig1_ward_l_p1_RHS_fig2}
    -\eqs[0.25]{ggg_1L_gloop_4g_fig1_ward_l_p1_RHS_fig3}
    \Bigg]
    \\&\qquad 
    \equiv \frac{1}{2}\Bigg[
    \eqs[0.25]{ggg_1L_gloop_4g_fig1_ward_l_p1_RHS_fig2}
    +\eqs[0.25]{ggg_1L_gloop_4g_fig1_ward_l_p1_RHS_fig3}
    \Bigg]
    \,,
\numberthis
\label{eq:ggg_gloop_ward_B_U_eq1}
\end{align*}
and
\begin{align*}
    &-\eqs[0.25]{ggg_1L_gloop_B_acw_diag1}
    +\frac{1}{2}\Bigg[
    \eqs[0.25]{ggg_1L_gloop_4g_fig3_ward_l_p1_RHS_fig2}
    -\eqs[0.25]{ggg_1L_gloop_4g_fig3_ward_l_p1_RHS_fig3}
    \Bigg]
    \\&\qquad 
    \equiv -\frac{1}{2}\Bigg[
    \eqs[0.25]{ggg_1L_gloop_4g_fig3_ward_l_p1_RHS_fig2}
    +\eqs[0.25]{ggg_1L_gloop_4g_fig3_ward_l_p1_RHS_fig3}
    \Bigg]
    \,.
    \numberthis
\label{eq:ggg_gloop_ward_B_U_eq2}
\end{align*}
The sum of integrands shown pictorially in eqs.~\eqref{eq:ggg_gloop_ward_B_U_eq1} and~\eqref{eq:ggg_gloop_ward_B_U_eq2}, which we denote by the kinematic function $\mathcal{U}_1$, is given by,
\begin{align}
\begin{split}
    &\mathcal{U}_{1}^{\nu\rho,abc}(q,k,l) =i\gs^3 \frac{C_A}{4}f^{abc} 
    \frac{1}{q^2(l-q)^2}\left[
    \frac{C^{\rho}{}_{\alpha\beta}(-q,k,q-k)f^{\nu;\alpha\beta}(l)}{k^2(q-k)^2}
    \right. \\&\qquad+\left.\frac{C^{\nu}{}_{\beta\alpha}(q-l,k-q,l-k)f^{\rho;\alpha\beta}(l)}{(l-k)^2(q-k)^2}
    \right]\,.
\end{split}
\label{eq:ggg_gloop_ward_B_U} 
\end{align}
The function $f$ above is defined as,
\begin{align}
    f^{\nu;\alpha\beta}(l) =
    l^\beta g^{\alpha\nu}+l^\alpha g^{\beta\nu} -2l^\nu g^{\alpha\beta}\,,
\end{align}
which is symmetric under exchange of the two Lorentz indices after the semicolon. Interestingly, the first (second) term in square brackets in eq.~\eqref{eq:ggg_gloop_ward_B_U} vanishes after performing a symmetrisation under the exchange $k\to q-k$ ($q\to q+l-k$). As usual, symmetric integration is implemented using a local infrared counterterm that integrates to zero,
\begin{align}
\begin{split}
        \Delta\,\mathcal{U}_1^{\nu\rho,abc}(q,k,l) 
        \equiv
        \mathcal{U}_1^{\nu\rho,abc}(q,k,l)
        + \delta_{\mathcal{U},1}^{\nu\rho,abc}
        (q,k,l) = 0\,,
\end{split}
\end{align}
with
\begin{align}
\begin{split}
    &\delta_{\mathcal{U},1}^{\nu\rho,abc}(q,k,l) =
    -i\gs^3\, \frac{C_A}{4}\,f^{abc} \frac{1}{q^2(l-q)^2}\frac{1}{2}\left[
    \frac{f^{\rho;}{}_{\alpha\beta}(q-2k)f^{\nu;\alpha\beta}(l)}{k^2(q-k)^2}
    \right. \\&\qquad +\left.
    \frac{f^{\nu;}{}_{\alpha\beta}(q+l-2k)f^{\rho;\alpha\beta}(l)}{(l-k)^2(q-k)^2}
    \right]\,.
\end{split}
\label{eq:delta_ggg_U}
\end{align}
This defines the second contribution to the local gluon triangle counterterm, eq.~\eqref{eq:delta_ggg}. \edit{Though $\delta_{\mathcal{U},1}^{\nu\rho}(q,k,l)$ integrates to zero, it has a residual UV divergence for large loop momentum $k$. This divergence is removed locally through the counterterm,}
\begin{align*}
    &\delta_{\mathcal{U},1 \text{ single-UV}}^{\nu\rho,abc}(q,k,l) =
    -i\gs^3\, \frac{C_A}{4}\,f^{abc} \frac{1}{q^2(l-q)^2}
    \biggl[
    g^{\nu\rho}\left(\frac{l\cdot (l+2q-4k)}{\prop{k}^2}
    - \frac{4\, k\cdot l \, k\cdot (l+2q)}{\prop{k}^3}
    \right)
    \\&\qquad
    +\frac{(5-4\e)\left(l^\nu l^\rho+ l^\nu q^\rho+l^\rho q^\nu\right)}{\prop{k}^2}
    -2(5-4\e)\left(l^\nu k^\rho+l^\rho k^\nu\right)\left(\frac{2\, k\cdot q}{\prop{k}^3}
    +\frac{1}{\prop{k}^2}\right)
    \\&\qquad
    - 4\,k\cdot l\,\frac{4(1-\e)k^\nu l^\rho+k^\rho l^\nu}{\prop{k}^3}
    \biggr]
    \,.\numberthis
    \label{eq:delta_ggg_U_UV}
\end{align*}

Next, the sum of the third and fourth graphs in the definition of the bubble contribution $\mathcal{B}_{g,1}$ in eq.~\eqref{eq:ggg_gloop_ward_B}, which we denote by $\mathcal{E}_1$, is given by the following integrand,
\begin{align}
    \begin{split}
        &\mathcal{E}_1^{\nu\rho,abc}(q,k,l) =
        \eqs[0.25]{ggg_1L_gloop_B_acw_diag2}
        -\eqs[0.25]{ggg_1L_gloop_B_cw_diag2} 
        \\&\qquad
        =
        i\gs^3 \frac{C_A}{2}f^{abc}\frac{1}{q^2(l-q)^2}\left[
        \frac{(q-k)^\rho k^\nu}{k^2(q-k)^2}
        +
        \frac{(l-k)^\rho (q-k)^\nu}{(l-k)^2(q-k)^2}
        \right]\,.
    \end{split}
\end{align}
Again, we notice that the first (second) term in square brackets is symmetric under the exchange $k\to q-k$ ($k\to q+l-k$), and symmetric integration yields,
\begin{align}
    \begin{split}
        &
        \Delta\,\mathcal{E}_1^{\nu\rho,abc}(q,k,l) \equiv
        \mathcal{E}_1^{\nu\rho,abc}(q,k,l)
        + \delta_{\mathcal{E},1}^{\nu\rho,abc}
        (q,k,l)
        \\&\qquad
        =
        i\gs^3 \frac{C_A}{2}f^{abc}\frac{1}{q^2(l-q)^2}\frac{1}{2}\left[
        \frac{(q-k)^\rho k^\nu+(q-k)^\nu k^\rho}{k^2(q-k)^2} \right.
        \\&\qquad+\left.
        \frac{(l-k)^\rho (q-k)^\nu+(l-k)^\nu (q-k)^\rho}{(l-k)^2(q-k)^2}
        \right]\,,
    \end{split}
\label{eq:ggg_gloop_B_sym}
\end{align}
where,
\begin{align}
\begin{split}
    &\delta_{\mathcal{E},1}^{\nu\rho,abc}(q,k,l)
    =i\gs^3 \frac{C_A}{2}f^{abc}\frac{1}{q^2(q-k)^2(l-q)^2}
    \frac{1}{2}\left[
    \frac{k^\rho q^\nu -k^\nu q^\rho}{k^2}
    \right.
    \\&\qquad+
    \left.
    \frac{(l-k)^\nu q^\rho-(l-k)^\rho q^\nu + 
    k^\nu l^\rho - k^\rho l^\nu}{(l-k)^2}
    \right]\,.
\end{split}
\label{eq:delta_ggg_E}
\end{align}
The UV divergence for large $k$ is cancelled locally using the following counterterm,
\begin{align}
\begin{split}
    &\delta_{\mathcal{E},1 \text{ single-UV}}^{\nu\rho,abc}(q,k,l)
    =i\gs^3 \frac{C_A}{4}f^{abc}\frac{1}{q^2(l-q)^2}
    \biggl[
    \frac{l^\nu q^\rho-l^\rho q^\nu}{\prop{k}^2}
    \\&\qquad
    -2\left(k^\nu q^\rho-k^\rho q^\nu\right)\left(\frac{k\cdot (l+2q)}{\prop{k}^3}+\frac{1}{\prop{k}^2}\right)
    \\&\qquad
    +\left(k^\nu l^\rho-k^\rho l^\nu\right)\left(\frac{2\,k\cdot (l+q)}{\prop{k}^3}+\frac{1}{\prop{k}^2}\right)
    \biggr]\,.
\end{split}
    \label{eq:delta_ggg_E_UV}
\end{align}

Superficially, the last two graphs in eq.~\eqref{eq:ggg_gloop_ward_B} look similar to the ghost loop contribution to the gluon triangle, eq.~\eqref{eq:ggg_ghloop}. It can be shown that the sum of these two graphs, which we represent by the function $\mathcal{C}_1^{\nu\rho,abc}(q,k,l)$, is proportional to $(l-q)^\nu$ or $q^\rho$, i.e. one of the ``external" gluons has longitudinal polarisation. In principle such terms can be shown to factorise from the hard-scattering tree-amplitude. However, when combined with the newly symmetrised expression in eq.~\eqref{eq:ggg_gloop_B_sym} and the ghost triangle, eq.~\eqref{eq:ggg_ghloop_l_p1}, the result is equal to a difference of two self-energy graphs with a ghost loop, as we will show below. Indeed, if such a singular contribution were to exist, even after integration, it would be in violation of the Ward-Slavnov-Taylor identity.

Explicitly, the sum of integrands reads,
\begin{align}
    \begin{split}
    &\mathcal{C}_1^{\nu\rho,abc}(q,k,l)  \equiv \eqs[0.25]{ggg_1L_gloop_B_cw_gh_LHS}
    +\eqs[0.25]{ggg_1L_gloop_B_acw_gh_LHS}
    \\&\qquad= -i\gs^3 \frac{C_A}{2}f^{abc}
    \frac{\mathcal{N}^{\nu\rho}_1 + \mathcal{N}^{\nu\rho}_2}{k^2q^2(l-k)^2(q-k)^2(l-q)^2}\,.
    \end{split}
    \label{eq:ggg_C}
\end{align}
Here, $\mathcal{N}^{\nu\rho}_1$ and $\mathcal{N}^{\nu\rho}_2$ denote the numerators of the first and second graphs in eq.~\eqref{eq:ggg_C}, which read
\begin{align}
\begin{split}
    \mathcal{N}^{\nu\rho}_1 &= k\cdot(l-k)(q-k)^\nu k^\rho\,,
    \\
    \mathcal{N}^{\nu\rho}_2 &= k\cdot(l-k)(l-k)^\nu(q-k)^\rho\,.
\end{split}
\end{align}
Their sum can be written as
\begin{align}
    \mathcal{N}^{\nu\rho}_1+\mathcal{N}^{\nu\rho}_2\equiv k\cdot(l-k)\left[(q-l)^\nu k^\rho +(l-k)^\nu q^\rho\right]\,.
\end{align}
Using the relation
\begin{align}
    2k\cdot(l-k) = l^2 - (l-k)^2-k^2\,,
\end{align}
and discarding terms of $\mathcal{O}(l^2)$, which give a vanishing contribution to the collinear limit $l\,||\,p_1$, we obtain,
\begin{align}
    \begin{split}
    &\mathcal{C}_1^{\nu\rho,abc}(q,k,l)  = \frac{1}{2}\Bigg[\eqs[0.25]{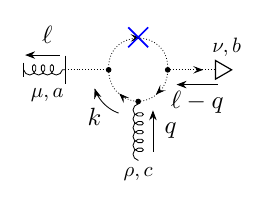}
    +\eqs[0.25]{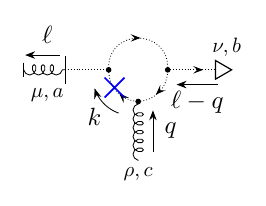}
    \\&\qquad 
    +\eqs[0.25]{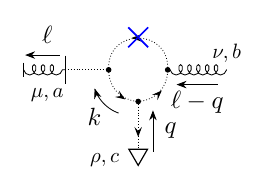}
    +\eqs[0.25]{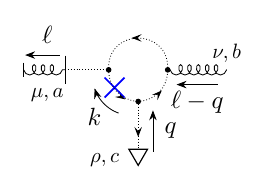}
    \Bigg]\,.
    \end{split}
    \label{eq:ggg_C_id}
\end{align}

Then, when we combine $\Delta\,\mathcal{E}_1$, defined in eq.~\eqref{eq:ggg_gloop_B_sym}, with $\mathcal{C}_1$ and the ghost triangle contribution, eq.~\eqref{eq:ggg_ghloop_l_p1}, we obtain,
\begin{align*}
    &(-l_\mu)\,\frac{-i}{q^2}\,\Gamma^{(1)\,\mu\nu\rho,abc}_{gh}(q,k,l)+
    \Delta\,\mathcal{E}_1^{\nu\rho,abc}
    (q,k,l)
    +\mathcal{C}_1^{\nu\rho,abc}
    (q,k,l)
    +\delta_{gh}^{\nu\rho,abc}(q,k,l)
    \\&\qquad
    =
    -i\gs f^{abc}\left[
    \frac{-i}{(l-q)^2}\,\Pi^{(1)\,\nu\rho}_{gh,\,sym}(-k,-q)
    -
    \frac{-i}{q^2}\,\Pi^{(1)\,\nu\rho}_{gh,\,sym}(q-k,q-l)
    \right]\,.
\numberthis
\label{eq:ggg_ghloop_l_p1_pol}
\end{align*}
Here, we have defined the local infrared counterterm,
\begin{align}
    \delta_{\Delta\,gh}^{\nu\rho,abc}(q,k,l) =
    -i\gs f^{abc} \frac{1}{2}\left[\frac{-i}{(l-q)^2}\,\delta_{gh}^{\rho\nu}(k,q)
    -\frac{-i}{q^2}\,\delta_{gh}^{\rho\nu}(q-k,q-l)\right]\,.
    \label{eq:delta_ggg_gh}
\end{align}
The terms $\delta_{gh}^{\rho\nu}(k,q)$ and $\delta_{gh}^{\rho\nu}(q-k,q-l)$ on the right-hand side are required to symmetrise bubble-type ghost loop contributions under the exchanges $k\to q-k$ and $k\to q+l-k$, respectively, according to the prescription given in eq.~\eqref{eq:gg_ghloop_sym}. The contributions to the three-loop amplitude in the $l\,||\,p_1$ limit given by the right-hand side of eq.~\eqref{eq:ggg_ghloop_l_p1_pol} factorise up to a shift term when combined with corrections to the external legs of the triangle subgraph, following the discussion after eq.~\eqref{eq:ggg_floop_l_p1_graph}. Again, we require a local counterterm to remove a left-over divergence in eq.\eqref{eq:delta_ggg_gh} in the single-UV region, which has the following form,
\begin{align}
    \begin{split}
        &\delta_{\Delta\,gh\text{ single-UV}}^{\nu\rho,abc}(q,k,l) =
        i\gs^3\,\frac{C_A}{4}\,f^{abc}\frac{1}{q^2(l-q)^2}\biggl[
        \frac{q^\nu l^\rho - q^\rho l^\nu}{\prop{k}^2}
        \\&\qquad
        + 2\left(k^\nu q^\rho - k^\rho q^\nu\right)
        \left(\frac{k\cdot(l+2q)}{\prop{k}^3}+\frac{1}{\prop{k}^2}\right)
        \\&\qquad
        + \left(l^\nu k^\rho - l^\rho k^\nu\right)
        \left(\frac{2\,k\cdot(l+q)}{\prop{k}^3}+\frac{1}{\prop{k}^2}\right)
        \biggr]\,.
    \end{split}
    \label{eq:delta_ggg_gh_UV}
\end{align}

The term $\mathcal{I}_{qq,1}$ on the right-hand side of eq.~\eqref{eq:ggg_gloop_l_p1} consists of graphs that lead to collinear insertions on the quark line, which can be shown to factorise independently. The function $\mathcal{I}_{\Delta,1}$ contains all gluon triangle graphs where one of the ``external" propagators $1/q^2$ or $1/(l-q)^2$ is cancelled. These terms vanish against ghost contributions to the $q\bar{q}g$-vertex, $(-l_\mu)\Gamma^{(2)\,\mu}_{qqg}$, either exactly or after a loop-momentum shift. The precise mechanism will be discussed in the next section.

We note that the integrands where the propagator $1/(q-k)^2$ is cancelled require special treatment through the addition of a local infrared counterterm $\delta^{\nu\rho,abc}_{\mathcal{D},1}$. As we will see, the modified integrand can be distributed among the functions $\mathcal{I}_{qq,1}$ and $\mathcal{I}_{\Delta,1}$.
To understand this point, we consider the difference,
\begin{align}
\begin{split}
    &\mathcal{D}_1^{\nu\rho,abc}(q,k,l) \equiv
    \eqs[0.25]{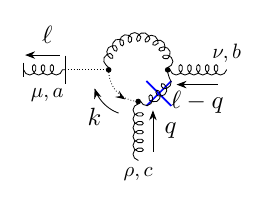}
    -\eqs[0.25]{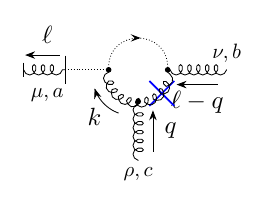}
    \\&\qquad =
    -i\gs^3 \frac{C_A}{2}f^{abc}
    \frac{C^{\alpha\nu\rho}(l-k,q-l,k-q)k_\alpha+C^{\beta\nu\rho}(k,q-k,-q)(l-k)_\beta }{k^2q^2(l-k)^2(l-q)^2}
    \\&\qquad =
    i\gs^3 \frac{C_A}{2}f^{abc}
    \left[\frac{1}{k^2(l-k)^2}\left[Q_0^{\nu\rho}(l,q) + O_0^{\nu\rho}(l,q)\right]
    +\frac{(l-k)^\nu l^\rho - k^\rho l^\nu}{k^2q^2(l-k)^2(l-q)^2} \right.
    \\&\qquad
    \left.
    +\frac{g^{\nu\rho}}{q^2(l-q)^2}\left(\frac{1}{(l-k)^2}-\frac{1}{k^2}\right)\right]\,,
    \label{eq:ggg_D}
\end{split}
\end{align}
Above, we recognise the familiar QCD Ward identity, eq.~\eqref{eq:gg_Ward}, up to terms in the integrand that are antisymmetric under the exchange $k \to l-k$. The term on the last line is a difference of two scaleless integrals, which vanish individually after integration over $k$. These terms are an impediment to local factorisation. However, we can remove these contributions altogether through the use of symmetric integration of the bubble integrand $\sim 1/k^2(l-k^2)$, equivalent to adding a counterterm of the form,
\begin{align}
    \delta^{\nu\rho,abc}_{\mathcal{D},1}(q,k,l) = \frac{1}{2}
    \left[
    \mathcal{D}_1^{\nu\rho,abc}(q,l-k,l)
    -\mathcal{D}_1^{\nu\rho,abc}(q,k,l)
    \right]\,,
    \label{eq:delta_ggg_D}
\end{align}
which is the last contribution to eq.~\eqref{eq:delta_ggg}.
This yields,
\begin{align}
\begin{split}
    &\mathcal{D}_1^{\nu\rho,abc}(q,k,l) +
    \delta^{\nu\rho,abc}_{\mathcal{D},1}(q,k,l) 
    \equiv 
    \eqs[0.23]{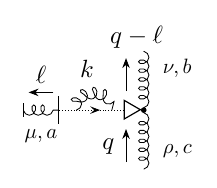}
    \\&\qquad 
    =
    i\gs^3\, \frac{C_A}{2}f^{abc}\frac{1}{k^2(l-k)^2}
    \left[Q_0^{\nu\rho}(l,q) + O_0^{\nu\rho}(l,q)\right]\,.
\end{split}
\label{eq:ggg_D_1_del}
\end{align}
As in eq.~\eqref{eq:qqg_1L_ward_ghost_id} symmetrisation of the bubble integrand under $k \to l-k$ is implied in the graphical notation on the right-hand side. 
The local infrared counterterm $\delta_{\mathcal{D},1}$ has the effect of removing terms from eq.~\eqref{eq:ggg_D} that are antisymmetric under the exchange $k \to l-k$, and is therefore valid only in the region $l\,||\,p_1$. %
To avoid spurious contributions to the $q\,||\,p_2$ region we use the prescription in eq.~\eqref{eq:ggg_mod_quark}, with a modified fermion propagator. As usual, the integrand-level modifications corresponding to $\delta_{\mathcal{D},1}$ generate additional UV divergences in the region where $k$ becomes large. While they integrate to zero, we require a counterterm to remove them locally,
\begin{align}
\begin{split}
    &\delta^{\nu\rho,abc}_{\mathcal{D},1 \text{ single-UV}}(q,k,l) = i \gs^3\,\frac{C_A}{2}\biggl[
    g^{\nu\rho}\left(\frac{l\cdot (l-2k)}{\prop{k}^2}-\frac{4(k\cdot l)^2}{\prop{k}^3}\right)
    \\&\qquad
    +\left(k^\nu l^\rho+k^\rho l^\nu\right)\left(\frac{2\,k\cdot l}{\prop{k}^3}
    +\frac{1}{\prop{k}^2}\right)
    - \frac{l^\nu l^\rho}{\prop{k}^2}
    \biggr]\,.
\end{split}
    \label{eq:delta_ggg_D_UV}
\end{align}

The ghost term in eq.~\eqref{eq:ggg_D_1_del}, proportional to $O_0^{\nu\rho}(l,q)$, leads to collinear insertions along the fermion line of the three-loop electroweak amplitude. By our definition, it therefore contributes to the function $\mathcal{I}_{qq,1}$ on the right-hand side of eq.~\eqref{eq:ggg_gloop_l_p1}, as follows,
\begin{align}
    \Delta\mathcal{I}^{\nu\rho,abc}_{qq,1}(q,k,l) = \mathcal{I}^{\nu\rho,abc}_{qq,1}(q,k,l)
    +i\gs^3\, \frac{C_A}{2}f^{abc}\frac{O_0^{\nu\rho}(l,q)}{k^2(l-k)^2}\,.
\end{align}
We represent the function $\Delta\mathcal{I}^{\nu\rho,abc}_{qq,1}$ graphically by,
\begin{align*}
    &\Delta\mathcal{I}^{\nu\rho,abc}_{qq,1}(q,k,l) =
    \eqs[0.25]{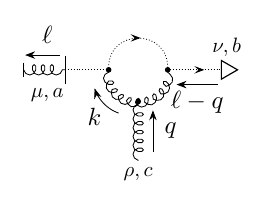}
    +\eqs[0.25]{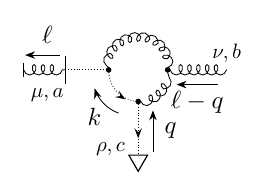}
    +\eqs[0.25]{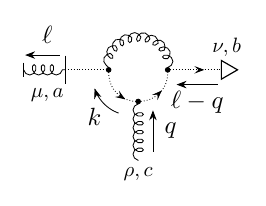}
    \\&\quad
    +\eqs[0.25]{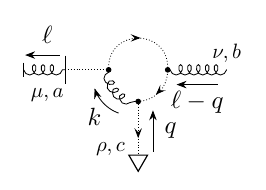}
    +\eqs[0.2]{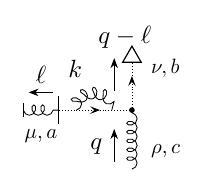}
    +\eqs[0.2]{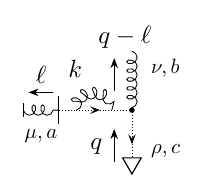}
    \,.
    \numberthis
    \label{eq:ggg_gloop_ward_I_qq}
\end{align*}
Note that this implicitly defines the unmodified function $\mathcal{I}^{\nu\rho,abc}_{qq,1}$ given by all but the last two terms.

Similarly, we add the contributions from the first term in eq.~\eqref{eq:ggg_D_1_del} to the function $\mathcal{I}_{\Delta,1}$, which defines the modified integrand,
\begin{align}
    \Delta\mathcal{I}^{\nu\rho,abc}_{\Delta,1}(q,k,l) = \mathcal{I}^{\nu\rho,abc}_{\Delta,1}(q,k,l)
    +i\gs^3\, \frac{C_A}{2}f^{abc}\frac{Q_0^{\nu\rho}(l,q)}{k^2(l-k)^2}\,,
\end{align}
and is represented by the following sum of graphs,
\begin{align*}
    &\Delta\mathcal{I}^{\nu\rho,abc}_{\Delta,1}(q,k,l) =
    \eqs[0.25]{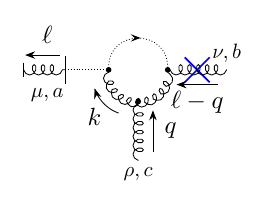}
    +\eqs[0.25]{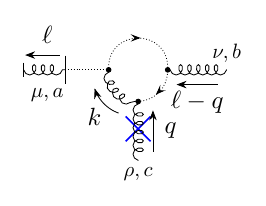}
    -\eqs[0.25]{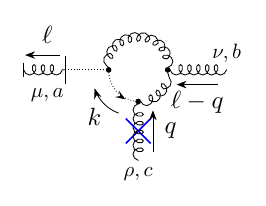}
    \\&\quad
    -\eqs[0.25]{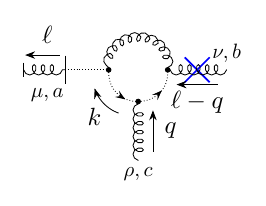}
    +\eqs[0.2]{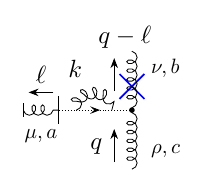}
    -\eqs[0.2]{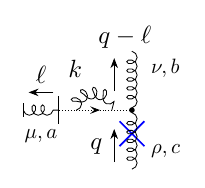}
    \,.
    \numberthis
    \label{eq:ggg_gloop_ward_I_Delta}
\end{align*}
The last two terms denote the scalar contributions proportional to $Q_0^{\nu\rho}(l,q)$ in eq.~\eqref{eq:ggg_D_1_del}. We have explicitly checked that the sum eqs.~\eqref{eq:ggg_gloop_ward_I_qq} and \eqref{eq:ggg_gloop_ward_I_Delta} agrees with the known Ward-Slavnov-Taylor identity.

Finally, the sum of ghost and gluon loop contributions to the gluon triangle is rendered finite in the single-UV region by subtracting the counterterm, 
\begin{align*}
         &\Gamma^{(1)\,\mu\nu\rho,abc}_{g\text{ single-UV}}(q,k,l)+\Gamma^{(1)\,\mu\nu\rho,abc}_{gh\text{ single-UV}}(q,k,l)
         \\&\qquad=
         -\gs^3\,\frac{C_A}{2}\,f^{abc} \frac{1}{(l-q)^2}
         \biggl[
         16(1-\e)\,k^\mu k^\nu k^\rho \left(\frac{2\, k\cdot (l+q)}{\prop{k}^4}+\frac{1}{\prop{k}^3}\right)
         \\&\qquad
         -8(1-\e)\,\frac{k^\nu k^\rho l^\mu + k^\mu k^\rho l^\nu +k^\mu k^\rho q^\nu+k^\mu k^\nu q^\rho}{\prop{k}^3}
         \\&\qquad
         +2\left(g^{\nu\rho}k^\mu+g^{\mu\rho}k^\nu+g^{\mu\nu}k^\rho\right)\left(\frac{k\cdot (l+q)}{\prop{k}^3}+\frac{1}{\prop{k}^2}\right)
         \\&\qquad
         +\frac{4(g^{\mu\nu}l^\rho+g^{\nu\rho}q^\mu)-3\left(g^{\nu\rho}l^\mu+g^{\mu\rho}l^\nu+g^{\mu\rho}q^\nu +g^{\mu\nu}q^\rho\right)}{\prop{k}^2}
         \\&\qquad
        + \frac{2\,k\cdot q\,g^{\nu\rho}k^\mu+2\,k\cdot l\,g^{\mu\nu}k^\rho}{\prop{k}^3}
         \biggr]\,.
\numberthis 
\label{eq:ggg_gloop+ghloop_single-UV}
\end{align*}

%% file: ghost.tex
\section{Local factorisation for ghost terms}
\label{sec:ghosts}

\edit{In section~\ref{sec:ward_2L} we established Ward identities for two-loop corrections to the quark-antiquark-gluon vertex, excluding gluon triangle subgraphs. In the single-collinear regions where the vertex is contracted by the longitudinal polarisation of the external gluon, the result can be expressed in terms of a difference of two-loop quark self-energy corrections plus contributions that cancel by loop momentum shifts and require shift regularisation. Consequently, for general three-loop electroweak amplitudes, scalar contributions to the Ward identities factorise up to shift-integrable Feynman graphs that can be cancelled locally through appropriate shift counterterms, c.f. section~\ref{sec:shift}. 

In this section, we extend this analysis by examining the ghost contributions that appear on the right-hand side of the two-loop QCD Ward identities, and investigate the interplay with modified gluon triangle subgraphs, discussed in section~\ref{sec:gluon_triangle}. As we will see below, ghost terms require shift regularisation and factorise locally only up to uncancelled loop polarisation terms, related to ghost self-energy corrections.}

In the previous section we developed graphical rules for the gluon triangle in the single-collinear limits. Of interest here are 1) the subgraphs that lead to collinear insertions on the fermion line, represented by the function $\Delta\mathcal{I}_{qq}$, and 2) the integrands $\Delta\mathcal{I}_\Delta$ where one of the gluon lines connecting directly to the fermion line (with momentum $l$ or $q-l$) is cancelled by Ward identities. The pictorial representations of these functions are given in eqs.~\eqref{eq:ggg_gloop_ward_I_qq} and~\eqref{eq:ggg_gloop_ward_I_Delta} for the $l\,||\,p_1$ limit. We will consider each of these contributions in turn below. We repeat that the bubble-type subgraphs $\mathcal{B}_g$, where one of the ``internal" $k$-dependent propagators is cancelled, are combined with ghost loop contributions and graphs with a quartic gluon vertex, and are regularised separately, c.f. the discussion after eq.~\eqref{eq:ggg_gloop_ward_B}.

The graphs contributing to $\Gamma^{(2)\,\mu}_{qqg}$ are all \acl{1pi}, though we will also have to consider diagrams without (two-loop) ladder structure. These involve disjoint insertions of the one-loop subgraphs on the same fermion line, satisfying the one-loop Ward identity, eq.~\eqref{eq:Ward_QCD_1L}. In particular, we are interested in diagrams with a three-gluon vertex, which we label by the superscript $(NL-3V)$,
\begin{align*}
\label{eq:qqg_NL-3V}
\numberthis
        &\gs\,\Gamma_{qqg}^{(2,NL-3V)\,\mu,c}(q,k,l) = 
        \eqs[0.3]{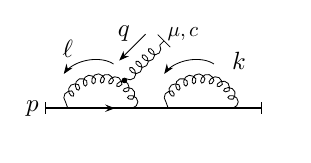}
        +\eqs[0.3]{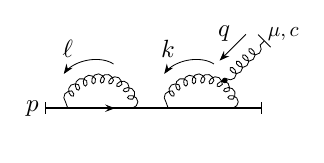}
        \\&\qquad =
        \gs^5\,
        \frac{C_AC_F}{2}\,\tq{c}
        \left[
        S^{(1)}(p+q,k)
        \frac{i}{\s{p}+\s{q}}
        W^{(1)\,\mu}(p,q,l)
        +W^{(1)\,\mu}(p,q,k)\frac{i}{\s{p}}S^{(1)}(p,l)
        \right]\,,
\end{align*}
where $S^{(1)}$
and $W^{(1)\,\mu}=Q^{(1)\,\mu}+O^{(1)\,\mu}$ were defined in eqs.~\eqref{eq:qq_1L_S}
and~\eqref{eq:qqg_1L_W}, respectively. 
We define the ghost contributions to $\Gamma_{qqg}^{(2,NL-3V)\,\mu}$ to be,
\begin{align}
\begin{split}
    O^{(2,NL-3V)\,\mu}(p,q,k,l) &=
    S^{(1)}(p+q,k)
    \frac{i}{\s{p}+\s{q}}
    O^{(1)\,\mu}(p,q,l)
    \\&
    +O^{(1)\,\mu}(p,q,k)\frac{i}{\s{p}}S^{(1)}(p,l)\,,
\end{split}
\label{eq:qqg_NL-3V_O}
\end{align}
where we have extracted the terms proportional to $O^{(1)\,\mu}$ in eq.~\eqref{eq:qqg_NL-3V}.

We let $O^{(2)\,\mu}$ be the sum of ghost terms from two-loop corrections to the fermion propagator involving one or more three-gluon vertices (except the gluon three-point function), defined in appendix~\ref{sec:Greens_2L}. We find it convenient to decompose $O^{(2)\,\mu}$ in terms of two colour coefficients as follows,
\begin{align}
\begin{split}
    &O^{(2)\,\mu,c}(p,q,k,l) = 
    \gs^5\tq{c} \frac{i}{\s{p}+\s{q}}
    \biggl[\frac{C_AC_F}{2}
    \biggl(O^{(2,NL-3V)\,\mu}+O^{(2,UL)\,\mu}
    +O^{(2,XL)\,\mu}\biggr)
    \\&\qquad
    +\frac{C_A^2}{4}\biggl(
    O^{(2,d3V)\,\mu}
    -O_2^{(2,UL)\,\mu} 
    - O^{(2,XL)\,\mu}
    \biggr)
    \biggr]
    \frac{i}{\s{p}}\,,
\end{split}
\label{eq:ghost_2L_sum}
\end{align}
where for readability we have suppressed the kinematic dependence on the right-hand side. Here, we consider the limit where the gluon with loop momentum $q$ becomes collinear to the incoming anti-quark and apply the approximation of eq.~\eqref{eq:g_k_p2}. \edit{Factorisation is guaranteed for both single-collinear regions, except for shift-integrable contributions that can be cancelled using local counterterms, as well as loop polarisation terms which we will discuss at the end of this section.} %

We contract eq.~\eqref{eq:ghost_2L_sum} with a longitudinal polarised gluon momentum $q_\mu$, and apply the QCD Ward identities, eqs.~\eqref{eq:qq_Ward},~\eqref{eq:gg_Ward} and~\eqref{eq:ghosts_ward_1L}, which yields,
\begin{align}
    q_\mu O^{(2)\,\mu,c} = \mathcal{O}^{(2)\,c}_2 - O^{(2)\,c\,\,\text{shift}}_{2} + O^{(2)\,c}_{\Delta,2}
    +O^{(2)\,c}_{NF,2}\,.
    \label{eq:ghost_2L}
\end{align}
Below, we explain each term on the right-hand side in turn, while their explicit derivation is provided in appendix~\ref{app:ghost}. 

The function $\mathcal{O}^{(2)\,c}_2$ is analogous to the ghost contribution to the Ward identity at one loop, c.f. {eqs.~\eqref{eq:ghosts_ward_1L}~-~\eqref{eq:qqg_1L_ward_ghost_id_2}}, consisting of terms where one of the outermost fermion propagators has been cancelled. The resulting expression, which is quite lengthy, is provided in diagrammatic form in eq.~\eqref{eq:ghost_2L_ext}. \edit{We analyse the remaining terms on the right-hand side of eq.~\eqref{eq:ghost_2L} in more detail, since \textit{a priori} they constitute an obstacle to local factorisation.}

In ref.~\cite{Anastasiou:2022eym} it was shown that in the single-collinear regions ghost contributions factorise independently from the two-loop electroweak amplitude, without extra modifications to the integrand. At three-loop order this is no longer the case, since we have additional non-factorisable shift contributions also for ghost terms, given by,
\begin{align*}
    &O^{(2)\,c\,\,\text{shift}}_{2}(p,q,k,l) 
    =
    i\gs^5 \,\frac{C_A^2}{4}\,\tq{c}\,
    \left[
    \s{l}(\s{p}+\s{l})\left(S^{(1)}(p+l,k)-S^{(1)}(p+l,k-l)\right) \right.
    \\&\qquad +\left.
    \left(S^{(1)}(p+l,k)-S^{(1)}(p+l,k+q-l)\right)
    (\s{p}+\s{l})(\s{q}-\s{l})
    \right]
    \frac{1}{l^2(l-q)^2(p+l)^2}
    \\&\qquad\equiv
    \frac{C_A}{2C_F}\Biggl[
    \Biggl(\eqs[0.25]{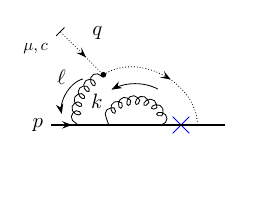}
    -(k \to k+q-l)\Biggr)
    \\&\qquad
    -\Biggl(
    \eqs[0.25]{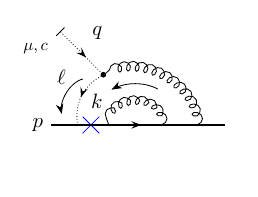}
    -(k\to k-l)\Biggr)
    \Biggr]\,.
\numberthis
\label{eq:ghost_2L_shift_2}
\end{align*}
Clearly, $O^{(2)\,c\,\,\text{shift}}_{2}$ vanishes after integrating out the one-loop fermion self-energy subgraph, i.e. it is one-loop shift-integrable in the variable $k$.
However, as for the scalar shift contributions to the Ward identity, this leads to non-factorisable graphs at the amplitude level and need to be subtracted in the \acs{ffs} scheme. \edit{We note that the analogous shift mismatch in the $l\,||\,p_1$ limit for self energy corrections to the antiquark is related by complex conjugation.}

The term $O^{(2)\,c}_{\Delta,2}$ in eq.~\eqref{eq:ghost_2L_sum} has the diagrammatic form,
\begin{align*}
    &O^{(2)\,c}_{\Delta,2}(p,q,k,l) = 
    \frac{C_A}{2C_F}\Biggl[\eqs[0.2]{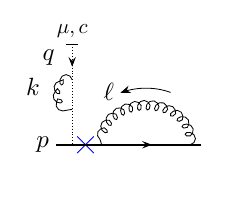}-\eqs[0.2]{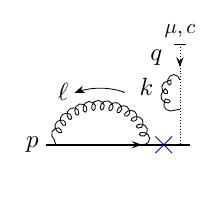}
    \Biggr]
    +\eqs[0.23]{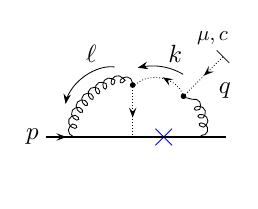}
    \\&\qquad
    -\eqs[0.23]{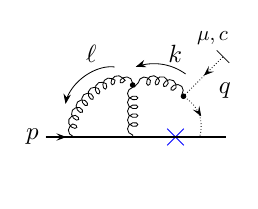}
    +\eqs[0.23]{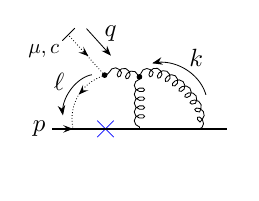}
    -\eqs[0.23]{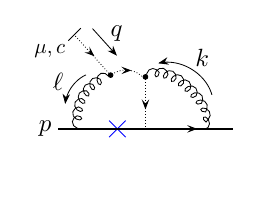}\,.
    \numberthis
\label{eq:ghost_2L_Delta}
\end{align*}
In appendix~\ref{app:ghost} we show that the contributions of the first two graphs proportional to the colour factor $C_AC_F/2$ cancel against similar graphs from the uncrossed ladder topology, and the remainder is proportional to $C_A^2/4$. Therefore, we multiply this combination by the non-standard colour factor $C_A/2C_F$. Equation~\eqref{eq:ghost_2L_Delta} combines with the $\mathcal{I}_{\Delta}$ contributions from the gluon three-point function, defined in eq.~\eqref{eq:ggg_gloop_ward_I_Delta}. Their sum cancels up to \textit{additional} non-factorisable one-loop shift-integrable terms, which we denote by $O^{(2)\,c\,\,\text{shift}}_{\Delta,2}$,
\begin{align}
    \label{eq:ghost_delta_shift_2}
    \begin{split}
    &O^{(2)\,c\,\,\text{shift}}_{\Delta,2}(p,q,k,l) =
    -\left[
    O^{(2)\,c}_{\Delta,2}(p,q,k,l)
    -\frac{C_A}{2}\,\Delta\mathcal{I}_{\Delta,2}^{\alpha\beta}(q,k,l)
    \frac{\gamma_\beta (\s{p}+\s{l})\gamma_\alpha}{(p+l)^2}
    \right]
    \\&\qquad =
    i\gs^5\,\frac{C_A^2}{4}\,\tq{c}\,
    \left[
    \frac{(\s{l}-\s{k})(\s{p}+\s{k})(\s{l}-\s{q})}{l^2(q-k)^2(l-k)^2(l-q)^2(p+k)^2}
    -\frac{S^{(1)}(p+q,k-q)}{l^2(l-q)^2}
    \right.\\&\qquad - \left.
    \frac{l_\mu W^{(1)\,\mu}(p+l,q-l,k-l)}{l^2(l-q)^2}
    \right]- (k\to k+q)
    \\&\qquad \equiv
    \frac{C_A}{2C_F}\eqs[0.2]{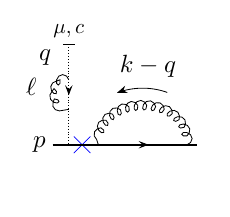}
    +\eqs[0.23]{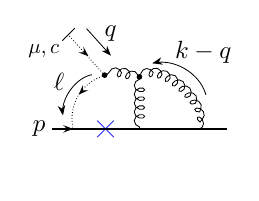}
    -\eqs[0.23]{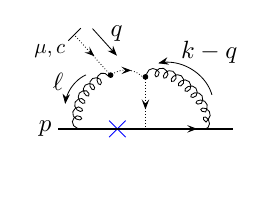}
   \\&\qquad 
   - (k\to k+q)\,.
   \end{split}
\end{align}

For the general electroweak amplitude, we again require a prescription to remove non-factorised ghost terms that cancel by loop momentum shifts. \edit{However, this do not seem to admit a representation in terms of regular Feynman diagrams in analogy to the discussion in section~\ref{sec:shift}.}
Instead, we remove ghost shift terms directly using the following counterterm,
\begin{align*}
    &\delta_{\text{shift},1}^{(3,O)}(p_1,p_2,q,k,l;
    \{q_1,\ldots,q_n\}) =
    \Biggl\{
    \Biggl[
    \eqs[0.27]{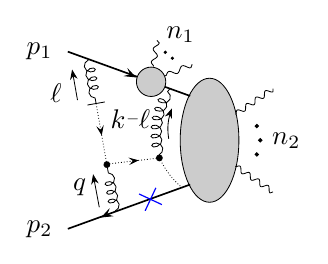}
    -\eqs[0.27]{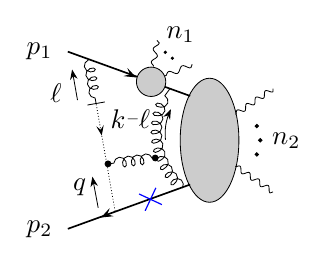} 
    \\&\quad 
     -\frac{C_A}{2C_F}
    \eqs[0.27]{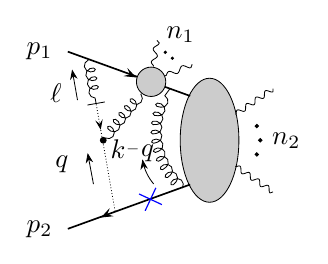}
    \Biggr] 
    - (k\to k+q)\Biggr\}
    -\Biggl(\frac{C_A}{2C_F}\eqs[0.27]{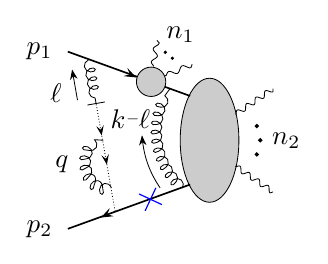}-(k\to k+l)\Biggr)
    \\&\quad 
    + \Biggl(\eqs[0.27]{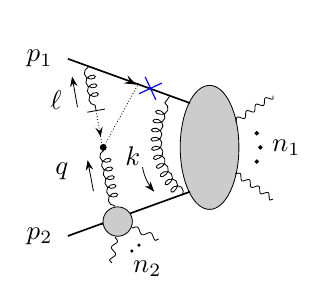}
    - (k\to l-q-k)     \Biggr)\,,
    \numberthis
    \label{eq:shift_O}
\end{align*}
which is the last remaining contribution to eq.~\eqref{eq:M3_shift_CNT}.
Again, the corresponding counterterm for the $q\,||\,p_2$ limit is analogous and we do not show it here. No fermion mass prescription (c.f. eqs.~\eqref{eq:shift_mod_quark} and~\eqref{eq:ggg_mod_quark}) is required in this case.

Next, we investigate the subgraphs contributing to $\mathcal{I}_{qq}$. It can be shown that the two graphs on the last line of eq.~\eqref{eq:ggg_gloop_ward_I_qq}, which correspond to the ghost contributions to the identity in eq.~\eqref{eq:ggg_D_1_del}, factorise independently when we sum over all such diagrams contributing to the relevant single-collinear region. As subgraphs of the two-loop corrections to the $q\bar{q}g$-vertex their contributions read,
\begin{align*}
        &-i \gs^5\frac{C_A^2}{4}\tq{c}\,
        \frac{O_0^{\mu\nu}(q,l)}{k^2(q-k)^2}
        \frac{\gamma_\beta (\s{p}+\s{l})\gamma_\alpha}{(p+l)^2}
        \equiv
        \eqs[0.2]{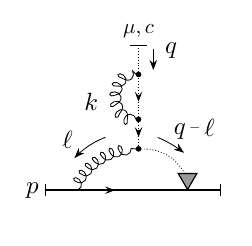}
        +\eqs[0.2]{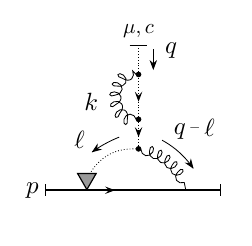}
        \numberthis
        \label{eq:qqqg_ggg_D_ward}
        \\&\quad
        =\eqs[0.2]{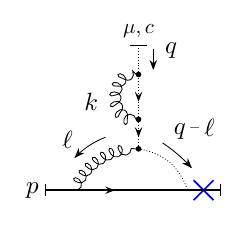}
        -\eqs[0.2]{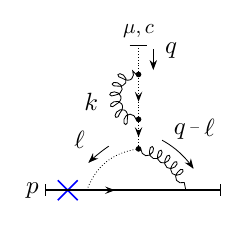}
        +\eqs[0.2]{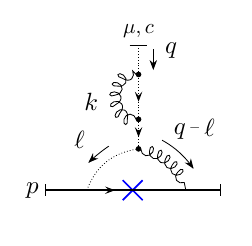}
        -\eqs[0.2]{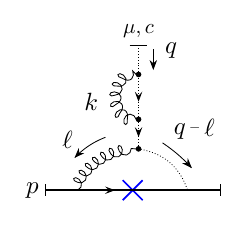}\,.
\end{align*}
Each collinear insertion leads to two terms according to the abelian Ward identity, eq.~\eqref{eq:qq_Ward}. We remark that the sum of graphs in eq.~\eqref{eq:qqqg_ggg_D_ward} where the fermion propagator $i/(\s{p}+\s{l})$ is cancelled is proportional to $\s{q}$, and therefore equivalent to the contraction of the scalar polarised gluon momentum with the $q\bar{q}g$-vertex. This is in complete analogy to the one-loop bubble identity, eq.~\eqref{eq:qqg_1L_ward_ghost_id}. To be specific, we obtain the relation,
\begin{align}
\label{eq:qqqg_ggg_D_ward_id}
\begin{split}
    &\eqs[0.2]{ggg_D_ward_RHS_fig3}
    -\eqs[0.2]{ggg_D_ward_RHS_fig2}
    = i\gs^5 \frac{C_A^2}{4}\tq{c}\,
    \frac{\s{q}}{k^2l^2(q-k)^2(l-q)^2}
    \\&\qquad
    =
    \frac{1}{2}\Bigg[
    \eqs[0.2]{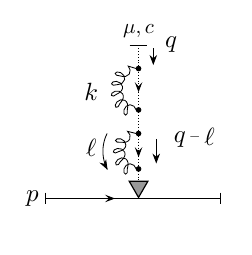}
    +\eqs[0.2]{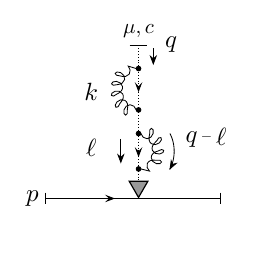}
    \Bigg]
    \equiv
    \eqs[0.2]{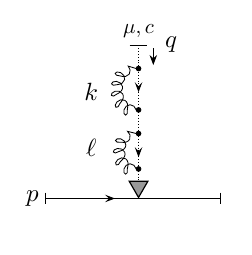}
\end{split}
\end{align}
This consists of a two-fold symmetrisation of the bubble integrands over their loop momentum flows, $k \to q-k$ and $\ell \to q-\ell$. 
This identity leads to a factorised integrand at the amplitude-level.  Collinear insertions on either side of an electroweak vertex lead to pairwise cancellations of singular integrands. The underlying mechanism of local factorisation is therefore the same as for the one-loop amplitude and we will not explain this further.

\subsection{Loop polarisations in ghost terms}

Next, consider the class of diagrams discussed in section~\ref{sec:shift_jet}, involving a (modified) one-loop $p_1$ jet function in the region $(1_l,H_k,H_q)$ with an outgoing longitudinal gluon with momentum $l$, connecting to a triple-gluon vertex. Factorisation of ghost terms follows directly from application of the one-loop Ward identities, eqs.~\eqref{eq:gg_Ward} and~\eqref{eq:qqg_1L_ward_ghost_id}, and is, up to the quark jet function, equivalent to factorisation of ghost terms for the two-loop amplitude (c.f. section~5.3 in ref.~\cite{Anastasiou:2022eym}),
\begin{align*}
    &\left.\lim_{l \to -z_{1,l}p_1}\, \frac{-ig_{\mu\nu}}{l^2}\,v(p_2)\widetilde{\mathcal{M}}^{(1)\,\mu,c}(p_1+l,p_2,q;\{q_1,\ldots,q_n\})
    \frac{(\s{p}_1+\s{l})}{(p_1+l)^2} \Delta \mathcal{J}_1^{(1)\,\nu,c} (p_1,l,k)
    u(p_1)\right|_{\text{ghost}} 
    \\&\quad
    =
    -\eqs[0.25]{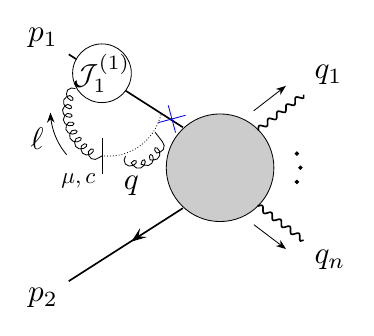}
    -\eqs[0.25]{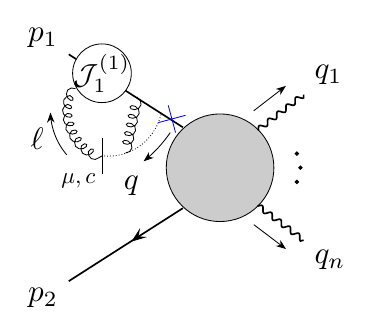}
    +\text{ finite}\,.
    \numberthis
\end{align*}
Terms where the $1/(\s{p}_1+\s{l})$ quark propagator is cancelled are not divergent in the $l\,||\,p_1$ limit and are not explicitly shown on the right-hand side. Modifications to the one-loop quark jet function, as summarised in section~\ref{sec:LP_1L}, eliminate non-factorising loop polarisation terms from the jet subgraph. %

\begin{figure}[h]
\centering
\begin{align*}
    \eqs[0.2]{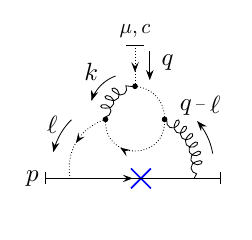}
    -\eqs[0.2]{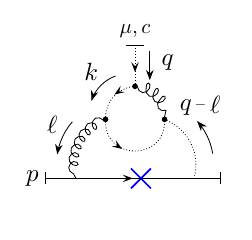}
    +\eqs[0.2]{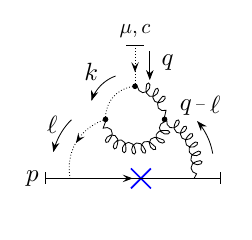}
    -\eqs[0.2]{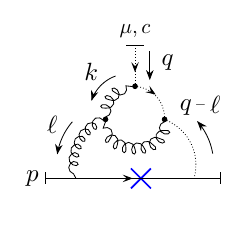}
\end{align*}
\caption{Subgraphs $\Gamma^{(2)\,c}_{NF,2}$ originating from the $q\,||\,p_2$ Ward identity that yield locally non-factorisable loop polarisation terms.
}
\label{fig:ggg_NF}
\end{figure}

Now, we analyse the remaining contributions to eq.~\eqref{eq:ggg_gloop_ward_I_qq}, which factorise up to loop polarisation terms that cannot be removed by symmetric integration. The problematic subgraphs in the $q\,||\,p_2$ region are shown in figure~\ref{fig:ggg_NF}, and we denote the corresponding integrand by $\Gamma^{(2)\,c}_{NF,2}$. Explicitly, we have,
\begin{align*}
\label{eq:qqg_ggg_NF}
\numberthis
    &\Gamma^{(2)\,c}_{NF,2}(p,q,k,l)
    = i\gs^5\,\frac{C_A^2}{4}\,\tq{c}\, \frac{1}{(l-k)^2}\frac{1}{2}\Bigg[
    \frac{4\s{k}-\s{q}}{l^2k^2(q-k)^2}
    -\frac{2(\s{l}-\s{q})}{l^2k^2(l-q)^2}
    \\&\quad
    +\frac{3\s{q}-4\s{k}}{k^2(q-k)^2(l-q)^2}
    +\frac{2\s{l}}{l^2(q-k)^2(l-q)^2}
    -\frac{\s{q}q^2}{l^2k^2(q-k)^2(l-q)^2}
    \Bigg]
    \,.
\end{align*}
To obtain the right-hand side we have written scalar products in the numerator in terms of the inverse propagators of the triangle integrand. The last term in square brackets is of $\mathcal{O}(q^2)$ and therefore finite in the $q\,||\,p_2$ limit, which means we can safely ignore it. Terms proportional to $\s{q}$ are unproblematic, since they are equivalent to a contraction of the longitudinal polarisation vector $q^\mu$ at the $q\bar{q}g$ vertex and factorise in analogy to the identity~\eqref{eq:qqg_1L_ward_ghost_id}. However, terms proportional to $\s{l}$ and $\s{k}$ spoil local factorisation since they correspond to an arbitrary polarisation.

Similarly non-factorising subgraphs appear on the right-hand side of eq.~\eqref{eq:ghost_2L}, denoted by $O^{(2)\,c}_{NF,2}$ and defined by,
\begin{align}
\begin{split}
    &O^{(2)\,c}_{NF,2}(p,q,k,l)\equiv\eqs[0.2]{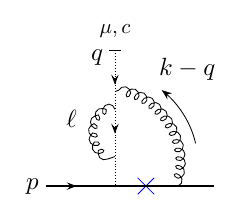}
    -\eqs[0.2]{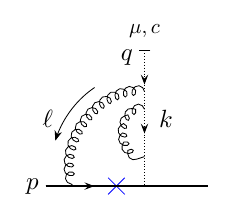}
    \\&\qquad= i\gs^3\,\frac{C_A}{2}\,\tq{c}
    \left[\frac{(\s{q}-\s{l})\,\widetilde{\Pi}(q-l,-k)}{l^2(l-q)^2}
    +\frac{\s{k}\,\widetilde{\Pi}(k,l)}{k^2(q-k)^2}\right]\,,
\end{split}
\label{eq:ghost_NF}
\end{align}
where the symmetrised ghost self-energy $\widetilde{\Pi}$ was defined in eq.~\eqref{eq:ghost_self_energy}. Again, these graphs constitute locally non-factorisable contributions and cannot be removed using symmetric integration due to different momentum dependence in the ghost self-energy subgraphs. \edit{Shifting the first term in square brackets in eq.~\eqref{eq:ghost_NF} by $k\to k-q$ and combining eqs.~\eqref{eq:ghost_NF} and~\eqref{eq:qqg_ggg_NF} one can show that the uncancelled non-factorising part is proportional to a difference of two bubble integrands. We note that these terms also spoil collinear factorisation in the double-UV region, where $k$ and $l$ correspond to large loop momenta but $q$ has scalar polarisation.}

%% file: app_framework.tex
\section{Collinear approximations}
\label{app:framework}
To show local factorisation, we perform a light-cone approximation of the metric tensor with respect to the direction specified by the momentum $k^\mu$ of the virtual electroweak boson or gluon, which becomes on-shell on the collinear \ac{ps} \cite{Anastasiou:2020sdt},
\begin{equation}
    g^{\mu \nu} =\frac{k^\mu \eta_i^\nu}{k \cdot \eta_i} +\frac{k^\nu\eta_i^\mu}{k \cdot \eta_i}   
    -\frac{\eta_i^2 k^\mu k^\nu
    + k^2 \eta_i^\mu \eta_i^\nu}{(k \cdot \eta_i)^2}
    + g_\perp^{\mu \nu}\,, \quad i=1,2\,,
    \label{eq:metric_decomp}
\end{equation}
where $\eta_i$ is an auxiliary vector chosen to have a large rapidity separation from $p_1$ ($p_2$) in the collinear limit $k\,||\,p_1$ ($k\,||\,p_2$), with $g_\perp^{\mu\nu}k_\nu =g_\perp^{\mu\nu}\eta_{i\,\nu}=0$, to avoid producing additional pinches. Since the virtual momentum goes on-shell in either collinear limit, we can drop the term proportional to $k^2=0$ in eq.~\eqref{eq:metric_decomp} without affecting the collinear behavior of the amplitude. In ref.~\cite{Anastasiou:2020sdt} we found it is useful to introduce an alternative version of eq.~\eqref{eq:metric_decomp} containing quadratic, instead of linear, denominators\footnote{The choice $\eta_1 = p_2$ ($\eta_2=p_1$) allows us to combine the integrands valid in the collinear regions $k\,||\,p_1$ ($k\,||\,p_2$) with the soft approximation into a single form factor subtraction term, c.f. sec.~3.1 of ref.~\cite{Anastasiou:2020sdt}.},
\begin{equation}
    g^{\mu \nu} =\frac{2k^\mu \eta_i^\nu}{d_i(k,\eta_i)} +\frac{2k^\nu\eta_i^\mu}{d_i(k,\eta_i)}   
    -\frac{4\eta_i^2 k^\mu k^\nu}{d_i^2(k,\eta_i)}
    + g_\perp^{\mu \nu}\,, \quad i=1,2\,,
    \label{eq:metric_decomp_2}
\end{equation}
with
\begin{align}
    d_1(k,\eta_1) = -(k-\eta_1)^2 +\eta_1^2\,, \quad
    d_2(k,\eta_2) = (k+\eta_2)^2 -\eta_2^2\,.
\end{align}
We note that eqs.~\eqref{eq:metric_decomp} and \eqref{eq:metric_decomp_2} agree in the strict collinear limits, where $k^2=0$. Similarly, for the jet-line momentum $k$ we perform a light-cone decomposition with respect to the collinear directions specified by the external momenta $p_1$ and $p_2$,
\begin{align}
    k^\mu = z_{i,k} \,p_i^\mu + \K{i}^\mu + \beta_{i,k} \,\eta_i^{\prime\mu}\,, \quad 0 < z_{i,k} < 1\,, \quad i=1,2\,,
    \label{eq:loop_lightcone}
\end{align}
with
\begin{align}
    z_{i,k} = \frac{k \cdot \eta^\prime_i}{\eta^\prime_i \cdot p_i}\,, \quad
    \beta_{i,k} = \frac{k \cdot p_i}{\eta^\prime_i \cdot p_i}\,.
\end{align}
Above, $\eta^\prime_i$ is an auxiliary light-like vector, $z_i$ is the momentum fraction of the jet-line in the direction of the external on-shell momentum $p_i$ and $\K{i}$ determines the transverse distance from the pinch singularity, $\K{i}\cdot \eta^\prime_i = 0$ and $\K{i}\cdot p_i^\mu=0$. We note that in general $\eta^\prime_i$ is not related to the auxiliary vector $\eta_i$ in eq.~\eqref{eq:metric_decomp} although in practice we set $\eta^\prime_1 =\eta_1 = p_2$ and $\eta^\prime_2 = \eta_2 = p_1$.

To identify the strength of the pinch singularities, we perform some elementary power counting by rescaling the components of soft and collinear momenta by a dimensionless parameter $\lambda$ and extracting the leading divergent behaviour of the amplitude in the limit $\lambda \to 0$. Soft lines have momenta that are vanishing in all four components, and scale as,
\begin{align}
    \ell_j \sim \lambda Q\,,
\end{align}
where $Q = \sqrt{s_{12}}$. Since a pinch surface $S$ is specified by algebraic conditions on loop momenta, it can be understood as an algebraic surface embedded within the loop momentum space of the Feynman graph~\cite{sterman_1993}. Therefore, it is useful to distinguish between \emph{intrinsic coordinates}, which parametrise the pinch surface, and \emph{normal coordinates}, which determine deviations from the pinch surface. For a soft divergence, the \ac{ps} has dimension zero, i.e. is given by a point in loop momentum space, and the normal coordinates are given by the components $\ell_j^\mu$. The components $z_{i,k}\,p_i^\mu$ correspond to points on the collinear \ac{ps}. The jet line components scale as,
\begin{align}
    \beta_{i,k} \sim \lambda Q\,, \quad \K{i} \sim \sqrt{\lambda} Q\,.
\end{align}

To leading order in the limit $k\,||\,p_i$, we can approximate any vector $v^\mu(p_i,z_{i,k})$ in the direction of $p_i$ by,
\begin{align}
    v^\mu(p_i,z_{i,k}) \sim \frac{v\cdot \eta_i}{d_i(k,\eta_i)}k^\mu = f(z_{i,k})p_i^\mu +\mathcal{O}{(\sqrt{\lambda})}\,.
    \label{eq:vector_decomp}
\end{align}
Again, we have replaced linear denominators for quadratic ones, $\eta_i \cdot k \to d_i$.
Together, eqs.~\eqref{eq:metric_decomp_2} and~\eqref{eq:vector_decomp} produce the correct behaviour in the corresponding collinear limits $k\,||\,p_1$ or $k\,||\,p_2$. 

%% file: app_jet.tex
\section{Counterterms for the one-loop jet function}
\label{app:jet}
At one-loop, the quark jet function reads,
\begin{align*}
    &\mathcal{J}_1^{(1)\,\mu,c}(p_1,k,l)u(p_1) = 
    \eqs[0.23]{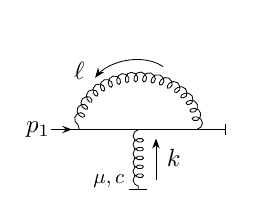}
    +\eqs[0.23]{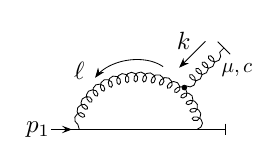}
    +\eqs[0.23]{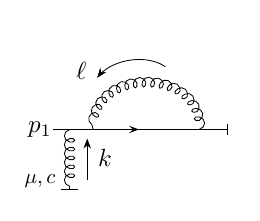}
    \\&\qquad
    = %
    \gs^3\,\tq{c}\bigg[\left(\frac{C_A}{2}-C_F\right)V^{(1)\,\mu}(p_1,k,l)
    +\frac{C_A}{2}W^{(1)\,\mu}(p_1,k,l) 
    \label{eq:jet_p1}
    \numberthis
    \\&\qquad 
    + C_F N_S^{(1)}(p_1+k,l)\gamma^\mu
    \bigg] u(p_1)\,,
\end{align*}
where we define $N_S^{(1)}$ to include the adjacent quark propagator,
\begin{align}
    N_S^{(1)}(p_1+k,l) = S^{(1)}(p_1+k,l)\frac{\s{p}_1+\s{k}}{(p_1+k)^2}
    \,.
\end{align}
In the definition of the one-loop jet function we exclude the quark propagator with momentum $p_1+k$. The one-loop two- and three-point functions $V^{(1)\,\mu}$, $W^{(1)\,\mu}$ and $S^{(1)}$ were defined in eqs.~\eqref{eq:qqg_1L_V},~\eqref{eq:qqg_1L_W} and~\eqref{eq:qq_1L_S}, respectively. At the integrand level, the quark self energy $N_S^{(1)}(p_1+k,l)$ has an enhanced collinear divergence in the region $(1_k,H_l)$ due to an additional power of the quark propagator with momentum $p_1+k$. Though this integrates to factorisable form, it spoils manifest local factorisation. As a consequence we can no longer use the collinear approximation of eq.~\eqref{eq:g_k_p1}, which is only valid at leading order in the collinear expansion, eq.~\eqref{eq:loop_lightcone}. As has been shown in ref.~\cite{Anastasiou:2022eym} this can be remedied by making the replacement,
\begin{align}
    N_S^{(1)}(p_1+k,l) \to \Delta N_S^{(1)}(p_1+k,l) 
    = \frac{(1-\e)}{(p_1+k+l)^2l^2}\,.
    \label{eq:S_mod}
\end{align}
where 
\begin{align}
    \int_l\, \left[N_S^{(1)}(p_1+k,l)-\Delta N_S^{(1)}(p_1+k,l)\right] =0\,.
\end{align}
The modified quark self-energy $\Delta N_S^{(1)}$ is obtained by symmetrising under the loop-momentum exchange $l \to -(p_1+k+l)$, which is equivalent to adding a counterterm 
$\delta_{S,1}$,
\begin{align}
    \Delta N_S^{(1)}(p_1+k,l) = 
    N_S^{(1)}(p_1+k,l)+\delta_{S,1}(p_1,k,l)\,,
\end{align}
with
\begin{align}
    \delta_{S,1}(p_1,k,l)
    = -(1-\e)\frac{(\s{p}_1+\s{k}+2\s{l})(\s{p}_1+\s{k})}{(p_1+k+l)^2l^2(p_1+k)^2}\,, \quad 
    \int_l\, 
    \delta_{S,1}
    =0\,.
\end{align}
Adopting the language of ref.~\cite{Anastasiou:2022eym}, we shall refer such modifications as \textit{symmetric integration}. Both the original and the modified quark self-energy function is finite in the region $(1_l,H_k)$. 

The modified quark jet function, which cancels the $k\,||\,p_1$ singularity and is free of loop polarisations in this region, reads,
\begin{align}
\begin{split}
    &\Delta \mathcal{J}_1^{(1)\,\mu,c}(p_1,k,l) =
    \left.\left[\mathcal{J}_1^{(1)\,\mu,c} + C_F\,\delta_{S,1}\gamma^\mu
    +\delta_{\perp,1}^{\mu,c}+\delta_{\mathcal{J},1}^{\mu,c}\right]\right.|_{\eta^\prime_1=p_2}\,,
\end{split}
    \label{eq:jet_mod_p1}
\end{align}
where for legibility we have suppressed the kinematic dependence of the functions on the right-hand side. The local infrared counterterms $\delta_{\perp,1}$ and $\delta_{\mathcal{J},1}$ remove loop polarisation terms from the one-loop jet integrand with,
\begin{align}
    \begin{split}
        &\delta_{\perp,1}^{\mu,c}(p_1,k,l) 
        = -\gs^2\,\tq{c}\,\frac{2(1-\e)}{l^2(p_1+l)^2}\s{l}_\perp
        \bigg[C_F\,\frac{(p_1+l)^\mu}{(p_1+k+l)^2}\\&\qquad
        -\frac{C_A}{2}\,\frac{2l^\mu}{(l-k)^2}
        +(l\to \tilde{l}\,)\bigg]\,,
    \end{split}
\label{eq:jet_1L_delperp}
\end{align}
and\footnote{Due to a sign error in the first line of eq.~(4.33) of ref.~\cite{Anastasiou:2022eym} the term $\delta\mathcal{J}^\mu(p_1,k,-l-p_1)$ multiplying $C_A$ in eq.~\eqref{eq:jet_1L_delJ} has opposite sign.}
\begin{align}
    \begin{split}
        &\delta_{\mathcal{J},1}^{\mu,c}(p_1,k,l) = \gs^3\,\tq{c}\,
        \bigg[\frac{C_A}{2}\left[{V}_k(k,-l-p_1) -{V}_k(k,l)\right]
        \\&\qquad+C_A\,\delta\mathcal{J}^\mu(p_1,k,-l-p_1) -C_F\,\delta\mathcal{J}^\mu(p_1,k,l)\bigg]\,,
    \end{split}
    \label{eq:jet_1L_delJ}
\end{align}
where
\begin{align}
    \delta\mathcal{J}^\mu(p_1,k,l)= \frac{2(1-\e)}{(p_1+l+k)^2}\left(
    \frac{2l^\mu+p_1^\mu+k^\mu}{l^2}-\frac{2(p_1+l)^\mu+k^\mu}{(p_1+l)^2}
    \right)\frac{\s{\eta}^\prime_1}{2p_1\cdot \eta^\prime_1}\,,
\end{align}
and $V_k^{(1)\,\mu}$ is the part of the the one-loop vertex function $V^{(1)}$, defined in eq.~\eqref{eq:QCD_vertex_1L_UV}, that produces a divergence when $l$ is hard and $k$ becomes collinear to $p_1$,
\begin{align}
    V_k^{(1)}(p_1,k,l)= -\frac{2(1-\e)}{(p_1+l+k)^2}\left[\frac{2(p_1+l)^\mu\s{l}}{l^2(p_1+l^2)}-\frac{\gamma^\mu}{l^2}\right]\,.
\end{align}
Here, we (implicitly) perform a light-cone decomposition of the loop momentum $l^\mu$ and the Dirac matrix $\s{l}$ in the $p_1$-direction according to eq.~\eqref{eq:loop_lightcone}. Finiteness in the region $(1_k,H_l)$ can be shown by decomposing also the Dirac matrix $\gamma^\mu$ into light-cone components as follows,
\begin{align}
    \gamma^\mu = \frac{p_1^\mu \s{\eta}^\prime_1}{p_1\cdot \eta^\prime_1} + \frac{\eta_1^{\prime\mu} \s{p}_1}{p_1\cdot \eta^\prime_1}+\gamma_\perp^\mu\,,
\end{align}
with $\gamma_\perp^\mu\,p_{1\,\mu} =\gamma_\perp^\mu\,\eta_{1\,\mu}^\prime =0$. 
The vector $\tilde{l}^\mu$ on the right-hand side of eq.~\eqref{eq:jet_1L_delperp} is equivalent to $l^\mu$ up to a reflection on the transverse plane $l_\perp \to -l_\perp$. The choice $\eta_1^\prime=p_2$ guarantees finiteness in the region $(2_k,H_l)$.

Counterterms for the $p_1$ jet function do not affect the $l\,||\,p_1$ region, which factorises (up to shift-integrable terms) and whose singularity is removed by a form-factor counterterm. Diagrams where the gluon line with momentum $k$ directly attaches to the incoming anti-quark leg also exhibit a collinear divergence in the limit $k\,||\,p_2$. Application of the QCD Ward identities on the (UV-regulated) quark jet integrand in the region $(2_k,H_l)$ produces a shift-mismatch, proportional to a (renormalised) on-shell one-loop self-energy correction. Though this shift-mismatch is equivalent to a scaleless integral which vanishes in dimensional regularisation, it spoils local factorisation. 
By construction, the jet counterterms eliminate this shift-mismatch locally, i.e. the modified jet integrand vanishes when contracted by $k^\mu \simeq z_{k,2}p_2^\mu$, except for non-abelian terms $O^\mu$ which generate the whole singularity in the region $(2_k,H_l)$,
\begin{align}
\begin{split}
    &k_\mu\,\Delta \mathcal{J}_1^{(1)\,\mu,c}(p_1,k,l)\equiv \eqs[0.25]{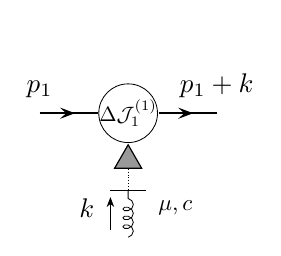} 
    \\&\qquad 
    \underset{k\,\simeq\, z_{k,2}p_2}{\to} 
    \eqs[0.24]{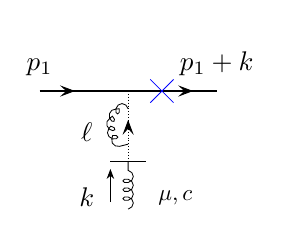}
    +\eqs[0.24]{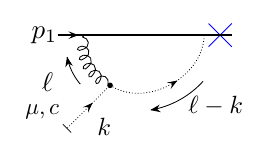}\,.
\end{split}
    \label{eq:jet_mod_p1_ward}
\end{align}
As has been shown in ref.~\cite{Anastasiou:2022eym} the sum of all ghost terms factorise in the two-loop amplitude.

For the analogous case for the $p_2$ jet function, which exhibits loop polarisation terms in the region $(2_k,H_l)$, the counterterms are obtained by a change of variables $p_1 \to -p_2$ and $\eta^\prime_1\to \eta^\prime_2$,
\begin{align}
    \Delta \mathcal{J}_2^{(1)\,\mu,c}(p_2,k,l) =
    \left.\Delta \mathcal{J}_1^{(1)\,\mu,c}(-p_2,k,l)\right|_{\eta^\prime_2=p_1}\,.
    \label{eq:jet_mod_p2}
\end{align}

%% file: app_greens.tex
\section{Two-loop two- and three-point Green's functions}
\label{sec:Greens_2L}
The uncrossed ladder contribution to the two-loop \acs{qcd} corrections to the quark self-energy, eq.~\eqref{eq:qq_QCD_2L} reads,
\begin{align}
    \Pi_{qq}^{(2,UL)} = \eqs[0.25]{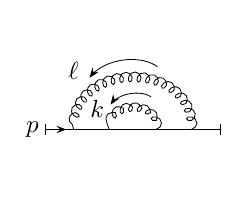} = i\gs^4\, C_F^2 S^{(2,UL)}(p,k,l)\,,
    \label{eq:qq_QCD_2L_UL}
\end{align}
where
\begin{align}
    S^{(2,UL)}(p,k,l) = 
    2(1-\e)\frac{(\s{p}+\s{l})S^{(1)}(p+l,k)(\s{p}+\s{l})}{ l^2 ((p+l)^2)^2}\,.
\end{align}
The crossed ladder contribution is given by,
\begin{align}
    \Pi_{qq}^{(2,XL)} = \eqs[0.25]{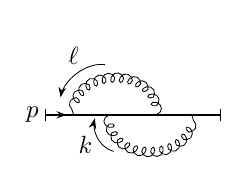} = i\gs^4\, C_F\,\left(C_F-\frac{C_A}{2}\right) S^{(2,XL)}(p,k,l)\,,
    \label{eq:qq_QCD_2L_XL}
\end{align}
with
\begin{align}
    S^{(2,XL)}(p,k,l) =
    \frac{\gamma_\mu (\s{p} + \s{k}) V^{(1)\,\mu}(p,k,l)}{k^2(p+k)^2}\,.
    \label{eq:S_QCD_2L_XL}
\end{align}
Finally, the three-gluon vertex contribution reads
\begin{align}
        \Pi_{qq}^{(2,3V)} = \eqs[0.25]{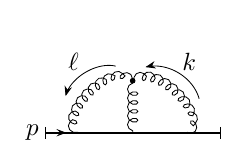} = -i\gs^4\, \frac{C_FC_A}{2}\, S^{(2,3V)}(p,k,l)\,,
        \label{eq:qq_QCD_2L_3V}
\end{align}
with kinematic part 
\begin{align}
    S^{(2,3V)}(p,k,l) = 
    \frac{\gamma_\mu (\s{p}+\s{k})W^{(1)\,\mu}(p,k,l)}{k^2(p+k)^2}\,,
    \label{eq:S_QCD_2L_3V}
\end{align}
which has the same structure as eq.~\eqref{eq:S_QCD_2L_XL}.

Similar to the self-energy graphs, we find it useful to split the $qq\gamma$-vertex into crossed and uncrossed ladder and three-gluon vertex contributions,
\begin{align}
    &\Gamma_{qq\gamma}^{(2)\,\mu} = 
    \Gamma_{qq\gamma}^{(2,UL)\,\mu}+ \Gamma_{qq\gamma}^{(2,XL)\,\mu} + \Gamma_{qq\gamma}^{(2,3V)\,\mu}\,.
    \label{eq:QED_vertex_2L}
\end{align}
The integrand for the uncrossed ladder contribution takes the form,
\begin{align}
        &\g\,\Gamma_{qq\gamma}^{(2,UL)\,\mu}(p,q,k,l) = \eqs[0.23]{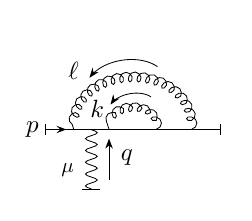}
        +\eqs[0.23]{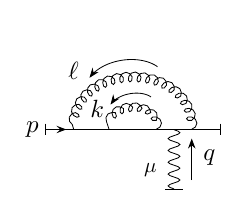}
        +\eqs[0.23]{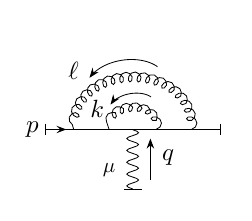} \nonumber \\
        &\qquad= i \g\, \gs^4\, C_F^2
        \,V^{(2,UL)\,\mu}(p,q,k,l)
        \,,
\label{eq:EW_vertex_2L_UL}
\end{align}
with $V^{(2,UL)\,\mu}\equiv\sum_{i=1}^3V_i^{(2,UL)\,\mu}$. The individual kinematic vertex functions are given explicitly by,
\begin{align}
    \begin{split}
        V^{(2,UL)\,\mu}_1
        &=-\frac{\gamma^\alpha (\s{p}+\s{l}+\s{q} )S^{(1)}(p+l+q,k)(\s{p}+\s{l}+\s{q} )\gamma^\mu(\s{p}+\s{l})\gamma_\alpha}{ l^2 ((p+l+q)^2)^2(p+l)^2}
        \label{eq:EW_vertex_2L_UL_V1}
    \end{split}\,,
    \\
    \begin{split}
        V^{(2,UL)\,\mu}_2
        &=-\frac{\gamma^\alpha (\s{p}+\s{l}+\s{q} ) \gamma^\mu(\s{p}+\s{l})S^{(1)}(p+l,k)(\s{p}+\s{l})\gamma_\alpha}{l^2 (p+l+q)^2((p+l)^2)^2} 
        \label{eq:EW_vertex_2L_UL_V2}
    \end{split}\,, \\
        \begin{split}
        V^{(2,UL)\,\mu}_3
        &=\frac{\gamma^\alpha (\s{p}+\s{l}+\s{q} ) V^{(1)\,\mu}(p+l,q,k)(\s{p}+\s{l})\gamma_\alpha}{l^2 (p+l+q)^2(p+l)^2}\,,
        \label{eq:EW_vertex_2L_UL_V3}
    \end{split}
\end{align}
where for the sake of readability we have suppressed the arguments on the left-hand side.
The crossed ladder diagrams are,
\begin{align}
        &\g\,\Gamma_{qq\gamma}^{(2,XL)\,\mu}(p,q,k,l) = \eqs[0.23]{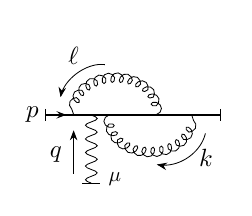}
        +\eqs[0.23]{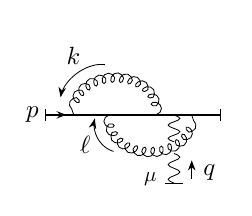}
        +\eqs[0.23]{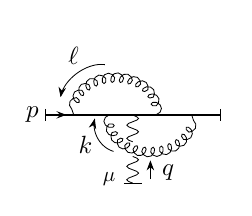} \nonumber \\
        &\qquad= i \g\, \gs^4\, C_F\left(C_F-\frac{C_A}{2}\right)
        V^{(2,XL)\,\mu}
        (p,q,k,l) 
        \,,
\label{eq:EW_vertex_2L_XL}
\end{align}
with $V^{(2,XL)\,\mu}\equiv\sum_{i=1}^3V_i^{(2,XL)\,\mu}$, where
\begin{align}
\begin{split}
    V^{(2,XL)\,\mu}_1
    &=\frac{V^{(1)\,\alpha}(p+l+q,-l,k)(\s{p} + \s{l} + \s{q}) \gamma^\mu  (\s{p} + \s{l}) \gamma_\alpha }{ l^2 (p+l)^2 (p+l+q)^2}
    \label{eq:EW_vertex_2L_XL_V1}
\end{split}\,,
\\
\begin{split}
    V^{(2,XL)\,\mu}_2&=\frac{\gamma_\alpha (\s{p} + \s{l} + \s{q}) \gamma^\mu  (\s{p} + \s{l})V^{(1)\,\alpha}(p,l,k)}{l^2 (p+l)^2(p+l+q)^2}\,.
    \label{eq:EW_vertex_2L_XL_V2}
\end{split}\\
\begin{split}
    V^{(2,XL)\,\mu}_3
    &=
    \frac{\gamma^\alpha (\s{p} + \s{k} + \s{q}) \gamma^\beta (\s{p} + \s{l} + \s{k} + \s{q})\gamma^\mu  (\s{p} + \s{k} + \s{l}) \gamma_\alpha (\s{p} + \s{l}) \gamma_\beta }{k^2 l^2 (p+l)^2 (p+k+l)^2 (p+k+l+q)^2 (p+k+q)^2}\,,
    \label{eq:EW_vertex_2L_XL_V3}
\end{split}
\end{align}
We emphasise that the second diagram on the right-hand side of eq.~\eqref{eq:EW_vertex_2L_XL}, corresponding to the kinematic function $V^{(2,XL)\,\mu}_2$ defined in eq.~\eqref{eq:EW_vertex_2L_XL_V2}, has a different momentum routing compared to the others. This proves a convenient choice for the single-UV region. However, the entire three-loop amplitude is symmetrised over the virtual gluon momenta, and we will present explicitly symmetrised versions of all double-UV counterterms.
This is especially useful for non-planar topologies, and lead to simplified numerators in the double-UV region.

Finally, we have the three-gluon vertex corrections,
\begin{align}%
        &\g\,\Gamma_{qq\gamma}^{(2,3V)\,\mu}(p,q,k,l) = \eqs[0.23]{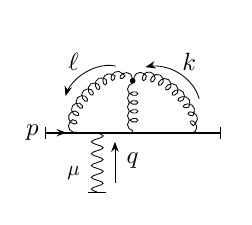}+\eqs[0.23]{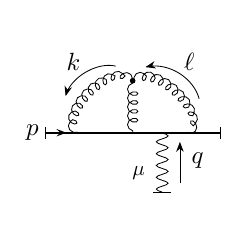}\nonumber \\
        &\qquad= -i \g\, \gs^4\, \frac{C_AC_F}{2} 
        V^{(2,3V)\,\mu}(p,q,k,l)
\label{eq:EW_vertex_2L_3V}
\end{align}
where again we split the vertex function into separate contributions as $V^{(2,3V)\,\mu} = V^{(2,3V)\,\mu}_{1}+ V^{(2,3V)\,\mu}_{2}$ where,
\begin{align}
\begin{split}
    V^{(2,3V)\,\mu}_{1}&=\frac{W^{(1)\,\alpha}(p+q+l,-l,k-l)(\s{p}+\s{l}+\s{q})\gamma^\mu(\s{p}+\s{l})\gamma_\alpha}{l^2(p+l)^2(p+l+q)^2}\,,
\label{eq:EW_vertex_2L_3V_V2}
\end{split}\\
\begin{split}
V^{(2,3V)\,\mu}_{2}&=\frac{\gamma_\alpha(\s{p}+\s{l}+\s{q})\gamma^\mu(\s{p}+\s{l})W^{(1)\,\alpha}(p,l,k)}{l^2(p+l)^2(p+l+q)^2}\,,
    \label{eq:EW_vertex_2L_3V_V1}
\end{split}
\end{align}
which has the same structure as eqs.~\eqref{eq:EW_vertex_2L_XL_V2} and~\eqref{eq:EW_vertex_2L_XL_V3}.

Next, the uncrossed ladder contributions to the two-loop \acs{qcd} corrections to the quark-antiquark-gluon vertex, eq.~\eqref{eq:qqg_QCD_2L} are given by,
\begin{align}
        &\gs\,\Gamma_{qqg}^{(2,UL)\,\mu,c}
        = \eqs[0.23]{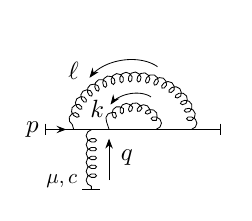}
        +\eqs[0.23]{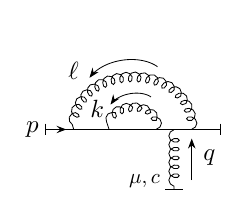}
        +\eqs[0.23]{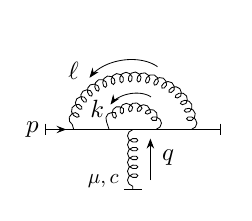} \nonumber \\
        &\quad= i \gs^5\, \tq{c} \left[
        C_F\left(C_F-\frac{C_A}{2}\right)\left(
        V^{(2,UL)\,\mu}_1
        +V^{(2,UL)\,\mu}_2
        \right)
        +\left(C_F -\frac{C_A}{2}\right)^2
        V^{(2,UL)\,\mu}_3
        \right]
        \,,
\label{eq:QCD_vertex_2L_UL}
\end{align}
where the kinematic functions $V^{(2,UL)\,\mu}_i$, $i\in\{1,2,3\}$, are defined in eqs.~\eqref{eq:EW_vertex_2L_UL_V1}-\eqref{eq:EW_vertex_2L_UL_V3}. The crossed ladder diagrams are
\begin{align}
        &\gs\,\Gamma_{qqg}^{(2,XL)\,\mu,c}
        = \eqs[0.23]{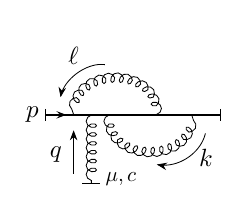}
        +\eqs[0.23]{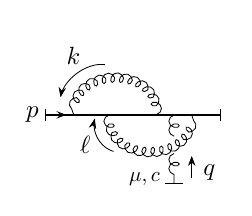}
        +\eqs[0.23]{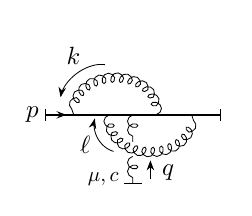} \nonumber \\
        &\qquad= i \gs^5\,\tq{c}\left[\left(C_F-\frac{C_A}{2}\right)^2\left(
        V^{(2,XL)\,\mu}_1
        +V^{(2,XL)\,\mu}_2
        \right) \right. 
        \label{eq:QCD_vertex_2L_XL}
        \\&\qquad\qquad\left.+\left(C_F-C_A\right)\left(C_F-\frac{C_A}{2}\right)
        V^{(2,XL)\,\mu}_3
        \right] \nonumber
        \,,
\end{align}
where $V^{(2,XL)\,\mu}_i$, $i\in\{1,2,3\}$, are defined in eqs.~\eqref{eq:EW_vertex_2L_XL_V1}-\eqref{eq:EW_vertex_2L_XL_V3}. Similarly,
\begin{align}
        &\gs\,\Gamma_{qqg}^{(2,3V)\,\mu,c} = 
        \eqs[0.23]{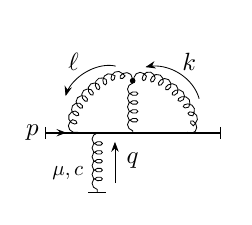}
        +\eqs[0.23]{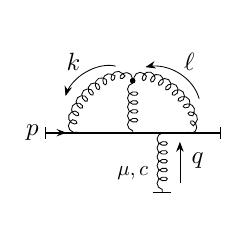}\nonumber \\
        &\qquad= -i \gs^5\, 
        \tq{c}\,
        \frac{C_A}{2}\,\left(C_F-\frac{C_A}{2}\right)
        \left(V^{(2,3V)\,\mu}_{1}
        +V^{(2,3V)\,\mu}_{2}\right)\,,
\label{eq:QCD_vertex_2L_3V}
\end{align}
with $V^{(2,3V)\,\mu}_{1}$ and $V^{(2,3V)\,\mu}_{2}$ defined in eqs.~\eqref{eq:EW_vertex_2L_3V_V1} and~\eqref{eq:EW_vertex_2L_3V_V2}.

If the external particle is a gluon, we have four additional topologies compared to the electroweak case. The four-gluon vertex contribution is
\begin{align}
\begin{split}
        &\gs\,\Gamma_{qqg}^{(2,4V)\,\mu,c} = 
        \eqs[0.23]{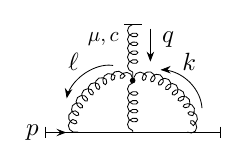}=i \gs^5\,
        \tq{c}\,
        \frac{C_A^2}{4}\,
        W^{(2,4V)\,\mu}\,,
\end{split}
\label{eq:QCD_vertex_2L_4V}
\end{align}
with
\begin{align}
    W^{(2,4V)\,\mu} = - \frac{\gamma_\nu (\s{p}+\s{k}+\s{q})\gamma_\sigma (\s{p}+\s{l})\gamma_\rho\left[g^{\mu\rho} g^{\nu\sigma} +g^{\mu\nu}g^{\rho\sigma}-2g^{\mu\sigma}g^{\nu\rho}\right]}{l^2k^2(p+l)^2(l-k-q)^2(p+k+q)^2}
    \,.
    \label{eq:QCD_vertex_2L_4V_W}
\end{align}
The case where the external gluon connects to an internal gluon line of the uncrossed ladder diagram, shown in eq.~\eqref{eq:qq_QCD_2L_UL}, is given by
\begin{align}
\begin{split}%
        &\gs\,\Gamma_{qqg}^{(2,UL-3V)\,\mu,c} = 
        \eqs[0.23]{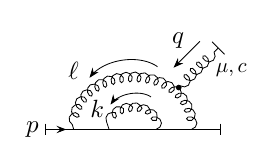}
        +\eqs[0.23]{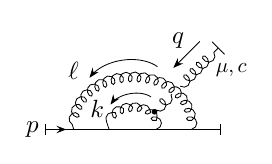} \\
        &\qquad= i \gs^5\, 
        \tq{c}
        \left[
        \frac{C_A C_F}{2}\,W^{(2,UL)\,\mu}_{1}
        +\frac{C_A}{2}\,\left(C_F-\frac{C_A}{2}\right)
        W^{(2,UL)\,\mu}_{2}
        \right]\,.
\end{split}
\label{eq:QCD_vertex_2L_UL-3V}
\end{align}
Analogous to eq.~\eqref{eq:qqg_1L_W} it is useful to decompose $W^{(2,UL)\,\mu}_{1}$ into a scalar contribution $Q^{(2,UL)\,\mu}$ and a contribution associated to ghosts $O^{(2,UL)\,\mu}$ as follows,
\begin{align}
    W^{(2,UL)\,\mu}_{1}(p,q,k,l) = 
    Q^{(2,UL)\,\mu}_1(p,q,k,l)
    +O^{(2,UL)\,\mu}_1(p,q,k,l)\,,
    \label{eq:W2_UL_1}
\end{align}
with
\begin{align}
    \label{eq:Q2_UL_1}
    Q^{(2,UL)\,\mu}_1(p,q,k,l) &= 2(1-\e)\frac{(q-2l)^\mu(\s{p}+\s{l})S^{(1)}(p+l,k)(\s{p}+\s{l})}{l^2((p+l)^2)^2(l-q)^2}\,, \\
    O^{(2,UL)\,\mu}_1(p,q,k,l) &=
    -\frac{1}{l^2((p+l)^2)^2(l-q)^2}\left[\gamma^\mu(\s{p}+\s{l})S^{(1)}(p+l,k)(\s{p}+\s{l})(\s{l}-2\s{q}) \right. \nonumber\\
    &\quad\left.+(\s{l}+\s{q})(\s{p}+\s{l})S^{(1)}(p+l,k)(\s{p}+\s{l})\gamma^\mu\right]
    \label{eq:O2_UL_1}
\end{align}
where $S^{(1)}$ is defined in eq.~\eqref{eq:qq_1L_S}. 

Similarly, $W^{(2,UL)\,\mu}_{2}$ can be written in terms of the one-loop function $W^{(1)\,\mu}$, defined in eq.~\eqref{eq:qqg_1L_W},
\begin{align}
    \begin{split}
        W^{(2,UL)\,\mu}_{2}(p,q,k,l) = -
        \frac{\gamma^\alpha (\s{p}+\s{l}+\s{q})W^{(1)\,\mu}(p+l,q,k)(\s{p}+\s{l})\gamma_\alpha}{l^2(p+l)^2(p+l+q)^2}\,.
    \end{split}
\label{eq:W2_UL_2}
\end{align}
Again, we can write, 
\begin{align}
    W^{(2,UL)\,\mu}_{2}(p,q,k,l) = 
    Q^{(2,UL)\,\mu}_2(p,q,k,l)
    +O^{(2,UL)\,\mu}_2(p,q,k,l)\,,
\end{align}
where $Q^{(2,UL)\,\mu}_2$ and $O^{(2,UL)\,\mu}_2$ are obtained by replacing $W^{(1)\,\mu}$ in eq.~\eqref{eq:W2_UL_2} by the one-loop scalar and ghost parts defined in eqs.~\eqref{eq:Q1} and \eqref{eq:O1}, respectively.

Next, we have the crossed-ladder-type diagrams with a three-gluon vertex,
\begin{align}
\begin{split}
        &\gs\,\Gamma_{qqg}^{(2,XL-3V)\,\mu,c} = 
        \eqs[0.23]{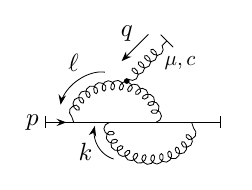}
        +\eqs[0.23]{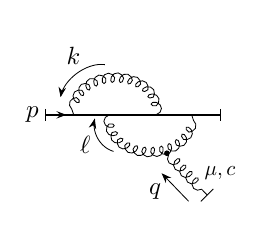} \\
        &\qquad= -i \gs^5\, 
        \tq{c}\,
        \frac{C_A}{2}\,\left(C_F-\frac{C_A}{2}\right)
        W^{(2,XL)\,\mu}(p,q,k,l)
        \,.
\end{split}
\label{eq:QCD_vertex_2L_XL-3V}
\end{align}
with $W^{(2,XL)\,\mu} \equiv W^{(2,XL)\,\mu}_{1}+W^{(2,XL)\,\mu}_{2}$. Again, we separate scalar and ghost contributions from the kinematic parts as
\begin{align}
    W^{(2,XL)\,\mu}_{i}(p,q,k,l) = 
    Q^{(2,XL)\,\mu}_i(p,q,k,l)
    +O^{(2,XL)\,\mu}_i(p,q,k,l)\,, \quad i \in \{1,2\}\,,
\end{align}
where
\begin{align}
    Q^{(2,XL)\,\mu}_1(p,q,k,l) &=
    - \frac{(q-2l)^\mu V^{(1)\,\alpha}(p+l,q-l,k)(\s{p}+\s{l})\gamma_\alpha}{l^2(l-q)^2(p+l)^2}\,, \label{eq:Q1_2L_XL}\\
    O^{(2,XL)\,\mu}_1(p,q,k,l) &= - \frac{1}{l^2(l-q)^2(p+l)^2}\left[ 
    (l+q)_\alpha V^{(1)\,\alpha}(p+l,q-l,k)(\s{p}+\s{l})\gamma^\mu
    \right. \nonumber\\
    &\quad\left.+ V^{(1)\,\mu}(p+l,q-l,k)(\s{p}+\s{l})(\s{l}-2\s{q})\right]\,,
    \label{eq:O1_2L_XL}
\end{align}
and
\begin{align}
    Q^{(2,XL)\,\mu}_2(p,q,k,l) &=
    - \frac{(q-2l)^\mu \gamma_\alpha(\s{p}+\s{l})V^{(1)\,\alpha}(p,l,k)}{l^2(l-q)^2(p+l)^2}\,,
    \label{eq:Q2_2L_XL}\\
    O^{(2,XL)\,\mu}_2(p,q,k,l) &= - \frac{1}{l^2(l-q)^2(p+l)^2}\left[ 
    \gamma^\mu(\s{p}+\s{l})V^{(1)\,\alpha}(p,l,k)(l-2q)_\alpha
    \right. \nonumber\\
    &\quad\left.+ (\s{l}+\s{q})(\s{p}+\s{l})V^{(1)\,\mu}(p,l,k)\right]\,.
    \label{eq:O2_2L_XL}
\end{align}
Above, we have used the one-loop vertex function $V^{(1)\,\mu}$ defined in eq.~\eqref{eq:qqg_1L_V}. In the coming sections, we will often use the shorthand $X^{(2,XL)\,\mu}= X_1^{(2,XL)\,\mu}+X_2^{(2,XL)\,\mu}$ also for the ghost and scalar contributions, $X\in \{Q,O\}$.

The last topology contains two three-gluon vertices and has the following contributions,
\begin{align*}
        &\gs\,\Gamma_{qqg}^{(2,d3V)\,\mu,c} = 
        \eqs[0.23]{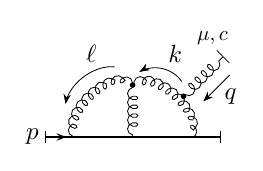}
        +\eqs[0.23]{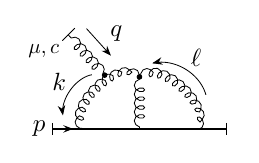}
        +\eqs[0.23]{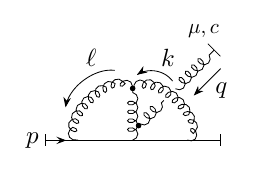}\\
        &\qquad= i \gs^5\, 
        \tq{c}\,
        \frac{C_A^2}{4}\,
        W^{(2,d3V)\,\mu}(p,q,k,l)
        \,. \numberthis
\label{eq:QCD_vertex_2L_d3V}
\end{align*}
where $W^{(2,d3V)\,\mu}=W^{(2,d3V)\,\mu}_{1}+W^{(2,d3V)\,\mu}_{2}$ is the sum of the first two diagrams shown on the right-hand side of eq.~\eqref{eq:QCD_vertex_2L_d3V}.
The third diagram has vanishing colour factor. We decompose the remaining contributions as usual into scalar and ghost parts,
\begin{align}
    W^{(2,d3V)\,\mu}_{i}(p,q,k,l) = 
    Q^{(2,d3V)\,\mu}_i(p,q,k,l)
    +O^{(2,d3V)\,\mu}_i(p,q,k,l)\,, \quad i \in \{1,2\}\,,
\end{align}
with
\begin{align}
    Q^{(2,d3V)\,\mu}_1(p,q,k,l) &=
    - \frac{(q-2k)^\mu \gamma_\alpha (\s{p}+\s{k})W^{(1)\,\alpha}(p,k,l)}{k^2(k-q)^2(p+k)^2}\,, 
    \label{eq:Q1_2L_d3V}\\
    O^{(2,d3V)\,\mu}_1(p,q,k,l) &= - \frac{1}{k^2(k-q)^2(p+k)^2}\left[ 
    \gamma^\mu (\s{p}+\s{k})W^{(1)\,\alpha}(p,k,l)(k-2q)_\alpha
    \right. \nonumber\\
    &\quad\left.+ (\s{k}+\s{q})(\s{p}+\s{k})W^{(1)\,\mu}(p,k,l)\right]\,,
    \label{eq:O1_2L_d3V}
\end{align}
and
\begin{align}
    Q^{(2,d3V)\,\mu}_2(p,q,k,l) &=
    - \frac{(q-2k)^\mu W^{(1)\,\alpha}(p+k,q-k,l+q-k) (\s{p}+\s{k})\gamma_\alpha}{k^2(k-q)^2(p+k)^2}\,, 
    \label{eq:Q2_2L_d3V}\\
    O^{(2,d3V)\,\mu}_2(p,q,k,l) &= - \frac{1}{k^2(k-q)^2(p+k)^2}\left[ 
    W^{(1)\,\mu}(p+k,q-k,l+q-k)(\s{p}+\s{k})(\s{k}-2\s{q})
    \right. \nonumber\\
    &\quad\left.+(k+q)_\alpha W^{(1)\,\alpha}(p+k,q-k,l+q-k)(\s{p}+\s{k})\gamma^\mu\right]\,,
    \label{eq:O2_2L_d3V}
\end{align}
Again, we will often write $X^{(2,d3V)\,\mu}= X_1^{(2,d3V)\,\mu}+X_2^{(2,d3V)\,\mu}$ also for the ghost and scalar contributions, $X\in \{Q,O\}$.

%% file: app_shift.tex
\section{Three-loop shift counterterms}
\label{app:shift}

In this appendix we provide additional technical information about shift counterterms, discussed in section~\ref{sec:shift}, and provide counterterms that remove shift-integrable contributions from the form factor counterterterms.

\subsection{One-loop shift integrability for the one-loop quark jet function}
\label{app:shift_jet}

In this appendix, we prove integrability in the $l\,||\,p_1$ limit of the set of integrands shown in fig.~\ref{fig:M3_shift_jet}, equal to the following sum of collinear insertions,
\begin{align*}
        &\sum_\text{insertions}
        \eqs[0.14]{l_p1}
        \otimes
        \eqs[0.2]{M1_NL_A_diag1_n}
        \equiv
        \eqs[0.2]{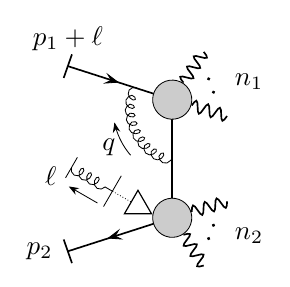}
        +\eqs[0.2]{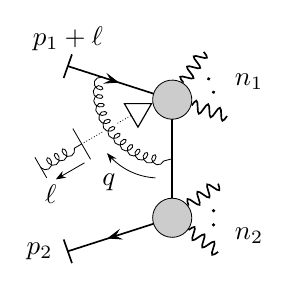}
        \\&\qquad
        +\eqs[0.2]{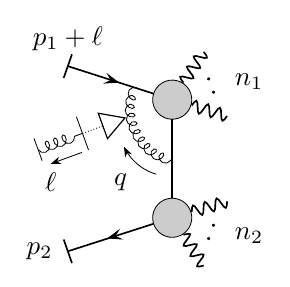}
        \numberthis
        \label{eq:M3_jet_diag1_sum_app}
        \end{align*}
Here, whenever a gluon with longitudinal polarisations flows into an (off-shell) tree-level sub-amplitude, it bisects the set of outgoing photon momenta as follows,
\begin{align*}
    &\eqs[0.2]{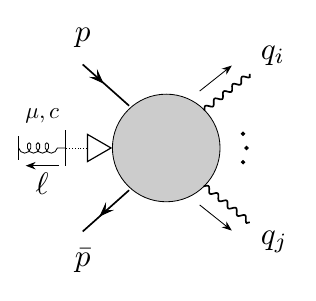}
    =
    \eqs[0.3]{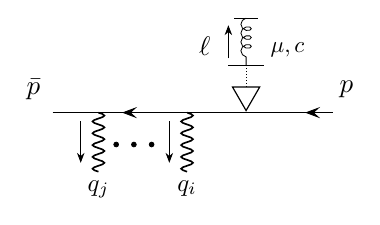}
    +\sum_{s=i}^{j}
    \eqs[0.3]{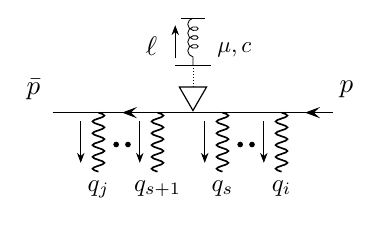}
    \\&\qquad 
    \numberthis
    \label{eq:born_qqg_coll}
    = \tq{c}\,\frac{i}{\s{\bar{p}}}(-l_\mu)\Bigg[
    \widetilde{\mathcal{M}}^{(0)}(p,\bar{p},l;
    \{q_i,\ldots,q_j\})\frac{\s{p}-\s{l}}{(p-l)^2}\gamma^\mu
    \\&\qquad+
    \sum_{s=i}^{j}\widetilde{\mathcal{M}}^{(0)\,\mu}_s(p,\bar{p},l;\{q_i,\ldots,q_s\},\{q_{s+1},\ldots,q_j\})\Bigg]\frac{i}{\s{{p}}}
    \\&\qquad 
    \tq{c}\Bigg[
    \frac{\s{l}-\s{\bar{p}}}{(l-\bar{p})^2}
    \widetilde{\mathcal{M}}^{(0)}_s(p,\bar{p},l;
    \{q_i,\ldots,q_j\})\,\frac{i}{\s{p}}
    -\frac{i}{\s{\bar{p}}}
    \widetilde{\mathcal{M}}^{(0)}_s(p,\bar{p},l;
    \{q_i,\ldots,q_j\})\frac{\s{{p}}-\s{l}}{(p-l)^2}
    \Bigg]
    \\&\qquad 
    \equiv\eqs[0.3]{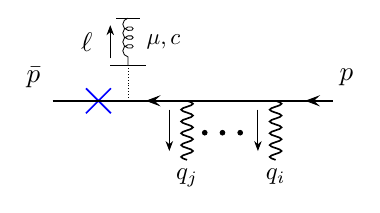}-\eqs[0.3]{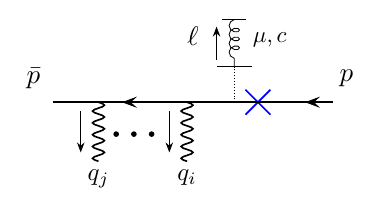}\,.
\end{align*}
Repeated applications of the abelian-type Ward identity, eq.~\eqref{eq:qq_Ward}, results in pairwise cancellation of all terms except two, shown graphically on the last line. The first 
term vanishes if $\bar{p}\equiv p_2$ 
due to the Dirac equation, c.f. eq.~\eqref{eq:qq_Ward_spinor}, corresponding to an insertion directly on the incoming antiquark 
leg. Applying eqs.~\eqref{eq:born_qqg_coll} and the QCD Ward identity for triple-gluon vertices, eq.~\eqref{eq:gg_Ward}, to the graphs on the right-hand side of eq.~\eqref{eq:M3_jet_diag1_sum_app} yields,
\begin{align*}
        &\sum_\text{insertions}
        \eqs[0.14]{l_p1}
        \otimes
        \eqs[0.2]{M1_NL_A_diag1_n}
        =-\eqs[0.2]{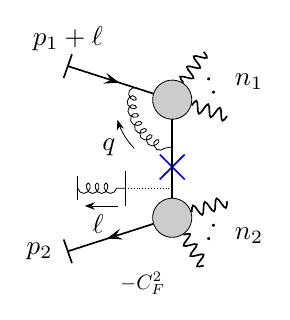}
        +\eqs[0.2]{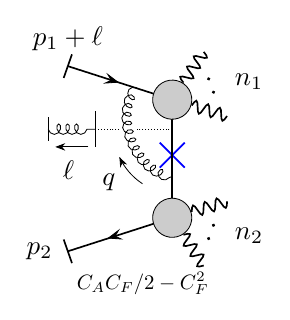}
        \\&\quad
        -\eqs[0.2]{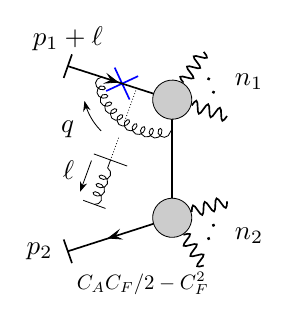}
        +\eqs[0.2]{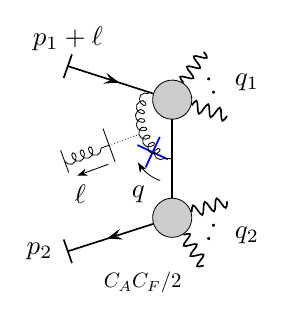}
        -\eqs[0.2]{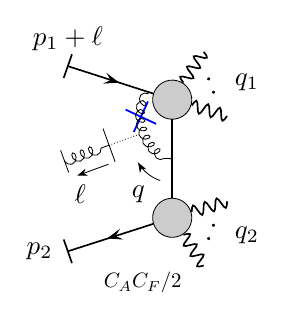}
        + \text{ghosts}\,.
        \numberthis
        \label{eq:M3_jet_diag1_ward_app}
\end{align*}
Colour factors are shown below each diagram for convenience. The first diagram after the equality cancels against the contribution of the second diagram proportional to $C_F^2$. On the second line, the part of the first diagram proportional to $C_AC_F/2$ cancels against the second diagram (insertion on the gluon line). The remainder consists of 1) an integrand proportional to $C_F^2$, which contributes to the singularity but is factorised from the hard one-loop sub-diagram, and 2) %
an integrand proportional to $C_AC_F/2$ contributing to both the second diagram on the second line and the last graph after the equality, which cancel locally up to a loop momentum shift $q \to q+l$ of the gluon line.

We can easily repeat the analysis above for the remaining integrands contributing to $\widetilde{\mathcal{M}}^{(1,C)}$, defined below eq.~\eqref{eq:M3_jet_sum}. This yields the shift counterterm shown in eq.~\eqref{eq:shift_jet_CNT}.

\subsection{Evaluation of shift integrands}
\label{app:shift_eval}
As an example, the first term in eq.~\eqref{eq:shift_NL} is written explicitly as,
\begin{align*}
    \eqs[0.25]{shift_NL_diag1_alt} 
    &=
    -\gs^4\,     
    \frac{1}{k^2(q-l)^2}\,
    \frac{-ig_{\mu\nu}}{l^2}\,
    \bar{v}(p_2)\,
    \overline{{\mathcal{M}}}^{(0)\,\mu}_{n_{123}+1,n_{1234}}(\xi_1;-p_2+l+\q{n_{123}+1}{n_{1234}},p_2,l)
    \\& \times
    \gamma^\beta
    {\mathcal{M}}^{(0)}_{n_{12}+1,n_{123}}(p_1+q-\q{1}{n_{12}},-p_1-q+\q{1}{n_{123}})
    \gamma_\beta
    \\& \times
    {\mathcal{M}}^{(0)\,\alpha}_{n_1+1,n_{12}}(p_1+l+k-\q{1}{n_1},-p_1-l+\q{1}{n_{12}},k)
    \gamma_\alpha
    \\& \times
    \mathcal{M}_{1,n_1}^{(0)}(p_1+l,-p_1-l+\q{1}{n_1})
    \gamma_\mu
    u(p_1)\,,
    \numberthis
    \label{eq:NL_shift_diag1}
\end{align*}
where $n_i$ denote the number of electroweak vertices at each tree-level sub-graph, with $n_{i\ldots j}\equiv n_i+\ldots +n_j$. Here, $ 0\leq n_1, n_3$ and $1\leq n_2,n_4$, where $n_{1234}\equiv n \geq 2$ is the total number of outgoing photons. On the first and second lines we have used momentum conservation,
\begin{align}
    p_1+p_2 = Q_{1,n}\,.
\end{align}
To improve readability, we have used the shorthand,
\begin{align}
  {\mathcal{M}}^{(0)}_{i,j}(p,\bar{p}) \equiv
  {\mathcal{M}}^{(0)}(p,\bar{p};\{q_i,\ldots, q_j\})\,,
\end{align}
and
\begin{align}
\begin{split}
    {\mathcal{M}}^{(0)\,\mu}_{i,j}(p,\bar{p},k) &\equiv
    \sum_{s=i}^{j}
    {\mathcal{M}}^{(0)\,\mu}_{s}(p,\bar{p},k;\{q_i,\ldots, q_s\},\{q_{s+1},\ldots, q_j\})
    \\&+  {\mathcal{M}}^{(0)}_{i,j}(p,\bar{p})
    \frac{\s{p}-\s{k}}{(p-k)^2}\gamma^\mu    \,.
\end{split}
\end{align}
Here, the functions $\mathcal{M}^{(0)}_{i,j}(p,\bar{p})$ and $\mathcal{M}^{(0)\,\mu}_{i,j}(p,\bar{p},k)$ are defined to include the adjacent (off-shell) fermion propagators,
\begin{align}
    \mathcal{M}^{(0)}_{i,j}(p,\bar{p}) \equiv
    iS_0(-\bar{p})
    \widetilde{\mathcal{M}}^{(0)}_{i,j}(p,\bar{p})
    iS_0(p)\,,
\end{align}
and
\begin{align}
    \mathcal{M}^{(0)\,\mu}_{i,j}(p,\bar{p},k) \equiv
    iS_0(-\bar{p})
    \widetilde{\mathcal{M}}^{(0)\,\mu}_{i,j}(p,\bar{p},k)
    iS_0(p)\,,
\end{align}
where $S_0$ denotes the fermion propagator defined in eq.~\eqref{eq:qq_Ward} and we set $iS_0(p)=1$ ($iS_0(-\bar{p})=1$) if $p^2 =0$ ($\bar{p}^2 =0$) is on-shell. As a tree-level subgraph, $\mathcal{M}^{(0)}_{i,j}(p,\bar{p})$ may contain no electroweak vertices, for which
\begin{align}
    {\mathcal{M}}_{i,j}(p,\bar{p}) \equiv iS_0(p)\,,
    \qquad
    Q_{i,j}\equiv q_i\,, \quad \text{if } j<i
    \,,
\end{align}
The amplitude $\overline{{\mathcal{M}}}^{(0)\,\mu}_{n_{123}+1,n_{1234}}$ on the first line of eq.~\eqref{eq:NL_shift_diag1} includes the graph where $l$ is adjacent to the incoming antiquark, and is regularised according to the prescription in eq.~\eqref{eq:shift_mod_quark} to avoid double-counting in different collinear regions,
\begin{align}
\begin{split}
    &\overline{{\mathcal{M}}}^{(0)\,\mu}_{n_{123}+1,n_{1234}}(\xi_1;-p_2+l+\q{n_{123}+1}{n_{1234}},p_2,l) 
    \\&\quad= 
    \gamma^\mu \bar{S}_0(l-p_2;p_1,\xi_1){{\mathcal{M}}}^{(0)}_{n_{123}+1,n_{1234}}(-p_2+l+\q{n_{123}+1}{n_{1234}},p_2,l)
    \\&\quad
    +{{\mathcal{M}}}^{(0)\,\mu}_{n_{123}+1,n_{1234}-1}(-p_2+l+\q{n_{123}+1}{n_{1234}},p_2,l)\,.
\end{split}
\end{align}

In eq.~\eqref{eq:NL_shift_diag1} the gluon propagator $-ig_{\mu\nu}/l^2$ is replaced by the approximation in eq.~\eqref{eq:g_k_p1} in the collinear region $(1_l,H_k,H_q)$.

\subsection{Form factor shift counterterms}
\label{sec:shift_FF}
The relevant shift counterterm $\mathcal{F}_{\text{shift},1}$ for the three-loop form factor follow straightforwardly from the results of 
section~\ref{sec:shift}.
We write the form factor shift counterterm as a sum of six terms,
\begin{align}
\begin{split}
    &\mathcal{F}_{\text{shift},1}^{(3)}(\xi_1;p_1,p_2,q,k,l;\{q_1,\ldots,q_n\})
    \\&\qquad=\sum_X\,\mathcal{F}_{\text{shift},1}^{(3,X)}(\xi_1;p_1,p_2,q,k,l,\{q_1,\ldots,q_n\})\,, \quad
    X\in \{\mathcal{J},UL,XL,3V,\Delta,O\}
    \,.
\end{split}
    \label{eq:FF_shift_CNT}
\end{align}
which follows the assignment explained after eq.~\eqref{eq:M3_shift_CNT}.
Again, it suffices to provide the counterterms valid for the region $(1_l,H_q,H_k)$. To avoid spurious contributions to overlapping collinear regions from a pinched fermion propagator adjacent to the incoming antiquark $p_2$, we implement the prescriptions (and use the graphical notation introduced) in eqs.~\eqref{eq:shift_mod_quark} and~\eqref{eq:ggg_mod_quark}. We remark that eq.~\eqref{eq:FF_shift_CNT} requires local ultraviolet regularisation.  The corresponding counterterms can be obtained straightforwardly using the method discussed in section~\ref{sec:UV}.

As in eq.~\eqref{eq:M3_shift_CNT}, the contribution $\mathcal{F}_{\text{shift},1}^{(3,\mathcal{J})}$ removes one-loop shift-integrable contributions from integrands with a one-loop jet subgraph,
\begin{align}
    \mathcal{F}_{\text{shift},1}^{(3,\mathcal{J})}= \frac{C_A}{2C_F}
    \eqs[0.21]{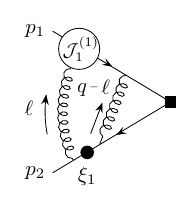}
    - (q\to q+l)\,.
    \label{eq:FF_shift_jet}
\end{align}
Here and below, we have suppressed the kinematic dependence for legibility. The square vertex in the graphical representation denotes the hard-scattering vertex (in this case, the Born amplitude) enclosed by a pair of Dirac projectors, eq.~\eqref{eq:P1projector}. The contribution to the $(UL)$, $(XL)$ and $(3V)$ sub-topologies read,
\begin{align}
    \begin{split}
        &\mathcal{F}_{\text{shift},1}^{(3,UL)}= 
        \frac{C_A}{2C_F}\Bigg(\eqs[0.21]{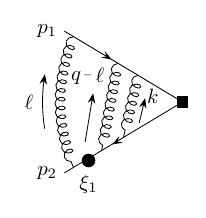} - (q\to q+l)\Bigg)
        +\Bigg[\Bigg(\frac{C_A}{2C_F}
        \eqs[0.21]{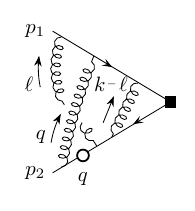}
        \\&\quad
        +\frac{C_A}{2C_F}\left(1-\frac{C_A}{2C_F}\right)
        \eqs[0.21]{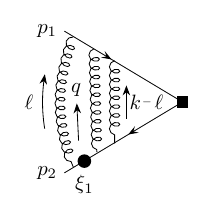} \Bigg)
        - (k\to k+l)\Bigg]\,,
    \end{split}
    \label{eq:FF_shift_UL}
\end{align}
\begin{align*}
        &\mathcal{F}_{\text{shift},1}^{(3,XL)}= 
        \frac{C_A}{2C_F}\Bigg(\eqs[0.21]{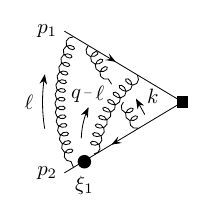} - (q\to q+l)\Bigg)
        +\Bigg\{\Bigg[\frac{C_A}{2C_F}
        \eqs[0.21]{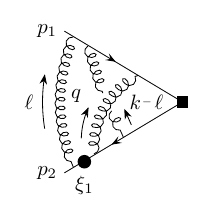}
        \\&\quad
        -C_A^2\Bigg(
        \eqs[0.21]{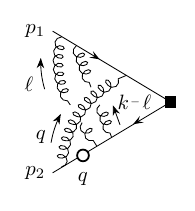}
        - (k\to k+l)\Bigg\}\,,
    \numberthis
    \label{eq:FF_shift_XL}
\end{align*}
and 
\begin{align}
        &\mathcal{F}_{\text{shift},1}^{(3,3V)}= 
        \frac{C_A}{2C_F}\Bigg(\eqs[0.21]{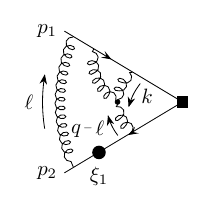}
        +\eqs[0.21]{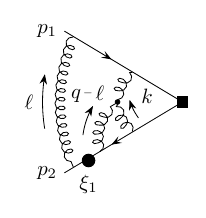}
        - (q\to q+l,k\to k-q)\Bigg)\,,
    \label{eq:FF_shift_3V}
\end{align}
respectively. Next, we have two-loop shift integrable terms due to gluon self-energy and vertex corrections, the form factor analogue of eq.~\eqref{eq:shift_DEL},
\begin{align*}
    &\mathcal{F}_{\text{shift},1}^{(3,\Delta)} = 
    i\gs^2\,\frac{C_A}{2C_F}\frac{\Gamma_{2}^\epsilon}{(q-k)^2(l-k)^2}
    \eqs[0.21]{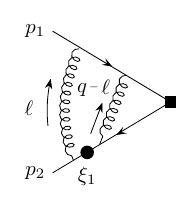} 
    - (k\to k+l,\, q\to q+l)\,.
\numberthis
\label{eq:FF_shift_DEL}
\end{align*}

Finally, we subtract ghost shift contributions using the following counterterm,
\begin{align*}
    &\mathcal{F}_{\text{shift},1}^{(3,O)}=
    \Biggl\{
    \Biggl[
    \eqs[0.21]{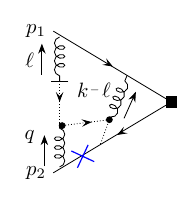}
    -\eqs[0.21]{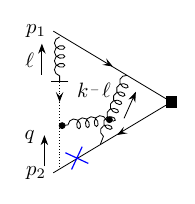} 
     -\frac{C_A}{2C_F}
    \eqs[0.21]{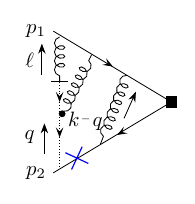}
    \Biggr] 
    - (k\to k+q)\Biggr\}
    \\&\quad 
    -\Biggl(\frac{C_A}{2C_F}\eqs[0.21]{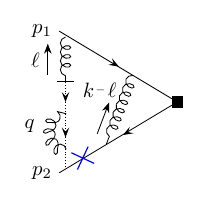}-(k\to k+l)\Biggr)
    + \Biggl(\eqs[0.21]{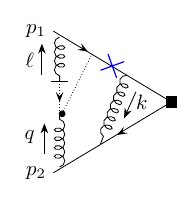}
    - (k\to l-q-k)     \Biggr)\,.
    \numberthis
    \label{eq:FF_shift_O}
\end{align*}

%% file: app_ghost.tex
\section{Ghost contributions to the QCD Ward identity for two-loop subgraphs}
\label{app:ghost}
In this appendix derive the ghost contributions to the QCD Ward identity for two-loop corrections (excluding gluon-self-energy subgraphs) to the fermion two-point function, eq.~\eqref{eq:ghost_2L}.

We begin by contracting the ghost terms for graphs without ladder structure, eq.~\eqref{eq:qqg_NL-3V_O}, with a longitudinal polarised gluon momentum $q_\mu$,
obtaining,
\begin{align*}
    &\gs^5\, \frac{C_AC_F}{2}\,\tq{c}\,
    \frac{i}{\s{p}+\s{q}}\,
    q_\mu
    O^{(2,NL-3V)\,\mu}(p,q,k,l)
    \,\frac{i}{\s{p}}
    \\&\qquad
    =\eqs[0.21]{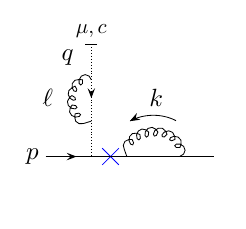}
    -\eqs[0.21]{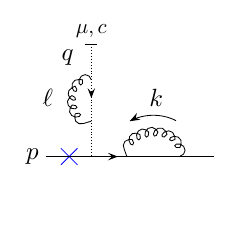}
    +\eqs[0.21]{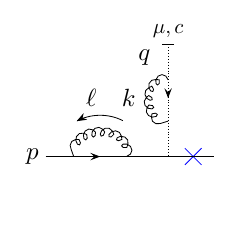}
    -\eqs[0.21]{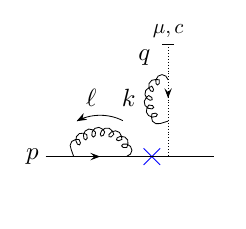}
    \\&\qquad
    +\eqs[0.3]{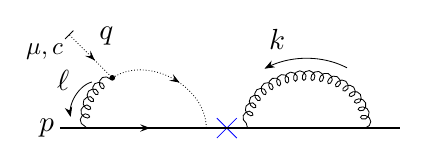}
    -\eqs[0.3]{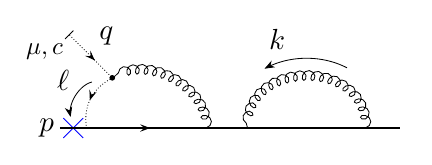}
    +\eqs[0.3]{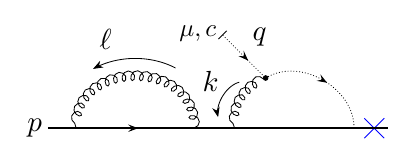}
    \\&\qquad
    -\eqs[0.3]{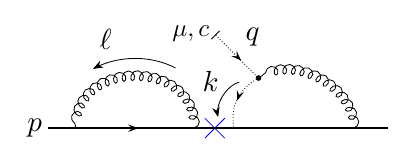}\,.
    \numberthis
    \label{eq:qqg_NL-3V_ward_ghost}
\end{align*}
The right-hand side is obtained directly from the one-loop ghost identities, eqs.~\eqref{eq:ghosts_ward_1L}, ~\eqref{eq:qqg_1L_ward_ghost_id} and~\eqref{eq:qqg_1L_ward_ghost_id_2}.

Next, we examine ghost contributions from the uncrossed-ladder topology, %
defined in eq.~\eqref{eq:QCD_vertex_2L_UL-3V}. The external gluon with loop momentum $q$ may either attach to the ``outer" or ``inner" loop, yielding different colour factors. For the former we obtain the relation, 
\begin{align}
\begin{split}
    &i\gs^5\, \tq{c}\, \frac{C_A C_F}{2}\, q_\mu O_1^{(2,UL)\,\mu} =
    \eqs[0.25]{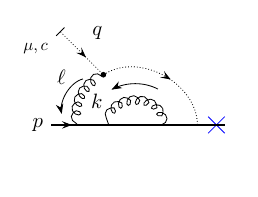}
    -\eqs[0.25]{qqg_QCD_2L_UL-3V_O1_ward_ghost_RHS_diag_1_2}
    \\&\qquad 
    +\eqs[0.25]{qqg_QCD_2L_UL-3V_O1_ward_ghost_RHS_diag_2_1}
    -\eqs[0.25]{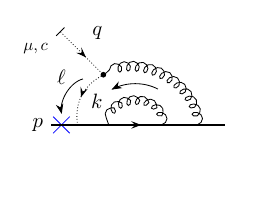}\,.
\end{split}
\label{eq:qqg_UL-3V_ward_ghost_O1}
\end{align}
Similarly, for a three-gluon vertex on the ``inner" loop we have,
\begin{align}
\begin{split}
    &i\gs^5\, \tq{c} \left(\frac{C_AC_F}{2}-\frac{C_A^2}{4}\right) q_\mu O_2^{(2,UL)\,\mu} =
    \eqs[0.2]{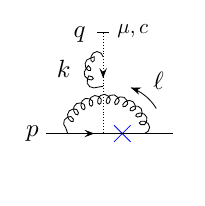}
    -\eqs[0.2]{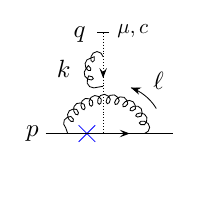}
    \\&\qquad
    +\eqs[0.25]{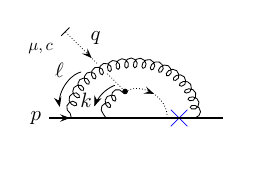}
    -\eqs[0.25]{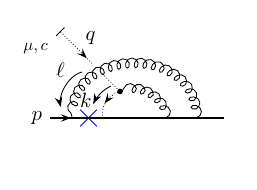}\,.
\end{split}
\label{eq:qqg_UL-3V_ward_ghost_O2}
\end{align}
The first two graphs are obtained by straightforward application of eqs.~\eqref{eq:qqg_1L_ward_ghost_id} and \eqref{eq:qqg_1L_ward_ghost_id_2}. We notice that the part of the first line proportional to $C_AC_F/2$ cancels against the first and fourth graphs on the second line of eq.~\eqref{eq:qqg_NL-3V_ward_ghost}. The graphs on the second line will be completely cancelled against contributions from the crossed ladder topology, as we will show next.

The crossed ladder graphs with a three-gluon vertex, defined in eq.~\eqref{eq:QCD_vertex_2L_XL-3V}, yield the ghost identity,
\begin{align}
\begin{split}
    &i\gs^5\, \tq{c} \left(\frac{C_AC_F}{2}-\frac{C_A^2}{4}\right) q_\mu O^{(2,XL)\,\mu} =
    \eqs[0.22]{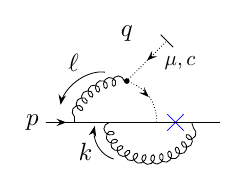}
    -\eqs[0.22]{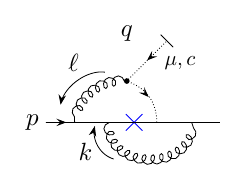}
    \\&\qquad 
    +\eqs[0.22]{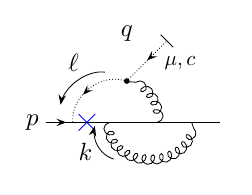}
    -\eqs[0.22]{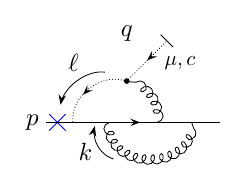}
    +\eqs[0.22]{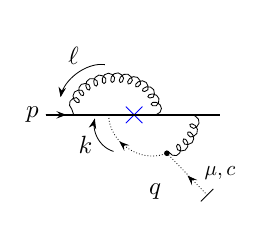}
    \\&\qquad 
    -\eqs[0.22]{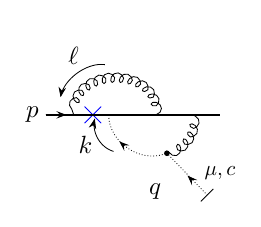} 
    +\eqs[0.22]{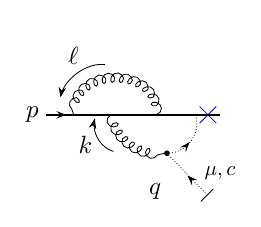}
    -\eqs[0.22]{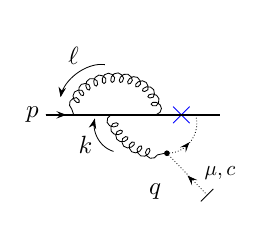}
    \,.
\end{split}
\label{eq:qqg_XL-3V_ward_ghost}
\end{align}
Examining the diagrams closely, we notice that the $C_AC_F/2$ part of the first graphs on the first and third lines of eq.~\eqref{eq:qqg_XL-3V_ward_ghost} cancel against the second and fourth graphs of the uncrossed ladder topology in eq.~\eqref{eq:qqg_UL-3V_ward_ghost_O1}, respectively. Similarly, the first diagram on the second line and the last diagram on the third line of eq.~\eqref{eq:qqg_XL-3V_ward_ghost} cancel completely against the diagrams on the second line of eq.~\eqref{eq:qqg_UL-3V_ward_ghost_O2}. Comparing the remaining contributions against the $(NL-3V)$ topology, we see that the $C_AC_F/2$ coefficient of the second graph on the first line as well as the third graph on the second line of eq.~\eqref{eq:qqg_XL-3V_ward_ghost} (where the fermion propagator $1/(\s{p}+\s{l}+\s{k})^2$ is cancelled by the Ward identity) vanish respectively against the first graphs on the third and last lines of eq.~\eqref{eq:qqg_NL-3V_ward_ghost}. 

Lastly, we take a look at the ghost contributions to the two-loop Ward identity for the $(d3V)$ topology of the QCD vertex, defined in eq.~\eqref{eq:QCD_vertex_2L_d3V}, which take the following form,
\begin{align*}
    &i\gs^5\, t^c \,\frac{C_A^2}{4}\,q_\mu O_1^{(2,d3V)\,\mu} =
    \eqs[0.22]{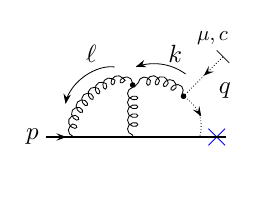}
    -\eqs[0.22]{qqg_QCD_2L_d3V_O1_ward_ghost_RHS_diag_1_2}
    +\eqs[0.25]{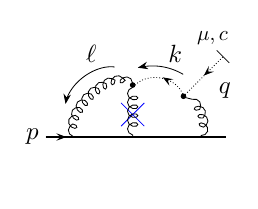}
    \\&\qquad 
    -\eqs[0.25]{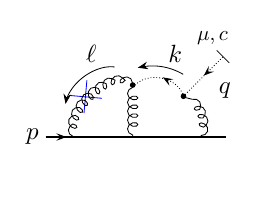}
    +\eqs[0.22]{qqg_QCD_2L_d3V_O1_ward_ghost_RHS_diag_3_1}
    -\eqs[0.22]{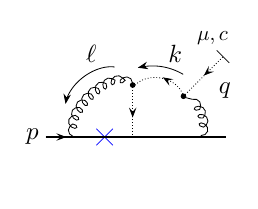}
    \\&\qquad
    +\eqs[0.22]{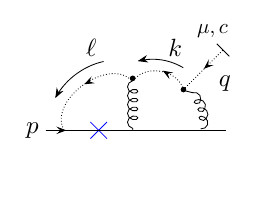}
    -\eqs[0.22]{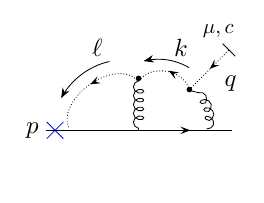}\,.
    \numberthis
\label{eq:qqg_d3V_ward_ghost_O1}
\end{align*}
and
\begin{align*}
    &i\gs^5\, t^c\, \frac{C_A^2}{4}\,q_\mu O_2^{(2,d3V)\,\mu} =
    \eqs[0.22]{qqg_QCD_2L_d3V_O2_ward_ghost_RHS_diag_1_1}
    -\eqs[0.22]{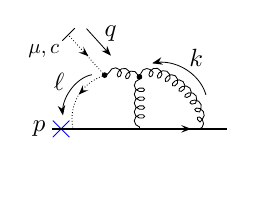}
    +\eqs[0.25]{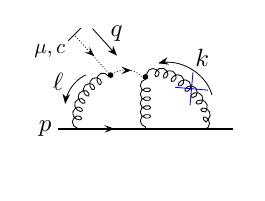}
    \\&\qquad 
    -\eqs[0.25]{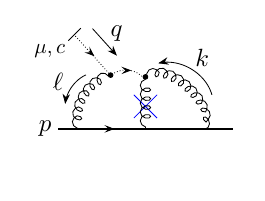}
    +\eqs[0.22]{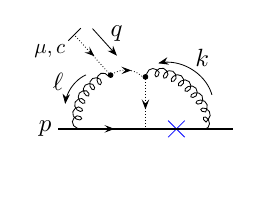}
    -\eqs[0.22]{qqg_QCD_2L_d3V_O2_ward_ghost_RHS_diag_3_2}
    \\&\qquad
    +\eqs[0.22]{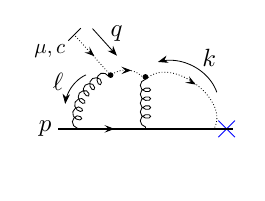}
    -\eqs[0.22]{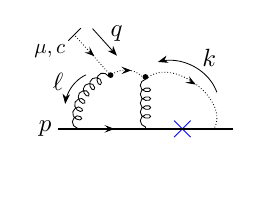}\,.
    \numberthis
\label{eq:qqg_d3V_ward_ghost_O2}
\end{align*}
We observe that the third graph on the second line and the first graph on the third line in eq.~\eqref{eq:qqg_d3V_ward_ghost_O1} as well as the second graph on the second line and last graph on the third line in eq.~\eqref{eq:qqg_d3V_ward_ghost_O2} satisfy, 
\begin{align}
\begin{split}
    &\eqs[0.22]{qqg_QCD_2L_d3V_O1_ward_ghost_RHS_diag_4_1}
    -\eqs[0.22]{qqg_QCD_2L_d3V_O1_ward_ghost_RHS_diag_3_2}
    \equiv
    \eqs[0.2]{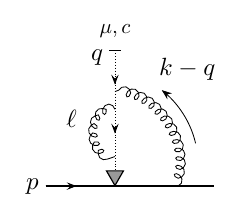}
    \\&\qquad
    =
    i \gs^5\, \frac{C_A^2}{4}\,\tq{c}\,
    \frac{-i}{\s{p}+\s{q}}\,
    \frac{\s{k}(\s{p}+\s{k})\s{k}}{l^2k^2(k-q)^2(l-k)^2(p+k)^2}
     \,\frac{-i}{\s{p}}\,,
\end{split}
\end{align}
and
\begin{align}
\begin{split}
    &\eqs[0.22]{qqg_QCD_2L_d3V_O2_ward_ghost_RHS_diag_3_1}
    -\eqs[0.22]{qqg_QCD_2L_d3V_O2_ward_ghost_RHS_diag_4_2}
    \equiv
    \eqs[0.2]{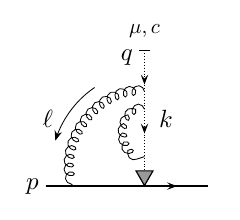}
    \\&\qquad
    =-i \gs^5 \frac{C_A^2}{4}\tq{c}
    \frac{-i}{\s{p}+\s{q}}\,
    \frac{(\s{q}-\s{l})(\s{p}+\s{l})(\s{q}-\s{l})}{l^2k^2(l-q)^2(l-k-q)^2(p+l)^2}
    \,\frac{-i}{\s{p}}\,.
\end{split}
\end{align}
In analogy to the one-loop ghost identity of eq.~\eqref{eq:qqg_1L_ward_ghost_id}, the convention is that we symmetrise over the loop momentum routing of the ghost self-energy subgraph in the pictorial representation on the right-hand sides. Again, we apply the abelian-type Ward identity for collinear insertions on the quark line to obtain,
\begin{align}
    \eqs[0.2]{qqg_2L_d3V_ghost_identity_diag2}
    =\eqs[0.2]{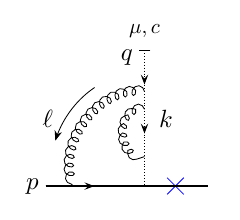}
    -\eqs[0.2]{qqg_2L_d3V_ghost_identity_diag2_2}\,,
    \label{eq:qqg_NF_ward_ghost_fig1}
\end{align}
and
\begin{align}
    \eqs[0.2]{qqg_2L_d3V_ghost_identity_diag1}
    =\eqs[0.2]{qqg_2L_d3V_ghost_identity_diag1_1}
    -\eqs[0.2]{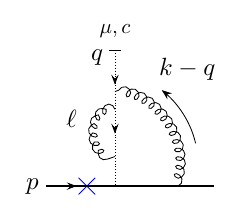}\,.
    \label{eq:qqg_NF_ward_ghost_fig2}
\end{align}
We denote the graphs where the ``internal" fermion propagator is cancelled, either $1/(\s{p}+\s{k})$ or $1/(\s{p}+\s{l})$, by $O^{(2)\,c}_{NF,2}$. The corresponding integrand, defined in eq.~\eqref{eq:ghost_NF}, contains loop polarisation terms that are an impediment to local factorisation and therefore the \ac{ffs} method at higher orders, c.f. the discussion at the end of sec.~\ref{sec:ghosts}.

Finally, we compare all remaining graphs with the colour coefficient $C_A^2/4$. The third graph on the first line of eq.~\eqref{eq:qqg_d3V_ward_ghost_O1} cancels identically against the part of the third crossed ladder graph on the second line of eq.~\eqref{eq:qqg_XL-3V_ward_ghost} proportional to $C_A^2/4$. Likewise, the first graph on the second line of eq.~\eqref{eq:qqg_d3V_ward_ghost_O2} cancels the $C_A^2/4$ part of the second crossed ladder graph on the first line of eq.~\eqref{eq:qqg_XL-3V_ward_ghost}. 

Additionally, for the first time to three-loop order, we also have a shift propagator for the ghost contributions to the QCD Ward identity. This shift term is the sum of the first graph on the second line in eq.~\eqref{eq:qqg_d3V_ward_ghost_O1}, the third graph on the first line of eq.~\eqref{eq:qqg_d3V_ward_ghost_O2}, and the $C_A^2/4$ part of the first crossed ladder graphs on the first and third  lines of eq.~\eqref{eq:qqg_XL-3V_ward_ghost}. This combination yields eq.~\eqref{eq:ghost_2L_shift_2}, represented in terms of planar graphs multiplied by the non-standard colour factor $C_A/(2C_F)$.

All graphs in eqs.~\eqref{eq:qqg_NL-3V_ward_ghost}~-~\eqref{eq:qqg_NF_ward_ghost_fig2} where an external quark propagator is cancelled, either $1/\s{p}$ or $1/(\s{p}+\s{q})$, enter the definition of $\mathcal{O}^{(2)\,c}_2$, corresponding to the first term on the right-hand side of eq.~\eqref{eq:ghost_2L}. Explicitly, we have,
\begin{align*}
   &
   \mathcal{O}^{(2)\,c}_2(p,q,k,l)=
    \eqs[0.19]{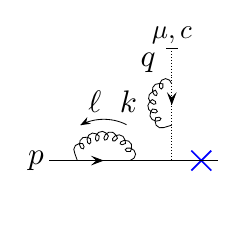}
    -\eqs[0.19]{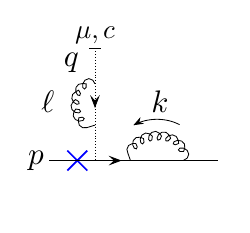}
    +\eqs[0.29]{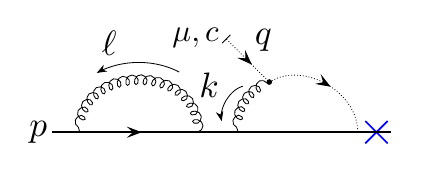}
    \\&\quad
    -\eqs[0.29]{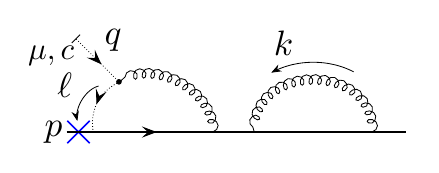}
    +\eqs[0.21]{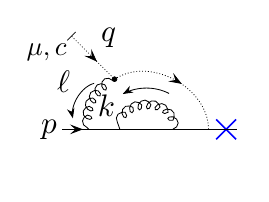}
    -\eqs[0.21]{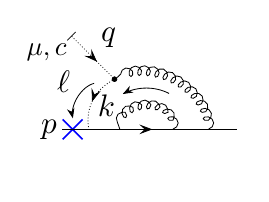}
    \\&\quad
    +\eqs[0.21]{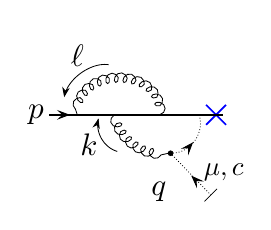}
    -\eqs[0.21]{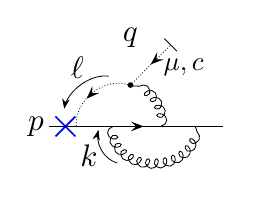}
    +\eqs[0.2]{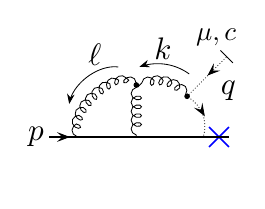}
     \\&\quad
    -\eqs[0.2]{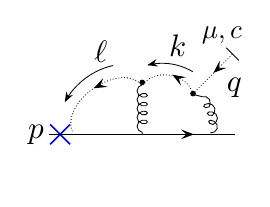}
    +\eqs[0.2]{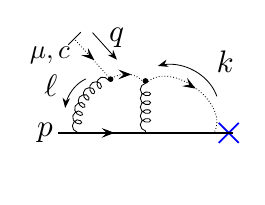}
    -\eqs[0.2]{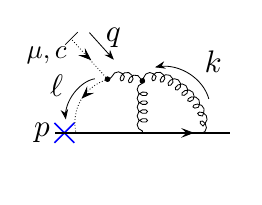}
    \\&\quad
    -\eqs[0.18]{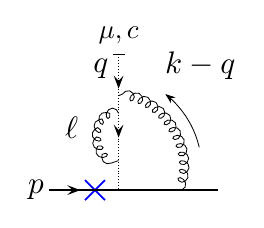}
    +\eqs[0.18]{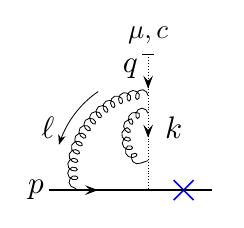}
\numberthis
\label{eq:ghost_2L_ext}
\end{align*}

The remaining, uncancelled graphs combine with ghost contributions to the two-loop Ward identity involving the gluon three-point function. This yields the term $O^{(2)\,c}_{\Delta,2}$, defined in eq.~\eqref{eq:ghost_2L_Delta}.

%% file: app_UV.tex
\section{Two-loop UV counterterms}
\label{app:uv}

In this appendix we provide explicit expressions for the single- and double-UV counterterms of two-loop subgraphs of the three-loop electroweak amplitude.

\subsection{\acs{uv} sub-divergence counterterms for two-loop subgraphs}

In this section, we provide the counterterms that remove divergences in the single-\acs{uv} regions $(H_{k\to\infty},H_l)$ and $(H_k,H_{l\to\infty})$ directly at the integrand level\footnote{In what follows the ``internal" loop momentum $k$ and $l$ flow through gluon lines, while $q$ denotes the momentum of an external off-shell gluon or photon.}. As we will see below, their integrands can be written in terms of the one-loop counterterms of section~\ref{sec:UV_1L}.
It is easy to verify that the approximations in the single-UV regions defined here satisfy the single-UV analog of the QCD Ward identity, eq.~\eqref{eq:Ward_QCD_2L}.

We begin with the single-UV approximations of the electroweak vertex. The uncrossed ladder contributions are given by,
\begin{align}
\begin{split}  
        &\g\, \Gamma_{qq\gamma \text{ single-UV}}^{(2,UL)\,\mu}(p,q,k,l) 
        =i \g\,  \gs^4\, C_F^2\, \sum_{i=1}^3V^{(2,UL)\,\mu}_{i \text{ single-UV}}(p,q,k,l)
        \\&\qquad\equiv \eqs[0.22]{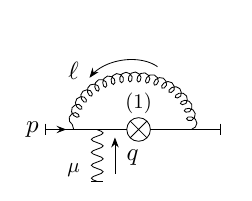}+\eqs[0.22]{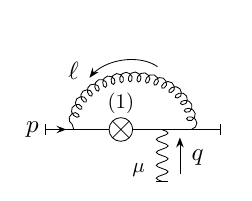}+\eqs[0.22]{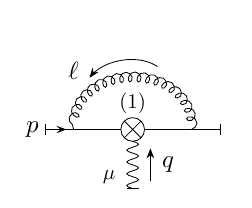}\,.
\end{split}
\label{eq:EW_vertex_2L_UL_sUV}
\end{align}
with
\begin{align}
    \begin{split}
        &V^{(2,UL)\,\mu}_{1 \text{ single-UV}}
        (p,q,k,l) 
        \\&\qquad\qquad=
        -\frac{\gamma^\alpha (\s{p} + \s{l} + \s{q}) S^{(1)}_\text{UV} (p+l+q,k)(\s{p} + \s{l} + \s{q}) \gamma^\mu (\s{p} + \s{l}) \gamma_\alpha}{l^2 (p+l)^2 ((p+l+q)^2)^2}\,,
    \\
        &V^{(2,UL)\,\mu}_{2 \text{ single-UV}}(p,q,k,l) =
        -\frac{\gamma^\alpha (\s{p} + \s{l} + \s{q}) \gamma^\mu (\s{p} + \s{l})  S^{(1)}_\text{UV} (p+l,k)(\s{p} + \s{l}) \gamma_\alpha}{l^2 ((p+l)^2)^2 (p+l+q)^2 }\,,
    \\
        &V^{(2,UL)\,\mu}_{3 \text{ single-UV}}(p,q,k,l) = 
        \frac{\gamma^\alpha (\s{p} + \s{l} + \s{q}) V^{(1)\,\mu}_\text{UV}(k)  (\s{p} + \s{l})  \gamma_\alpha}{l^2 (p+l)^2 (p+l+q)^2}\,.
    \end{split}
\label{eq:EW_vertex_2L_UL_V_UV}
\end{align}
Here, we have simply replaced the one-loop functions $V^{(1)\,\mu}$ and $S^{(1)}$ in eqs.~\eqref{eq:EW_vertex_2L_UL_V1} and~\eqref{eq:EW_vertex_2L_UL_V2} with their \acs{uv}-expansions, $V^{(1)\,\mu}_\text{UV}$ and $S^{(1)}_\text{UV}$ defined in eqs.~\eqref{eq:V_1L_UV} and~\eqref{eq:S_1L_UV}. %
We remark that by simplifying numerators we could find alternative expressions that are equivalent up to terms which are finite in the UV limits. However, this spoils the application of the one-loop Ward identities, and therefore local factorisation in the limit where $q$ becomes collinear to an external fermion. 

It is straightforward to see that the crossed ladder and three-gluon vertex contributions to the single-UV limits combine to,
\begin{align}
    &\g\,
    \Gamma_{qq\gamma, \text{ single-UV}}^{(2,XL+3V)\,\mu}
    \equiv \eqs[0.25]{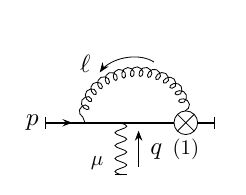}
    +\eqs[0.25]{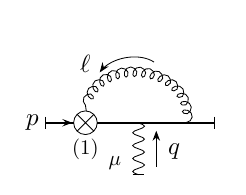} \nonumber \\
    \begin{split}
    &\qquad=
    -i\g\, \gs^2\, 2C_F\,
    \frac{\Gamma_{qqg,\text{ UV}}^{(1)\,\alpha}(k)
    (\s{p} + \s{l} + \s{q}) \gamma^\mu  (\s{p} + \s{l}) \gamma_\alpha}{{ l^2 (p+l)^2 (p+l+q)^2}}
    \,,
    \label{eq:EW_vertex_2L_XL+3V_sUV}
    \end{split}
\end{align}
where the UV counterterm $\Gamma_{qqg,\text{ UV}}^{(1)\,\alpha}$ for the one-loop QCD vertex was defined in eq.~\eqref{eq:QCD_vertex_1L_UV}. Here, the superscript $(XL+3V)$ denotes the sum of crossed ladder and three-gluon vertex contributions to the elctroweak vertex. We note that the third crossed ladder diagram shown in eq.~\eqref{eq:EW_vertex_2L_XL} does not contribute to the single-UV region, as can be shown by simple power counting that the corresponding vertex function $V^{(2,XL)\,\mu}_{3}$, defined in eq.~\eqref{eq:EW_vertex_2L_XL_V3}, is finite when either loop momentum becomes large.

The single-\acs{uv} approximations of the two-loop self-energy corrections contributing to $\Pi_{qq}^{(2)}$ can be written in terms of the UV-approximated vertex functions, up to terms that vanish upon integration. This is in analogy to the one-loop identity eq.~\eqref{eq:EW_selfenergy_1L_UV} for $S^{(1)}_\text{UV}$. The counterterm for the uncrossed ladder diagram reads,
\begin{align}
\begin{split}
    \Pi_{qq \text{ single-UV}}^{(2,UL)}(p,k,l) =
    \eqs[0.27]{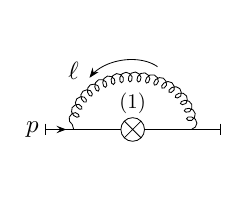} 
    =i\gs^4\, C_F^2\, S^{(2,UL)}_\text{single-UV}(p,k,l)\,,
\end{split}
\end{align}
with
\begin{align*}
    \numberthis
    \label{eq:S2_UL_single-UV}
    &S^{(2,UL)}_\text{single-UV}(p,k,l) = 
    2(1-\e)\frac{\gamma^\alpha (\s{p}+\s{l})S_\text{UV}^{(1)}(p+l,k)(\s{p}+\s{l})\gamma^\alpha}{l^2((p+l)^2)^2}
    \\&\qquad\equiv 
    \frac{2(1-\e)}{(k^2-\m^2)^2}
    \frac{\gamma^\mu(\s{p}+\s{l}+\s{q})\s{k}(\s{p}+\s{l})\gamma_\mu}{l^2(p+l)^2(p+l+q)^2}
    +(p+l+q)_\mu \sum_{i=1}^3V^{(2,UL)\,\mu}_{i \text{ single-UV}}(p,q,k,l)\,.
\end{align*}
The first term after the second equality diverges linearly in the UV region where $k$ is large, but again corresponds to an odd integral in $k$ and therefore vanishes.

The crossed ladder and three-gluon vertex self-energy contributions are singular single-UV region when either gluon loop momentum become large. We can combine them to obtain,
\begin{align*}
    \numberthis
    &
    \Pi_{qq \text{ single-UV}}^{(2,XL+3V)}(p,k,l)
    \equiv
    \eqs[0.25]{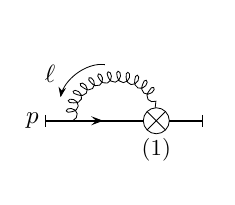} 
    +\eqs[0.25]{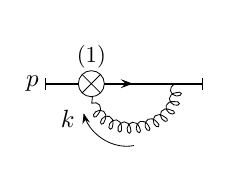} 
    \\&\qquad=i\gs^4 
    \left[
    C_F\left(C_F-\frac{C_A}{2}\right)S^{(2,XL)}_\text{single-UV}(p,k,l)
    -\frac{C_F C_A}{2}S^{(2,3V)}_\text{single-UV}(p,k,l)\right]
    +(k\leftrightarrow l)
    \,,
\end{align*}
where we define
\begin{align}
\begin{split}
    &S^{(2,XL)}_\text{single-UV}(p,k,l)
    = \frac{ V^{(1)\,\alpha}_\text{UV}(k)(\s{p}+\s{l})\gamma_\alpha}{l^2(p+l)^2} 
    \equiv
    (p+l+q)_\mu V^{(2,XL)\,\mu}_{1 \text{ single-UV}}(p,q,k,l)
    \,,
\end{split}
\label{eq:S2_XL_single-UV}
\end{align}
and
\begin{align}
\begin{split}
    &
    S^{(2,3V)}_\text{single-UV}(p,k,l)
    = \frac{W^{(1)\,\alpha}_\text{UV}(k)(\s{p}+\s{l})\gamma_\alpha}{l^2(p+l)^2} 
    \equiv
    (p+l+q)_\mu
    V^{(2,3V)\,\mu}_{1 \text{ single-UV}}(p,q,k,l)
    \,.
\end{split}
\label{eq:S2_3V_single-UV}
\end{align}
The vertex functions in the expressions on the second line of eqs.~\eqref{eq:S2_XL_single-UV} and~\eqref{eq:S2_3V_single-UV} are given by,
\begin{align}
\begin{split}
    V^{(2,XL)\,\mu}_{1 \text{ single-UV}}(p,q,k,l)
    \equiv V^{(2,XL)\,\mu}_{2 \text{ single-UV}}(p,q,k,l) =
    \frac{V^{(1)\,\alpha}_\text{UV}(k)(\s{p} + \s{l} + \s{q}) \gamma^\mu  (\s{p} + \s{l}) \gamma_\alpha }{ l^2 (p+l)^2 (p+l+q)^2} \,,
\end{split}
\label{eq:EW_vertex_2L_XL_V_UV}
\end{align}
and
\begin{align}
\begin{split}
    V^{(2,3V)\,\mu}_{1 \text{ single-UV}}(p,q,k,l)
    \equiv V^{(2,3V)\,\mu}_{2 \text{ single-UV}}(p,q,k,l)
    &=\frac{W^{(1)\,\alpha}_\text{UV}(k)(\s{p}+\s{l}+\s{q})\gamma^\mu(\s{p}+\s{l})\gamma_\alpha}{l^2(p+l)^2(p+l+q)^2}\,.
\end{split}
\label{eq:EW_vertex_2L_3V_V_UV}
\end{align}
Again, we have simply replaced the one-loop functions $V^{(1)\,\mu}$ and $W^{(1)\,\mu}$ by their UV-approximations $V^{(1)\,\mu}_\text{UV}$ and $W^{(1)\,\mu}_\text{UV}$, defined in eqs.~\eqref{eq:V_1L_UV} and~\eqref{eq:W_1L_UV}, respectively, without simplifying numerators. 

We note that together, eqs.~\eqref{eq:EW_vertex_2L_UL_sUV} and~\eqref{eq:EW_vertex_2L_XL+3V_sUV} respect local collinear factorisation by virtue of the abelian one-loop Ward identity, eq.~\eqref{eq:qq_Ward}, and its UV equivalent, eq.~\eqref{eq:Ward_QED_1L_UV},
\begin{align}
    q_\mu \Gamma_{qq\gamma, \text{ single-UV}}^{(2)\,\mu}(p,q,k,l) = 
    \Pi_{qq \text{ single-UV}}^{(2)}(p,k,l)
    -\Pi_{qq \text{ single-UV}}^{(2)}(p+q,k,l)\,,
    \label{eq:Ward_QED_2L_sUV}
\end{align}
where $\Gamma_{qq\gamma, \text{ single-UV}}^{(2)\,\mu}$ is the sum of eqs.~\eqref{eq:EW_vertex_2L_UL_sUV} and~\eqref{eq:EW_vertex_2L_XL+3V_sUV}.

Next, we discuss the single-UV limits of the two-loop corrections to the quark-antiquark-gluon vertex, introduced in appendix~\ref{sec:Greens_2L}.
By simple power counting it is easy to see that the four-gluon vertex contribution $\Gamma_{qqg}^{(2,4V)\,\mu}(p,q,k,l)$, defined in eq.~\eqref{eq:QCD_vertex_2L_4V}, is finite when either loop momentum $k$ or $l$ becomes large. For the $(UL)$, $(XL)$ and $(3V)$ topologies we have already provided the UV-approximated kinematic functions in eqs.~\eqref{eq:EW_vertex_2L_UL_V_UV},~\eqref{eq:EW_vertex_2L_XL_V_UV} and~~\eqref{eq:EW_vertex_2L_3V_V_UV} when we discussed the electroweak vertex (only the colour factors change). For each of the remaining topologies, we obtain the approximated integrands, 
\begin{align}
    \Gamma_{qqg}^{(2,X)\,\mu,c} \to
    \Gamma_{qqg \text{ single-UV}}^{(2,X)\,\mu,c}\,,\quad
    X\in\{UL-3V,XL-3V,d3V\}\,,
\end{align}
by replacing the one-loop functions $S^{(1)}$, $V^{(1)\,\mu}$ and $S^{(1)}$ in the numerators by their UV counterparts $S^{(1)}_\text{UV}$, $V^{(1)\,\mu}_\text{UV}$ and $S^{(1)}_\text{UV}$ defined section~\ref{sec:UV_1L}. For example, for the three-point function $\Gamma_{qqg}^{(2,UL-3V)\,\mu,c}$ defined in eq.~\eqref{eq:QCD_vertex_2L_UL-3V}, the single-UV limit yields,
\begin{align}
\begin{split}%
        &\Gamma_{qqg \text{ single-UV}}^{(2,UL-3V)\,\mu,c} =
        i \gs^5\, \tq{c}
        \left[
        \frac{C_A C_F}{2}W^{(2,UL)\,\mu}_{1 \text{ single-UV}}
        +\frac{C_A}{2}\left(C_F-\frac{C_A}{2}\right)
        W^{(2,UL)\,\mu}_{2 \text{ single-UV}}
        \right]\,.
\end{split}
\label{eq:QCD_vertex_2L_UL-3V_UV}
\end{align}
Here, $W^{(2,UL)\,\mu}_{1 \text{ single-UV}}$ and $W^{(2,UL)\,\mu}_{2 \text{ single-UV}}$ are obtained from eqs.~\eqref{eq:W2_UL_1} - \eqref{eq:W2_UL_2} by making the following replacements,
\begin{align}
    \lim_{k\to \infty} S^{(1)}(p+l,k) \sim S^{(1)}_\text{UV}(p+l,k)\,, \quad
    \lim_{k\to \infty} W^{(1)\,\mu}(p+l,q,k) \sim W^{(1)\,\mu}_\text{UV}(k)\,,
\end{align}
with $S^{(1)}_\text{UV}$ and $W^{(1)\,\mu}_\text{UV}$ defined in eqs.~\eqref{eq:S_1L_UV} and~\eqref{eq:W_1L_UV}, respectively.

Similarly, the two-loop shift terms $\Pi^{(2,X)\,\text{shift}}_{qq}$ with $X\in\{UL,XL,3V\}$, introduced in section~\ref{sec:ward_2L}, are singular in the single-UV regions. For the uncrossed ladder contribution, the singularities are subtracted with the counterterm,
\begin{align*}
   &\Pi_{qq \text{ single-UV}}^{(2,UL)\,\text{shift}}(p,q,k,l) = 
   i\gs^4\, \frac{C_AC_F}{2}\left[S^{(2,UL)}_{\text{single-UV}}(p+q,k,l-q)-S^{(2,UL)}_{\text{single-UV}}(p+q,k,l)\right]
   \\&\qquad-i\gs^2\left(C_F-\frac{C_A}{2}\right)
   \frac{\gamma^\alpha(\s{p}+\s{l}+\s{q})\Pi_{qq\,\text{UV}}^{(1)\,\text{shift}}(p+l,q,k)(\s{p}+\s{l})\gamma_\alpha}{l^2(p+l)^2(p+l+q)^2}\,,
   \numberthis
\label{eq:shift_2L_UL_single-UV}
\end{align*}
with $S^{(2,UL)}_{\text{single-UV}}$ and $\Pi_{qq\,\text{UV}}^{(1)\,\text{shift}}$ defined in eqs.~\eqref{eq:S2_UL_single-UV} and~\eqref{eq:shift_1L_UV}, respectively. The crossed ladder and triple-gluon vertex contributions yield, respectively,
\begin{align}
\begin{split}
    &
    \Pi_{qq \text{ single-UV}}^{(2,XL)\,\text{shift}}(p,q,k,l) =
    i\gs^4\, \frac{C_A}{2}\left(C_F-\frac{C_A}{2}\right)
    \left[
    S^{(2,XL)}_\text{single-UV}(p+q,k,l-q)\right.
    \\& \qquad \left. 
    -S^{(2,XL)}_\text{single-UV}(p+q,k,l)
    +(k\leftrightarrow l)\right]\,,
\end{split}
\label{eq:shift_2L_XL_single-UV}
\end{align}
and
\begin{align}
\begin{split}
    &
    \Pi_{qq \text{ single-UV}}^{(2,3V)\,\text{shift}}(p,q,k,l) =
    -i\gs^4\,\frac{C_A^2}{4}
    \left[
    S^{(2,3V)}_\text{single-UV}(p+q,k,l-q)\right.
    \\& \qquad \left. 
    -S^{(2,3V)}_\text{single-UV}(p+q,k,l)
    +(k\leftrightarrow l)\right]\,.
\end{split}
\label{eq:shift_2L_3V_single-UV}
\end{align}
Diagrammatically, we can represent the sum of eqs.~\eqref{eq:shift_2L_XL_single-UV} and~\eqref{eq:shift_2L_3V_single-UV} by
\begin{align}
\begin{split}
    \Pi_{qq \text{ single-UV}}^{(2,XL+3V)\,\text{shift}}(p,q,k,l)
    &\equiv\frac{C_A}{2C_F}\Bigg[
    \eqs[0.22]{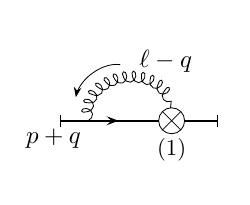}
    -\eqs[0.22]{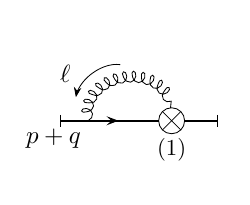}
    \Bigg]
    \\&
    +(k\leftrightarrow l)
    \,.
\end{split}
\end{align}
We emphasise that each of the shift counterterms integrates to zero, which also holds in the single-UV (and double-UV) regions,
\begin{align}
    \int_{k,l}\,\Pi_{qq \text{ single-UV}}^{(2,X)\,\text{shift}}(p,q,k,l) = 0\,,\qquad
    X\in\{UL,XL,3V\}\,.
\end{align}

As we have explained in section~\ref{sec:ghosts}, we obtain additional shift-integrable terms associated to ghosts, $O^{(2)\,\text{shift}}_{2}$ and $O^{(2)\,\text{shift}}_{\Delta,2}$ defined in eqs.~\eqref{eq:ghost_2L_shift_2} and~\eqref{eq:ghost_delta_shift_2}, respectively. These require the following counterterms for the single-UV regions,
\begin{align}
    \begin{split}
        &O^{(2)\,\text{shift}}_{2, \text{ single-UV}}(p,q,k,l) =
        -i\gs^3\,\frac{C_A}{2}\,
        \frac{1}{l^2(l-q)^2(p+l)^2}
        \left[
        \s{l}(\s{p}+\s{l})\,\Pi^{(1)\,\text{shift}}_{\,\text{UV}}(l,k) \right.
        \\&\qquad +\left.
        \Pi^{(1)\,\text{shift}}_{\,\text{UV}}(l-q,k)(\s{p}+\s{l})(\s{q}-\s{l})
        \right]\,,
    \end{split}
\label{eq:ghost_2L_shift_2_single-UV}
\end{align}
and
\begin{align*}
        &O^{(2)\,\text{shift}}_{\Delta,2 \text{ single-UV}}(p,q,k,l) =
        i\gs^5\,C_A^2\,\frac{(1-\e)}{\prop{l}^2}
        \left(\frac{\s{p}+\s{k}+\s{q}}{k^2(p+k+q)^2}
        -\frac{\s{p}+\s{k}}{(q-k)^2(p+k)^2}\right)
        \\&\qquad
        -i\gs^3\,\frac{C_A}{2}\,\frac{\Pi^{(1)\,\text{shift}}_{\,\text{UV}}(q,k)}{l^2(l-q)^2}\,.
\numberthis
\label{eq:ghost_delta_shift_2_single-UV}
\end{align*}

\subsection{Double-\acs{uv} counterterms at two loops.}
After treating the single-UV region in which the two-loop amplitude diverges when one of the loop momenta approaches infinity, we turn our attention to the double-UV limit, in which both loop momenta $k$, $l$ become infinitely large.
Below, we will find it particularly convenient to symmetrise the kinematic functions\footnote{Antisymmetric terms in the loop momenta $k,l$ vanish and the numerator $\mathcal{N}^\mu_{qqg \text{ double-UV}}$ belonging to the double-UV approximation of the three-point function, $\Gamma^{(2)\,\mu}_{qqg \text{ double-UV}}$, satisfies $\mathcal{N}^\mu_{qqg \text{ double-UV}} = f_1^\mu\s{k}+f_2^\mu\s{l}+f_3^\mu\gamma^\mu$, where $f_i^\mu = f_i^\mu(k,l)$.} in the loop-momenta $k$ and $l$.

We begin with the uncrossed ladder vertex functions $V_{i}^{(2,UL)\,\mu}$, defined in eqs.~\eqref{eq:EW_vertex_2L_UL_V1}-\eqref{eq:EW_vertex_2L_UL_V3}. Though strictly speaking a symmetrisation is not required since the integrand is simple enough, we find it convenient in the application of Ward identities in the collinear regions. Since $V_1^{(2,UL)\,\mu}$ and $V_2^{(2,UL)\,\mu}$ have the same colour coefficient both as part of the electroweak and QCD vertex, we find it useful to combine them,
\begin{align}
\begin{split}
    &\mathscr{S}_{k,l}\,\left(V_{1 \text{ double-UV}}^{(2,UL)\,\mu}(k,l)+V_{2 \text{ double-UV}}^{(2,UL)\,\mu}(k,l)\right) \\
    &\qquad =
    \frac{2(1-\e)^2}{\prop{l}^2}\left[
    \frac{4l^\mu \s{l}}{\prop{l}^2}\left(
    \frac{1}{k^2-\m^2}-\frac{1}{(l+k)^2-\m^2}
    \right. \right.\\ &\qquad \left.\left.
    -\frac{2k\cdot l}{\prop{k}^2}
    \left(1 - \frac{2k\cdot l}{k^2-\m^2}\right)
    \right)
    + \frac{2k^\mu\s{l}+2l^\mu\s{k}}{\prop{l}\prop{k}}
    \right. \\ &\qquad \left.
    \times
    \left(
    \frac{1}{k^2-\m^2}
    \left(1 - \frac{2k\cdot l}{k^2-\m^2}\right)
    -\frac{1}{(l+k)^2-\m^2}
    \right)
    \right. \\ &\qquad \left.
    -\gamma^\mu \left(
    \frac{1}{\prop{l}\prop{k}}
    -\frac{2k\cdot l}{\prop{l}\prop{k}^2}
     \left(1 - \frac{2k\cdot l}{k^2-\m^2}\right)
    \right.\right. \\ &\qquad \left.\left.
    +\frac{1}{\prop{k}\prop{(l+k)}}
     -\frac{1}{\prop{l}\prop{(l+k)}}
    \right)
    \right]
    \\&\qquad
    +2(1-\e)\frac{V^{(1)\,\mu}_\text{UV}(l)}{\prop{k}^2}
    + (k \leftrightarrow l)\,.
\end{split}
\end{align}
Here, we have introduced a symmetrisation operator $\mathscr{S}_{k,l}$ as follows,
\begin{align}
    \mathscr{S}_{l,k}\,\mathcal{I}(l,k)= 
    \frac{1}{2}\left(
    \mathcal{I}(l,k) + (l \leftrightarrow k)
    \right)\,.
    \label{eq:sym_int}
\end{align}
where $\mathcal{I}$ is a kinematic function of the loop momenta $k$ and $l$.
The remaining double-UV approximation reads
\begin{align}
\begin{split}
    &\mathscr{S}_{k,l}\,V_{3 \text{ double-UV}}^{(2,UL)\,\mu}(k,l)
    \\&\qquad =
    \frac{1-\e}{k^2-\m^2}
    \left[V^{(1)\,\mu}_\text{UV}(l)\left(\frac{1}{(l+k)^2-\m^2}
    -\frac{1}{k^2-\m^2}\right)
    +\frac{2k^\mu k_\alpha V^{(1)\,\alpha}_\text{UV}(l)}{\prop{k}^2}
    \right. \\ &\qquad \left.
    -\frac{4(1-\e)(l+k)^\mu}{\prop{k}\prop{(l+k)}}
    \left(
    \frac{1}{(l+k)^2-\m^2}
    \left(\frac{\s{k}}{k^2-\m^2}
    +\frac{\s{l}}{l^2-\m^2}\right)
    \right. \right.\\\ &\qquad \left.\left.
    -\frac{\s{k}}{\prop{l}\prop{k}}
    \right)
    \right]
    + (k \leftrightarrow l)\,.
\end{split}
\end{align}
Then, the double-UV counterterm for the two-loop uncrossed ladder contribution to the two-point function can be written as,
\begin{align}
    &\mathscr{S}_{k,l}\,S^{(2,UL)}_\text{double-UV}(p,k,l)
    = -p_\mu\left(\mathscr{S}_{k,l}\, V^{(2,UL)\,\mu}_\text{double-UV}(k,l)\right)
    \nonumber\\ &\qquad
    +\left\{
    \frac{2(1-\e)^2}{\prop{l}^2}\left[
        \left(\frac{\s{l}}{l^2-\m^2}
    + \frac{\s{k}}{k^2-\m^2}\right)
    \left(
    \frac{1}{k^2-\m^2}
    - \frac{2l\cdot k}{\prop{k}^2} 
    \right. \right. \right. \\&\qquad \left.\left.\left. 
    -\frac{1}{(l+k)^2-\m^2}
    \right)
    -\frac{\s{l}}{\prop{k}^2}
    +\frac{4(l\cdot k)^2 \s{l}}{\prop{l}\prop{k}^3}
    \right]
    + (k \leftrightarrow l)
    \right\}\,.\nonumber
\end{align}

For the crossed ladder contributions we have
\begin{align}
\begin{split} %
    &
    \mathscr{S}_{k,l}\,
    \left(
        V_{1 \text{ double-UV}}^{(2,XL)\,\mu}(k,l)+V_{2 \text{ double-UV}}^{(2,XL)\,\mu}(k,l)
    \right) \\
    &\qquad =
    \frac{2(1-\e)}{(l^2-\m^2)(k^2-\m^2)}\left[
    \frac{4(l+k)^\mu (\s{l}+\s{k})}{\prop{l}\prop{k}\prop{(l+k)}}
    \right.\\
    &\qquad\left.
    + \frac{4\e\,l^\mu }{\prop{l}\prop{(l+k)}}\left(
    \frac{\s{l}}{l^2-\m^2}
    + \frac{\s{k}}{k^2-\m^2}
    \right)
    \right.\\
    &\qquad\left.
    +\frac{\e \gamma^\mu}{\prop{l}\prop{k}}
    + \frac{4l^\mu(\s{l}+\s{k})}{\prop{l}^2}\left(
    \frac{1}{(l+k)^2-\m^2}
    - \frac{1}{k^2-\m^2}
    \right)
    \right.\\
    &\qquad\left.
    -\frac{4(1+\e)l^\mu\s{l}}{\prop{l}^2\prop{k}}
    +\frac{8 k\cdot l\, l^\mu \s{k}}{\prop{l}^2\prop{k}^2}
    \right]
    +(k \leftrightarrow l)\,,
\end{split}
\end{align}
and
\begin{align}%
    \begin{split}
        &
        \mathscr{S}_{k,l}\,
        V_{3 \text{ double-UV}}^{(2,XL)\,\mu}(k,l)
    \\&\qquad
    =
    -\frac{2(1-\e)}{\prop{k}\prop{l}}\left[
    \frac{2\e\,l^\mu \s{k}}{\prop{l}\prop{k}\prop{(l+k)}}
    \right. \\
    &\qquad\left.
    +\frac{2(l+k)^\mu}{\prop{k}\prop{(l+k)}}
    \left(
    \frac{\s{l}+\s{k}}{l^2-\m^2}
    -\frac{2\s{l}+2(1+\e)\s{k}}{(l+k)^2-\m^2}
    \right)
    \right. \\
    &\qquad\left.
    -\frac{(2+\e)\gamma^\mu}{k^2-\m^2}\left(
    \frac{1}{2\prop{l}}
    -\frac{1}{(l+k)^2-\m^2}
    \right)
    \right]
    +(k \leftrightarrow l)\,.
    \end{split}
\end{align}
Then, the corresponding crossed self-energy contribution can be written as
\begin{align}
    \begin{split}
        &\mathscr{S}_{k,l}\,S^{(2,XL)}(p,k,l)=
        -p_\mu\left(\mathscr{S}_{k,l}\,V^{(2,XL)\,\mu}_{\text{double-UV}}(k,l)\right)
        \\&\qquad
        -\frac{2(1-\e)}{\prop{l}\prop{k}}\left[
        \frac{\s{l}+\s{k}}{l^2-\m^2}\left(
        \frac{1+\e}{k^2-\m^2}
        -\frac{2}{(l+k)^2-\m^2}
        \right)
        \right. \\
        &\qquad\left.
        -\frac{2\e\,\s{l}}{\prop{l}\prop{(l+k)}}
        - \frac{4l\cdot k \s{l} }{\prop{l}^2\prop{k}}
        + (k \leftrightarrow l)
        \right]\,.
    \end{split}
\end{align}

Next, we turn our attention to the three-gluon vertex contribution, whose three-point function in the double-UV limit reads,
\begin{align}
    \begin{split}
        &\mathscr{S}_{k,l}\,V^{(2,3V)\,\mu}(k,l)=
        \frac{2(1-\e)}{\prop{l}\prop{k}}\left[
        \frac{4l^\mu(\s{l}+\s{k})}{\prop{l}^2\prop{k}}
        \right. \\&\qquad \left.
        -\frac{(\s{l}-\s{k})(l-k)^\mu-2l^\mu \s{l}}{\prop{l}\prop{k}\prop{(l-k)}}
        +\frac{4l^\mu(\s{l}-\s{k})}{\prop{l}^2\prop{(l-k)}}
        \right. \\ & \qquad \left.
        -\frac{\gamma^\mu}{k^2-\m^2}\left(
        \frac{2}{(l-k)^2-\m^2}+\frac{1}{l^2-\m^2}
        \right)
        +\frac{2V^{(1)\,\mu}_\text{UV}(k)}{\prop{l}^2}
        \right. \\ & \qquad \left.
        -\frac{2l^\mu l_\alpha V^{(1)\,\alpha}_\text{UV}(k)}{\prop{l}^3}
        \right]
        + (k \leftrightarrow l)\,.
    \end{split}
\end{align}
The corresponding self-energy contribution can be written as
\begin{align}
    \begin{split}
        &\mathscr{S}_{k,l}\,S^{(2,3V)}(p,k,l)=
        -p_\mu\left(\mathscr{S}_{k,l}\,V^{(2,3V)\,\mu}_{\text{double-UV}}(k,l)\right)
        \\&\qquad
        +\frac{2(1-\e)}{\prop{l}\prop{k}}\left[
        \frac{\s{l}-\s{k}}{(l-k)^2-\m^2}\left(
        \frac{1}{l^2-\m^2}-\frac{1}{k^2-\m^2}
        \right)
        \right. \\ &\qquad \left.
        -\frac{\s{l}+\s{k}}{\prop{l}\prop{k}}
        + \frac{4l\cdot k \s{l}}{\prop{k}\prop{l}^2}
        + (k \leftrightarrow l)
        \right]\,.
    \end{split}
\end{align}

If the external particle is a gluon, we have additional non-abelian topologies which contribute to the double-UV limits. We begin with the four-gluon vertex function, defined in eq.~\eqref{eq:QCD_vertex_2L_4V_W}. It has a particularly simple form if we symmetrise in $k$ and $l$,
\begin{align}
        &\mathscr{S}_{k,l}\,W^{(2,4V)\,\mu}_\text{double-UV}(k,l)
        =- \frac{2(1-\e)}{\prop{l}\prop{k}}
        \left[
        \frac{2l^\mu \s{k}}{\prop{l}\prop{k}\prop{(l-k)}}
        \nonumber\right. \\ &\qquad \left.
        + \frac{\gamma^\mu}{l^2-\m^2}\left(
        \frac{1}{k^2-\m^2}-\frac{2}{(l-k)^2-\m^2}
        \right)
        \right] + (k \leftrightarrow l)\,.
\end{align}

Next, we discuss the uncrossed-ladder-type topology with a three-gluon vertex, denoted by $\Gamma^{(2, UL-3V)}$, consisting of two diagrams shown on the right-hand side of eq.~\eqref{eq:QCD_vertex_2L_UL-3V}. 
The kinematic function $W_1^{(2,UL)}$ belonging to the first diagram in eq.~\eqref{eq:QCD_vertex_2L_UL-3V} has the following expansion in the double-UV limit,
\begin{align*}
        &\mathscr{S}_{k,l}\,Q^{(2,UL)\,\mu}_{1,\text{ double-UV}}(k,l) = 
        \frac{4(1-\e)^2 l^\mu}{\prop{l}^3}\left[
        \frac{1}{(l+k)^2-\m^2}\left(
        \frac{\s{l}}{l^2-\m^2}+\frac{\s{k}}{k^2-\m^2}
        \right)
        \right. \\ &\qquad \left.
        -\frac{1}{\prop{k}^2}\left(\s{k}-\frac{2l\cdot k\s{l}}{l^2-\m^2}\right)
        \left(1-\frac{2l\cdot k}{k^2-\m^2}\right)
        +\frac{\s{l}}{k^2-\m^2}\left(
        \frac{1}{k^2-\m^2}-\frac{1}{l^2-\m^2}
        \right)
        \right]
        \\ &\qquad 
        +(k \leftrightarrow l)\,. \numberthis
\end{align*}
Likewise, the contribution associated to ghosts has the double-UV counterterm,
\begin{align*}
    &\mathscr{S}_{k,l}\,O^{(2,UL)\,\mu}_{1,\text{ double-UV}}(k,l) = \frac{1-\e}{\prop{l}^2}\Biggl\{
    \frac{2l^\mu\s{k}-2k^\mu\s{k}}{\prop{l}\prop{k}}
    \Biggl[\frac{1}{k^2-\m^2}\Biggl(
    1-\frac{2l\cdot k}{k^2-\m^2}
    \Biggr) 
    \\ & \qquad 
    -\frac{1}{(l+k)^2-\m^2}\Biggr]
    + \gamma^\mu \Biggl[
    \frac{1}{(l+k)^2-\m^2}\Biggl(
    \frac{1}{l^2-\m^2}-\frac{1}{k^2-\m^2}
    \Biggr)
    + \frac{2}{\prop{k}^2}
     \\ & \qquad 
    -\frac{1}{\prop{l}\prop{k}}
    + \frac{2l \cdot k}{\prop{l}\prop{k}^2}\Biggl(
    1-\frac{2l\cdot k}{k^2-\m^2}
    \Biggr)
    \Biggr]
    \Biggr\}
    \numberthis
    \\ & \qquad 
    +(k \leftrightarrow l)\,.
\end{align*}
The term $W_2^{(2,UL)}$ belonging to the second diagram in eq.~\eqref{eq:QCD_vertex_2L_UL-3V}, has the following double-UV scalar and ghost contributions,
\begin{align}
\begin{split}
    &\mathscr{S}_{k,l}\,Q^{(2,UL)\,\mu}_{2,\text{ double-UV}}(k,l) =\frac{4(1-\e)^2 k^\mu}{\prop{l}^2\prop{k}}
    \left[
    \frac{1}{(l+k)^2-\m^2}\left(\frac{\s{l}}{l^2-\m^2}+\frac{\s{k}}{k^2-\m^2}\right)
    \right. \\ & \qquad \left.
    - \frac{\s{k}}{\prop{k}\prop{(l+k)}}
    \right]
    + (1-\e)\left(\frac{Q_\text{UV}^{(1)\,\mu}(k)}{\prop{l}^2}
    -\frac{2l^\mu l_\alpha Q_\text{UV}^{(1)\,\alpha}(k)}{\prop{l}^3}
    \right)
    +(k \leftrightarrow l)\,,
\end{split}
\end{align}
and
\begin{align*}
    &\mathscr{S}_{k,l}\,O^{(2,UL)\,\mu}_{2,\text{ double-UV}}(k,l) =
    \frac{(1-\e)}{\prop{l}^2\prop{k}}
    \left[
    \frac{4(k+l)^\mu \s{l}}{\prop{k}\prop{(l+k)}}
    \right. \\ & \qquad \left.
    -\frac{4l^\mu\s{l}}{\prop{l}\prop{k}}
    +\frac{\gamma^\mu}{k^2-\m^2}
    \right]
    + (1-\e)\left(\frac{O_\text{UV}^{(1)\,\mu}(k)}{\prop{l}^2}
    -\frac{2l^\mu l_\alpha O_\text{UV}^{(1)\,\alpha}(k)}{\prop{l}^3}
    \right)
     \\ & \qquad 
    +(k \leftrightarrow l)\,, \numberthis
\end{align*}
where $Q^{(1)}_\text{UV}$ and $O^{(1)}_\text{UV}$ are defined in eqs.~\eqref{eq:Q_1L_UV} and~\eqref{eq:O_1L_UV}.

Similarly, the scalar and ghost parts of the uncrossed-ladder-type contributions read,
\begin{align}
    \begin{split}
        &\mathscr{S}_{k,l}\,Q^{(2,XL)\,\mu}_{\text{ double-UV}}(k,l) =
        \frac{2(1-\e)}{\prop{l}^2}\left\{
        \frac{2(l+k)^\mu(\s{l}+\s{k})}{\prop{k}^2 \prop{(l+k)}}
        \right. \\ & \qquad \left.
        + \frac{8l\cdot k l^\mu \s{k}}{\prop{l}\prop{k}^3}
        - \frac{4\e l^\mu}{k^2-\m^2}
        \left[
        \frac{\s{l}}{\prop{l}\prop{k}}
        \right. \right.\\ & \qquad \left.\left.
        -\frac{1}{(l+k)^2-\m^2}\left(\frac{\s{l}}{l^2-\m^2}
        +\frac{\s{k}}{k^2-\m^2}
        \right)\right]
        \right. \\ & \qquad \left.
        -\frac{4l^\mu(\s{l}+\s{k})}{\prop{l}\prop{k}}
        \left(
        \frac{1}{k^2-\m^2}-\frac{1}{(l+k)^2-\m^2}
        \right)
        \right\} + (k \leftrightarrow l)\,,
    \end{split}
\end{align}
and
\begin{align*}
        &\mathscr{S}_{k,l}\,O^{(2,XL)\,\mu}_{\text{ double-UV}}(k,l) =
        \frac{2(1-\e)}{\prop{l}^2}\left\{
        \frac{(l+k)^\mu(\s{l}+\s{k})}{\prop{k}^2 \prop{(l+k)}}
        \right. \\ & \qquad \left.
        + \frac{l^\mu \s{k}-k^\mu \s{l}}{\prop{l}\prop{k}}\left(\frac{1}{k^2-\m^2}-\frac{1}{(l+k)^2-\m^2}\right)
        \right. \\ & \qquad \left.
        - \frac{2l^\mu \s{l}}{\prop{l}\prop{k}^2}
        + \frac{1}{2}\gamma^\mu
        \left[
        \frac{1}{(l+k)^2-\m^2}\left(
        \frac{1}{l^2-\m^2}-\frac{1}{k^2-\m^2}
        \right)
        \right.\right. \\ & \qquad \left.\left.
        +\frac{1}{k^2-\m^2}
        \left(
        \frac{2l\cdot k}{\prop{l}\prop{k}}
        \left(1-\frac{2l\cdot k}{k^2-\m^2}\right)
        +\frac{3}{k^2-\m^2}
        \right.\right. \right.\\ & \qquad \left.\left.\left.
        -\frac{1}{l^2-\m^2}
        \right)
        \right]
        \right\}+ (k \leftrightarrow l)\,,
        \numberthis
\end{align*}
where we have defined, $X^\mu=X_1^\mu + X_2^\mu$ for $X\in\{Q^{(2,XL)},O^{(2,XL)}\}$. In a similar fashion, the symmetrised double-UV counterterms for the double three-gluon vertex part is
\begin{align}
    \begin{split}
        &\mathscr{S}_{k,l}\,Q^{(2,d3V)}_\text{double-UV}(k,l) =
        \frac{8(1-\e)k^\mu}{\prop{l}\prop{k}^2}\left[
        \frac{\s{l}-\s{k}}{\prop{l}\prop{k}}
        \right. \\ & \qquad \left.
        +\frac{\s{l}-\s{k}}{(l-k)^2-\m^2}\left(
        \frac{1}{l^2-\m^2}-\frac{1}{k^2-\m^2}
        \right)
        + \frac{2l\cdot k \s{l}}{\prop{l}^2\prop{k}}
        \right]
        \\ & \qquad 
        + (k \leftrightarrow l)\,,
    \end{split}
\end{align}
and
\begin{align*}
        &\mathscr{S}_{k,l}\,O^{(2,d3V)}_\text{double-UV}(k,l) =
        \frac{\gamma^\mu}{\prop{l}\prop{k}^2}\left(
        \frac{2}{(l-k)^2-\m^2}-\frac{3}{l^2-\m^2}
        \right. \\ &\qquad \left.
        -4(1-\e)\frac{(l\cdot k)^2}{\prop{l}^2\prop{k}}
        \right)
        -\frac{1-\e}{\prop{k}^2}\left\{
        \frac{4l^\mu\s{l}}{\prop{l}^3}
        \right. \\ &\qquad \left.
        +\frac{2(k^\mu \s{l}-l^\mu \s{k})}{\prop{l}\prop{k}}
        \left(
        \frac{1}{l^2-\m^2}
        -\frac{1}{(l-k)^2-\m^2}
        \right)
        \right. \\ &\qquad \left.
        +\frac{2(k-2l)^\mu \s{l}}{\prop{l}^2\prop{(l-k)}}
        -\gamma^\mu\left[
        \frac{1}{(l-k)^2-\m^2}\left(\frac{1}{l^2-\m^2}+\frac{1}{k^2-\m^2}\right)
        \right. \right.\\ &\qquad \left.\left.
        -\frac{1}{\prop{l}\prop{k}}\left(1+\frac{2l\cdot k}{l^2-\m^2}\right)
        \right]
        \right\} + (k \leftrightarrow l)\,, \numberthis
\end{align*}
where $X^\mu=X_1^\mu + X_2^\mu$ for $X\in\{Q^{(2,d3V)},O^{(2,d3V)}\}$.

Finally, we provide the (symmetrised) double-UV approximations for the shift terms, including ghost contributions. The uncrossed ladder contribution, given in eq.~\eqref{eq:shift_2L_UL_graph}, yields the counterterm,
\begin{align}
\begin{split}
    &\mathscr{S}_{k,l}\,\Pi_{qq \text{ double-UV}}^{(2,UL)\,\text{shift}}(q,k,l) = i\gs^4\,(1-\e)^2\left[C_AC_F\, s^{(UL)}_1(q,k,l) 
    \right. \\&\qquad \left.
    +
    C_A\left(C_F-\frac{C_A}{2}\right)s^{(UL)}_2(q,k,l) \right]
    + (k\leftrightarrow l)\,,
\end{split}
\end{align}
with coefficients $s^{(UL)}_1$ and $s^{(UL)}_2$, corresponding to loop momentum shifts of the ``outer" and ``inner" loops, given by
\begin{align*}
        &s^{(UL)}_1(q,k,l) = \frac{6 k\cdot q\,\s{k}}{\prop{k}^4}\left[\frac{1}{l^2-\m^2}-\frac{1}{(k+l)^2-\m^2}
        -\frac{2k\cdot l}{\prop{l}^2}\left(1-\frac{2k\cdot l}{l^2-\m^2}\right)
        \right]
        \\&\quad
        +\frac{1}{\prop{k}^3}
        \left[
        \frac{\s{q}}{(k+l)^2-\m^2}
        -\frac{2k\cdot q\,(\s{k}-2\s{l})}{l^2-\m^2}
        \left(
        \frac{1}{l^2-\m^2}
        -\frac{1}{(k+l)^2-\m^2}
        \right) \right.
        \\&\quad\left.
        -\frac{2(k+l)\cdot q\,\s{k}}{\prop{(k+l)}^2}
        -\frac{4k\cdot l\,(k\cdot l\,\s{q} + 4 k\cdot q\,\s{l})}{\prop{l}^3}
        \right]
        -\frac{1}{\prop{k}\prop{l}^2}
        \left[
        \frac{2(k+l)\cdot q\,\s{k}}{\prop{(k+l)}^2}
        \right.\\&\quad\left.
        + \frac{1}{l^2-\m^2}
        \left(\s{q} +\frac{2 l\cdot q\,\s{l}}{(k+l)-\m^2}\right)
        \right]
        +\frac{1}{\prop{k}^2\prop{l}^2}\left(
        \s{q}+2\frac{k\cdot l \,\s{q}+k\cdot q \,\s{l}}{l^2-\m^2}
        \right)\,,
        \numberthis
\end{align*}
and
\begin{align*}
    &s^{(UL)}_2(q,k,l) =
    -\frac{2k\cdot q}{\prop{k}^3\prop{l}^2}
    \left(\frac{4k\cdot l\, \s{l}}{l^2-\m^2}
    -3\s{k}\right)
    \\&\quad
    +\frac{1}{\prop{k}\prop{l}^2\prop{(k+l)}}
    \left(\s{q}-\frac{2{(k+l)\cdot q\,}\s{k}}{(k+l)^2-\m^2}\right)
    \\&\quad
    +\frac{1}{\prop{k}^2\prop{l}^2}
    \left[
    2k\cdot q\left(\frac{\s{l}}{l^2-\m^2}
    -\frac{\s{k}}{(k+l)^2-\m^2}\right)
    -\s{q}
    \right]
    \\&\quad
    -\frac{2{(k+l)\cdot q\,}\s{k}}{\prop{k}^3\prop{(k+l)}^2}
    \,.
    \numberthis
\end{align*}
The double-UV counterterm for the crossed-ladder shift term, defined in eq.~\eqref{eq:shift_2L_XL}, is given by,
\begin{align}
\begin{split}
    &\mathscr{S}_{k,l}\,\Pi_{qq \text{ double-UV}}^{(2,XL)\,\text{shift}}(q,k,l) = i\gs^4\,
    2C_A\left(C_F-\frac{C_A}{2}\right)(1-\e)\,s^{(XL)}(q,k,l)
    \\&\qquad+ (k\leftrightarrow l)\,,
\end{split}
\end{align}
with
\begin{align*}
    &s^{(XL)}(q,k,l) = \frac{2{k\cdot q}}{\prop{k}^3\prop{l}}\left(
    2\,\frac{(1+\e)\s{k}+\s{l}}{(k+l)^2-\m^2}
    -\frac{2(\s{k}+\s{l})
   +(1+2\e)\s{k} %
    }{l^2-\m^2}
    +\frac{4k\cdot l\, \s{l}}{\prop{l}^2}%
    \right)
    \\&\quad
    +\frac{1}{\prop{k}^2\prop{l}^2}\left(
    (1
    +\e %
    )\s{q} + 2\,\frac{k\cdot q\,(\s{k}+(3+\e)\s{l})-2l\cdot q\,\s{k}}{(k+l)^2-\m^2}
    \right)
    \\&\quad
    +\frac{1}{\prop{k}\prop{l}^2\prop{(k+l)}}
    \left(4(k+l)\cdot q\,\frac{(1+\e)\s{l}+\s{k}}{(k+l)^2-\m^2}-(2+\e)\s{q}\right)
    \numberthis
\end{align*}
Similarly, for the $(3V)$-shift term, defined in eq.~\eqref{eq:shift_2L_3V}, we have,
\begin{align}
\begin{split}
    &\mathscr{S}_{k,l}\,\Pi_{qq \text{ double-UV}}^{(2,3V)\,\text{shift}}(q,k,l) = i\gs^4\,
    C_A^2(1-\e)\,s^{(3V)}(q,k,l)
    \\&\qquad+ (k\leftrightarrow l)\,,
\end{split}
\end{align}
where the kinematic function $s^{(3V)}$ reads,
\begin{align*}
    &s^{(3V)}(q,k,l)=
    -\frac{2k\cdot q}{\prop{k}^3\prop{l}}\left(\frac{2\s{l}-3\s{k}
    }{l^2-\m^2}
    -\frac{2(\s{l}-\s{k})}{(l-k)^2-\m^2}
    +\frac{4k\cdot l\, \s{l}}{\prop{l}^2}%
    \right)
    \\&\quad
    -\frac{1}{\prop{k}^2\prop{l}^2}\left(\s{q}
    +\frac{2k\cdot q\,(\s{l}-\s{k})}{(l-k)^2-\m^2}\right)\,.\numberthis
\end{align*}

In addition to the ``standard" shift contributions to the quark self-energy, the $q\,||\,p_2$ Ward identities generate a shift mismatch in terms associated to ghosts,
c.f. eqs.~\eqref{eq:ghost_2L_shift_2} and~\eqref{eq:ghost_delta_shift_2}. These are rendered finite in the double-UV regions using the following counterterms,
\begin{align*}
        &\mathscr{S}_{k,l}\,O^{(2)\,\text{shift}}_{2, \text{ double-UV}}(p,q,k,l) =
        i\gs^5\, \frac{C_A^2}{4}\frac{1-\e}{\prop{l}\prop{k}}
        \Biggl[\frac{4k\cdot q\,\s{k}}{\prop{k}^2\prop{l}}%
        \\&\quad
        +\frac{\s{q}}{l^2-\m^2}\left(\frac{1}{(l+k)^2-\m^2}+\frac{1}{(l-k)^2-\m^2}
        + \frac{2}{k^2-\m^2}\left(\frac{2 (k\cdot l)^2}{\prop{l}\prop{k}}
        -1\right)
        \right)
        \\&\quad
        +\frac{\s{k}\s{l}(\s{p}+\s{q})-\s{p}\s{l}\s{k}}{\prop{l}^2}\left(
        \frac{1}{(l+k)^2-\m^2}-\frac{1}{(l-k)^2-\m^2}
        \right)
        \\&\quad
        + \frac{1}{(l-k)^2-\m^2}\left(\frac{(l-k)\cdot q}{(l-k)^2-\m^2}\left(\frac{\s{k}}{l^2-\m^2}-\frac{\s{l}}{k^2-\m^2}\right)\right.
        \\&\quad
        -\left.\frac{2 k\cdot q}{\prop{k}\prop{l}}\right)
        \Biggr]
        +(k\leftrightarrow l)\,,
\numberthis
\label{eq:ghost_2L_shift_2_double-UV}
\end{align*}
and
\begin{align*}
        &O^{(2)\,\text{shift}}_{\Delta,2 \text{ double-UV}}(q,k,l) = i\gs^5\, \frac{C_A^2}{2}
        \left[
        \frac{2(1-\e)k \cdot q\, \s{k}}{\prop{k}^2\prop{l}^2}
        \left(\frac{1}{(l-k)^2-\m^2}+\frac{2}{k^2-\m^2}\right)
        \right.\\&\quad
        -\left.
        \frac{(l-k)\cdot q\left(2(1-\e)\s{k}+\s{l}\right)}{\prop{k}\prop{l}^2\prop{(l-k)}^2}
        +\frac{k\cdot q\,\s{l}}{\prop{k}^2\prop{l}^2\prop{(l-k)}}
        \right.\\&\quad
        -\left.\frac{(1-\e)\s{q}}{\prop{k}\prop{l}^2}\left(
        \frac{1}{k^2-\m^2}+\frac{1}{(l-k)^2-\m^2}
        \right)
        \right]
        \numberthis
        \label{eq:ghost_delta_shift_2_double-UV}
\end{align*}
\edit{We note the dependence of eq.~\eqref{eq:ghost_2L_shift_2_double-UV} on the external fermion momentum $p$, which is unusual for a shifted UV counterterm.}

Finally, we provide the double-UV subtraction term for self-energy corrections to the fermion propagator with a gluon-triangle subgraph, defined in eq.~\eqref{eq:qqg_ggg_mod},
\begin{align}
\Delta \Gamma_{ggg\text{ double-UV}}^{(1)\,\rho,c}(k,l) = \gs^4\, \left(2\, n_f\,s^\mu_{n_f}(q,k,l)+\frac{C_A}{2}\,s^\mu_{A}(q,k,l)\right)
\label{eq:qqg_ggg_mod_double-UV}
\end{align}
with $s^\mu_{n_f}$ and $s^\mu_{A}$ are decomposed as,
\begin{align}
    s^\mu_{X}(k,l) = k^\mu\,s_{X,k}(k,l) +l^\mu \,s_{X,l}(k,l) +\gamma^\mu \,s_{X,\gamma}(k,l)\,, \quad X\in\{n_f,A\}\,,
\end{align}
where
\begin{align*}
    s_{n_f,k}(k,l)&= \frac{4\s{k}-2\s{l}}{\prop{k}\prop{l}^3}\left(\frac{1}{k^2-\m^2}-\frac{1}{(l-k)^2-\m^2}\right)-\frac{4\s{k}}{\prop{k}^3\prop{l}^2} 
    \\&
    +\frac{4\,k\cdot l(2\s{k}-\s{l})}{\prop{k}^3\prop{l}^3}
    +\frac{16(k\cdot l)^2\s{k}}{\prop{k}^4\prop{l}^3}
    \\&
    +\frac{2(1-\e)\s{l}}{\prop{k}^2\prop{l}^2\prop{l-k}}\,,
    \\
    s_{n_f,l}(k,l)&= \frac{2(1-\e)\s{l}-2\s{k}}{\prop{k}\prop{l}^3}\left(\frac{1}{k^2-\m^2}-\frac{1}{(l-k)^2-\m^2}\right) -\frac{4\,k\cdot l\,\s{k}}{\prop{k}^3\prop{l}^3}\,,
    \\s_{n_f,\gamma}(k,l)&=
    -\frac{1}{\prop{k}\prop{k}^2}\left(\frac{1}{l^2-\m^2}+\frac{1}{(l-k)^2-\m^2}\right)
    +\frac{2}{\prop{k}^2\prop{l}^2} 
    \\&
    + \frac{1}{\prop{k}^3\prop{(l-k)}}
    -\frac{2\,k\cdot l}{\prop{k}^2\prop{l}^3}\left(1+\frac{2\,k\cdot l}{k^2-\m^2}\right)\,.
    \numberthis
    \label{eq:qqg_ggg_mod_double-UV_nf}
\end{align*}
and
\begin{align*}
    s_{A,k}(k,l)&= 2\,\frac{(3-2\e)\s{l}+8(1-\e)(\s{l}-\s{k})}{\prop{k}\prop{l}^3}\left(
    \frac{1}{k^2-\m^2}-\frac{1}{(l-k)^2-\m^2} +\frac{2\,k\cdot l}{\prop{k}^2}\right)
    \\&
    +\frac{16(1-\e)\s{k}}{\prop{k}^3\prop{l}^2}\left(1-\frac{4\,(k\cdot l)^2}{\prop{k}\prop{l}}\right)
    \\&
    -\frac{2(11-8\e)\s{l}}{\prop{k}^2\prop{l}^2\prop{(l-k)}}\,,
    \\
    s_{A,l}(k,l)&= -2\,\frac{\s{k}+9(1-\e)\s{l}}{\prop{k}\prop{l}^3}\left(\frac{1}{k^2-\m^2}-\frac{1}{(l-k)^2-\m^2}\right)
    -\frac{4\,k\cdot l \s{k}}{\prop{k}^3\prop{l}^3}
    \\&
    +2\,\frac{\s{k}+4(1-\e)\s{l}}{\prop{k}^2\prop{l}^2\prop{(l-k)}}
    \,,
    \\s_{A,\gamma}(k,l)&=-\frac{2}{\prop{l}^3}\left(\frac{1}{k^2-\m^2}-\frac{1}{(l-k)^2-\m^2}\right)
    -\frac{5}{\prop{k}^2\prop{l}^2}
    \\&
    +\frac{1}{\prop{k}\prop{l}\prop{(l-k)}}\left(\frac{7}{l^2-\m^2}+\frac{2}{k^2-\m^2}\right)
    \\&
    -\frac{2\,k\cdot l}{\prop{k}\prop{l}^2}\biggl(\frac{1}{\prop{k}\prop{l}}+\frac{1}{\prop{l}\prop{(l-k)}}
    \\&
    +\frac{1}{\prop{k}\prop{(l-k)}}+\frac{2\,k\cdot l}{\prop{k}^2\prop{l}}\biggr)
    \,.
    \numberthis
    \label{eq:qqg_ggg_mod_double-UV_CA}
\end{align*}
We note that the integrand-level modifications to the gluon three-point function, defined through additive counterterms according to eq.~\eqref{eq:ggg_CNT}, do not affect the amplitude in the double-UV region.